\documentclass[numbib]{imaiai}

\usepackage{amsthm, amsfonts, amssymb, amsmath} 
\usepackage{enumerate}  
\usepackage{mathtools}
\usepackage{natbib}
\mathtoolsset{showonlyrefs}

\def\mathbbm{\mathrm}  

\theoremstyle{remark}
\newtheorem{assumption}{\sc Assumption}[section]

\newcounter{myalgctr}
\numberwithin{myalgctr}{section}

\numberwithin{equation}{section}

\def\tcr{\textcolor{red}}

\usepackage{xcolor}
\usepackage{graphicx}  
\usepackage{flafter}  
\usepackage{bm} 
\usepackage{xr}
\usepackage{xr-hyper}
\usepackage{colortbl}
\usepackage{arydshln}
\usepackage{booktabs}
\definecolor{dullmagenta}{rgb}{0.4,0,0.4}   
\definecolor{darkblue}{rgb}{0,0,0.4}
\definecolor{coquelicot}{rgb}{0.20, 0.12, 0.72}
\definecolor{navyblue}{rgb}{0,0,0.5}
\definecolor{coquelicot}{rgb}{0.90, 0.42, 0.72}
\definecolor{burntorange}{rgb}{0.8, 0.33, 0.0}
\definecolor{burntblue}{RGB}{0, 114, 206}

\def\crev{\color{black}}
\def\crevmag{\color{black}}
\def\crevjb{\color{black}}
\def\crevyz{\color{black}}

\def\tcr{\textcolor{black}}

\def\cred{\color{black}}

\def\tco{\textcolor{black}}
\def\jelena{\textcolor{black}}

\def\mod{\textcolor{black}} 

\newcommand\independent{\protect\mathpalette{\protect\independenT}{\perp}}
\def\independenT#1#2{\mathrel{\rlap{$#1#2$}\mkern2mu{#1#2}}}

\usepackage{epstopdf}
\usepackage{multirow}
\usepackage[colorlinks=true, pdfstartview=FitV,linkcolor=blue, citecolor=blue, urlcolor=blue, filecolor=blue]{hyperref} 
\AtBeginDocument{
  \label{CorrectFirstPageLabel}
  
}

\makeatletter
\def\namedlabel#1#2{\begingroup
    #2%
    \def\@currentlabel{#2}%
    \phantomsection\label{#1}\endgroup
}
\makeatother
\newcommand{\vertiii}[1]{{\vert\kern-0.25ex\vert\kern-0.25ex\vert #1
    \vert\kern-0.25ex\vert\kern-0.25ex\vert}}

\usepackage[toc,page]{appendix}

\def\bx{\mathbf{x}}
\def\bX{\mathbf{X}}

\def\bz{\mathbf{z}}
\def\bZ{\mathbf{Z}}
\def\bu{\mathbf{u}}

\def\bv{\mathbf{v}}
\def\bV{\mathbf{V}}

\def\K{\mathbb{K}}
\def\E{\mathbb{E}}

\def\P{\mathbb{P}}

\def\bxv{\overrightarrow{\bx}}
\def\bXv{\overrightarrow{\bX}}

\def\balpha{\boldsymbol{\alpha}}
\def\bbeta{\boldsymbol{\beta}}

\def\bnabla{\boldsymbol{\nabla}}

\def\bSigma{\boldsymbol{\Sigma}}

\def\bbetahat{\widehat{\bbeta}}

\def\Deltahat{\widehat{\Delta}}
\def\deltahat{\widehat{\delta}}

\def\mhat{\widehat{m}}

\newcommand\ind{\protect\mathpalette{\protect\independenT}{\perp}}
\def\independenT#1#2{\mathrel{\rlap{$#1#2$}\mkern4mu{#1#2}}}

\def\Lsc{\mathcal{L}}
\def\Usc{\mathcal{U}}

\def\bzeta{\boldsymbol{\zeta}}

\def\Var{\mbox{Var}}

\def\bzero{\mathbf{0}}

\def\bR{\mathbf{R}}

\def\bS{\mathbf{S}}

\def\S{\mathbb{S}}

\def\bDelta{\boldsymbol{\Delta}}

\def\Vtil{\widetilde{\mathbb{V}}}

\def\bgamma{\boldsymbol{\gamma}}
\def\bdelta{\boldsymbol{\delta}}

\def\C{\mathbb{C}}

\def\R{\mathbb{R}}

\def\Nsc{\mathcal{N}}

\def\S{\mathbb{S}}

\def\Nsc{\mathcal{N}}

\def\Z{\mathbb{Z}}

\def\Lsc{\mathcal{L}}

\def\bSigma{\boldsymbol{\Sigma}}

\def \hs2{\hspace{2mm}}

\numberwithin{table}{section}
\numberwithin{equation}{section}

\definecolor{jcolor}{RGB}{041,122,000}
\definecolor{darkred}{RGB}{100,000,000}
\definecolor{purple}{RGB}{200,000,200}

\def\boxit#1{\vbox{\hrule\hbox{\vrule\kern6pt  \vbox{\kern6pt#1\kern6pt}\kern6pt\vrule}\hrule}}

\def\bS{\mathbf{S}}

\def\pitil{\widetilde{\pi}}
\def\pihat{\widehat{\pi}}
\def\mhat{\widehat{m}}

\def\bgammahat{\widehat{\bgamma}}

\def\bR{\mathbf{R}}

\def\pitil{\widetilde{\pi}}

\def\thetahat{\widehat{\theta}}
\def\psitil{\widetilde{\psi}}
\def\Vtil{\widetilde{V}}

\usepackage{abstract}

\begin{document}

\title{Double Robust Semi-Supervised Inference for the Mean: Selection Bias under MAR Labeling with Decaying Overlap\thanks{Accepted by {\it Information and Inference: A Journal of the IMA}.}}

\shorttitle{Semi-Supervised Inference under MAR with Decaying Overlap} 
\shortauthorlist{Zhang, Chakrabortty, and Bradic} 

\author{{\sc Yuqian Zhang}\thanks{All authors contributed equally in this work.}\thanks{\crevyz{Zhang was previously at the Department of Mathematics, University of California San Diego during the initial preparation of this work.}}\\[2pt]
Institute of Statistics and Big Data\\
Renmin University of China, Beijing 100872, China\\
{\email{yuqianzhang@ruc.edu.cn}}\\[4pt]
{\sc Abhishek Chakrabortty}\footnotemark[2]\\[2pt]
Department of Statistics\\
Texas A\&M University, College Station, TX 77843, USA\\
{\email{abhishek@stat.tamu.edu}}\\[4pt]
{\sc AND}\\[4pt]
{\sc Jelena Bradic}\footnotemark[2]\\[2pt]
Department of Mathematics and Halicioglu Data Science Institute\\
University of California, San Diego, La Jolla, CA 92093, USA\\
{\email{jbradic@ucsd.edu}}
}
\maketitle

\begin{abstract}
Semi-supervised (SS) inference has received much attention in recent years. Apart from a moderate-sized labeled data, $\mathcal L$, the SS setting is characterized by an additional, \emph{much larger sized}, unlabeled data, $\mathcal U$.  The setting of  $|\mathcal U|\gg|\mathcal L|$,  makes  SS inference unique and different from the standard missing data problems,  owing to natural violation of the so-called ``positivity'' or ``overlap'' assumption. However, most of the SS literature implicitly assumes $\mathcal L$ and $\mathcal U$ to be equally distributed,  i.e., no selection bias in the labeling.
 Inferential challenges in missing at random (MAR) type labeling allowing for selection bias, are inevitably exacerbated by the decaying nature of the  propensity score (PS). We address this gap  for a prototype problem, the estimation of  the  response's mean. We propose a double robust  SS (DRSS)   mean estimator and give a complete characterization of its asymptotic properties. The proposed estimator is consistent as long as either the outcome or the PS model is correctly specified. When both   models are correctly specified, we provide  inference results with a non-standard consistency rate that depends on the smaller size $|\mathcal L|$. The results are also extended to causal inference with imbalanced treatment groups. Further, we provide several novel choices of models and estimators of the decaying  PS,  including a novel offset logistic model and a stratified labeling model. We present  their  properties  under both high and low dimensional settings. These may be of independent interest.   Lastly, we present  extensive simulations  and also a real data application.
\end{abstract}

\section{Introduction}\label{sec:intro}

\tcr{Inference in semi-supervised (SS) settings has received substantial attention in recent times. Unlike traditional statistical learning settings that are usually either supervised or unsupervised, an SS setting
represents a confluence of these two settings. 
%
A typical SS setting has two types of available data:} apart from a small or moderate-sized \tcr{\emph{labeled} (or supervised)} data $\Lsc = (Y_i,\bX_i)_{i=1}^n$, one has access to a \emph{much larger} sized \tcr{\emph{unlabeled} (or unsupervised)} data $\Usc =(\bX_i)_{i=n+1}^N$ \tcr{with} 
 $N \gg n$. Here, $Y_i\in\R$ and $\bX_i\in\R^p$ denote the outcome of interest and a covariate vector \tcr{(possibly high dimensional)}, respectively. To integrate the notation, \tcr{we use $R_i\in\{0,1\}$ to denote the missingness/labeling indicator and} use
 $\S=\mathcal L\cup\mathcal U=(R_i,R_iY_i,\bX_i)_{i=1}^N$ to denote \tcr{the full data, a collection of} $N$ i.i.d. \tcr{(independent and identically distributed)} observations \tcr{of $(R,RY,\bX)$, where} 
 throughout this paper, we let $(R,RY,\bX)$ denote an independent copy of $(R_i,R_iY_i,\bX_i)$.

 \tcr{SS settings arise naturally} whenever the covariates are easily available \tcr{for a large cohort (so that $\Usc$ is plentiful),} but the \tcr{corresponding} response is expensive \tcr{and/or} difficult to obtain due to various practical constraints (thus limiting the size of $\Lsc$), a frequent scenario in modern studies involving large databases \tcr{in the `big data' era. Examples of such settings are ubiquitous across various scientific disciplines, including machine learning problems like speech recognition, text mining etc. \tco{\citep{zhu2005semi,chapelle2006semi}}, as well as more recent (and relevant to our work) biomedical applications, like electronic health records (EHR) and integrative genomic studies} \jelena{\citep{chakrabortty2018efficient, tony2020semisupervised}}. 
  \tcr{It is important to note that while SS settings can be viewed as a missing data problem of sorts, the fact that $|\Usc| \gg |\Lsc|$ is a {\em key distinguishing feature} of SS settings (for instance, $|\Lsc|$ could be of the order of hundreds, while $|\Usc|$ could be in the order of tens of thousands!). This condition, a natural consequence of the underlying practical situations leading to these data,
  implies that the proportion of labeled observations in SS settings converges to 0 as the sample sizes $|\Lsc|, |\Usc| \rightarrow \infty$. This makes SS settings unique and fundamentally {\em different} from any standard missing data problem where this proportion is always assumed to be bounded away from 0, a condition also known as the \emph{positivity} (or \emph{overlap}) assumption in the missing data literature \citep{imbens2004nonparametric,tsiatis2007semiparametric}, 
  which is \emph{naturally violated} here.}


    \tcr{Most of the} SS literature, however, implicitly assumes that $\bX$ is equally distributed in $\mathcal L$ and $\mathcal U$ samples, that is, a missing completely at random (MCAR) \tcr{setting,} where $R \independent (Y,\bX)$, and the goal is to improve efficiency over an (already valid) \tcr{supervised} estimator \tcr{based on} 
    $\mathcal L$.
     A biased, covariate-dependent, missing at random (MAR) type labeling \tcr{mechanism} has not been studied much, \tcr{although they are much more realistic in practice, especially in biomedical applications (including the examples discussed earlier) where \emph{selection bias} is common. For instance, in EHR data, relatively `sicker' patients may often be more likely to be labeled, especially if the labeling is for a disease response.} 
     We work in \tcr{this type of a \emph{`decaying'} MAR domain,} which we name \tcr{\emph{MAR-SS} for} short, \tcr{  under the typical}  
     ``ignorability'' assumption:  $$R~\independent~ Y ~|~ \bX,$$
     \tcr{thereby allowing for a selection bias in the process.}
     \tcr{It is important to note that} the \tcr{traditional} MAR setting amongst the missing data literature is typically studied together with an overlap (positivity) condition that  \jelena{bounds  away the propensity score (PS) $\E(R|\bX)$ uniformly   from zero }\citep{bang2005doubly}.
   \tcr{Compared to such MAR settings, our MAR-SS setting is \jelena{significantly} more challenging due to the inevitably decaying nature of the PS. \tcr{We also interchangeably refer to this setting as \emph{decaying overlap.}} 
   \jelena{As} $N \gg n$ here, positivity  is \jelena {automatically excluded,  thus leading} to a \emph{non-standard} asymptotic regime.}

   \paragraph*{Subtleties} {\cred To work} \jelena{with} {\cred such  unbalanced labeling, we denote the \tco{PS} as $\pi_N(\bX):=\E(R|\bX) \equiv \P(R=1|\bX)$ and let {\crev $\bar\pi_N := \E(R) \equiv \P(R = 1)$}. It is important to note that {\crev in order} to allow a non-degenerate \tco{PS} with $\E(R)\to 0$ as $N\to\infty$, we {\em must} allow $R$, $\pi_N(\bX)$ and $\bar\pi_N$ to {\em depend} on $N$ (otherwise forcing $n/N\to0$ would lead to a degenerate situation with $\E(R)=0$ and $\E(R|\bX)=0$ almost surely (a.s.)). Hence, both $\{R_{N,i}\}_{N,i}$ and $\{\pi_N(\bX_i)\}_{N,i}$ form triangular arrays.
   We suppress the dependence of $R_N$ on $N$ throughout for notational simplicity. }

  \tcr{Under such a \emph{decaying \jelena{MAR-SS} setting}, we study the fundamental problem of} estimation and inference towards the mean \tcr{response, defined as:}
  $$\theta_0~:=~\E(Y).$$
   %
   {\cred The mean estimation problem above is a canonical problem in classical missing data \jelena{as well as causal inference literature}, and we consider it here mainly as a prototype problem. The bigger purpose of this paper is to provide a deeper understanding of this \emph{MAR-SS} setting and all its subtleties, where the main challenge is to allow for the uniform decay of the PS with the sample size and handle the non-standard asymptotics that arises inevitably. Moreover, unlike ``traditional'' SS settings (with MCAR), the goal here is {\em not} to ``improve'' over a supervised estimator from $\Lsc$ (which is no longer valid under selection bias) but rather {\em develop from scratch a consistent and rate-optimal estimator along with inferential tools for it.} The contributions of this work therefore constitutes advances both in the literature of classical missing data and causal inference as well as that of traditional SS inference.}
   \tcr{We first provide an overview of the existing literature(s), followed by a summary of our contributions.}

\subsection{Related Literature}\label{sec:literatre}

{\crevmag SS} literature on 
\tcr{prediction problems} is vast, \tcr{typically under the name of semi-supervised learning;} see \cite{zhu2005semi} and \cite{chapelle2006semi} for a review.  \tcr{SS inference} has attracted a lot of \tcr{recent} attention.
\cite{zhang2019semi} and \cite{zhang2022high} proposed SS mean estimators.  \tcr{The} estimators in \cite{zhang2022high} can be roughly seen as a special (MCAR) case of \tcr{the} \tco{MAR-SS} \tcr{setting here}. 
\cite{azriel2016semi} and \cite{chakrabortty2018efficient} tackled the SS linear regression \tcr{problem}, \tcr{while \cite{kawakita2013semi} considered likelihood based SS inference}. \cite{tony2020semisupervised} studied 
\tcr{SS inference} of the explained variance in high dimensional linear regression. However, they all require a MCAR assumption, i.e., $R\independent(Y,\bX)$.  MCAR   is practically too strong, and these estimators lead to doubtful results once the dependency of $R$ on \tcr{$\bX$} occurs {\crevmag causing a selection bias in the labeling}. 

Works that remove some of the MCAR restrictions have been proposed recently.
A special stratified labeling  in a SS framework  was studied in \cite{gronsbell2020efficient}  with a focus on prediction \tcr{performance measures}. 
\tcr{Stratified labeling was also studied in \cite{hong2020inference}, though their setting is very specific in that their only source of randomness arises from the treatment assignment.}
{\crev\cite{ryan2015semi} considered {\crevmag regression problems under a SS framework with covariate shift}{\crevjb{\crevyz;} however,}
  certain manifold conditions {\crevmag were} {\crevjb imposed {\crevmag therein} that we avoid altogether.}
  {\crevmag
  More recently,}
  \cite{liu2020doubly}
  {\crevjb used {\crevyz an approach based on} density ratios for the same setup, albeit working with {\crevyz semi-nonparametric models} and {\crevyz a} non-decaying PS.}
%
%
}
{\tcr {\crevjb A problem setup closest to us was considered only in }\cite{kallus2020role}. Their main focus, however, was on treatment effects estimation and efficiency theory when surrogate variables occur in the MAR setting with positivity.
  They do} provide
  a semiparametric efficiency bound {\crevjb under a decaying PS setting}.
{\crevjb However,} we provide a \emph{complete characterization} (see Sections \ref{sec:knownpi}-\ref{sec:unknownpi}) of the {\crevyz estimators'} 
asymptotic properties {\crevyz and the inferential tools} \tcr{(see Section \ref{sec:varestimation})}, \tcr{and} {\crevyz {\crevmag further,} such results are provided} under much weaker conditions. \tcr{For instance,} we only require $N\bar\pi_N\to\infty$ \tcr{(while they require $N \bar\pi_N^2 \rightarrow \infty$)} \tco{and we allow an unbounded support for $\bX$, which is essentially violated under the uniformly bounded density ratio condition $\bar\pi_N/\pi_N(\bX) <C$ assumed in \cite{kallus2020role}.
}
\tcr{Moreover, the authors therein} did not provide any results \tcr{and/or methodology} on the decaying  \tcr{PS's estimation} 
\tcr{which is an essential component of the problem here}.  

Our work is \tcr{also} naturally connected to the rich missing data \tcr{(and causal inference)} literature \tcr{on semi-parametric methods,} and especially to \tcr{so-called doubly robust (DR)} inference;  see \cite{robins1994estimation}, \cite{robins1995semiparametric}, \cite{bang2005doubly}, \cite{tsiatis2007semiparametric}, \cite{kang2007demystifying}, and \cite{graham2011efficiency} for a review.  High-dimensional DR equivalents have been presented recently as well; see  for example \cite{belloni2014inference,farrell2015robust,chernozhukov2018double,smucler2019unifying,bradic2019sparsity}.
They work on a low-dimensional parameter estimation problem that involves high-dimensional nuisance parameters. On the other hand, \cite{semenova2017estimation} and \cite{chakrabortty2019high} work on problems where the parameters of interest themselves are high-dimensional.  However, the \tco{positivity} assumption is always assumed. Our work is a direct extension of the above literature where we now include \tco{a decaying PS},
and therefore a setting of imbalanced  treatment mechanisms.


{\cred Another related setting to our decaying \jelena{PS} setting is the so-called ``limited overlap'' setting.
A few notable prior works on}
limited overlap  include  \cite{crump2009dealing,khan2010irregular,yang2017asymptotic,rothe2017robust,visconti2018handling} among others, where a truncation of \tcr{the} PS is introduced and a restricted analysis to the portions of the treatment groups \tcr{such} that overlap \jelena{holds} is performed.
\tcr{The ``limited overlap'' condition is also weaker than the usual overlap condition, but very \emph{different} from our \tco{decaying PS}
situation.}
The limited overlap  allows the \tco{PS} to approach  zero on some 
specific regions in the support of $\bX$, while we allow $\E(R|\bX)$ to shrink to zero \tcr{(with $N$)} uniformly \tcr{in $\bX$}.
Moreover, they assume that $\E(R|\bX)$ is independent of $N$. By allowing $R$ to   depend  on $N$, we allow $\P_R$ and $\P_{R|\bX}$ to depend on $N$ so  that $\bar\pi_N=\E(R)\to0$  is permissible \tcr{(a necessity under our settings of interest),} much unlike the  existing limited overlap \tco{literature.}

  \subsection{Our Contributions}

  Contributions of our work are three fold: on (i) {\it double robust estimation with decaying \tco{PS}}, (ii) {\it estimation of decaying \tco{PS}}, and 
  (iii) {\it average treatment effect \tco{(ATE)} estimation with imbalanced groups}.

 \paragraph*{Double robust estimation with decaying propensity}
We believe this work fills \jelena{in} an \jelena{important} gap in both the SS literature  and the missing data literature. A selection bias in the labeling \jelena{mechanism} is allowed, therefore parting  with the SS literature. A \tco{PS} is allowed to decay to zero uniformly, consequently enriching the MAR literature. \tco{We propose a {\it double robust  semi-supervised} \jelena{(DRSS)} mean estimator \tcr{(see Sections \ref{sec:knownpi}-\ref{sec:unknownpi})}, which can be viewed as an adaptation of the standard DR estimator \citep{robins1994estimation} \jelena{to} our MAR-SS setting. Theorem \ref{t4}, \tcr{our main result for this part}, provides a full characterization of the \jelena{DRSS} estimator and its \jelena{asymptotic} ex\tcr{p}ansion when at least one of the nuisance functions is correctly specified.} \tcr{Throughout, our results} \jelena{bring in a new set of rate-adjusted} \tcr{high-level} \jelena{estimation error conditions} \tcr{on the nuisance estimators}  that are \tcr{agnostic to their mode of construction}.
    When both \tcr{nuisance} 
    models
    are correctly specified, we derive the asymptotic normality \tcr{of our estimator} if a product rate condition for the estimation errors is further assumed, with an asymptotic variance reaching the semi-parametric efficiency bound \jelena{derived} in \cite{kallus2020role}. \jelena{W}\tcr{e also} \jelena{construct a corresponding confidence interva\tcr{l}} \tcr{(see Section \ref{sec:varestimation})} \jelena{that adapt\tcr{s} to the rate of decay of the PS. Adaptivity here implies that the confidence sets are wider for the cases of faster decay without cha\tcr{n}ging the estimators themselves}. 
    The analys\tcr{e}s and the methods are considerably more involved here compared to the standard problems, due to the decaying nature of the PS.    For example, we establish that the rate of convergence is no longer govern\tcr{ed} by $N$ solely; \tcr{rather the} effective rate is identified to be \tcr{in terms of} $Na_N$, where \tco{$a_N^{-1}=\E\{\pi_N^{-1}(\bX)\}$. In high dimensions, \tcr{and using standard parametric nuisance models,} the product rate condition required for the asymptotic normality is $s_ms_\pi\{\log(p)\}^2=o(Na_N)$, where $s_m$ and $s_\pi$ are the sparsity levels of the (linear/logistic) nuisance functions $m(\bX)=\E(Y|\bX)$ and $\pi_N(\bX)=\E(R|\bX)$, respectively. When $a_N\asymp1$, such a condition coincides with the usual product condition \citep{chernozhukov2018double} where the positivity condition is assumed.}  \jelena{However, whenever $a_N\to \tcr{0}$, 
    the condition is stricter in order to compensate \tcr{for} the decay of \tcr{the} PS.} 

 \paragraph*{Estimation of \tcr{the} decaying propensity}
\jelena{A} key challenge \tcr{for any methodological development in our MAR-SS setting} is the modeling of the decaying PS. We propose several choices \tcr{and associated results} in this regard, including (i) \emph{stratified labeling} \tcr{(see Section \ref{stratified})} as well as (ii) a novel \emph{\tco{offset based} imbalanced logistic regression model} \tcr{(see Section \ref{logistic}),} \tcr{under \emph{both} low and high dimensional settings.}
  \tcr{The first approach}\jelena{, (i),} \tcr{is often practically relevant in the presence of \emph{apriori} information available on a stratifying variable. The second approach,} \jelena{(ii),}  \tcr{on the other hand, is applicable quite generally and constitutes a natural extension of logistic models to our case of a decaying PS. Related to the latter model,} imbalanced classification in low-dimensions was \tcr{recently} studied by
  \cite{owen2007infinitely} and \cite{wang2020logistic}.  Our offset based model is closely related to their diverging intercept model, \tcr{and yet has distinct methodological advantages (see Appendix \ref{sec:offset} and Remark \ref{remark:offset} therein).}
  We provide  theoretical results \jelena{about} estimation rates and other properties of these models under both high and low dimensional settings. These results may be more generally useful and are of independent interest; for example, our results on estimation of decaying PS under a logistic model in high dimensions \tcr{are the \emph{first} such results to our knowledge.} 
  \tco{We demonstrate that, for a sub-Gaussian $\bX$, the estimation error of $\pi_N(\cdot)$ is $O_p(\sqrt{s_\pi\log(p)/(N\bar\pi_N)})$, where $s_\pi$ is the sparsity level of the logistic \tcr{model} parameter. Such a result is non-trivial  as, \jelena{per} Theorem \ref{thm:high-dim}, an appropriate choice of the regularization parameter is \jelena{non-standard} with $\lambda_N\asymp\sqrt{\bar\pi_N\log(p)/N}$.}
  \tco{We also obtain \tcr{a regular and asymptotically linear} \jelena{(RAL)}
  expansion for the \jelena{estimator} of \tcr{the} 
  logistic \tcr{regression} parameter \tcr{in the low-dimensional case; see Theorem \ref{thm:ex3}.}   \jelena{Moreover, we showcase that the estimator reaches the asymptotic variance} as  \jelena{established} in \cite{wang2020logistic} for low-dimensional \jelena{problems}.  \jelena{For the cases where the outcome model is misspecified, we}
  further construct an adjusted \jelena{RAL} exp\jelena{a}nsion \jelena{of our} DRSS estimator.}    \tcr{Lastly, in Appendix \ref{sec:MCAR}, we also consider the special case of the MCAR model and the corresponding results in that setting.}

\paragraph*{Average treatment effect \tcr{(ATE)} estimation with imbalanced groups}
Drawing on a natural connection between the causal inference and missing data \tcr{settings} (see the discussions in Section 1.1 of \cite{chakrabortty2019high} for instance) we extend our results to a corresponding \tco{ATE} estimation \tcr{problem}. Our results  allow for an extremely imbalanced treatment \jelena{or} \tcr{control} groups, in that $\bar\pi_N=\P(R=1)\to0$ (or \tcr{alternatively,} 
$\bar\pi_N\to1$) as $N\to\infty$.

\tco{We establish a \jelena{RAL} expansion for the \tcr{proposed} ATE estimator with a non-stand\jelena{a}rd consistency rate, $O_p(1/\sqrt{N\bar\pi_N})$,  \jelena{where without loss of generality we  assume}   $\bar\pi_N\to0$.
\jelena{
A sufficient condition for \tcr{the expansion's} validity is correctness of the model \tcr{for} the treatmen\tcr{t} 
group\tcr{'s} outcome as well as \tcr{that of} the PS model. Notably, the control group's outcome and PS models \emph{can} be (even both) misspecified if $\bar\pi_N\to0$ fast enough.} Such a condition is different from \tcr{most of} the \jelena{recent results, such as} \cite{farrell2015robust} and \cite{chernozhukov2018double},  \jelena{where} the nuisance functions in both of the groups need to be correctly specified for valid inference results.
 It is also different from the recent  work of \cite{smucler2019unifying} and \cite{tan2020model}, where they used specific parametric working models and they required at least one of the nuisance functions to be correctly specified for both of the groups.}

\tco{The PS setting, the parameter of interest, and the methodology are also different from the limited overlap literature, e.g., \citep{crump2009dealing}. As shown in \cite{khan2010irregular}, the information bound for the ATE estimation is $0$ if only under the ignorability assumption and a.s., $\pi_N(\cdot)\in(0,1)$. As a result, a common approach in the limited overlap literature is to \jelena{re-target the parameter of interest by considering}  a ``shifted'' ATE \jelena{induced by the} truncati\jelena{on} \tco{of} the PS \citep{crump2009dealing}.
In our paper, we show that it is in fact possible to estimate the ATE directly when we have additional information that the inverse PS has well-behaved tails, e.g., $\pi_N(\cdot)$ follows an offset logistic model and $\bX$ is sub-Gaussian; see Theorems \ref{t4}, \ref{thm:ex3}, and \ref{thm:high-dim}. }

\subsection{Organization} 
The rest of this paper is organized as follows. In Section \ref{setup}, we formulate the decaying \tco{PS} setting and the mean estimation problem. In Section \ref{sec:EY}, we construct the mean estimators  and provide a  complete characterization of the asymptotic properties of our proposed estimators. In Section \ref{sec:knownpi}, we first consider a special case that the \tco{PS} function is known, and in Section \ref{sec:unknownpi}, we consider a general unknown $\pi_N(\cdot)$ case and provide results on our final proposed \jelena{DRSS} estimator. \tco{In Section \ref{pihat_N}, we analyze three specific decaying \tco{PS} models, including \tco{an} offset logistic model under both low and high dimensions (Section \ref{logistic}), a stratified labeling model (Section \ref{stratified}), and a MCAR model (Section \ref{MCAR}). For each of the \tco{PS} models, we propose a corresponding \tco{PS} estimator and establish their asymptotic properties as well as the asymptotic results of the \jelena{DRSS} estimators based on them. In Section \ref{ATE}, we extend our results to an ATE estimation problem where the treatment groups are extremely imbalanced. } Simulation results and an application to a NHEFS data are presented in Section \ref{numerical}, followed by a concluding discussion in Section \ref{discussion}. \tco{Further discussions, additional theoretical and numerical results, as well as the proofs of the main results are provided in the \hyperref[supp_mat]{Supplement} \tcr{(Appendices \ref{sec:offset}-\ref{Ssec:proof})}.}

\subsection{Notation}

 \tco{We use the following notation throughout.} Let $\P(\cdot)$ \tcr{and} $\E(\cdot)$ denote the probability measure and expectation characterizing the joint distribution of the underlying (possibly unobserved) random vector $\Z := (R,Y,\bX)$, respectively\tco{, where $R\in\{0,1\}$, $Y\in\R$, and $\bX\in\R^p$}.
Let $\P_{\bX}$ denote the marginal distribution of $\bX$.
For any $r > 0$, let $ \| f(\cdot) \|_{r,\P} := \{\E | f(\bZ)|^r\}^{1/r}$ and $ \| f(\cdot) \|_{r,\P_{\bX}} := \{\E_{\bX}| f(\bX)|^r\}^{1/r}$. \tco{For any vector $\bz\in\R^p$, we denote $\bz(j)$ as the j-th coordinate of $\bz$.}
 For $r\geq1$, 
 \jelena{define the $l_r$-norm of a vector $\bz$ with}
$\|\bz\|_r:=(\sum_{j=1}^p|\bz(j)|^r)^{1/r}$\tco{, \tcr{$\|\bz\|_0 := |\{j:\bz(j)\neq0\}|$},} and $\|\bz\|_\infty:=\max_j|\bz(j)|$.
For a matrix $A\in\R^{p\times p}$, $\|A\|_{\tcr{r}}:=\sup_{\bz\neq\boldsymbol{0}}\|A\bz\|_{\tcr{r}}/\|\bz\|_{\tcr{r}}$ and $\lambda_{\min}(A)$ denotes the smallest eigenvalue of $A$. 
 \tco{For sequences $a_N$ and $b_N$, we denote $a_N\asymp b_N$ if there exists constants $c,C,N_0>0$ such that $cb_N<a_N<Cb_N$ for all $N>N_0$. \tcr{Lastly,} we define the logit function as $\mbox{logit}(u):=\log\{u/(1-u)\}$ for any $u\in(0,1)$.  }

 \section{Problem setup}\label{setup}

Let the entire dataset be denoted \tcr{as:} 
$\mathbb S:= \{\bZ_i= (R_i,R_iY_i,{\bX}_i ),i=1,\dots, N \}.$
The dimension of the covariates $p$ can be either fixed or growing with $N$ in that $p=p_N\to\infty$ as $N\to\infty$. We assume the following ignorability condition \tcr{throughout.} 

\begin{assumption}[Ignorability \jelena{or} \tcr{MAR condition}]\label{cond:ignorability}
{\crevjb $R, Y$ and $\bX$ are such that} $R~\independent~ Y~|~{\bX}$.
\end{assumption}
The ignorability condition is standard in the missing data literature \citep{bang2005doubly, tsiatis2007semiparametric}.
\tco{Let $m(\bx):=\E(Y|\bX=\bx)$ and $\pi_N(\bx):=\E(R|\bX=\bx)$ denote the conditional mean of Y and the \jelena{conditional} \tco{PS}, respectively.
%
{\crev Assume $\E\{\pi_N^{-1}(\bX)\}<\infty$ for each $N$.} We then define {\crevjb a positive sequence of real numbers},  $a_N$ as:} 
%
\begin{equation}
a_N^{-1}~:=~\E\{\pi_N^{-1}(\bX)\}, \label{def:a_N}
\end{equation}
\tcr{{\crevjb which}  plays a key role in determining the rates of any inverse-probability weighting type estimator}.
{\crevjb We are primarily interested in the case of $a_N\to0$, although our results hold more broadly.}
{\crevmag We assume $\E\{\pi_N^{-1}(\bX)\}<\infty$ for each $N$, i.e. $a_N > 0$ for each $N$, but we {\it still} allow $a_N \to 0$ as $N \to \infty$.}
The value $a_N$ shrinks when the distribution of $\pi_N(\cdot)$ has too much mass concentrated around 0; see Remark \ref{remark:NaN} for more details.
Notice that the usual positivity (overlap) condition, $\pi_N(\bX)>c>0$, is NOT assumed throughout the paper, and we allow a uniformly decaying \tco{PS} in that $\pi_N(\bx)\to0$ \tcr{as $N \to \infty$,} for every $\bx$ in the support $\mathcal X$ {\crevmag of $\bX$}.

\begin{example}[Offset based \tco{PS} model]\label{e1}
{\crevjb A}{\crevyz s} \tcr{an illustration of {\crevjb a}  decaying PS model {\crevyz with a dependence of $\pi_N(\bX)$ onto $N$},} we {\crevjb consider}  a \tcr{general} \emph{offset based \tco{PS} model} as follows
$$
\pi_N(\bX)~=~g(f(\bX)+\log(\bar\pi_N)),\quad\mbox{with some}\;\;f:\R^p\to\R\;\;\mbox{and}\;\;g(u)~:=~\frac{\exp(u)}{1+\exp(u)},
$$
where $\log(\bar\pi_N)$ {\crevjb is considered as}  an ``offset''. \tcr{The model above constitutes a fairly general way of incorporating the {\crevjb uniformly}  decaying  PS {\crevjb model}. Further details on} the rationale behind and the analysis of \tcr{such} an \tcr{offset} model \tcr{are} 
discussed in \tcr{Section \ref{logistic}}{\crevyz,} {\crevjb where}
 we allow a linear $f$ with {\crevjb a} sub-Gaussian $\bX${\crevyz,} {\crevjb thus clearly violating} the positivity condition.
 {\crevjb Uniform decay of  $\pi_N(\bX)$ can also be seen whenever} $\bX$ has a compact support $\mathcal X$, {\crevjb leading to} $c_1\bar\pi_N\leq\pi_N(\bx)\leq c_2\bar\pi_N$ for all $\bx\in\mathcal X$ with constants $0<c_1<c_2$.
\end{example}


\paragraph*{\tcr{Preliminaries: Identification and alternative representations}}  We have the following three alternative representations \jelena{or identifications} of $\theta_0=\E(Y)$ \tcr{based on the observable variables and some unknown (but estimable) nuisance functions, i.e.\tco{,} $m(\bX)$ and $\pi_N(\bX)$.}
\begin{enumerate}
\item[(Reg)] Regression based  representation: $\theta_0=\E\{m(\bX)\}$.
\item[(IPW)] Inverse probability weighting  representation: $\theta_0=\E\{\pi_N^{-1}(\bX)RY\}$.
\item[(DR)]  Doubly robust   representation: $\theta_0=\E[m(\bX)+\pi_N^{-1}(\bX)\{RY-Rm(\bX)\}]$.
\end{enumerate}
A natural estimator of $\theta_0$  would be the empirical mean of the observed responses,
$\bar Y_\mathrm{labeled}:={\sum_{i=1}^NR_iY_i}/{\sum_{i=1}^NR_i}.$  Under \tcr{a} MCAR setting, $\bar Y_\mathrm{labeled}$ is \tcr{a} consistent estimator.
However, under \tcr{the} MAR setting, $\bar Y_\mathrm{labeled}$ is no longer a consistent estimator; $\bar Y_\mathrm{labeled}\xrightarrow{p}\E(Y|R=1)\neq\E(Y)$ in general.
According to the above representations, with $\mhat(\cdot)$ and $\pihat_N(\cdot)$  estimating $m(\cdot)$ and $\pi_N(\cdot)$, respectively, we could consider
  $\thetahat_\mathrm{Reg}:=\tcr{N^{-1}}\sum_{i=1}^N\mhat(\bX_i)$ and $\thetahat_\mathrm{IPW}:= \tcr{N^{-1}}\sum_{i=1}^NR_iY_i\pihat_N^{-1}(\bX_i)$.
  For the sake of simplicity, \tco{
here }we consider an ideal case that $\mhat(\cdot)$ and $\pihat_N(\cdot)$ are trained on another additional set so that $(\mhat(\cdot),\pihat_N(\cdot))\independent(\bX_i)_{i=1}^N$. 

It is \tcr{then} not hard to show \tcr{that}
\begin{align*}
\thetahat_\mathrm{Reg}-\theta_0&~=~O_p\left(\|\mhat(\bX)-m(\bX)\|_{1,\P_\bX}+N^{-1/2}\right),\\
\thetahat_\mathrm{IPW}-\theta_0&~=~O_p\left(\|1-\pi_N(\bX)/\pihat_N(\bX)\|_{2,\P_\bX}+N^{-1/2}\right).
\end{align*}
\tco{Hence, \tcr{the} Reg and IPW  estimators are not even consistent when the corresponding nuisance \jelena{model} is misspecified. Even when the corresponding nuisances are correctly specified,  \jelena{estimators} directly depend on the estimation error of $\mhat(\cdot)$ and $\pihat_N(\cdot)$, respectively, which \jelena{are not} $\sqrt N$-consistent \tcr{(nor $\sqrt{N \bar\pi_N}$-consistent)} in the high-dimensional  \jelena{or non-parametric} settings.}

\tco {The DR representation of $\theta_0$, viewed as a combination of the Reg and IPW representations \citep{accomando1974optimal},
leads to double robustness. 
DR estimators are consistent as long as at least one of the models are correctly specified (this property is called ``double robustness'', see, e.g., Theorem 2 of \cite{farrell2015robust}). \jelena{W}hen both models are correctly specified, the estimation errors of the DR estimators depend on the product of estimation errors of the nuisance functions; this property is called ``rate double robustness\jelena{,}'' as defined in Definition 2 of \cite{smucler2019unifying}.}
Moreover, DR estimators are known to be semi-parametric\jelena{ally} optimal when both 
models are correct \citep{bang2005doubly}, \jelena{as well as } first order insensitive to the estimation errors of the nuisance functions \tco{\citep{chernozhukov2018double}; see the discussions in \cite{chakrabortty2019high}.} 
In Section \ref{sec:EY}, we propose estimators based on the \jelena{above} DR  \jelena{representation}.

\section{Semi-supervised \tco{inference under a MAR-SS setting}}\label{sec:EY}

\subsection[Known PS]{Known \tco{PS} $\pi_N(\cdot)$}\label{sec:knownpi}

We first consider an oracle case \jelena{where} the \tcr{PS}\jelena{,}
$\pi_N(\cdot)$\jelena{,} is known. {In other words, the missing mechanism is designed and controlled by the researcher. This is also closely related to the randomized controlled trials in causal inference literature.} Based on the DR representation, we consider the following \tcr{SS} estimator:
\begin{equation}\label{theta:knownpi}
\tilde\theta~:=~ N^{-1}\sum_{i=1}^{ N}\mhat(\bX_i)+ N^{-1}\sum_{i=1}^{ N}\frac{R_i}{\pi_N(\bX_i)}\{Y_i-\mhat(\bX_i)\},
\end{equation}
where $\mhat(\bX_i)$ is a \emph{cross-fitted} estimator established as follows: 1) for any fixed $\K\geq2$,  let $\{\mathcal I_k\}_{k=1}^{\mathbb K}$ be a random partition of $\mathcal I:=\{1,\dots, N\}$; 2) for each $k\leq\K$, obtain the estimator $\mhat(\cdot;\S_{-k})$ using the training set $\S_{-k}:=\{\bZ_i:i\in\mathcal I\setminus\mathcal I_k\}$, \tco{where for typical supervised methods, $\mhat(\cdot;\S_{-k})$ only depends on the labeled observations, $\{\bZ_i:i\in\mathcal I\setminus\mathcal I_k,R_i=1\}$};
3) for each $i=1,\dots,N$, let $\mhat(\bX_i):=\mhat(\bX_i;\mathbb S_{-k(i)})$, where $k(i)$ denotes the unique $k$ such that $i\in\mathcal I_k$.
The proposed   $\tilde\theta$ can be seen as a debiased   $\thetahat_\mathrm{Reg}$ estimator, where the misspecification or estimation bias of $\mhat(\cdot)$ is removed by the knowledge of $\pi_N(\cdot)$. On the other hand, $\thetahat_\mathrm{IPW}$ is a special case of $\tilde\theta$ with  $\pihat_N(\cdot)=\pi_N(\cdot)$ and $\mhat(\cdot)\equiv0$. However, $\tilde\theta$ with a ``good'' \tcr{estimator for} 
the outcome model improves the efficiency of the IPW estimator; see e.g., Remark \ref{efficiency}.
The cross-fitting is vital for the bias correction; see discussions in \cite{chernozhukov2018double} and \cite{chakrabortty2019high}. By the cross-fitting construction, \tco{$\mhat(\cdot;\mathbb S_{-k(i)})\independent\bZ_i$} for each $i\leq N$.
As a result,
$$\E_{\bX}\left[\mhat(\bX)+\frac{R}{\pi_N(\bX)}\{Y-\mhat(\bX)\}\right]~=~\E_{\bX}\left[\mhat(\bX)+\frac{\pi_N(\bX)}{\pi_N(\bX)}\{Y-\mhat(\bX)\}\right]~=~\theta_0,$$
and hence the proposed estimator $\tilde\theta$ is unbiased for $\theta_0$, even if \tco{$m(\cdot)$} is misspecified. We denote $\mu(\cdot)$ as a ``limit'' \tcr{(potentially \emph{misspecified})} of $\mhat(\cdot)$, \tcr{i.e.\tco{,} in general,} $\mu(\cdot)\neq m(\cdot)$ is \tcr{allowed.} 


\begin{assumption}[Basic assumption]\label{a1}
(a)  $\bZ$ has finite 2nd moments and $\boldsymbol\Sigma\equiv\Var(\bX)$ is positive definite. (b)  Let $\E[\{Y-m(\bX)\}^2|\bX=\bx]\geq\sigma_{\zeta,1}^2>0$ and $\E[\{Y-\mu(\bX)\}^2|\bX=\bx]\leq\sigma_{\zeta,2}^2<\infty$ for all $\bx$ in the support   $\mathcal X$   of $\mathbb P_{\bX}$. Moreover, $\Var(Y)\leq\sigma_{\zeta,2}^2$.
\end{assumption}
\begin{assumption}[Tail condition]\label{a4}
Let
$a_N^{-1}\E\left[\psi_{\mu,\pi}^2(\bZ)\mathbbm{1}\left\{|\psi_{\mu,\pi}(\bZ)|>c \sqrt{N/a_N}
\right\}\right]\to0,$
for any $c>0$ as $ N\to\infty$, \jelena{where recall that $a_N$} \tco{ is defined in \eqref{def:a_N},}
\jelena{and} with $\psi_{\mu,\pi}(\bZ)$  as:
 \begin{align}
\psi_{\mu,\pi}(\bZ)~:=~\mu(\bX)+\frac{R}{\pi_N(\bX)}\{Y-\mu(\bX)\}-\theta_0
~=~Y-\theta_0+\left\{\frac{R}{\pi_N(\bX)}-1\right\}\{Y-\mu(\bX)\}.
\label{psi1}
\end{align}
\end{assumption}

\begin{remark}[Discussion on Assumptions \ref{a1} and \ref{a4}]\label{r1}

\tcr{Assumption \ref{a1} imposes some mild moment conditions; similar versions can be found in \cite{zhang2019semi,zhang2022high}.}
Assumption \ref{a4} is needed only for the asymptotic normality and is satisfied if
1) $\pi_N(\cdot)$ follows an offset  propensity  model as in Example \ref{e1} with sub-Gaussian $f(\bX)$ (see \tcr{Section \ref{logistic}} where  we analyzed a  special case of the offset  model); 2) $\E\{|Y-\mu(\bX)|^{2+\delta}|X\}< C$, $\E(|Y-\theta_0|^{2+\delta})<C$ with constants $\delta,C>0$; and 3) $N\bar\pi_N\to\infty$ as $N\to\infty$. A  sufficient condition for Assumption \ref{a4} \tcr{is given in} 
in Assumption \ref{a4'}.
\end{remark}


\tcr{In the result below, we analyze the theoretical properties of $\tilde\theta$ including \jelena{its} consistency, convergence rate, asymptotic normality and robustness properties.}

\begin{theorem}\label{t3}
Let Assumptions \ref{cond:ignorability} and \ref{a1} hold. Let  $Na_N \rightarrow \infty$ as $N \rightarrow \infty$. Let $\mu(\cdot)$ be
a well-defined limit of
 the cross-fitted  $\mhat(\cdot)$,  that satisfy:
\begin{equation}\label{cmuN}
\E_{\bX}\left[ \frac{a_N}{\pi_N(\bX)} \{\mhat(\bX; \mathbb S_{-k}) - \mu(\bX) \}^2 \right] ~=~ O_p(c_{\mu,N}^2), \mbox{ with sequence } c_{\mu, N} ~=~ o(1),
\end{equation}
for $k\leq\K$.
Then,
\begin{align*}
& \tilde\theta - \theta_0 ~=~ N^{-1} \sum_{i=1}^N \psi_{\mu,\pi}(\bZ_i) + O_p\left( \frac{c_{\mu,N}}{\sqrt{Na_N}}\right) \;\; \mbox{and} \;\; V_N(\mu)~:=~\Var\{\psi_{\mu,\pi}(\bZ)\} ~\asymp~ a_N^{-1},
\end{align*}
where
$\psi_{\mu,\pi}(\bZ_i)$
is defined in \eqref{psi1}. Alternatively,   we also have the following asymptotically linear representation:
\begin{align*}
& \tilde\theta - \theta_0 ~=~ N^{-1} \sum_{i=1}^N \psitil_{\mu}(\bZ_i) + O_p\left( \frac{c_{\mu,N}}{\sqrt{Na_N}} + \frac{1}{\sqrt{N}}\right) \;\; \mbox{and} \;\; \Vtil_N(\mu):=\Var\{\psitil_{\mu}(\bZ)\} \asymp a_N^{-1},
\end{align*}
where
  $\psitil_\mu(\bZ) := R/\pi_N(\bX) \{ Y - \mu(\bX)\} - \E [R/\pi_N(\bX) \{ Y - \mu(\bX)\}]$ and $\E\{\psitil_\mu(\bZ)\} = 0$.
Additionally, as long as Assumption \ref{a4} holds, we have:
$$(Na_N)^{1/2}(\tilde\theta - \theta_0) ~=~ O_p(1), \;\; \mbox{and} \;\; N^{1/2}V_N^{-1/2}(\mu)(\tilde\theta - \theta_0)~\xrightarrow{d}~\mathcal N(0,1).$$
Moreover, if  $a_N\to0$ as $N\to\infty$, then,
$$N^{1/2}\Vtil_N^{-1/2}(\mu)(\tilde\theta - \theta_0) ~\xrightarrow{d}~ \mathcal N(0,1), \;\; \mbox{and} \;\; \frac{V_N(\mu)}{\Vtil_N(\mu)} ~=~ 1 + O(a_N).$$
\end{theorem}
\begin{remark}[\tco{Discussion on condition \eqref{cmuN}}]\label{remark:cmuN}
As per Theorem \ref{t3},  consistency  and  asymptotic normality of $\tilde\theta$ depend on  \eqref{cmuN}, a condition that involves
  1) the convergence rate of $\mhat(\cdot)$ towards some $\mu(\cdot)$,  depending on the (expected) labeled sample size $(N\bar\pi_N)$, and 2) the tail of $\pi_N^{-1}(\bX)$, \tco{that is, how much of the mass of the distribution of $\pi_N(\bX)$ \jelena{concentrates} around zero.}
For a special case of $\pi_N(\bX)\equiv\bar\pi_N$, MCAR, \eqref{cmuN}  is equivalent to $\|\mhat(\cdot;\S_{-k})-\mu(\cdot)\|_{2,\P_\bX}=o_p(1)$ coinciding with \cite{zhang2022high}. On the other hand,  when $\pi_N(\cdot)$ follows the offset  model (Example \ref{e1}) with sub-Gaussian $f(\bX)$, we have $a_N\asymp\bar\pi_N$, and   \eqref{cmuN} holds once $\E_{\bX}\{|\mhat(\bX;\S_{-k})-\mu(\bX)|^{2+\delta}\}=o_p(1)$ with   $\delta>0$.

\end{remark}
\begin{remark}[\tco{Efficiency of $\tilde\theta$ and the choice of $\mu(\cdot)$}]\label{efficiency}
Although the choice of $\mhat(\cdot)$ is arbitrary as long as it converges to some $\mu(\cdot)$  as in \eqref{cmuN}, the efficiency of  $\tilde\theta$ does depend on the limit $\mu(\cdot)$, and hence also  on the choice of $\mhat(\cdot)$. For a simple case  of $\mhat(\bx)=\mu(\bx)=0$ for all $\bx\in\mathcal X$, $\tilde\theta$  can be written as
$\tilde\theta~=~ N^{-1}\sum_{i=1}^{ N}R_iY_i/\pi_N(\bX_i),$
which coincides with  the IPW estimator, \jelena{an estimator independent of} $m(\cdot)$.
However, an appropriate estimator $\mhat(\cdot)$ will provide a better efficiency for $\tilde\theta$. The optimal choice of $\mu(\cdot)$ that minimizes the asymptotic variance $V_N(\mu)$ is $\mu(\cdot)=m(\cdot)$ indicating that the outcome model is correctly specified. 
\end{remark}
\begin{remark}[Intuition behind the {\crev influence functions (IFs)}]\label{remark:IFs}
\tcr{Two separate \tco{IFs} $\psi_{\mu,\pi}(\bZ)$ and $\psitil_\mu(\bZ)$ appear in the expansions of $\tilde\theta$ in Theorem \ref{t3}. The first 
IF}  $\psi_{\mu,\pi}(\bZ)$ is an ``accurate influence function'' in that $\tilde\theta-\theta_0= {\crevjb T_1}+o_p((Na_N)^{-1/2})$ with ${\crevjb T_1} =N^{-1}\sum_{i=1}^N\psi_{\mu,\pi}(\bZ_i)\asymp(Na_N)^{-1/2}$.
 {\crevjb Whenever}  $a_N\to0$ and $N\to\infty$, the second IF $\psitil_\mu(\bZ)$ captures the main contribution of $\psi_{\mu,\pi}(\bZ)$. It only involves   the labeled samples and hence one can clearly see that the rate of $\tilde\theta$ is effectively determined by the smaller sized, labeled data only. When the outcome model is correctly specified, the second IF $\psitil_\mu(\bZ)$ coincides with the  efficient \tco{IF} of \cite{kallus2020role}; \jelena{see  Theorem 4.1 therein}. 
\end{remark}
\begin{remark}[\tco{Convergence rate and ``effective sample size''}]\label{remark:NaN} \tco{Suppose the conditions in Theorem \ref{t3} hold, then $\tilde\theta$ is a $(Na_N)^{1/2}$-consistent estimator for $\theta_0$. The value $Na_N$ can be seen as an ``effective sample size''  having a similar role as the sample size in supervised learning. Bellow is a discussion on the value $Na_N$. By Jensen's inequality, $Na_N\leq N\bar\pi_N$, where the difference between the two rates is related to the tail of $\pi_N^{-1}(\bX)$. Here, $N\bar\pi_N$ is the expected sample size \jelena{as} $N\bar\pi_N=\E(n)$, where $n:=\sum_{i=1}^NR_i$. 
Therefore, the effective sample size, $Na_N$, depends on 1) how much of the mass of the distribution of $\pi_N(\bX)$ concentrates around 0 and
2) the (expected) size of the labeled sample. MCAR is a special case \jelena{with} $\pi_N(\cdot)$ \jelena{being} a constant and therefore $Na_N=N\bar\pi_N$. In another example, the offset based model in Example \ref{e1} and Section \ref{logistic}, \tcr{we have} $a_N\asymp\bar\pi_N$ for sub-Gausian $\bX$; see Theorems \ref{thm:ex3} and \ref{thm:high-dim}.}
\end{remark}

\subsection[{Unknown \tco{PS} and the general version of \tcr{the DRSS} estimator}]{Unknown \tco{PS} $\pi_N(\cdot)$ and the general version of \tcr{the DRSS} estimator}\label{sec:unknownpi}

With $\pi_N(\cdot)$ being unknown in general {observational studies},  we propose our final estimator, a \emph{doubly robust semi-supervised} \jelena{(DRSS)} estimator \tcr{of the mean $\theta_0$, given by:}
\begin{equation}\label{theta:unknownpi}
\thetahat_\mathrm{DRSS}~:=~ N^{-1}\sum_{i=1}^{ N}\mhat(\bX_i)+ N^{-1}\sum_{i=1}^{ N}\frac{R_i}{\pihat_N(\bX_i)}\{Y_i-\mhat(\bX_i)\},
\end{equation}
where $\pihat_N(\bX_i)$ is a cross-fitted estimator of $\pi_N(\bX_i)$ constructed similarly as $\mhat(\bX_i)$, \tcr{as discussed below \eqref{theta:knownpi} in Section \ref{sec:knownpi}}.   The proposed estimator \eqref{theta:unknownpi} is a plug-in version of \eqref{theta:knownpi}.
%
We denote with $e_N(\cdot)$ a  ``limit'' of $\pihat_N(\cdot)$, which is \tcr{possibly \emph{misspecified}, i.e.\tco{,} $e_N(\cdot)$  is} not necessarily  the same as $\pi_N(\cdot)$. Define the following generalization of \eqref{psi1}, i.e., a DR score (influence) function:
\begin{align}
\psi_{\mu,e}(\bZ)~:=~&\mu(\bX)+\frac{R}{e_N(\bX)}\{Y-\mu(\bX)\}-\theta_0~=~Y-\theta_0+\left\{\frac{R}{e_N(\bX)}-1\right\}\{Y-\mu(\bX)\}.\label{def:psimue}
\end{align}
We have the following asymptotic results under the two cases: (a) both \tco{$\pi_N(\cdot)$ and $m(\cdot)$} are correctly specified; (b) one of \tco{$\pi_N(\cdot)$ and $m(\cdot)$} is correctly specified.

\begin{theorem}\label{t4}
Let Assumptions \ref{cond:ignorability} and \ref{a1} hold and let $Na_N\to\infty$, as $N\to\infty$. Suppose the cross-fitted versions of $\mhat(\cdot)$ and \tcr{$\pihat_N(\cdot)$} have well-defined \tcr{(possibly misspecified)} limits $\mu(\cdot)$ and $e_N(\cdot)$, respectively,
such that \eqref{cmuN} holds
 for $k\leq\K$ as well as
\begin{align}
& \E_{\bX}\left[ \frac{a_N}{\pi_N(\bX)} \left\{1 - \frac{e_N(\bX)}{\pihat_N(\bX;\mathbb S_{-k})} \right\}^2\right] ~=~ O_p(c_{e,N}^2) \;\; \mbox{with sequence} \;\; c_{e, N} ~=~ o(1),\label{drthm:ratecond2} \\
& \E_{\bX}\{\mhat(\bX; \mathbb S_{-k}) - \mu(\bX) \}^2 ~=~ O_p(r_{\mu,N}^2) \;\; \mbox{with sequence} \;\; r_{\mu,N} ~=~ o(1),\;\;\mbox{and}\label{drthm:ratecond3}\\
& \E_{\bX}\left\{1 - \frac{e_N(\bX)}{\pihat_N(\bX; \mathbb S_{-k})} \right\}^2 ~=~ O_p(r_{e,N}^2) \;\; \mbox{with sequence} \;\; r_{e,N} ~=~ o(1).\label{drthm:ratecond4}
\end{align}
The properties of $\thetahat_\mathrm{DRSS}$ under different cases are as follows:

\vspace{0.5em}
(a) Suppose both $\mu(\cdot)=m(\cdot)$ and $e_N(\cdot)=\pi_N(\cdot)$ hold. Then, as $N\to\infty$, $\thetahat_\mathrm{DRSS}$ satisfies the following asymptotic linear expansion:
$$\thetahat_\mathrm{DRSS} - \theta_0~=~N^{-1} \sum_{i=1}^N \psi_{\mu,e}(\bZ_i) + O_p\left( \frac{c_{\mu,N}}{\sqrt{Na_N}} + \frac{c_{e,N}}{\sqrt{Na_N}} + r_{\mu,N}r_{e,N}\right),$$
and $V_N(\mu,e) \asymp a_N^{-1}$, where $V_N(\mu,e) := \Var\{\psi_{\mu,e}(\bZ)\}$. Hence, as long as the product rate $r_{\mu,N}r_{e,N}$ from \eqref{drthm:ratecond3} and \eqref{drthm:ratecond4} further satisfies $r_{\mu,N}r_{e,N} = o(1/\sqrt{Na_N})$, and Assumption \ref{a4} holds, we have:
\begin{equation}
(Na_N)^{1/2}(\thetahat_\mathrm{DRSS} - \theta_0) ~=~ O_p(1), \;\; \mbox{and} \;\; N^{1/2}V_N^{-1/2}(\mu,e)(\thetahat_\mathrm{DRSS} - \theta_0)~\xrightarrow{d}~\mathcal N(0,1).\label{norm:main}
\end{equation}

(b) Suppose now that either $\mu(\cdot)=m(\cdot)$ or $e_N(\cdot)=\pi_N(\cdot)$ holds.  Moreover, if $e_N(\cdot)\neq\pi_N(\cdot)$, we assume $c\leq\pi_N(\bX)/e_N(\bX)\leq C$ \tco{a.s.} for some constants $c,C>0$. Then, as $N\to\infty$, $\thetahat_\mathrm{DRSS}$ satisfies the following asymptotic linear expansion:
\begin{align}
& \thetahat_\mathrm{DRSS} - \theta_0~=~ N^{-1} \sum_{i=1}^N \psi_{\mu,e}(\bZ_i) + O_p\left( \frac{c_{\mu,N}}{\sqrt{Na_N}} + \frac{c_{e,N}}{\sqrt{Na_N}} + r_{\mu,N}r_{e,N}\right) + \Deltahat_N,\nonumber
\end{align}
with $ \Deltahat_N $ satisfying \eqref{Deltahat} or \eqref{Deltahat2}:
\begin{align}
& \Deltahat_N ~:=~  N^{-1} \sum_{i=1}^N \left\{\frac{R_i}{\pi_N(\bX_i)} - \frac{R_i}{\pihat_N(\bX_i)}\right\}\{\mu(\bX_i) - m(\bX_i)\} \quad \mbox{if} \;\; e_N(\cdot) ~=~ \pi_N(\cdot),\label{Deltahat}\\
&\Deltahat_N ~:=~  N^{-1} \sum_{i=1}^N  \left\{\frac{R_i}{\pi_N(\bX_i)} - \frac{R_i}{e_N(\bX_i)}\right\}\{\mhat(\bX_i) - m(\bX_i)\} \quad \mbox{if} \;\; \mu(\cdot) ~=~ m(\cdot)\label{Deltahat2}.
\end{align}
Suppose for case 
\tco{\eqref{Deltahat}}, $\| m(\cdot) - \mu(\cdot)\|_{2,\P_{\bX}} < C$, while for case 
\tco{\eqref{Deltahat2}}, $\| 1 - \pi_N(\cdot)/e_N(\cdot) \|_{2,\P_{\bX}} <C$, with a constant $C<\infty$.
Then, $\thetahat_\mathrm{DRSS}$ satisfies:
\begin{align*}
\thetahat_\mathrm{DRSS} - \theta_0&~=~ O_p\left(\frac{1+c_{\mu,N}+c_{e,N}}{\sqrt{Na_N}} + r_{\mu,N}r_{e,N}+r_{e,N}\mathbbm{1}\{\mu(\cdot) \neq m(\cdot)\} + r_{\mu,N}\mathbbm{1}\{e_N(\cdot) \neq \pi_N(\cdot)\}\right).
\end{align*}
\end{theorem}

A few remarks \jelena{pertaining} to the estimation rates are presented next.
\begin{remark}[\tco{Conditions in Theorem \ref{t4}}]\label{remark:conds}
\tco{Here we discuss the rate conditions \eqref{cmuN}, \eqref{drthm:ratecond2}, \eqref{drthm:ratecond3}, and \eqref{drthm:ratecond4} required in Theorem \ref{t4}. The rate \eqref{drthm:ratecond3} is a standard estimation error of the outcome model; see for example,  \cite{zhang2019semi}. The other rates, \eqref{cmuN}, \eqref{drthm:ratecond2}, and \eqref{drthm:ratecond4}, are rescaled or \jelena{self-normalized}   versions of  conditions in \cite{chernozhukov2018double}. \jelena{T}hey   are \jelena{needed} as the price of violating the positivity condition.
The rate \eqref{drthm:ratecond4},  a rescaled version of the usually considered $\E_{\bX}\{\pihat_N(\bX)-e_N(\bX)\}^2$,  is a change needed to properly address \tco{a} decaying \tco{PS estimator}.
Then, \eqref{cmuN} and \eqref{drthm:ratecond2} can be seen as  \jelena{self-normalized} versions \jelena{ with the normalization factor being}  $\omega(\bX):=a_N/\pi_N(\bX)$. Notice that $\E\{\omega(\bX)\}=1$, \jelena{so these weights $\omega(\cdot)$ can be viewed as reweighing or redistribution factor}. \jelena{Then}, the estimation errors of $\pihat_N(\bX)$ and $\mhat(\bX)$ at $\bX$\jelena{,} with a smaller PS\jelena{,} contribute more to rates \eqref{drthm:ratecond3} and \eqref{drthm:ratecond4}.
The rates of the reweighed versions, $c_{\mu,N}$ and $c_{e,N}$ in \eqref{cmuN} and \eqref{drthm:ratecond2}, only need to be $o(1)$; whereas $r_{\mu,N}$ and $r_{e,N}$ in \eqref{drthm:ratecond3} and \eqref{drthm:ratecond4} appear in the final rate for $\thetahat_\mathrm{DRSS}$.
In high dimensions, assume $\pi_N(\cdot)$ follows an offset based model as in Example \ref{e1}. Suppose $m(\cdot)$ and $f(\cdot)$ in Example \ref{e1} are linear with sparsity levels $s_m$ and $s_\pi$, respectively. Then, for sub-Gaussian $\bX$, we demonstrate in Theorem \ref{thm:high-dim} that $a_N\asymp\bar\pi_N$ \jelena{as long as} $r_{e,N}=\sqrt{s_\pi\log(p)/(N\bar\pi_N)}$ and $r_{\mu,N}=\sqrt{s_m\log(p)/(N\bar\pi_N)}$}, \jelena{therefore coming close to the simplest missingness pattern, that of MCAR. }
\end{remark}

\begin{remark}[\tco{Double robustness, \tcr{rates} and efficiency}]\label{remark:DR}
Here, we discuss the double robustness and the efficiency of the proposed estimator.
{\it Whenever \tco{$\pi_N(\cdot)$ and $m(\cdot)$} are  correctly specified}, the asymptotic normality with a \jelena{rate of consistency} $(Na_N)^{-1/2}$ is guaranteed if a product rate condition $r_{\mu,N}r_{e,N} = o(1/\sqrt{Na_N})$ is satisfied.  We can see that our product rate condition is an analog of the usual product rate condition in the literature \tco{\citep{chernozhukov2018double},} 
if \tcr{the} sample size is replaced with $N a_N$, the ``effective sample size'' \tcr{in our case}; see Remark \ref{remark:NaN}.     In addition, when the asymptotic normality occurs, our estimator reaches the semi-parametric efficiency bound proposed in \cite{kallus2020role} when $\bar\pi_N\to0$ as $N\to\infty$.
{\it When one of \tco{$\pi_N(\cdot)$ and $m(\cdot)$} is misspecified}, we obtain a consistency rate of $O_p(r_{e,N})$ if \tco{$\pi_N(\cdot)$} is correctly specified, whereas the rate is  $O_p(r_{\mu,N})$ if \tco{$m(\cdot)$} is correctly specified.
Therefore, the consistency rate of $\thetahat_\mathrm{DRSS}$ directly depends on the estimation error rate of the correct model. As a special case, $\thetahat_\mathrm{DRSS}$ is consistent as long as the correct model is consistently estimated. \tco{Additionally}, we can see that $\thetahat_\mathrm{DRSS}$ can still be $(Na_N)^{1/2}$-consistent as long as the correct model is estimated with an error rate $O_p(Na_N)^{-1/2}$\tco{, which is reachable in low dimensions. For instance,
for a (correctly specified) low-dimensional offset logistic PS model as introduced in Section \ref{low-dim}, as shown in Theorem \ref{thm:ex3},  \jelena{not only do we reach} the error rate $O_p(Na_N)^{-1/2}$ but \jelena{are able to}  construct a RAL expansion for $\thetahat_\mathrm{DRSS}$.}
\end{remark}

\begin{remark}[\tco{Unbounded support for $\bX$}] \label{remark:support}
\tco{We do \jelena{not} enforce a bounded support for $\bX$, which is typically an assumption assumed (implicitly) in missing data and causal inference literature. For instance, suppose  $\pi_N(\cdot)$ follows an (offset based) logistic model as in Example \ref{e1}. Both the usual positivity condition $\P(\pi_N(\bX)>c>0)=1$ in the standard missing data literature \citep{imbens2004nonparametric,tsiatis2007semiparametric,imbens2015causal} and the uniform bounded density ratio condition, $\bar\pi_N/\pi_N(\bX) <C$, in \cite{kallus2020role}, which tack\tcr{le}s a MAR-SS setting, essentially require a compact support for $\bX$. However, our results only require a sub-Gaussian $\bX$ as in Theorems \ref{thm:ex3} and \ref{thm:high-dim}.
 }
\end{remark}


\begin{remark}[\tco{Asymptotic linearity and $(Na_N)^{1/2}$-consistency under misspecification}]\label{remark:RAL_mis}
Mor\tcr{eo}ver, in Section \ref{pihat_N}, we demonstrate that $\thetahat_\mathrm{DRSS}$ can still be asymptotically normal even if \tco{$m(\cdot)$} is misspecified. Such an asymptotic normality is constructed based on a careful analysis \tcr{to obtain} 
the regular and asymptotically linear (RAL) expansion and the \tco{IF} for the additional error term $\Deltahat_N$  \tcr{in} \eqref{Deltahat}, in that
$$
\Deltahat_N~=~N^{-1}\sum_{j=1}^N\mathrm{IF}_\pi(\bZ_j)+o_p\left((Na_N)^{-1/2}\right),
$$
for some $\mathrm{IF}_\pi(\cdot)$ with $\E\{\mathrm{IF}_\pi(\bZ)\}=0$ and $\E\{\mathrm{IF}_\pi^2(\bZ)\}\asymp a_N^{-1}$. The final \tcr{IF} 
of $\thetahat_\mathrm{DRSS}$ involves the extra IF contributed from the estimation error of $\pihat_N(\cdot)$. Consequently, the RAL expansion and the asymptotic normality of $\thetahat_\mathrm{DRSS}$ are also affected accordingly. Using the above expansion for $\Deltahat_N$ and the general expansion of $\thetahat_\mathrm{DRSS}$ from Theorem \ref{t4}, we have a RAL expansion of $\thetahat_\mathrm{DRSS}$ as: 
\begin{align*}
\thetahat_\mathrm{DRSS}-\theta_0~=~& N^{-1} \sum_{i=1}^N \psi_{\mu,e}(\bZ_i) + O_p\left( \frac{c_{\mu,N}}{\sqrt{Na_N}} + \frac{c_{e,N}}{\sqrt{Na_N}} + r_{\mu,N}r_{e,N}\right)+\Deltahat_N\\
~=~&N^{-1}\sum_{i=1}^N\{\psi_{\mu,e}(\bZ_i)+\mathrm{IF}_\pi(\bZ_i)\}+o_p\left((Na_N)^{-1/2}\right).
\end{align*}
The function $\Psi(\bZ):=\psi_{\mu,e}(\bZ)+\mathrm{IF}_\pi(\bZ)$ is the final \emph{adjusted} IF of $\thetahat_\mathrm{DRSS}$ with $\E\{\Psi(\bZ)\}=0$ and $\Var\{\Psi(\bZ)\}\asymp a_N^{-1}$. Consequently, we also have:
\begin{equation}\label{norm:mod}
N^{1/2}[\Var\{\Psi(\bZ)\}]^{-1/2}(\thetahat_\mathrm{DRSS}-\theta_0)~\xrightarrow{d}~\mathcal N(0,1).
\end{equation}
\end{remark}

{\crev
\begin{remark}[Estimation when both models are misspecified]\label{remark:wrong}
In this remark, we briefly discuss a few aspects of the proposed estimator's behavior when both nuisance models are misspecified.

Firstly, as one of the anonymous reviewers pointed out, we do not need at least one nuisance model to be correctly specified; in fact, the identification of the mean parameter only requires
\begin{align}\label{eq:dr'}
\E\left[\left\{1-\frac{\pi_N(\bX)}{e_N(\bX)}\right\}\{\mu(\bX)-m(\bX)\}\right]=0.
\end{align}
For instance, \eqref{eq:dr'} holds when $\mu(\bx)=m(\bx)$ for all $\bx\in\mathcal X_1$ and $e_N(\bx)=\pi_N(\bx)$ for all $\bx\in\mathcal X_2${\crevyz, for some $\mathcal X_1$ and $\mathcal X_2$} with $\mathcal X_1\cup\mathcal X_2=\mathcal X$. {\crevyz This is indeed weaker than requiring one of $\pi_N(\cdot)$ and $m(\cdot)$ to be correctly specified, and, of course, there are also other ways possible to achieve the weaker condition \eqref{eq:dr'}.} However, for the sake of interpretation, we  {\crevjb consider} the {\crevyz simpler and} sufficient  {\crevjb (}but not necessary{\crevjb)} condition that either $\mu(\cdot)=m(\cdot)$ or $e_N(\cdot)=\pi_N(\cdot)$ throughout the paper;  {\crevjb this is not unlike the missing data or causal inference literature}
, e.g., \cite{tsiatis2007semiparametric,tan2020model,rubin1974estimating,robins1995semiparametric,bang2005doubly,chernozhukov2018double,farrell2015robust}.

Secondly, if \eqref{eq:dr'} does not hold and both models have non-ignorable misspecification errors  {\crevjb with}  $\mu(\cdot)-m(\cdot)\asymp1$ and $1-\pi_N(\cdot)/e_N(\cdot)\asymp1$, we do not expect any consistent estimate for $\theta_0$  {\crevjb to exist}   in general. On the other hand, if at least one model is  misspecified but with a \emph{decaying} misspecification error, we can still obtain consistent estimates. For instance,
consider a ``weakly misspecified'' linear outcome model  {\crevjb with}
\begin{align*}
m(\bX)=\bX^T\bbeta_0+\xi(\bX),
\end{align*}
where $\xi(\bX)=\xi_N(\bX)$ is the misspecification error whose distribution is dependent on $N$.   {\crevjb The case of misspecification of the PS model follows analogously.}
 {\crevjb If{\crevyz,} f}or the sake of simplicity, we assume that $\E\{\xi(\bX)\}=0$ and $\E\{\bX\xi(\bX)\}=\bzero${\crevyz,}
   {\crevjb then whenever}   the linear model is approximately correct with $\E\{\xi^2(\bX)\}=o(1)$ as $N\to\infty$, the proposed DRSS mean estimator is still consistent; see, e.g.,
\cite{belloni2014inference,belloni2014high,belloni2012sparse}, where the authors also only require the true nuisance functions to be close enough to the linear approximations.
Moreover, another indirect situation with ``weakly misspecified'' models is when the nuisance function is ``weakly sparse''; see, e.g., Section 4.3 of \cite{negahban2012unified} {\crevmag and Example 10 of \cite{smucler2019unifying}}. In general, let $\E\{1-\pi_N(\bX)/e_N(\bX)\}^2\asymp\bar r_{e,N}^2=O(1)$ and $\E\{\mu(\bX)-m(\bX)\}^2\asymp\bar r_{\mu,N}^2=O(1)$. The additional bias of {\crevyz our} DRSS {\crevyz mean estimator originating} from {\crevyz model} misspecification is
\begin{align*}
\mathrm{Bias}_\mathrm{DR}:=\E\{\psi_{\mu,e}(\bZ)\}-\theta_0=\E\left[\{1-\pi_N(\bX)/e_N(\bX)\}\{\mu(\bX)-m(\bX)\}\right]=O(\bar r_{e,N}\bar r_{\mu,N})=o(1),
\end{align*}
as long as either $\bar r_{e,N}=o(1)$ or $\bar r_{\mu,N}=o(1)$. 
{\crevyz For comparison,} if we consider {\crevyz the corresponding regression-based estimator} $\thetahat_\mathrm{Reg}$ or {\crevyz the IPW estimator} $\thetahat_\mathrm{IPW}$ (see definitions in Section \ref{setup}), the additional biases {\crevyz of these estimators} from {\crevyz model} misspecification are{\crevyz, respectively,}
\begin{align*}
\mathrm{Bias}_\mathrm{Reg}:=\E\{\mu(\bX)\}-\theta_0=O(\bar r_{\mu,N}),\;\;\mathrm{Bias}_\mathrm{IPW}:=\E\{\pi_N(\bX)m(\bX)/e_N(\bX)\}-\theta_0=O(\bar r_{e,N}).
\end{align*}
That is, $\thetahat_\mathrm{Reg}$ is consistent if $\bar r_{\mu,N}=o(1)$; $\thetahat_\mathrm{IPW}$ is consistent if $\bar r_{e,N}=o(1)$. We can see $\thetahat_\mathrm{DRSS}$ is more robust than $\thetahat_\mathrm{Reg}$ and $\thetahat_\mathrm{IPW}$ as we allow both $\bar r_{e,N}=o(1)$ and $\bar r_{\mu,N}=o(1)$. In addition, $\thetahat_\mathrm{DRSS}$ also has a faster consistency rate when both misspecification error{\crevjb s}   shrink; see empirical comparisons of the mean estimators in Section \ref{sec:both-mis}.
\end{remark}
}

\subsection{Asymptotic variance estimation}\label{sec:varestimation}

In this section, we consider the estimation of \tcr{the asymptotic variances} $V_N(\mu)$ in Theorem \ref{t3} (\jelena{with} $\pi_N(\cdot)$  known) and $V_N(\mu,e)$ in Theorem \ref{t4} (\jelena{with} $\pi_N(\cdot)$ unknown and both \tco{$m(\cdot)$ and $\pi_N(\cdot)$} are correctly specified). \tcr{These facilitate inference on $\theta_0$ (via confidence intervals, hypothesis tests etc.) using $\tilde\theta$ and $\thetahat_\mathrm{DRSS}$.}
We assume the following tail condition.
\begin{assumption}[Tail condition]\label{a4'}
With $ N\to\infty$, for a constant $\delta>0$, let
$$
N^{-\delta/2}a_N^{1+\delta/2}\E\{|\psi_{\mu,\pi}(\bZ)|^{2+\delta}\}~\to~0.
$$
\end{assumption}
The Assumption \ref{a4'} is a sufficient condition for Assumption \ref{a4}.
Under the setting in Theorem \ref{t3} and part (a) of Theorem \ref{t4}, we have:
$$
N^{1/2}V_N^{-1/2}(\mu)(\tilde\theta - \theta_0)~\xrightarrow{d}~\mathcal N(0,1),\qquad N^{1/2}V_N^{-1/2}(\mu,e)(\thetahat_\mathrm{DRSS} - \theta_0)~\xrightarrow{d}~\mathcal N(0,1).
$$
\jelena{W}e propose the following plug-in estimates, $\widehat V_N(\mu)=\widehat V_N(\mhat, \bar\pi_N, \tilde\theta)$ and $\widehat V_N(\mu,e)=\widehat V_N(\mhat, \pihat_N,\thetahat_\mathrm{DRSS})$, where 
\begin{align*}
\widehat V_N(a,b,c)~:=~&N^{-1}\sum_{i=1}^{N}\left[a(\bX_i)-c+\frac{R_i}{b(\bX_i)}\{Y_i-a(\bX_i)\}\right]^2.
\end{align*}
\begin{theorem}\label{thm:var}
(a) Let Assumptions in Theorem \ref{t3} hold. Then, as $N\to\infty$,
$
\widehat V_N(\mu)=V_N(\mu)\{1+o_p(1)\}.
$
(b) Let  Assumptions  (a) of Theorem \ref{t4} hold. Further let Assumption \ref{a4'} hold and
\begin{equation}\label{cond:drift2}
\E\left[\frac{a_N}{\pi_N(\bX)}\left\{1-\frac{\pi_N(\bX)}{\pihat_{\tcr{N}}(\bX;\mathbb S_{-k})}\right\}^2\{\mhat(\bX;\mathbb S_{-k})-m(\bX)\}^2\right]~=~o_p(1).
\end{equation}
Then, as $N\to\infty$,
$
\widehat V_N(\mu,e)=V_N(\mu,e)\{1+o_p(1)\}.
$
\end{theorem}
\tco{
Notice that we only require a $o_p(1)$ condition in \eqref{cond:drift2}. \tcr{S}uch a condition can be satisfied as long as we have upper bounds for \tcr{the} $(2+c)$-th moment of the estimation errors and the tail of $\pi_N^{-1}(\bX)$ is well-behaved. Under a standard positivity condition, when $\mu(\cdot)=m(\cdot)$, \eqref{cond:drift2} only requires $r_{\mu,N}=o(1)$, which  \jelena{would have been already assumed for consistency}.}

\tco{Under the conditions in Theorem \ref{thm:var},} asymptotic\tcr{ally valid $100(1-\alpha)\%$} confidence intervals \tcr{(CIs)} for $\tilde\theta$ and $\thetahat_\mathrm{DRSS}$ at any significance level $\alpha$ \tcr{can now be obtained as:} 
\begin{align}
\mathrm{CI}(\tilde\theta)~:=~&\left(\tilde\theta-N^{-1/2}\widehat V_N^{1/2}(\mu)z_{1-\alpha/2},\;\tilde\theta+N^{-1/2}\widehat V_N^{1/2}(\mu)z_{1-\alpha/2}\right),\nonumber\\
\mathrm{CI}(\thetahat_\mathrm{DRSS})~:=~&\left(\thetahat_\mathrm{DRSS}-N^{-1/2}\widehat V_N^{1/2}(\mu,e)z_{1-\alpha/2},\;\thetahat_\mathrm{DRSS}+N^{-1/2}\widehat V_N^{1/2}(\mu,e)z_{1-\alpha/2}\right),\label{CI}
\end{align}
where $z_{1-\alpha/2}$ is the $(1-\alpha/2)$-quantile of a standard normal distribution. \tco{As shown in Theorems \ref{t3} and \ref{t4}, $V_N(\mu)\asymp a_N^{-1}$ and $V_N(\mu,e)\asymp a_N^{-1}$. Hence, the length of \tcr{the} proposed confidence intervals are of the order $(Na_N)^{-1/2}$. }

\jelena{It is important to note, that these confidence intervals are valid when both the outcome and propensity score models are correctly specified.  Whenever the outcome model is misspecified, they need further adjustment based on an adjusted RAL expansion as discussed in Remark \ref{remark:RAL_mis}. }Based on the adjusted IF, $\Psi(\bZ)$, \tcr{therein,} one can estimate the asymptotic variance $\Var\{\Psi(\bZ)\}$ using a plug-in estimate $N^{-1}\sum_{i=1}^N\widehat{\Psi}^2(\bZ_i;\thetahat_\mathrm{DRSS})$, where $\widehat{\Psi}(\cdot;\thetahat_\mathrm{DRSS})$ is a consistent estimator of $\Psi(\cdot)$, \tcr{and obtain the corresponding \emph{adjusted} confidence intervals.} \tco{We also illustrate the numerical performance of the\tcr{se} adjusted confidence intervals in Appendix \ref{sim:modCI}}  \jelena{of the \hyperref[supp_mat]{Supplement}}.


\section{Decaying \tco{PS} models}\label{pihat_N}

In Section \ref{sec:EY}, we proposed a DR estimator $\thetahat\tcr{_\mathrm{DRSS}}$ of $\theta_0=E(Y)$. Such an estimator is based on an outcome estimator $\mhat(\cdot)$ and a \tco{PS} estimator $\pihat_N(\cdot)$. Due to the decaying nature of the \tco{PS}, the estimation of $\pi_N(\cdot)$ itself is also an interesting and challenging problem.
In this section, we illustrate three decaying \tco{PS} models: (i) \tco{an} \jelena{\emph{offset logistic model}} 
\tcr{(Section \ref{logistic})}, (ii) a \emph{stratified labeling model} \tcr{(Section \ref{stratified})}, and (iii) a \emph{MCAR labeling model} \tcr{(presented in Appendix \ref{sec:MCAR} of the \hyperref[supp_mat]{Supplement} in the interest of space)}. {These are just some natural examples of modeling a decaying PS -- our main results are completely general.}

We propose \tco{PS} estimators under each of the three models and establish detailed asymptotic results, especially for the \jelena{offset} logistic model (in both low and high dimensions).
Moreover, as discussed in Remark \tco{\ref{remark:RAL_mis}},
for \tcr{a} misspecified \tco{$m(\cdot)$}, based on a case by case study of $\pi_N(\cdot)$, we further construct an adjusted RAL expansion of $\thetahat_\mathrm{DRSS}$ and hence provide an asymptotic normality with \tco{an adjusted} asymptotic variance.

\tcr{We begin with a brief remark here that addresses this particular point for the MCAR model, though all other results and detailed discussions for the MCAR setting are deferred to Appendix \ref{sec:MCAR} of the \hyperref[supp_mat]{Supplement} for brevity.}

\begin{remark}[\tcr{Adjusted IF and general RAL expansion under} MCAR]\label{rem:mcar}
For a simple case of {\crevyz a} MCAR labeling mechanism, \tcr{i.e.\tco{,} $\pi_N(\bX) \equiv \bar\pi_N$}, we establish a similar result \tco{as in \cite{zhang2022high}} 
under fairly general conditions. {\crevmag In Theorem \ref{thm:ex1} of the \hyperref[supp_mat]{Supplement},} \jelena{we provide} \tcr{an adjusted {\crevmag (and general)} RAL expansion of $\thetahat_\mathrm{DRSS}$ {\crevmag under the MCAR setting,} allowing for misspecification of $\mhat(\cdot)$.} The \tcr{adjusted IF} 
takes the form of $\Psi(\bZ) := \psi_{\mu, \pi}(\bZ)+ \mathrm{IF}_{\pi}(\bZ)$, {\crevyz where $\psi_{\mu,\pi}(\bZ)$ is {\crevmag as} defined in \eqref{psi1}} {\crevmag and} 
\[
\mathrm{IF}_{\pi}(\bZ)~:=~ \left(\frac{R - \bar\pi_N}{\bar\pi_N}\right)\Delta_{\mu},\quad \mbox{\crevmag with}\quad \Delta_{\mu} ~:=~ \E \{ \mu(\bX) - m(\bX) \}{\crevyz,}
\]
and a fast {\crevmag consistency} rate \tcr{of} \tcr{$(N a_N)^{-1/2}$}  {\crevmag for $\thetahat_\mathrm{DRSS}$} 
is achievable. We {\crevmag further} show \tcr{that}
\begin{align}\label{var_compare}
\Var\{\psi_{\mu,\pi}(\bZ)\}~=~\Var\{\Psi(\bZ)\} + (\bar\pi_N^{-1}-1)\Delta_\mu^2 ~\geq~ \Var\{\Psi(\bZ)\};
\end{align}
see Appendix \ref{sec:MCAR} in the \hyperref[supp_mat]{Supplement} for more details and formal statements. {\crevyz Recall that, as shown in Theorem \ref{t3}, $\psi_{\mu,\pi}(\bZ)$ is the IF of $\tilde\theta$, the mean estimator constructed based on the \emph{true} PS $\bar\pi_N$. Meanwhile, under the MCAR setting, $\Psi(\bZ)$ is the IF of the DRSS mean estimator $\thetahat_\mathrm{DRSS}$, which is based on an \emph{estimated} constant PS; see Theorem \ref{thm:ex1} of the \hyperref[supp_mat]{Supplement}. Hence, $\Var\{\psi_{\mu,\pi}(\bZ)\}$ and $\Var\{\Psi(\bZ)\}$ are the asymptotic variances of $\tilde\theta$ and $\thetahat_\mathrm{DRSS}$, respectively.} {\crev
From the above inequality, we {\crevmag can thus conclude} 
that $\thetahat_\mathrm{DRSS}$ is {\crevmag asymptotically} more efficient than $\tilde\theta$.} This {\crevmag therefore} suggests that, under the MCAR setting, \emph{even if} $\bar\pi_N$ is known, it is \emph{s\tcr{ti}ll} worth \tcr{estimating} 
$\bar\pi_N$ instead of directly plug\tcr{ging} in the true value $\bar\pi_N$, as long as $\Delta_\mu\neq0$. {\crevyz Formal results of the proposed DRSS mean estimator under the MCAR setting, along with additional connections with the semi-supervised estimators of \cite{zhang2019semi} and \cite{zhang2022high} can be found in Appendix \ref{sec:MCAR} of the \hyperref[supp_mat]{Supplement}.}
\end{remark}

\subsection{\jelena{Offset} logistic regression}\label{logistic}

In this section, we propose a 
parametric logistic model for extremely unbalanced outcomes, i.e.\tco{,} $\bar\pi_N=\P(R=1)\to0$ as $N\to\infty$, 
{\crev where we consider a {\crevyz PS model ({\crevmag assumed} correctly specified) with the form:} 
}
\begin{equation}\label{offset_model}
\pi_N(\bX) ~=~ \bar\pi_N \frac{\exp(\bXv^T\bgamma_0)}{1+ \bar\pi_N \exp(\bXv^T\bgamma_0 )}~=~\frac{\exp\{\bXv^T\bgamma_0+\log(\bar\pi_N)\}}{1+ \exp\{\bXv^T\bgamma_0+\log(\bar\pi_N) \}},
\end{equation}
\tcr{where $\bXv := (1,\bX^T)^T$} and \tcr{the parameter} $\bgamma_0\in\R^{p+1}$  possibly depend\jelena{s} on $N$ \jelena{with} \tco{ $\|\bgamma_0\|_2<C$ \tcr{for} 
some constant $C>0$}. This model is  fairly natural and \jelena{allows for }a general way to incorporate the decaying nature of the labeling fraction. At the same time, it ensures that the dependence of $\pi_N(\bX)$ on $\bX$ is not distorted by the decaying nature of $\bar\pi_N$.
Model \eqref{offset_model} could also be viewed as a logistic   model with $\log(\bar\pi_N)$ (a diverging negative intercept) as an \emph{offset}.
If a standard logistic model is used \citep{owen2007infinitely, wang2020logistic} \tcr{instead, i.e.\tco{,} we let}
\begin{align}
\tco{\pi_N(\bX)~=~g(\bXv^T\bbeta),\quad\mbox{where}\;\;g(u)~:=~\frac{\exp(u)}{1+\exp(u)},}
 \label{divmod-def}
\end{align}
 then under some standard conditions whenever an extreme imbalance exists, $\exp(-\bbeta(1)) \asymp \bar\pi_N^{-1} \rightarrow \infty$ whenever $N \to \infty$\jelena{; see Appendix \ref{sec:offset} and Remark \ref{remark:offset} in the \hyperref[supp_mat]{Supplement} for further details. }This provides a clear justification for our offset  model \eqref{offset_model} where we precisely extract out   $\log\tcr{(\bar\pi_N)}$ as an offset to be estimated separately and plugged in apriori to the  likelihood equation\jelena{. In this way, we}  are able to treat the \jelena{auxiliary} intercept  and the slope  \jelena{as}   well-behaved, i.e.\tco{,} finite and independent \jelena{or bounded in $N$}.


\begin{remark}[Connections with density ratio estimation]\label{rem:densityratio}
There is an intricate connection between  the offset model \eqref{offset_model} and a model for density ratios usually used in the covariate shift \tcr{literature} where \tco{$R_i$s} are treated as fixed (or condition\tcr{ed} on) and $\P_{\bX}\neq\P_{\bX|R=1}$ is allowed \citep{kawakita2013semi,liu2020doubly}.
Observe that
$$\mbox{logit}\{\pi_N(\bX)\}~=~\log(\bar\pi_N) -\log(1-\bar\pi_N)-\log\{\Lambda_N(\bX)\},$$
where  $\Lambda_N(\bX):={f(\bX|R=0)}/{f(\bX|R=1)}$ and $f(\cdot|R=\cdot)$ is the conditional density  of $\bX$ \tcr{given} 
$R$. However, direct estimation of density ratios  is often arduous. The above representation, however, suggests that the \emph{same} model can be fitted by a simple   logistic regression  of  $R|\bX$, and further using $\log\{\bar\pi_N/(1-\bar\pi_N)\}$ as an \emph{offset}\jelena{.}
 Therefore,  missing data literature related to density ratios can   now \jelena{be} enriched with an effective estimation of the decaying \tco{PS;}
see Section 4 of \cite{kallus2020role} where semi-parametric efficiency is established but no estimator is discussed. In some sense, such {a} density ratio estimator   is \tcr{also} optimal \citep{Qin_1998}. \tcr{For more discussions on these connections, see Remark \ref{rem:densityratio-supp} in the \hyperref[supp_mat]{Supplement}.}
 \end{remark}

\subsubsection{Low-dimensional offset logistic regression}\label{low-dim}

We {\crevyz first consider a low-dimensional setting where} 
{\crev $p$ is fixed}.
We propose a \tco{PS} estimator $\pihat_N(\cdot)$ for  the offset model \eqref{offset_model}, based on the full sample $\S$ and use \jelena{its} cross-fitted version \tco{(based on a subsample $\S_{-k}$)} to construct the DR mean estimator $\thetahat_\mathrm{DRSS}$.

We construct $\pihat_N(\cdot)$ based on an  apriori ch\tcr{o}sen 
estimate $\pihat_N:=N^{-1}\sum_{i=1}^NR_i$. Let $\widehat{\bgamma}$ be the minimizer of $\ell_N(\bgamma;\pihat_N)$, where
\begin{equation}\label{loglike}
\ell_N(\bgamma;a)~:=~-N^{-1}\sum_{i=1}^N\left[R_i\bXv_i^T\bgamma-\log\{1+ a\exp(\bXv_i^T\bgamma)\}\right],
\end{equation}
\tco{where recall that $\bXv=(1,\bX)^T$.}
Then, the \tco{PS}   \tcr{estimate} \jelena{,} $\pihat_N(\cdot)$\jelena{,} can be obtained by plugging  $\pihat_N$  into   \eqref{offset_model}, \tcr{as} \jelena{follows:}
\begin{equation}\label{eq:pi}
\pihat_N(\bX)~:=~\frac{\pihat_N\exp(\bXv^T\widehat{\bgamma})}{1+\pihat_N\exp(\bXv^T\widehat{\bgamma})}.
\end{equation}
Here, for any $a\in(0,1]$, $\ell_N(\bgamma;a)$ is the negative log-likelihood under the offset based model, up to a term $-N^{-1}\sum_{i=1}^NR_i\log(a)$ that is independent of $\bgamma$. Existence and uniqueness of $\widehat{\bgamma}$ is discussed \tcr{in detail} in Remark \ref{remark:existence} in the \hyperref[supp_mat]{Supplement}. 
\tcr{But}\jelena{,} \tcr{it is worth mentioning} that the results of \cite{owen2007infinitely}  showcasing the existence of the  MLE for the model \eqref{divmod-def} can be extended to guarantee \tcr{the} existence of $\widehat{\bgamma}$ \tcr{as well}.

The following theorem provides asymptotic results for $\bgammahat$ and $\pihat_N(\cdot)$, as well as an adjusted RAL expansion of the DR\tcr{SS}  estimator $\thetahat_\mathrm{DRSS}$  \jelena{in low-dimensional setting with $p$ being fixed} and  when \tco{$m(\cdot)$} is possibly misspecified. \jelena{For this result alone we consider the following conditions on the design: $\E\{\exp(t\|\bX\|_2)\}<\infty$ for any $t>0$, $\lambda_{\min}\left[\E\{\bXv\bXv^T\dot{g}(\bXv^T\bgamma_0)\}\right]>0$, where \tco{$g(\cdot)$} was defined in \eqref{divmod-def} \tco{and $\dot{g}(\cdot)=g(\cdot)\{1-g(\cdot)\}$ is the derivative of $g(\cdot)$}.}
\begin{theorem}\label{thm:ex3}
Let $N\bar\pi_N\to\infty$ as $N\to\infty$, and $\|\bgamma_0\|_2<C<\infty$ \jelena{where $\bgamma_0$ was defined in \eqref{offset_model}}. \tco{Suppose that $\|[\E\{\dot{g}(\bXv^T\bgamma_0)\bXv\bXv^T\}]^{-1}\|_2<C$ with some constant $C>0$.}
\tco{Then, as $N\to\infty$,
\begin{align*}
&\widehat{\bgamma}-\bgamma_0~=~N^{-1}\sum_{i=1}^N\mathrm{IF}_{\bgamma}(\bZ_i)+\bR_{N},\quad\mbox{with}\;\;\|\bR_{N}\|_2~=~o_p\left((N\bar\pi_N)^{-1/2}\right),\\
&\mathrm{IF}_{\bgamma}(\bZ)~:=~\mathcal J^{-1}(\bgamma_0,\bar\pi_N)\{R_i-g(\bXv^T\bgamma_0+\log(\bar\pi_N))\}\bXv-(\bar\pi_N^{-1}R-1)\mathbf{e}_1,
\end{align*}
where $\mathbf{e}_1:=(1,0,\dots,0)^T\in\R^{p+1},$} 
$\mathcal J(\bgamma_0,\bar\pi_N):=\E\{\bXv\bXv^T\dot{g}(\bXv^T\bgamma_0+\log(\bar\pi_N))\}$\tco{,} and $\|\widehat{\bgamma}-\bgamma_0\|_2=O_p((N\bar\pi_N)^{-1/2})$. Further, we also have the following error rates:
\begin{align}
&\|\pi_N^{-1}(\bX)\|_{r,\P_\bX}~\asymp~\bar\pi_N^{-1}\quad\forall r>0,\quad\mbox{and hence}\;\;a_N~\asymp~\bar\pi_N,\nonumber\\
&\left\|1-\frac{\pi_N(\bX)}{\pihat_N(\bX)}\right\|_{2,\P_\bX}~=~O_p\left((N\bar\pi_N)^{-1/2}\right),\label{2norm:pi}\\
&\E_\bX\left[\frac{a_N}{\pi_N(\bX)}\left\{1-\frac{\pi_N(\bX)}{\pihat_N(\bX)}\right\}^2\right]~=~O_p\left((N\bar\pi_N)^{-1}\right)~=~o_p(1).\label{2norm:pi*aN}
\end{align}
If \tcr{we} further assume that $\|m(\cdot)-\mu(\cdot)\|_{2+c,\P_\bX}<\infty$, then we have a RAL expansion of the term $\Deltahat_N$ defined in \eqref{Deltahat} as follows:
\begin{align}
\Deltahat_N~:=~&N^{-1}\sum_{i=1}^N\mathrm{IF}_\pi(\bZ_i)+o_p\left((N\bar\pi_N)^{-1/2}\right),\quad\mbox{where}\label{IF:Deltahat}\\
\mathrm{IF}_{\pi}(\bZ) ~:=~& \E\left[\{1-\pi_N(\bX)\}\{\mu(\bX)-m(\bX)\}\bXv^T\right]\mathcal J^{-1}(\bgamma_0,\bar\pi_N)\bXv\{R-\pi_N(\bX)\}.\label{def:IFpi}
\end{align}
Moreover, if we assume $\|\mhat(\cdot)-\mu(\cdot)\|_{2+c,\P_\bX}=o_p(1)$ (we suppressed the dependency of $\mhat(\cdot)$ on $k$ as in Theorem \ref{t4}), then we have the following rate:
\begin{equation}\label{2norm:m*aN}
\E_\bX\left[\frac{a_N}{\pi_N(\bX)}\{\mhat(\bX)-\mu(\bX)\}^2\right]~=~o_p(1)\jelena{,}
\end{equation}
and \jelena{with it }  a RAL expansion of  $\thetahat_\mathrm{DRSS}$ as:
\begin{align}
& \thetahat_\mathrm{DRSS} - \theta_0 ~=~ N^{-1} \sum_{i=1}^N \Psi(\bZ_i) + o_p\left(  \frac{1}{\sqrt{N\bar\pi_N}}\right), \;\; \mbox{where} \;\Psi(\bZ) ~:=~ \psi_{\mu, \pi}(\bZ) + \mathrm{IF}_{\pi}(\bZ),\label{modRAL}
\end{align}
 and \tco{$\psi_{\mu, \pi}(\bZ)$} is defined in \eqref{psi1}. \jelena{Lastly, }
 $\E\{\Psi(\bZ)\} = 0$, $\E\{ \Psi^2(\bZ)\}=O(\bar\pi_N^{-1})$.
\end{theorem}

The displays \eqref{2norm:pi}, \eqref{2norm:pi*aN} and \eqref{2norm:m*aN} are the conditions we need to \jelena{guarantee the assumptions of } Theorem \ref{t4}, while the result \eqref{IF:Deltahat} on $\Deltahat_N$ helps characterize the full RAL expansion of $\thetahat_\mathrm{DRSS}$ under misspecification of $\mhat(\cdot)$. Lastly, notice that we do not assume $\pi_N(\bX)/\bar\pi_N$ to be bounded bellow \tco{a.s.}, which is a condition required in \cite{kallus2020role}.

\begin{remark}[\tco{Necessity of the RAL \tcr{expansion}\jelena{'s} modification}]\label{remark:RAL_logsitic}
When $\bar\pi_N\to0$, we observe that part of the additional IF\tcr{, $\mathrm{IF}_{\pi}(\bZ)$, in} \tco{\eqref{def:IFpi}}  has the following property:
$$\E\left[\{1-\pi_N(\bX)\}\{\mu(\bX)-m(\bX)\}\bXv\right]~=~\E\left[\{\mu(\bX)-m(\bX)\}\bXv\right]+O_p(\bar\pi_N).$$
If the outcome model is fitted by a linear model \jelena{whose} limit has a linear form $\mu(\bX)=\bXv^T\bbeta^*$, with $\bbeta^*:=\{\E(\bXv\bXv^T)\}^{-1}\E(\bXv Y)$, then,
$$\E\left[\{\mu(\bX)-m(\bX)\}\bXv\right]={\bbeta^*}^T\E(\bXv\bXv^T)-\E(\bXv Y)=0,$$ \jelena{indicating that} the RAL \tcr{expansion}\jelena{'s}   modification is unne\tcr{c}essary when $\bar\pi_N\to0$ and $\mhat(\cdot)$ converges to the linear projection $\mu(\cdot)$. \jelena{Here, $\mu(\cdot)\neq m(\cdot)$.}The same argu\tcr{m}ent holds if one perform\tcr{s} a linear transformation on some basis function $\{\phi_j(\bX)\}_{j=1}^d$ with a fixed $d<\infty$. However, when $d$ grows with $N$ in that $d/N\to c\in(0,1)$, a least squares estimator leads to a latent misspecification
i.e., the limit $\mu(\cdot)\neq m(\cdot)$ even if $m(\cdot)$ is indeed linear on $\{\phi_j(\bX)\}_{j=1}^d$. Hence, an adjusted RAL would be more appropriate if the outcome model is linear with a growing degree of freedom; see Appendix \ref{sim:modCI} of the \hyperref[supp_mat]{Supplement} for corresponding simulation results.
\end{remark}

{\crev
\begin{remark}[Comparison with alternative PS estimators based on under-sampling of the unlabeled group]\label{remark:under-sampling}
Under the decaying PS model, a possible alternative to our offset logistic regression estimator could be the so-called ``under-sampled'' estimators, as studied by \cite{wang2020logistic}  {\crevjb in low-dimensional setting}{\crevyz s}, where the observations from the large unlabeled data are under-sampled in some way to create a more ``balanced'' setting. Since the under-sampled data is biased (as the under-sampling is done only for one group, i.e., the unlabeled group), additional bias correction or weight adjustment is needed{\crevyz; see the discussion before Section 3.1 of \cite{wang2020logistic}}. Furthermore, {\crevyz as they have also discussed in their Remarks 3 and 4,} even after such techniques are applied, the resulting PS estimator may still be less efficient than the full-data-based estimator, such as our proposed offset-based formulation.
 {\crevjb {\crevyz The h}igh-dimensional setting {\crevmag for such methods} has yet to be studied in depth{\crevyz;} however{\crevyz,} we expect to see additional bias accumulation due to regularization effects. Our results in {\crevyz high dimensions} {\crevmag for the offset logistic model} are non-trivial where non-standard rates are proposed {\crevmag and these serve as important extensions of the existing high dimensional literature;}}  see Theorem \ref{thm:high-dim} and Remark \ref{remark:nsr} below.


Additionally, the loss of efficiency due to the under-sampling may also affect the finite sample  {\crevjb performance} or even the asymptotic efficiency of the final mean estimator. When the nuisance models are all correctly specified, the efficiency loss will not affect the final estimator's asymptotic efficiency but will likely impact the estimator's finite sample properties. More importantly, when the outcome model is misspecified, the adjusted RAL expansion of the mean estimator is directly dependent on the estimation error and the IF of the PS estimator; see our results in Theorem \ref{thm:ex3} and Remark \ref{remark:RAL_mis}. Hence, a less efficient PS estimator will result in a less efficient mean estimator.
\end{remark}
}

\subsubsection{High-dimensional \tco{offset} logistic regression}\label{high-dim}
Next, we consider \tcr{a} high-dimensional setting with {\crevyz $p$ possibly depending on $N$ and} $p \to \infty$ {\crevyz as $N$ grows}.
The problem here is challenging as  together with $p\to\infty$,  the labels are extremely imbalanced in that $\bar\pi_N=\P(R=1)\to0$.
Unlike before, an adjusted RAL expansion
for the case \tcr{when} 
\tco{$m(\cdot)$} is misspecified is now not available, as we are no longer able to obtain a  parametric rate for the \tco{PS} estimation.
 In this section, we provide the consistency rate $r_{e,N}$ \tcr{in} \eqref{drthm:ratecond4} for an \jelena{offset,} sparse\jelena{,} logistic \tco{PS} model and establish asymptotic results for $\thetahat_\mathrm{DRSS}$ when both \tco{$m(\cdot)$ and $\pi_N(\cdot)$} are correctly specified.


Consider the same parametric \jelena{offset} model \eqref{offset_model}, except here we allow   $p\to\infty$ as $N\to\infty$.  \jelena{In this subsection, we assume }\tco{the parameter $\bgamma_0$ to be sparse with $s:=\|\bgamma_0\|_0$ denoting its sparsity level.}
Let $\pihat_N:=N^{-1}\sum_{i=1}^NR_i$ and for every $\bgamma\in\R^{p+1}$ and $a\in(0,1]$,  recall  $\ell_N(\bgamma;a)$  defined in \eqref{loglike}.
Let $\widehat{\bgamma}$ be a minimizer of the convex program:
\begin{equation}\label{gamma_hd}
\arg\min_{\bgamma\in\R^{p+1}}\left\{\ell_N(\bgamma;\pihat_N)+\lambda_N\|\bgamma\|_1\right\},
\end{equation}
with a sequence $\lambda_N>0$. Then, $\pi_N(\bX)$ can be estimated similarly as in \eqref{eq:pi} by
$
\pihat_N(\bX)
:=g(\bXv^T\widehat{\bgamma}+\log(\pihat_N)).
$
We establish the theoretical properties of our estimators $\widehat{\bgamma}$ and $\hat\pi_N(\cdot)$ in 3 parts: 1) establish a \tcr{restricted strong convexity} (RSC) property; 2) control the \jelena{$l$}\tcr{$_{\infty}$ norm of the} gradient \tcr{of the loss} at the true parameter, \tcr{i.e.}\tco{,} $\|\nabla_{\bgamma}\ell_N(\bgamma_0;\pihat_N)\|_\infty$; \tcr{and} 3) obtain the final probabilistic bounds on the error rates of our estimator.

\paragraph*{\jelena{\bf RSC property for the offset logistic model}} We first analyze the RSC property of our \tcr{high dimensional} \jelena{offset} logistic \tcr{model}. 
Under our imbalanced \jelena{treatment} setting, we  show that the RSC condition holds with a parameter of the order of $\bar\pi_N \to 0$ (rather than a constant bounded away from 0), once the RSC condition holds for a balanced logistic model with some constant $\kappa>0$.
For any $\bDelta,\bgamma\in\R^{p+1}$, define the following:
\begin{equation}\label{deltaL}
\delta\ell(\bDelta;a;\bgamma)~:=~\ell_N(\bgamma+\bDelta;a)-\ell_N(\bgamma;a)-\bDelta^T\nabla_{\bgamma}\ell_N(\bgamma;a).
\end{equation}
We say the restricted strong convexity (RSC) property holds for $\delta\ell(\bDelta;a;\bgamma_0)$ with parameter $\kappa$ on a given set $A$ if
\begin{equation}\label{RSC}
\delta\ell(\bDelta;a;\bgamma_0)~\geq~\kappa\|\bDelta\|_2^2,\qquad\mbox{for all}\;\;\bDelta\in A.
\end{equation}
We have the following \emph{deterministic} result.

\begin{lemma}\label{lemma:RSC}
For any $a\in(0,1]$,
$$\delta\ell(\bDelta;a;\bgamma_0)~\geq~ a\delta\ell(\bDelta;1;\bgamma_0).$$
Hence, for a given set $A$ and for any given realization of the data, if the RSC property holds for $\delta\ell(\bDelta;1;\bgamma_0)$ with parameter $\kappa$ on \jelena{a} set $A$, then the RSC property also holds for $\delta\ell(\bDelta;a;\bgamma_0)$ with parameter $a\kappa$ on $A$.
\end{lemma}
Notice that
\begin{align*}
\delta\ell(\bDelta;1;\bgamma_0)~=~&\ell_N(\bgamma_0+\bDelta;1)-\ell_N(\bgamma_0;1)-\bDelta^T\nabla_{\bgamma}\ell_N(\bgamma_0;1)\\
~=~&\ell_N^\mathrm{bal}(\bgamma_0+\bDelta)-\ell_N^\mathrm{bal}(\bgamma_0)-\bDelta^T\nabla_{\bgamma}\ell_N^\mathrm{bal}(\bgamma_0),\quad\mbox{where}\\
\ell_N^\mathrm{bal}(\bgamma)~:=~&-N^{-1}\sum_{i=1}^N[R_i^*\bXv^T\bgamma-\log\{1+\exp(\bXv^T\bgamma)\}],\;\;\forall\bgamma\in\R^{p+1},
\end{align*}
with $(R_i^*)_{i=1}^N$ being i.i.d. random variables generated from Bernoulli$(g(\bXv^T\bgamma_0))$. Here, $\ell_N^\mathrm{bal}(\bgamma)$ is the negative log-likelihood function under a \emph{balanced} logistic model with the true parameter $\bgamma_0$.
By Lemma \ref{lemma:RSC}, we relate the RSC property of our \tcr{imbalanced} model to a standard balanced logistic model.
 The RSC property for a balanced logistic model has been studied in \cite{negahban2010unified}, among others. We also present a more general version in this paper; see Lemma \ref{RSC:general}.


\paragraph*{\jelena{\bf Gradient control}} Now, we control the \tcr{$l_{\infty}$ norm of the} gradient, $\|\nabla_{\bgamma}\ell_N(\bgamma_0;\pihat_N)\|_\infty$, and the following lemma demonstrate\jelena{s} that the rate \tcr{of} $\|\nabla_{\bgamma}\ell_N(\bgamma_0;\pihat_N)\|_\infty=O_p(\{N^{-1}\bar\pi_N\log(p)\}^{1/2})$.

\begin{lemma}\label{lemma:gradient}
Let  $\bXv^T\bgamma_0$ be a sub-Gaussian  random variable and $\bXv$  a marginal sub-Gaussian random vector, in that    $\|\bXv^T\bgamma_0\|_{\psi_2}\leq\sigma_{\bgamma_0}<\infty$ and $\max_{1\leq j\leq p+1}\|\bXv(j)\|_{\psi_2}\leq\sigma<\infty$, respectively. Then, for any $t_1,t_2\geq0$ and $t_2<N\bar\pi_N/9$,
\begin{align*}
\|\nabla_{\bgamma}\ell_N(\bgamma_0;\pihat_N)\|_\infty&~\leq~ C_1(\bar\pi_N+\bar\pi_N^{1/2})\sqrt\frac{\{t_1+\log(p+1)\}}{N}+C_4\left\{\sqrt\frac{t_2\bar\pi_N}{N}+\frac{t_2}{N}\right\}\\
&\qquad+(C_2+C_3\bar\pi_N)\frac{\sqrt{\log(2N)}\{t_1+\log(p+1)\}}{N},
\end{align*}
\tco{with probability at least $1-6\exp(-t_1)-2\exp(-t_2)$}\jelena{. The}
 constants $C_1,C_2,C_3,C_4>0$ independent of $N$ \jelena{are defined through equations}  \eqref{constants1}-\eqref{constants2}  in the \hyperref[supp_mat]{Supplement}.
\end{lemma}

Define $S:=\{j\leq p+1:\bgamma_0(j)\neq0\}$, $s=|S|$ and the \emph{cone set}:
\begin{equation}
\C_\delta(S;3)~:=~\{\bDelta\in\R^{p+1}:\|\bDelta_{S^c}\|_1\leq3\|\bDelta_S\|_1,\;\;\|\bDelta\|_2=\delta\},
\end{equation}
where $\bDelta_S=\{\bDelta(j)\}_{j\in S}$ and $\bDelta_{S^c}=\{\bDelta(j)\}_{j\not\in S}$.
Define the \emph{critical tolerance}:
$$\delta_N:=\inf\left\{\delta>0:\delta\geq2\lambda_Ns^{1/2}\tilde\kappa^{-1},\;\;\mbox{RSC holds for $\ell_N(\cdot;\pihat_N)$ with parameter $\tilde\kappa$ over}\;\;\C_\delta(S;3)\right\}.$$
Then, any optimal solution $\widehat{\bgamma}=\widehat{\bgamma}_{\lambda_N}$ to the convex program \eqref{gamma_hd} satisfies $\|\widehat{\bgamma}-\bgamma_0\|_2\leq\delta_N$.

\paragraph*{\jelena{\bf Probabilistic bounds}}
\tcr{Finally, we now obtain the probabilistic bounds and convergence rates for $\widehat{\bgamma}$, and subsequently $\pihat_N(\cdot)$, in the following result.}
\begin{theorem}\label{thm:high-dim}
Assume $\log(p)\log(N)=O(N\bar\pi_N)$ and $s\log(p)=o(N\bar\pi_N)$ as $N,p\to\infty$\tco{, where $s:=\|\bgamma_0\|_0$}. Assume conditions in Lemma \ref{lemma:gradient}. Suppose the RSC property holds for $\delta\ell(\bDelta;1;\bgamma_0)$ with parameter $\kappa>0$ on the set $\overline{\C}(S;3) := \{\bDelta\in\R^{p+1}:\|\bDelta_{S^c}\|_1\leq3\|\bDelta_S\|_1,\;\;\|\bDelta\|_2\leq1\}$, with probability at least $1-\alpha_N$, where $\alpha_N=o(1)$.
Let $$M_N~:=~C_5\sqrt\frac{\bar\pi_N\log(p+1)}{N}+C_6\frac{\sqrt{\log(2N)}\log(p+1)}{N},$$
with some constants $C_5,C_6>0$. For any $\lambda_N$ satisfying $2(1+c)M_N\leq\lambda_N\leq9\kappa\bar\pi_Ns^{-1/2}$ with  $c>0$, whenever $N\bar\pi_N>9c\log(p+1)$,
$$\|\widehat{\bgamma}-\bgamma_0\|_2~\leq~\frac{1}{9}\lambda_Ns^{1/2}\bar\pi_N^{-1}\kappa^{-1},\quad\mbox{with probability at least}\;\;1-8(p+1)^{-c}-\alpha_N.$$
Further assume that $\|\bXv^T\bv\|_{\psi_2}\leq\sigma\|\bv\|_2$ for any $\bv\in\mathbb R^{p+1}$. Then, for any $r>0$, \tco{with some $\lambda_N\asymp\sqrt{\bar\pi_N\log(p)/N}$,}
\begin{align}
&\|\pi_N^{-1}(\bX)\|_{r,\P_\bX}~\asymp~\bar\pi_N^{-1}\quad\forall r>0,\quad\mbox{and hence}\;\;a_N~\asymp~\bar\pi_N,\nonumber\\
&\left\|1-\frac{\pi_N(\cdot)}{\pihat_N(\cdot)}\right\|_{r,\P_\bX}~=~O_p\left(\sqrt\frac{s\log(p)}{N\bar\pi_N}\right)\quad\forall r>0,\label{rate:pi}\\
&\E_{\bX}\left[ \frac{a_N}{\pi_N(\bX)} \left\{1 - \frac{\pi_N(\bX)}{\pihat_N(\bX)} \right\}^2\right]~=~O_p\left(\frac{s\log(p)}{N\bar\pi_N}\right)~=~o_p(1).\label{rate:pi&aN}
\end{align}
Moreover, if $\|\mhat(\cdot)-m(\cdot)\|_{2+c,\P_\bX}=o_p(1)$ with constant $c>0$, then,
$$\E_{\bX}\left[ \frac{a_N}{\pi_N(\bX)} \{\mhat(\bX) - m(\bX) \}^2 \right]~=~o_p(1).$$
\end{theorem}

\begin{remark}[\mod{Non-standard rates}]\label{remark:nsr}
The implication of Theorem \ref{thm:high-dim}  is that for
 some $\lambda_N\asymp\{N^{-1}\bar\pi_N\log(p)\}^{1/2}$,
\begin{equation}\label{rate:gamma}
\|\widehat{\bgamma}-\bgamma_0\|_2~=~O_p\left(\sqrt\frac{s\log(p)}{N\bar\pi_N}\right).
\end{equation}
As long as $\bar\pi_N\to0$,  the rate $\lambda_N\asymp\{N^{-1}\bar\pi_N\log(p)\}^{1/2}$, is \emph{faster} than the usual rate \tcr{of} $\{N^{-1}\log(p)\}^{1/2}$ used for \tcr{tuning parameter choice in} a \tcr{standard (i.e.\tco{,} balanced)} \tco{$\ell_1$}-penalized logistic regression.  This in turn implies \emph{slower} than usual rate of convergence in \eqref{rate:gamma}, as $N a_N$ is much smaller than $N$. This is also reflected in the error rates of the conditional  propensity \tcr{score}, \tcr{in} \eqref{rate:pi}. The ``effective sample size'' \tcr{here} is $N a_N$ rather than $N$\mod{, thus leading to \emph{non-standard} rates. The results above may therefore be seen as a \emph{generalization} of standard high-dimensional logistic regression models (i.e., where positivity holds) to the case of a decaying PS. To our knowledge, these rates are novel for high-dimensional settings.}
\end{remark}



\begin{remark}[\tco{Marginal \tcr{versus} ``Joint'' sub-Gaussianity}]
In \tcr{Th}eorem \ref{thm:high-dim}, we obtained a non-asymptotic upper bound for $\|\bgammahat-\bgamma_0\|_2$ that only requires a marginal sub-Gaussianity of $\bXv$, that \tcr{is} $\max_{1\leq j\leq p+1}\|\bXv(j)\|_{\psi_2}\leq\sigma<\infty$. Unfortunately, to show \eqref{rate:pi} and \eqref{rate:pi&aN}, we do require a \tcr{``joint''} 
sub-Gaussianity of $\bXv$ \tcr{in} that $\|\bXv^T\bv\|_{\psi_2}\leq\sigma\|\bv\|_2$ for any $\bv\in\mathbb R^{p+1}$. In high-dimensions, the \tcr{joint} 
 sub-Gaussianity is stronger than the marginal sub-Gaussianity \tcr{in} that \tcr{the latter} 
 enfor\tcr{c}es a weak\tcr{er} dependency \tcr{among} 
 the covariates; see Section 4 of \cite{kuchibhotla2018moving} for more details.
\end{remark}
\tcr{Note that} in \tcr{Th}eorem \ref{thm:high-dim}, we \tcr{\emph{only}} assume the RSC property for a classical balanced logistic regression model, which is standard in \tcr{the} high-dimensional regression (and classification) literature. As shown in Proposition 2 of \cite{negahban2010unified}, with probability at least $1-2\exp(-c_1N)$,
\begin{align}
\delta\ell(\bDelta;1;\bgamma_0)&~=~\ell_N^\mathrm{bal}(\bgamma_0+\bDelta)-\ell_N^\mathrm{bal}(\bgamma_0)-\bDelta^T\nabla_{\bgamma}\ell_N^\mathrm{bal}(\bgamma_0)\nonumber\\
&~\geq~ c_2\|\bDelta\|_2\left\{\|\Delta\|_2-c_3\sqrt\frac{\log(p+1)}{N}\|\Delta\|_1\right\}, \quad\forall\|\Delta\|_2\leq1,\label{RSC:finite}
\end{align}
with some constants $c_1,c_2,c_3>0$, and hence, the RSC property holds for $\delta\ell(\bDelta;1;\bgamma_0)$ with some $\kappa>0$ on the set $\overline{\C}(S;3)$. The conditions required in \cite{negahban2010unified} \tcr{essentially amount to:} 
$s\log(p)=o(N)$, the intercept term $\bgamma_0(1)=0$, $\bX$ is a jointly sub-Gaussian with mean zero, and $\lambda_{\min}\{\mathrm{Cov}(\bX)\}\geq c>0$. Similar conditions are also required in Example 9.17 and Theorem 9.36 \tcr{of} \cite{wainwright2019high}. In the following Lemma \ref{RSC:general}, we propose a 
\tcr{user-}friendly version of RSC condition results for a balanced logistic regression problem that only \tcr{require} a marginal sub-Gaussianity \tcr{of $\bX$} and an additional $(2+c)$-th moment condition $\sup_{\|\bv\|_2\leq1}\|\bXv^T\bv\|_{2+c,\P_\bX}\leq M<\infty$. 
\tco{In addition}, we do not enforce a mean zero $\bX$, \tcr{and we do not require} 
a zero intercept term \tcr{in the logistic model either}.
\begin{lemma}\label{RSC:general}
Assume the smallest eigenvalue $\lambda_{\min}\{\E(\bXv\bXv^T)\}\geq\kappa_l>0$, a $(2+c)$-th moment condition $\sup_{\|\bv\|_2\leq1}\|\bXv^T\bv\|_{2+c,\P_\bX}\leq M<\infty$, a $c$-th moment condition $\|\bXv^T\bgamma_0\|_{c,\P_\bX}\leq\mu_c<\infty$ and the marginal sub-Gaussianity $\sup_{1\leq j\leq p+1}\|\bXv(j)\|_{\psi_2}\leq\sigma<\infty$. Then, with probability at least $1-2\exp(-c_1N)$, \eqref{RSC:finite} holds, with constants $c_1,c_2,c_3>0$.  If
  we further assume that $s\log(p)=o(N)$, then, for large enough $N$, there exists a constant $\kappa>0$ such that, with probability at least $1-2\exp(-c_1N)$,
\begin{equation}\label{RSC:asymp}
\delta\ell(\bDelta;1;\bgamma_0)~\geq~\kappa\|\Delta\|_2^2,\quad\forall\Delta\in\overline{\C}(S;3).
\end{equation}
\end{lemma}

Although Lemma \ref{RSC:general} is based on the logistic loss function, it in fact applies to any loss function $\ell_N^\mathrm{bal}(\cdot)$ based on the maximum likelihood of a balanced generalized linear model.


\subsection{Stratified labeling}\label{stratified}

We consider \tcr{here} a stratified labeling mechanism. Here, the labeling \tcr{indicator} $R$  depends on $\bX$, but does so only through an intermediate stratification in $\bX$. \tcr{Such mechanisms are often of practical relevance in biomedical studies when \emph{prior} information is available on stratification through another observed variable.} Specifically, let $\delta\in\{0,1\}$ denote an observed random stratum indicator  and  assume that $R \ind \bX | \delta$.
Note that nothing changes \jelena{ if we were to move from binary to} finitely many strata, and \tcr{while we stick to a binary $\delta$ here for simplicity}, our work can be easily extended to a multiple-stratum situation. Let $\pi_{j,N} := \P(R = 1 | \delta = j, \bX) \equiv \P(R = 1 | \delta = j)$ for each $j = 0,1$. We assume $\delta$ is a ``well behaved'' indicator whose distribution is independent \tcr{of} 
$N$ and \jelena{itself} satisfies \jelena{ the overlap condition}\tco{
$$c~<~p_{\delta}(\bx) ~:=~ \P(\delta = 1 | \bx) ~<~1-c,\quad\text{for all}\;\;\bx\in\mathcal X,$$
with a constant $0<c<1/2$ independent of $N$.}
Then, we have:
$$
\pi_N(\bX)~=~ \pi_{1,N} p_{\delta}(\bX) + \pi_{0,N} \{1-p_{\delta}(\bX)\}.
$$
As long as $\delta$ is observed, then $\pi_{j,N}$ for each $j$ can be estimated very easily and at a rate $O_p((N\bar\pi_N)^{-1/2})$. Moreover, when $\bar\pi_N\to0$ as $N\to\infty$, \tco{$p_{\delta}(\bX)$ can be estimated at \jelena{a parametric $N^{-1/2}$ rate} if the model is parametric, or a\jelena{t a} rate slower than $N^{-1/2}$ but still as a function of $N$ (rather than $N\bar\pi_N$) if a non-parametric estimator is performed}. Therefore, we will continue to have a fast enough rate for $\pihat_N(\cdot)$ under this setting, so that the error term \tcr{$\Deltahat_N$ in} \eqref{Deltahat} \tcr{can} potentially ha\tcr{ve} a rate:
$$
\Deltahat_N~=~O_p(r_{e,N})~=~O_p\left((N\bar\pi_N)^{-1/2}\right).
$$
In this section, we propose a \tco{PS} estimator based on the stratified labeling \tcr{model above}, \tcr{and provide a full characterization of its properties as well as}
a RAL expansion \tcr{for the error} $\Deltahat_N$.

With a slight abuse of notation, we define $\bZ_i=(R_i,R_iY_i,\delta_i,\bX_i)$ in this section, 
{\crev and let $\S=(\bZ_i)_{i=1}^N$, $\S_{-k}=\{\bZ_i:i\in\mathcal I\setminus\mathcal I_k\}$}. Suppose $\widehat p_\delta(\cdot)$ is an estimator of $p_\delta(\cdot)$, and let $\widehat p_\delta(\bX_i):=\widehat p_\delta(\bX_i;\S_{-k(i)})$ be a \tcr{corresponding} cross-fitted \tcr{version of this estimator.} 

 Define $\pihat_1^{-k}=:\sum_{i\not\in I_k}\delta_iR_i/\sum_{i\not\in I_k}\delta_i$ and $\pihat_0^{-k}:=\sum_{i\not\in I_k}(1-\delta_i)R_i/\sum_{i\not\in I_k}(1-\delta_i)$ to be the cross-fitted estimators of $\pi_{1,N}$ and $\pi_{0,N}$, respectively. The \tcr{PS $\pi_N(\cdot)$} 
 is then estimated by:
$$\pihat_N(\bX_i)~:=~\pihat_1^{-k(i)}\widehat p_\delta(\bX_i)+\pihat_0^{-k(i)}\{1-\widehat p_\delta(\bX_i)\}.$$
\begin{theorem}\label{thm:ex2}
Assume $\bar\pi_N\to0$ and $N\bar\pi_N\to\infty$ as $N\to\infty$.
Suppose
\begin{equation}\label{rdeltaN}
\|\widehat p_\delta(\cdot)-p_\delta(\cdot)\|_{2,\mathbb P_\bX}=O_p(r_{\delta,N}),\qquad\mbox{for some sequence}\;\;r_{\delta,N}=o(1).
\end{equation}
Then, for each $k\leq\K$,
$$
\E_\bX\left\{1-\frac{\pi_N(\bX)}{\pihat_N(\bX)}\right\}^2~=~O_p\left(r_{\delta,N}^2+N\bar\pi_N\right).
$$
Besides, assume $\|\mu(\cdot)-m(\cdot)\|_{\infty,\mathbb P_\bX}<\infty$ and \eqref{drthm:ratecond3}. Then, the overall RAL ex\tcr{p}ansion of our \tcr{DRSS} estimator $\thetahat_\mathrm{DRSS}$ under \tcr{a} stratified labeling \tcr{model as above} is:
$$
\thetahat_\mathrm{DRSS}-\theta_0~=~N^{-1}\sum_{i=1}^N \Psi(\bZ_i) + O_p\left((N\bar\pi_N)^{-1}+N^{-1/2}+r_{\mu,N}(N\bar\pi_N)^{-1/2}+r_{\delta,N}\right),
$$
where $\Psi(\bZ) := \psi_{\mu, \pi}(\bZ) + \mathrm{IF}_{\pi}(\bZ)$ and $\E\{\Psi(\bZ)\} = 0$ with $\psi_{\mu, \pi}(\bZ)$ as defined in \eqref{psi1}  and
\begin{align*}
 \mathrm{IF}_\pi(\bZ)&~:=~\left\{\frac{\delta R}{p_\delta}-\pi_{1,N}\right\}\E_\bX\left[\frac{p_\delta(\bX)}{\pi_N(\bX)}\{\mu(\bX)-m(\bX)\}\right]\\
&\qquad+\left\{\frac{(1-\delta)R}{1-p_\delta}-\pi_{0,N}\right\}\E_\bX\left[\frac{1-p_\delta(\bX)}{\pi_N(\bX)}\{\mu(\bX)-m(\bX)\}\right],
\end{align*}
where $p_\delta=\E\{p_\delta(\bX)\}=\E(\delta)$. If we further assume   $r_{\delta,N}=o((N\bar\pi_N)^{-1/2})$, then
\begin{equation}\label{RAL:stratified}
\thetahat_\mathrm{DRSS}-\theta_0~=~N^{-1}\sum_{i=1}^N \Psi(\bZ_i)+o_p\left((N\bar\pi_N)^{-1/2}\right).
\end{equation}
\end{theorem}
Note that Theorem \ref{thm:ex2} still holds if $\pi_1$ and $\pi_0$ are estimated without cross-fitting in that $\pihat_1:=\sum_{i=1}^N\delta_iR_i/\sum_{i=1}^N\delta_i$ and $\pihat_0:=\sum_{i=1}^N(1-\delta_i)R_i/\sum_{i=1}^N(1-\delta_i)$.

\begin{example}\label{ex:2}
Here we illustrate a simple logistic model for $p_\delta(\cdot)$ and investigate the conditions we need for $r_{\delta,N}$ \tcr{to be} $o_p((N\bar\pi_N)^{-1/2})$, so that the RAL expansion \eqref{RAL:stratified} holds. For a fixed dimensional $\bX$, let $\widehat p_\delta(\cdot)$ be the MLE of the logistic model. \tcr{T}hen $r_{\delta,N}=O(N^{-1/2})=o((N\bar\pi_N)^{-1/2})$ as long as $\bar\pi_N\to0$. 
\tcr{A}s for a high-dimensional $\bX$, \tcr{consider a} sparse \tcr{logistic} model  \tcr{for $p_\delta(\cdot)$, and} let $\widehat p_\delta(\cdot)$ be the logistic estimator 
\tcr{based on} a Lasso penalty. \tcr{T}hen, $r_{\delta,N}=O((s_\delta\log(p)/N)^{1/2})$, where $s_\delta$ is the sparsity level of the logistic \tcr{model}\jelena{s} parameter. Hence, $r_{\delta,N}=o_p((N\bar\pi_N)^{-1/2})$ if $s_\delta\bar\pi_N\log(p)=o(1)$ as $N\to\infty$.
\end{example}

\begin{remark}[Comparisons with other works]
\tcr{We note that s}imilar, \tcr{yet} different, problems are studied in \cite{gronsbell2020efficient} and \cite{hong2020inference}. They both work on decaying stratified labeling propensity s\tcr{co}re models, but with different types of stratified labeling m\tcr{e}chanisms and parameters of interest \tcr{compared to} our setting. \cite{hong2020inference} assume a deterministic $\delta$ given $\bX$. Essentially, they require $Y\independent R|\delta$, \tcr{so} that $\delta$ can be seen as a univariate confounder with a finite support. On the other hand, both \cite{gronsbell2020efficient} and our work  allow additional randomness \tcr{in} 
$\delta$. Besides, \cite{hong2020inference} work on a ``finite-population'' ATE estimation problem, where the treatment assignment is the only source of randomness. \cite{gronsbell2020efficient} focus on the estimation of the regression parameters and prediction performance \tcr{measures}, for low-dimensional covariates and a binary outcome problem; \tcr{and} we are mainly working on the estimation \tcr{of} the mean response \tcr{while} allowing \tcr{for} high-di\tcr{men}sional covar\tcr{ia}tes and real valued outcomes.
\end{remark}

\section{Average treatment effect estimation with imbalanced treatment groups}\label{ATE}
One important application of our proposed method \jelena{of} Section \ref{sec:EY} is the \jelena{\tcr{popular} causal inference problem of 
ATE estimation and hypothesis testing. Our method is particularly suited when extremely {\it imbalanced} treatment groups occur. \tcr{The c}ausal inference literature \tcr{typically} access\tcr{es} ATE inference by imposing an overlap condition by which $\P(c<\E(R|\bX)<1-c)=1$ for some constant $c>0$. Here, $R\in\{0,1\}$ is a \tcr{binary}  treatment indicator.  In contrast, we show that our method identifies and performs inference about \tcr{the} ATE without requiring an overlap condition.}  We extend our results for the \tcr{MAR-SS} 
setting \jelena{of} \tcr{Section \ref{sec:EY}} to a causal inference setting while allowing a decaying \tco{PS} in that $\bar\pi_N:=\E(R)\to0$ (or \tcr{alternatively,} 
$\bar\pi_N\to1$) as $N\to\infty$. To the best of our knowledge, no previous work has addressed such  an extremely imbalanced treatment groups setting \tcr{in the context of ATE estimation}.


We formulate the problem setup first. Suppose we have i.i.d. samples $\S:=(R_i,Y_i,\bX_i)_{i=1}^N$ with $(R,Y,\bX)$ being an independent copy of $(R_i,Y_i,\bX_i)$. Here, $R=R_N\in\{0,1\}$ is a treatment indicator that, similarly as in Section \ref{sec:EY}, is allowed to   depend  on $N$, i.e., $R=R_N$. As before, \jelena{$\bX\in\R^p$ denotes the covariate vector  while $Y=\mathrm Y(R)$ now denotes the observed potential outcome.}  Here,   $\mathrm Y(1)$ denotes the potential outcome if the individual \jelena{have been} treated and $\mathrm Y(0)$ denotes the potential outcome if the individual \jelena{haven't been} treated. For each individual, only one of the potential outcomes $\mathrm Y(R)$ is observable. \jelena{Consistency of the potential outcomes is assumed throughout:} $Y=\mathrm Y(R)=R\mathrm Y(1)+(1-R)\mathrm Y(0)$; see \cite{rubin1974estimating} and \cite{imbens2015causal}.

Now we define the parameter of interest, $\theta_\mathrm{ATE}:=\theta^1-\theta^0$ to be the \tco{ATE}  of $R$ on $Y$, where with a slight abuse of notation \jelena{we denote with} $\theta^1:=\E\{\mathrm Y(1)\}$ and $\theta^0:=\E\{\mathrm Y(0)\}$. \jelena{Moving forward we assume} the \tcr{usual \emph{unconfoundedness condition} \citep{imbens2004nonparametric,tsiatis2007semiparametric}:}  
$${\cred \{\mathrm Y(0),\mathrm Y(1)\}}~\independent~ R~|~\bX.$$
 Then, $\theta^1=\E\{m_1(\bX)\}$ and $\theta^0=\E\{m_0(\bX)\}$, where $m_r(\bX):=\E(Y|R=r,\bX) {\cred \equiv \E\{Y(r)|\bX\}}$ \jelena{denotes the conditional outcome model, and $r \in \{0,1\}$}.  With extremely imbalanced groups, without loss of generality, we assume $\bar\pi_N=\P(R=1)\to0$, i.e., most of the individuals are likely to be in the control group.

The estimation of $\theta^1$ is the same as the mean estimation \tcr{problem} in the MAR\tcr{-SS} setting, if we set $\bZ_i$\jelena{s to be }$(R_i,R_iY_i,\bX_i)$. Similarly,  $\theta^0$ can be \jelena{identified} as a mean \jelena{with} $\bZ_i$\jelena{s being} $(1-R_i,(1-R_i)Y_i,\bX_i)$. Now, as in \eqref{theta:unknownpi}, $\theta^1$ and $\theta^0$ can be estimated by:
\begin{align}
&\thetahat^1~:=~ N^{-1}\sum_{i=1}^N\left[\mhat_1(\bX_i)+\frac{R_i}{\pihat_N(\bX_i)}\{Y_i-\mhat_1(\bX_i)\}\right],\nonumber\\
&\thetahat^0~:=~ N^{-1}\sum_{i=1}^N\left[\mhat_0(\bX_i)+ \frac{1-R_i}{1-\pihat_N(\bX_i)}\{Y_i-\mhat_0(\bX_i)\}\right],\label{theta0}
\end{align}
\tco{where, for each $k\leq\K$ and $i\in\S_k$, $\mhat_1(\bX_i)=\mhat_1(\bX_i;\S_{-k})$, $\mhat_0(\bX_i)=\mhat_0(\bX_i;\S_{-k})$, and $\pihat_N(\bX_i)=\pihat_N(\bX_i;\S_{-k})$ are cross-fitted estimators  of $m_1(\bX_i)$, $m_0(\bX_i)$, and $\pi_N(\bX_i)$, respectively. Here, $\S_{-k}=\{\bZ_i:i\in\mathcal I\setminus\mathcal I_k\}$ is defined analogously as discussed below \eqref{theta:knownpi} in Section \ref{sec:knownpi}, and for $r\in\{0,1\}$, $\mhat_r(\cdot)$ is constructed based on $\{\bZ_i:i\in\S_{-k},R_i=r\}$.} \jelena{Hence,} $\theta_\mathrm{ATE}$ can be estimated by \tcr{the \emph{DRSS ATE estimator}}:
\begin{equation}\label{ATEhat}
\thetahat_\mathrm{ATE}~:=~\thetahat^1-\thetahat^0.
\end{equation}
The asymptotic properties of $\thetahat^1$ follow directly from Theorem \ref{t4}. The following theorem provides \tcr{the} asymptotic results \tcr{for}
$\thetahat^0$. \tco{For the sake of a better interpretability, in the following theorem, we suppose $c\bar\pi_N<\pi_N(\bX)$ with some constant $c>0$ and hence we have $a_N\asymp\bar\pi_N$.}
\begin{theorem}\label{t6}
{\crev Let Assumption \ref{cond:ignorability} hold and Assumption \ref{a1} hold for $m_0(\cdot)$ and $\mu_0(\cdot)$, let} $N\to\infty$, $\bar\pi_N\to0$ and $N\bar\pi_N\to\infty$. \tco{Suppose for all $\bx\in\mathcal X$, $c\bar\pi_N<\pi_N(\bx), e_N(\bx)<C\bar\pi_N$,}
for some $c,C>0$.
Let \tco{$\varepsilon:=Y-Rm_1(\bX)-(1-R)m_0(\bX)$,}
assume $\|\varepsilon\|_{2,\P}<\infty$,
\tco{$\|m_0(\cdot)-\mu_0(\cdot)\|_{2,\P_\bX}<C<\infty$,}
$\Var\{m_0(\bX)\}<\infty$ as well as
\begin{align} \label{4.4}
&\|\mhat_0(\cdot)-\mu_0(\cdot)\|_{2,\P_X}~=~O_p(r_{\mu,0,N}),\quad\sup_{\bx\in\mathcal X}\left|\frac{\pihat_N(\bx)- e_N(\bx)}{\bar\pi_N}\right|~=~O_p(r_{e,N}),
\end{align}
for  \jelena{a sequence of positive numbers} $r_{\mu,0,N}=o(1)$ and $r_{e,N}=o(1)$.
Then,
\begin{align*}
 \thetahat^0-\theta^0&~=~N^{-1}\sum_{i=1}^N\psi_0(\bZ_i)+O_p(N^{-1/2}\bar\pi_N^{1/2}r_{\mu,0,N}+N^{-1/2}\bar\pi_Nr_{e,N}+\bar\pi_Nr_{e,N}r_{\mu,0,N})\\
 &  \qquad +\mathbbm1\{m_0(\cdot)\neq\mu_0(\cdot)\}O_p(\bar\pi_Nr_{e,N})+\mathbbm1\{ e_N(\cdot)\neq\pi_N(\cdot)\}O_p(\bar\pi_Nr_{\mu,0,N}),
\end{align*}
where
\begin{align}
 \psi_0(\bZ)~:=~& \mu_0(\bX) - \theta^0 + \frac{1-R}{1- e_N(\bX)} \{ Y - \mu_0(\bX)\}\\
~=~&\frac{ e_N(\bX)-R}{1- e_N(\bX)}\{m_0(\bX)-\mu_0(\bX)\}+m_0(\bX)-\theta^0+\frac{\varepsilon(1-R)}{1- e_N(\bX)}, \nonumber
\end{align}
with $\E\{\psi_0(\bZ)\}=\mathbbm1\{ e_N(\cdot)\neq\pi_N(\cdot),\;\mu_{{\crev 0}}(\cdot)\neq m_{{\crev 0}}(\cdot)\}O_p(\bar\pi_N)$ and $\Var\{\psi_0(\bZ)\}=O_p(N^{-1})$. Hence,
$$
\thetahat^0-\theta^0~=~N^{-1}\sum_{i=1}^N\psi_0(\bZ_i)+o_p(N^{-1/2}),\quad\mbox{with }\;\;\E\{\psi_0(\bZ)\}~=~0,
$$
once $\bar\pi_Nr_{e,N}r_{\mu,0,N}=o(N^{-1/2})$, $\bar\pi_Nr_{e,N}=o(N^{-1/2})$ if \tco{$m_{{\crev 0}}(\cdot)$} is misspecified, $\bar\pi_Nr_{\mu,0,N}=o(N^{-1/2})$ if \tco{$\pi_N(\cdot)$} is misspecified
and at least one of \tco{$m_{{\crev 0}}(\cdot)$ and $\pi_N(\cdot)$} is correctly specified. If both \tco{$m_{{\crev 0}}(\cdot)$ and $\pi_N(\cdot)$} are misspecified, then we have \tco{$\thetahat^0-\theta^0=O_p(\bar\pi_N+N^{-1/2})$}.
\end{theorem}

\begin{remark}[\tco{Comparison with \tcr{the naive}  estimator}]\label{label:comp-naive}
Now we consider the comparison of the doubly robust estimator $\thetahat^0$ with the empirical average of the response over the control group $\bar Y_0:=\sum_{i=1}^N(1-R_i)Y_i/\sum_{i=1}^N(1-R_i)$. The empirical average $\bar Y_0$ can be seen as a \tcr{sp}ecial case of \tcr{the} estimator \tco{\eqref{theta0}} 
(without cross-fitting) in that $\pihat_N(\bX)=N^{-1}\sum_{i=1}^N(1-R_i)$ and $\mhat_0(\bX)=0$. Notice that when $\bar\pi_N\to0$,
\begin{align*}
&\bar Y_0~=~\E\{m_0(\bX)|R=0\}+O_p(N^{-1/2}),\quad\mbox{with}\\
&\theta^0-\E\{m_0(\bX)|R=0\}~=~[\E\{m_0(\bX)|R=1\}-\E\{m_0(\bX)|R=0\}]\bar\pi_N~=~O(\bar\pi_N),
\end{align*}
and hence \tco{$\bar Y_0-\theta^0=O_p(\bar\pi_N+N^{-1/2})$},  
which coincides with the case that both \tco{$m(\cdot)$ and $\pi_N(\cdot)$} are misspecified in Theorem \ref{t6}.
\end{remark}


\begin{corollary}\label{c4.2}
Let  \tcr{the a}ssumptions  of  Theorem \ref{t6} hold
{\crev and the assumptions of Theorems \ref{t4} (b) hold for $m_1(\cdot)$, $\mu_1(\cdot)$, and $\mhat_1(\cdot)$}. Assume that at least one of $ e_N(\cdot)=\pi_N(\cdot)$ and $\mu_1(\cdot)=m_1(\cdot)$ holds, and let
$$\|\mhat_1(\bX;\S_{-k})-\mu_1(\cdot)\|_{2,\P_\bX}~=~O_p(r_{\mu,1,N}),\quad\mbox{with }r_{\mu,1,N}~=~o(1).$$
Then,
\begin{align*}
\thetahat_\mathrm{ATE} - \theta_\mathrm{ATE}&~=~N^{-1}\sum_{i=1}^N\psi_1(\bZ_i)+ \Deltahat_N  + O_p(r_{e,N}r_{\mu,1,N}+\bar\pi_Nr_{e,N}r_{\mu,0,N}) \\
&\qquad+\mathbbm1\{m_0(\cdot)\neq\mu_0(\cdot)\}O_p(\bar\pi_Nr_{e,N})+\mathbbm1\{ e_N(\cdot)\neq\pi_N(\cdot)\}O_p(\bar\pi_Nr_{\mu,0,N})\\
&\qquad+\mathbbm1\{ e_N(\cdot)\neq\pi_N(\cdot),\;\mu_{{\crev 0}}(\cdot)\neq m_{{\crev 0}}(\cdot)\}O_p(\bar\pi_N)+o_p\left((N\bar\pi_N)^{-1/2}\right),
\end{align*}
where
$$
\psi_1(\bZ)~:=~ \mu_1(\bX) - \theta^1 + \frac{R}{ e_N(\bX)} \{ Y - \mu_1(\bX)\},
$$
with $\E\{\psi_1(\bZ)\}=0$, $\E\{\psi_1^2(\bZ)\}\asymp\bar\pi_N^{-1}$, and
\begin{align*}
&\Deltahat_N :=  \frac{1}{N} \sum_{i=1}^N \left\{\frac{R_i}{\pi_N(\bX_i)} - \frac{R_i}{\pihat_N(\bX_i)}\right\}\{\mu_1(\bX_i) - m_1(\bX_i)\}=O_p(r_{e,N}) \;\; \mbox{if} \;\;  e_N(\cdot) = \pi_N(\cdot), \\
&\Deltahat_N := \frac{1}{N} \sum_{i=1}^N  \left\{\frac{R_i}{\pi_N(\bX_i)} - \frac{R_i}{ e_N(\bX_i)}\right\}\{\mhat_1(\bX_i) - m_1(\bX_i)\}=O_p(r_{\mu,1,N}) \;\; \mbox{if} \;\; \mu_1(\cdot) = m_1(\cdot).
\end{align*}
Moreover,  if $r_{e,N}r_{\mu,1,N}=o_p((N\bar\pi_N)^{-1/2})$,
\begin{align*}
\thetahat_\mathrm{ATE} - \theta_\mathrm{ATE}~=~N^{-1}\sum_{i=1}^N\psi_1(\bZ_i)+o_p\left((N\bar\pi_N)^{-1/2}\right) + \Deltahat_N,
\end{align*}
as long as one of the following holds:
{\crevyz
\begin{enumerate}
\item[(a)] both $m_{{\crev 0}}(\cdot)$ and $\pi_N(\cdot)$ are correctly specified, and $r_{e,N}r_{\mu,0,N}=o(N^{-1/2}\bar\pi_N^{-3/2})$; or
\item[(b)] $\pi_N(\cdot)$ is correctly specified, $m_{{\crev 0}}(\cdot)$ is misspecified, and $r_{e,N}=o(N^{-1/2}\bar\pi_N^{-3/2})$; or
\item[(c)] $m_{{\crev 0}}(\cdot)$ is correctly specified, $\pi_N(\cdot)$ is misspecified, and $r_{\mu,0,N}=o(N^{-1/2}\bar\pi_N^{-3/2})$; or
\item[(d)] both $m_{{\crev 0}}(\cdot)$ and $\pi_N(\cdot)$ are misspecified, and $N\bar\pi_N^3=o(1)$.
\end{enumerate}}
\end{corollary}

{\crev When $\bar\pi_N\to0$, the control group has a much larger sample size compared with the treatment group. As a result, the estimation of $\theta^0$ is relatively ``simpler'' than $\theta^1$. As shown in Theorem \ref{t4}, the consistency of $\thetahat^1$ requires at least one nuisance model to be correctly specified. On the other hand, $\thetahat^0$ is consistent even when both nuisance models are misspecified; see Theorem \ref{t6}. In fact, since most of the samples are in the control group and we can observe most of the potential outcomes $Y_i(0)$, the missing data problem is trivial here, and a naive sample mean estimator over the control group is also consistent as discussed in Remark \ref{label:comp-naive}. Although obtaining a consistent estimator for $\theta^0$ is a trivial problem, the consistency rate still depends on the correctness of the nuisance models. As a result, the RAL expansion of the whole ATE parameter still requires a number of conditions corresponding to the control arm's estimation; see Corollary \ref{c4.2}. The smaller the probability $\bar\pi_N$ is, the weaker conditions we need for the model correctness and estimation errors of the control arm. When $N\bar\pi_N^3=o(1)$, we can get the RAL expansion for the ATE parameter even when $m_{{\crev 0}}(\cdot)$ and $\pi_N(\cdot)$ are both misspecified; however, when $\bar\pi_N$ is not small enough, we still need either $m_{{\crev 0}}(\cdot)$ or $\pi_N(\cdot)$ to be correctly specified to get the RAL expansion of the whole ATE parameter.}

\section{Numerical studies}\label{numerical}

\subsection{Simulation studies}\label{sim}

We illustrate the performance of our \tco{DRSS} estimators through extensive simulations under various data generating processes (DGPs). We first provide our main simulation results in Section \ref{sim:main}, where the double robustness (in the sense of consistency or inference) \tcr{shows up} in different misspecification settings. Then, in Section \ref{sim:stratified}, we show the simulation results under a special stratified labeling \tco{PS} model that \tcr{was} discussed in Section \ref{stratified}. We further focus on sparse linear models in high dimensions, and provide  results under different sparsity levels in Section \ref{sim:hd}. 

\subsubsection{Main \tcr{simulation} results}\label{sim:main}

We consider the following choices of parameters $p$, $N$ and $\bar\pi_N$:
$$p~\in~\{10,500\},\quad(N,\bar\pi_N)~\in~\{(10000,0.01),\; (50000,0.01),\; (10000,0.1)\}.$$
We generate i.i.d. Gaussian covariates  $\bX_i\sim^\mathrm{iid}\mathcal N_p(\boldsymbol{0},\boldsymbol{I}_p)$ and  residuals $\varepsilon_i\sim^\mathrm{iid}\mathcal N(0,1)$. \tcr{Given $\bX_i$, we generate} $R_i|\bX_i\sim\mathrm{Bernoulli}(\pi_N(\bX_i))$, with the following \tco{PS} models:
\begin{enumerate}
\item[P1.] (Constant \tco{PS}) $\pi_N(\cdot)\equiv\bar\pi_N$.
\item[P2.] (\jelena{Offset} logistic \tco{PS}) $\pi_N(\bx)=g(\bxv^T\bgamma_0+\log(\bar\pi_N))$, where $g(\cdot)$ is defi\tcr{ne}d in \eqref{divmod-def}.
\end{enumerate}
We consider the following outcome models \tcr{for $Y_i$ given $\bX_i$}:
\begin{enumerate}
\item[O1.] (Linear outcome) $Y_i=\bXv_i^T\bbeta_0+\varepsilon_i$.
\item[O2.] (Quadratic outcome) $Y_i=\bXv_i^T\bbeta_0+\sum_{j=1}^{p+1}\balpha_0(j)\bXv_i(j)^2+\varepsilon_i$.
\end{enumerate}
The parameter values are \tcr{chosen as:}
\begin{align*}
&\bbeta_0~=~(-0.5,1,1,1,\boldsymbol{0}_{1\times(p-3)})^T,\quad\bgamma_0~=~(\bgamma_0(1),1,\boldsymbol{0}_{1\times(p-1)})^T,\quad\balpha_0~=~(0,1,1,1,\boldsymbol{0}_{1\times(p-3)})^T,
\end{align*}
where $\bgamma_0(1)$ is chosen so that $\E(R)=\bar\pi_N$ for each $\bar\pi_N$. The following DGPs are considered:
Setting a: P1+O1, Setting b: P1+O2, Setting c: P2+O1, and Setting d: P2+O2.
For each DGP, we compare the performance of the following estimators:
(1) A naive mean estimator over the labeled samples $\bar Y_\mathrm{labeled}:=\sum_{i=1}^NR_iY_i/\sum_{i=1}^NR_i$;
(2) An oracle case of the mean estimator $\thetahat_\mathrm{DRSS}$ in \eqref{theta:unknownpi} 
\tcr{with} $\pi_N(\cdot)$ and $m(\cdot)$ treated as known;
(3)  The proposed \tcr{DRSS} mean estimator $\thetahat_\mathrm{DRSS}$ in \eqref{theta:unknownpi}, with $\K=5$.


\definecolor{newblue}{RGB}{46, 77, 167}

\begin{table}[h!]
\centering
\caption{\underline{Simulation setting a} with $p=10. $ Bias: emp\tcr{i}rical bias; RMSE: root mean square error; Length: average length of the $95\%$ confidence intervals; Coverage: average coverage of the $95\%$ confidence intervals; ESD: emp\tcr{i}rical standard deviation; ASD: average of estimated standard deviations. The {\color{newblue}\bf blue} color in the tables denotes the ``smallest'' and correctly specified parametric model for each of the settings.} \label{table:a_p10}
\vspace{0.5em}
\resizebox{!}{2.8in}{%
\begin{tabular}{llccccccc}
\hline
\toprule
$\pihat_N(\cdot)$&$\mhat(\cdot)$&Bias&RMSE&Length&Coverage&ESD&ASD\\
\hline
\multicolumn{2}{c}{} & \multicolumn{6}{c}{ \cellcolor{gray!50} $ N=10000,\bar\pi_N=0.01~(N\bar\pi_N=100)$}\\
\cline{3-8}
\multicolumn{2}{c}{$\bar Y_\mathrm{labeled}$}&0.013&0.204&0.789&0.932&0.203&0.201\\
\multicolumn{2}{c}{oracle}&0.003&0.106&0.397&0.942&0.106&0.101\\
\cdashline{3-8}
\multirow{4}{*}{{\color{newblue!90} \bf constant}}&\color{newblue} \bf LS&\color{newblue} 0.003&\color{newblue} 0.115&\color{newblue} 0.436&\color{newblue} 0.938&\color{newblue} 0.115&\color{newblue} 0.111\\
&poly&0.002&0.127&0.482&0.934&0.127&0.123\\
&RF&0.008&0.155&0.604&0.926&0.155&0.154\\
&RKHS&0.009&0.145&0.547&0.936&0.145&0.139\\
\cdashline{3-8}
\multirow{4}{*}{logistic}&LS&0.005&0.127&0.534&0.972&0.127&0.136\\
&poly&0.003&0.144&0.600&0.972&0.144&0.153\\
&RF&0.003&0.152&0.762&0.990&0.152&0.194\\
&RKHS&0.007&0.161&0.698&0.976&0.161&0.178\\
\hline
\multicolumn{2}{c}{} &\multicolumn{6}{c}{  \cellcolor{gray!50}  $ N=50000,\bar\pi_N=0.01~(N\bar\pi_N=500)$}\\
\cline{3-8}
\multicolumn{2}{c}{$\bar Y_\mathrm{labeled}$}&0.008&0.092&0.352&0.950&0.092&0.090\\
\multicolumn{2}{c}{oracle}&0.002&0.045&0.179&0.948&0.045&0.046\\
\cdashline{3-8}
\multirow{4}{*}{{\color{newblue}\bf constant}}&{\color{newblue} \bf LS}&\color{newblue} 0.003&\color{newblue} 0.045&\color{newblue} 0.182&\color{newblue} 0.956&\color{newblue} 0.045&\color{newblue}  0.046\\
&poly&0.003&0.046&0.185&0.952&0.046&0.047\\
&RF&0.004&0.056&0.218&0.942&0.056&0.056\\
&RKHS&0.003&0.056&0.213&0.942&0.056&0.054\\
\cdashline{3-8}
\multirow{4}{*}{logistic}&LS&0.003&0.046&0.189&0.960&0.046&0.048\\
&poly&0.003&0.047&0.192&0.962&0.046&0.049\\
&RF&0.002&0.052&0.227&0.974&0.052&0.058\\
&RKHS&0.001&0.054&0.223&0.960&0.054&0.057\\
\hline
\multicolumn{2}{c}{} & \multicolumn{6}{c}{  \cellcolor{gray!50}  $ N=10000,\bar\pi_N=0.1 ~(N\bar\pi_N=1000)$}\\
\cline{3-8}
\multicolumn{2}{c}{$\bar Y_\mathrm{labeled}$}&0.002&0.066&0.260&0.958&0.066&0.066\\
\multicolumn{2}{c}{oracle}&-0.001&0.037&0.146&0.958&0.037&0.037\\
\cdashline{3-8}
\multirow{4}{*}{{\color{newblue} \bf constant}}&{\color{newblue} \bf LS}&\color{newblue} -0.001&\color{newblue} 0.038&\color{newblue} 0.147&\color{newblue} 0.956&\color{newblue} 0.038&\color{newblue} 0.038\\
&poly&-0.001&0.038&0.148&0.954&0.038&0.038\\
&RF&0.000&0.041&0.164&0.958&0.041&0.042\\
&RKHS&-0.001&0.042&0.163&0.938&0.042&0.042\\
\cdashline{3-8}
\multirow{4}{*}{logistic}&LS&0.000&0.038&0.150&0.960&0.038&0.038\\
&poly&0.000&0.038&0.150&0.956&0.038&0.038\\
&RF&-0.001&0.039&0.167&0.964&0.039&0.043\\
&RKHS&-0.001&0.041&0.166&0.950&0.041&0.042\\
\bottomrule
\end{tabular}
}
\end{table}

\begin{table}[h!]
\centering
\caption{\underline{Simulatio\tcr{n} setting b} with $p=10$. \tcr{The rest of the caption details remain the same as those in Table \ref{table:a_p10}.}}\label{table:b_p10}
\vspace{0.5em}
\resizebox{!}{2.8in}{%
\begin{tabular}{llccccccc}
\toprule
$\pihat_N(\cdot)$&$\mhat(\cdot)$&Bias&RMSE&Length&Coverage&ESD&ASD\\
\hline
\multicolumn{2}{c}{} & \multicolumn{6}{c}{  \cellcolor{gray!50}  $ N=10000,\bar\pi_N=0.01 ~(N\bar\pi_N=100)$}\\
\cline{3-8}
\multicolumn{2}{c}{$\bar Y_\mathrm{labeled}$}&0.008&0.334&1.251&0.936&0.334&0.319\\
\multicolumn{2}{c}{oracle}&0.001&0.103&0.410&0.946&0.103&0.105\\
\cdashline{3-8}
\multirow{4}{*}{{\color{newblue}\bf constant}}&LS&0.009&0.311&1.167&0.938&0.311&0.298\\\
&{\color{newblue} \bf poly}&\color{newblue}0.002&\color{newblue}0.124&\color{newblue}0.496&\color{newblue}0.950&\color{newblue}0.125&\color{newblue}0.127\\
&RF&0.002&0.253&0.939&0.934&0.253&0.239\\
&RKHS&0.004&0.245&0.930&0.928&0.245&0.237\\
\cdashline{3-8}
\multirow{4}{*}{logistic}&LS&0.154&0.420&1.540&0.950&0.391&0.393\\
&poly&0.001&0.143&0.615&0.968&0.144&0.157\\
&RF&0.075&0.319&1.222&0.960&0.310&0.312\\
&RKHS&0.107&0.327&1.216&0.950&0.310&0.310\\
\hline
\multicolumn{2}{c}{} & \multicolumn{6}{c}{  \cellcolor{gray!50}   $ N=50000,\bar\pi_N=0.01 ~(N\bar\pi_N=500)$}\\
\cline{3-8}
\multicolumn{2}{c}{$\bar Y_\mathrm{labeled}$}&0.005&0.138&0.558&0.958&0.138&0.142\\
\multicolumn{2}{c}{oracle}&0.001&0.046&0.184&0.952&0.046&0.047\\
\cdashline{3-8}
\multirow{4}{*}{{\color{newblue} \bf constant}}&LS&0.008&0.119&0.478&0.952&0.119&0.122\\
&{\color{newblue}\bf poly}&\color{newblue}0.000&\color{newblue}0.047&\color{newblue}0.189&\color{newblue}0.952&\color{newblue}0.047&\color{newblue}0.048\\
&RF&0.001&0.076&0.304&0.942&0.076&0.077\\
&RKHS&0.001&0.076&0.306&0.952&0.076&0.078\\
\cdashline{3-8}
\multirow{4}{*}{logistic}&LS&0.032&0.128&0.505&0.954&0.124&0.129\\
&poly&0.000&0.048&0.196&0.956&0.048&0.050\\
&RF&0.009&0.079&0.320&0.958&0.079&0.082\\
&RKHS&0.014&0.080&0.324&0.960&0.079&0.083\\
\hline
\multicolumn{2}{c}{} & \multicolumn{6}{c}{  \cellcolor{gray!50}   $N=10000,\bar\pi_N=0.1 ~(N\bar\pi_N=1000)$}\\
\cline{3-8}
\multicolumn{2}{c}{$\bar Y_\mathrm{labeled}$}&0.001&0.107&0.411&0.944&0.107&0.105\\
\multicolumn{2}{c}{oracle}&0.000&0.045&0.176&0.946&0.045&0.045\\
\cdashline{3-8}
\multirow{4}{*}{{\color{newblue}\bf constant}}&LS&0.000&0.091&0.353&0.958&0.091&0.090\\
&{\color{newblue} \bf poly}&\color{newblue} 0.000&\color{newblue}0.045&\color{newblue}0.177&\color{newblue}0.944&\color{newblue}0.046&\color{newblue}0.045\\
&RF&-0.001&0.057&0.228&0.952&0.057&0.058\\
&RKHS&-0.001&0.058&0.232&0.952&0.058&0.059\\
\cdashline{3-8}
\multirow{4}{*}{logistic}&LS&0.012&0.093&0.363&0.958&0.092&0.093\\
&poly&0.000&0.046&0.179&0.944&0.046&0.046\\
&RF&0.003&0.058&0.233&0.956&0.058&0.059\\
&RKHS&-0.005&0.058&0.237&0.956&0.058&0.061\\
\bottomrule
\end{tabular}
}
\end{table}

\begin{table}[h!]
\centering
\caption{\underline{Simulatio\tcr{n} setting c} with $p=10$. The rest of the caption details remain the same as those in Table \ref{table:a_p10}.} \label{table:c_p10}
\vspace{0.5em}
\resizebox{!}{2.8in}{%
\begin{tabular}{llccccccc}
\toprule
$\pihat_N(\cdot)$&$\mhat(\cdot)$&Bias&RMSE&Length&Coverage&ESD&ASD\\
\hline
\multicolumn{2}{c}{} & \multicolumn{6}{c}{  \cellcolor{gray!50}   $N=10000,\bar\pi_N=0.01 ~(N\bar\pi_N=100)$}\\
\cline{3-8}
\multicolumn{2}{c}{$\bar Y_\mathrm{labeled}$}&0.980&0.999&0.783&0.002&0.194&0.200\\
\multicolumn{2}{c}{oracle}&0.003&0.106&0.397&0.942&0.106&0.101\\
\cdashline{3-8}
\multirow{4}{*}{constant}&LS&0.005&0.151&0.434&0.850&0.151&0.111\\
&poly&-0.004&0.194&0.480&0.792&0.194&0.122\\
&RF&0.570&0.610&0.596&0.108&0.218&0.152\\
&RKHS&0.403&0.439&0.543&0.210&0.175&0.139\\
\cdashline{3-8}
\multirow{4}{*}{{\color{newblue}\bf logistic}}&{\color{newblue}\bf LS}&\color{newblue}0.015&\color{newblue}0.234&\color{newblue}0.884&\color{newblue}0.952&\color{newblue}0.234&\color{newblue}0.226\\
&poly&-0.002&0.365&1.083&0.922&0.365&0.276\\
&RF&-0.171&0.705&1.708&0.950&0.685&0.436\\
&RKHS&-0.140&0.675&1.525&0.938&0.661&0.389\\
\hline
\multicolumn{2}{c}{} & \multicolumn{6}{c}{  \cellcolor{gray!50}   $ N=50000,\bar\pi_N=0.01 ~(N\bar\pi_N=500)$}\\
\cline{3-8}
\multicolumn{2}{c}{$\bar Y_\mathrm{labeled}$}&0.968&0.972&0.350&0.000&0.090&0.089\\
\multicolumn{2}{c}{oracle}&0.000&0.070&0.281&0.968&0.070&0.072\\
\cdashline{3-8}
\multirow{4}{*}{constant}&LS&0.001&0.063&0.181&0.868&0.063&0.046\\
&poly&0.001&0.070&0.183&0.812&0.070&0.047\\
&RF&0.341&0.351&0.215&0.004&0.085&0.055\\
&RKHS&0.240&0.251&0.211&0.028&0.072&0.054\\
\cdashline{3-8}
\multirow{4}{*}{{\color{newblue}\bf logistic}}&{\color{newblue} \bf LS}&\color{newblue}0.000&\color{newblue}0.072&\color{newblue}0.297&\color{newblue}0.964&\color{newblue}0.073&\color{newblue}0.076\\
&poly&0.000&0.076&0.307&0.956&0.076&0.078\\
&RF&-0.016&0.121&0.491&0.956&0.120&0.125\\
&RKHS&-0.015&0.123&0.464&0.938&0.122&0.118\\
\hline
\multicolumn{2}{c}{} & \multicolumn{6}{c}{  \cellcolor{gray!50}   $N=10000,\bar\pi_N=0.1 ~(N\bar\pi_N=1000)$}\\
\cline{3-8}
\multicolumn{2}{c}{$\bar Y_\mathrm{labeled}$}&0.820&0.822&0.244&0.000&0.063&0.062\\
\multicolumn{2}{c}{oracle}&0.001&0.052&0.200&0.946&0.052&0.051\\
\cdashline{3-8}
\multirow{4}{*}{constant}&LS&0.001&0.046&0.142&0.868&0.046&0.036\\
&poly&0.001&0.051&0.143&0.864&0.051&0.036\\
&RF&0.241&0.248&0.156&0.006&0.058&0.040\\
&RKHS&0.149&0.158&0.154&0.102&0.053&0.039\\
\cdashline{3-8}
\multirow{4}{*}{{\color{newblue}\bf logistic}}&{\color{newblue} \bf LS}&\color{newblue}0.001&\color{newblue}0.052&\color{newblue}0.203&\color{newblue}0.936&\color{newblue}0.052&\color{newblue}0.052\\
&poly&0.001&0.053&0.206&0.930&0.053&0.053\\
&RF&-0.009&0.070&0.294&0.970&0.069&0.075\\
&RKHS&-0.005&0.069&0.277&0.942&0.069&0.071\\
\bottomrule
\end{tabular}
}
\end{table}

\begin{table}[h!]
\centering
\caption{\underline{Simulatio\tcr{n} setting d} with $p=10$. The rest of the caption details remain the same as those in Table \ref{table:a_p10}.} \label{table:d_p10}
\vspace{0.5em}
\resizebox{!}{2.8in}{%
\begin{tabular}{llccccccc}
\toprule
$\pihat_N(\cdot)$&$\mhat(\cdot)$&Bias&RMSE&Length&Coverage&ESD&ASD\\
\hline
\multicolumn{2}{c}{} & \multicolumn{6}{c}{  \cellcolor{gray!50}   $ N=10000,\bar\pi_N=0.01 ~(N\bar\pi_N=100)$}\\
\cline{3-8}
\multicolumn{2}{c}{$\bar Y_\mathrm{labeled}$}&1.885&1.934&1.617&0.002&0.432&0.412\\
\multicolumn{2}{c}{oracle}&0.008&0.168&0.625&0.952&0.168&0.160\\
\cdashline{3-8}
\multirow{4}{*}{constant}&LS&-0.929&1.040&1.148&0.226&0.469&0.293\\\
&poly&0.002&0.189&0.492&0.808&0.189&0.125\\
&RF&0.317&0.443&1.033&0.752&0.309&0.264\\
&RKHS&0.392&0.472&1.022&0.664&0.263&0.261\\
\cdashline{3-8}
\multirow{4}{*}{{\color{newblue}\bf logistic}}&LS&0.378&1.386&3.996&0.910&1.335&1.019\\
&{\color{newblue}\bf poly}&\color{newblue}0.012&\color{newblue}0.255&\color{newblue}1.011&\color{newblue}0.932&\color{newblue}0.255&\color{newblue}0.258\\
&RF&0.026&0.573&1.828&0.954&0.573&0.466\\
&RKHS&-0.014&0.570&1.727&0.950&0.570&0.441\\
\hline
\multicolumn{2}{c}{} & \multicolumn{6}{c}{  \cellcolor{gray!50}   $ N=50000,\bar\pi_N=0.01 ~(N\bar\pi_N=500)$}\\
\cline{3-8}
\multicolumn{2}{c}{$\bar Y_\mathrm{labeled}$}&1.901&1.910&0.728&0.000&0.192&0.186\\
\multicolumn{2}{c}{oracle}&-0.002&0.074&0.287&0.944&0.074&0.073\\
\cdashline{3-8}
\multirow{4}{*}{constant}&LS&-0.951&0.972&0.473&0.000&0.202&0.121\\
&poly&-0.002&0.076&0.189&0.798&0.076&0.048\\
&RF&0.074&0.123&0.323&0.816&0.099&0.082\\
&RKHS&0.163&0.191&0.328&0.500&0.100&0.084\\
\cdashline{3-8}
\multirow{4}{*}{{\color{newblue}\bf logistic}}&LS&0.078&0.485&1.661&0.924&0.479&0.424\\
&{\color{newblue}\bf poly}&\color{newblue}-0.002&\color{newblue}0.082&\color{newblue}0.318&\color{newblue}0.940&\color{newblue}0.082&\color{newblue}0.081\\
&RF&0.002&0.181&0.594&0.924&0.181&0.152\\
&RKHS&0.003&0.178&0.557&0.936&0.178&0.142\\
\hline
\multicolumn{2}{c}{} & \multicolumn{6}{c}{  \cellcolor{gray!50}  $ N=10000,\bar\pi_N=0.1 ~(N\bar\pi_N=1000)$}\\
\cline{3-8}
\multicolumn{2}{c}{$\bar Y_\mathrm{labeled}$}&1.371&1.377&0.471&0.000&0.120&0.120\\
\multicolumn{2}{c}{oracle}&-0.001&0.053&0.225&0.980&0.053&0.057\\
\cdashline{3-8}
\multirow{4}{*}{constant}&LS&-0.746&0.756&0.342&0.000&0.122&0.087\\
&poly&-0.001&0.052&0.172&0.914&0.052&0.044\\
&RF&0.010&0.065&0.227&0.924&0.064&0.058\\
&RKHS&0.053&0.084&0.231&0.838&0.065&0.059\\
\cdashline{3-8}
\multirow{4}{*}{{\color{newblue}\bf logistic}}&LS&0.019&0.234&0.951&0.944&0.233&0.243\\
&{\color{newblue}\bf poly}&\color{newblue}-0.001&\color{newblue}0.054&\color{newblue}0.226&\color{newblue}0.976&\color{newblue}0.054&\color{newblue}0.058\\
&RF&-0.005&0.100&0.379&0.940&0.099&0.097\\
&RKHS&0.000&0.094&0.358&0.940&0.094&0.091\\
\bottomrule
\end{tabular}
}
\end{table}

\begin{table}[h!]
\centering
\caption{\underline{Simulatio\tcr{n} setting a} with $p=500$.
The rest of the caption details remain the same as those in Table \ref{table:a_p10}.}\label{table:a_p500}
\vspace{0.5em}
\resizebox{!}{1.7in}{%
\begin{tabular}{llccccccc}
\toprule
$\pihat_N(\cdot)$&$\mhat(\cdot)$&Bias&RMSE&Length&Coverage&ESD&ASD\\
\hline
\multicolumn{2}{c}{} & \multicolumn{6}{c}{  \cellcolor{gray!50}  $ N=10000,\bar\pi_N=0.01 ~(N\bar\pi_N=100)$}\\
\cline{3-8}
\multicolumn{2}{c}{$\bar Y_\mathrm{labeled}$}&-0.003&0.195&0.788&0.960&0.195&0.201\\
\multicolumn{2}{c}{oracle}&-0.001&0.103&0.400&0.954&0.103&0.102\\
\cdashline{3-8}
\multirow{2}{*}{{\color{newblue}\bf constant}}&{\color{newblue} \bf Lasso}&\color{newblue}0.000&\color{newblue}0.119&\color{newblue}0.478&\color{newblue}0.950&\color{newblue}0.119&\color{newblue}0.122\\
&poly-Lasso&-0.002&0.120&0.493&0.950&0.120&0.126\\
\cdashline{3-8}
\multirow{2}{*}{log-Lasso}&Lasso&0.000&0.120&0.487&0.960&0.120&0.124\\
&poly-Lasso&-0.002&0.121&0.502&0.952&0.121&0.128\\
\hline
\multicolumn{2}{c}{} & \multicolumn{6}{c}{  \cellcolor{gray!50}  $ N=50000,\bar\pi_N=0.01 ~(N\bar\pi_N=500)$}\\
\cline{3-8}
\multicolumn{2}{c}{$\bar Y_\mathrm{labeled}$}&0.003&0.093&0.352&0.944&0.093&0.090\\
\multicolumn{2}{c}{oracle}&0.001&0.044&0.178&0.948&0.044&0.046\\
\cdashline{3-8}
\multirow{2}{*}{{\color{newblue}\bf constant}}&{\color{newblue} \bf Lasso}&\color{newblue}0.001&\color{newblue}0.046&\color{newblue}0.184&\color{newblue}0.952&\color{newblue}0.046&\color{newblue}0.047\\
&poly-Lasso&0.001&0.046&0.185&0.952&0.046&0.047\\
\cdashline{3-8}
\multirow{2}{*}{log-Lasso}&Lasso&0.001&0.046&0.185&0.948&0.046&0.047\\
&poly-Lasso&0.001&0.046&0.186&0.952&0.046&0.047\\
\hline
\multicolumn{2}{c}{} & \multicolumn{6}{c}{  \cellcolor{gray!50}  $ N=10000,\bar\pi_N=0.1 ~(N\bar\pi_N=1000)$}\\
\cline{3-8}
\multicolumn{2}{c}{$\bar Y_\mathrm{labeled}$}&-0.003&0.063&0.260&0.958&0.063&0.066\\
\multicolumn{2}{c}{oracle}&-0.002&0.037&0.147&0.956&0.037&0.037\\
\cdashline{3-8}
\multirow{2}{*}{{\color{newblue}\bf constant}}&{\color{newblue} \bf Lasso}&\color{newblue}-0.002&\color{newblue}0.038&\color{newblue}0.149&\color{newblue}0.960&\color{newblue}0.038&\color{newblue}0.038\\
&poly-Lasso&-0.002&0.038&0.149&0.960&0.038&0.038\\
\cdashline{3-8}
\multirow{2}{*}{log-Lasso}&Lasso&-0.002&0.038&0.149&0.956&0.038&0.038\\
&poly-Lasso&-0.002&0.038&0.149&0.960&0.038&0.038\\
\bottomrule
\end{tabular}
}
\end{table}

\begin{table}[h!]
\centering
\caption{\underline{Simulatio\tcr{n} setting b} with $p=500$.
The rest of the caption details remain the same as those in Table \ref{table:a_p10}.}\label{table:b_p500}
\vspace{0.5em}
\resizebox{!}{1.7in}{%
\begin{tabular}{llccccccc}
\toprule
$\pihat_N(\cdot)$&$\mhat(\cdot)$&Bias&RMSE&Length&Coverage&ESD&ASD\\
\hline
\multicolumn{2}{c}{} & \multicolumn{6}{c}{  \cellcolor{gray!50} $ N=10000,\bar\pi_N=0.01 ~(N\bar\pi_N=100)$}\\
\cline{3-8}
\multicolumn{2}{c}{$\bar Y_\mathrm{labeled}$}&0.003&0.306&1.244&0.948&0.306&0.317\\
\multicolumn{2}{c}{oracle}&-0.003&0.107&0.411&0.934&0.107&0.105\\
\cdashline{3-8}
\multirow{2}{*}{{\color{newblue}\bf constant}}&Lasso&0.004&0.296&1.220&0.966&0.296&0.311\\
&{\color{newblue}\bf poly-Lasso}&\color{newblue}0.000&\color{newblue}0.174&\color{newblue}0.668&\color{newblue}0.954&\color{newblue}0.175&\color{newblue}0.171\\
\cdashline{3-8}
\multirow{2}{*}{log-Lasso}&Lasso&0.008&0.297&1.241&0.966&0.298&0.317\\
&poly-Lasso&0.002&0.175&0.680&0.954&0.175&0.173\\
\hline
\multicolumn{2}{c}{} & \multicolumn{6}{c}{  \cellcolor{gray!50}   $ N=50000,\bar\pi_N=0.01 ~(N\bar\pi_N=500)$}\\
\cline{3-8}
\multicolumn{2}{c}{$\bar Y_\mathrm{labeled}$}&-0.004&0.147&0.554&0.926&0.147&0.141\\
\multicolumn{2}{c}{oracle}&-0.002&0.054&0.215&0.954&0.054&0.055\\
\cdashline{3-8}
\multirow{2}{*}{{\color{newblue}\bf constant}}&Lasso&-0.005&0.129&0.483&0.928&0.129&0.123\\
&{\color{newblue}\bf poly-Lasso}&\color{newblue}-0.002&\color{newblue}0.051&\color{newblue}0.195&\color{newblue}0.934&\color{newblue}0.051&\color{newblue}0.050\\
\cdashline{3-8}
\multirow{2}{*}{log-Lasso}&Lasso&-0.004&0.129&0.484&0.926&0.129&0.124\\
&poly-Lasso&-0.002&0.051&0.196&0.936&0.051&0.050\\
\hline
\multicolumn{2}{c}{} & \multicolumn{6}{c}{  \cellcolor{gray!50}   $ N=10000,\bar\pi_N=0.1 ~(N\bar\pi_N=1000)$}\\
\cline{3-8}
\multicolumn{2}{c}{$\bar Y_\mathrm{labeled}$}&-0.001&0.104&0.411&0.948&0.104&0.105\\
\multicolumn{2}{c}{oracle}&0.001&0.042&0.175&0.964&0.043&0.045\\
\cdashline{3-8}
\multirow{2}{*}{{\color{newblue}\bf constant}}&Lasso&-0.001&0.091&0.357&0.944&0.092&0.091\\
&{\color{newblue}\bf poly-Lasso}&\color{newblue}0.001&\color{newblue}0.043&\color{newblue}0.178&\color{newblue}0.960&\color{newblue}0.043&\color{newblue}0.046\\
\cdashline{3-8}
\multirow{2}{*}{log-Lasso}&Lasso&-0.001&0.091&0.357&0.948&0.091&0.091\\
&poly-Lasso&0.001&0.043&0.179&0.962&0.043&0.046\\
\bottomrule
\end{tabular}
}
\end{table}

\begin{table}[h!]
\centering
\caption{\underline{Simulatio\tcr{n} setting c} with $p=500$.
The rest of the caption details remain the same as those in Table \ref{table:a_p10}.}\label{table:c_p500}
\vspace{0.5em}
\resizebox{!}{1.7in}{%
\begin{tabular}{llccccccc}
\toprule
$\pihat_N(\cdot)$&$\mhat(\cdot)$&Bias&RMSE&Length&Coverage&ESD&ASD\\
\hline
\multicolumn{2}{c}{} & \multicolumn{6}{c}{  \cellcolor{gray!50}  $ N=10000,\bar\pi_N=0.01 ~(N\bar\pi_N=100)$}\\
\cline{3-8}
\multicolumn{2}{c}{$\bar Y_\mathrm{labeled}$}&0.977&0.997&0.782&0.006&0.198&0.199\\
\multicolumn{2}{c}{oracle}&-0.003&0.160&0.612&0.970&0.160&0.156\\
\cdashline{3-8}
\multirow{2}{*}{constant}&Lasso&0.275&0.320&0.480&0.412&0.164&0.123\\
&poly-Lasso&0.324&0.368&0.496&0.320&0.173&0.127\\
\cdashline{3-8}
\multirow{2}{*}{{\color{newblue} \bf log-Lasso}}&{\color{newblue} \bf Lasso}&\color{newblue}0.099&\color{newblue}0.210&\color{newblue}0.563&\color{newblue}0.782&\color{newblue}0.186&\color{newblue}0.144\\
&poly-Lasso&0.117&0.229&0.602&0.778&0.197&0.154\\
\hline
\multicolumn{2}{c}{} & \multicolumn{6}{c}{  \cellcolor{gray!50}   $ N=50000,\bar\pi_N=0.01 ~(N\bar\pi_N=500)$}\\
\cline{3-8}
\multicolumn{2}{c}{$\bar Y_\mathrm{labeled}$}&0.975&0.979&0.350&0.000&0.087&0.089\\
\multicolumn{2}{c}{oracle}&-0.003&0.071&0.282&0.956&0.071&0.072\\
\cdashline{3-8}
\multirow{2}{*}{constant}&Lasso&0.115&0.132&0.184&0.370&0.065&0.047\\
&poly-Lasso&0.130&0.146&0.185&0.306&0.067&0.047\\
\cdashline{3-8}
\multirow{2}{*}{{\color{newblue} \bf log-Lasso}}&{\color{newblue} \bf Lasso}&\color{newblue}0.022&\color{newblue}0.072&\color{newblue}0.241&\color{newblue}0.884&\color{newblue}0.069&\color{newblue}0.061\\
&poly-Lasso&0.026&0.075&0.243&0.886&0.070&0.062\\
\hline
\multicolumn{2}{c}{} & \multicolumn{6}{c}{  \cellcolor{gray!50}  $ N=10000,\bar\pi_N=0.1 ~(N\bar\pi_N=1000)$}\\
\cline{3-8}
\multicolumn{2}{c}{$\bar Y_\mathrm{labeled}$}&0.822&0.824&0.245&0.000&0.065&0.062\\
\multicolumn{2}{c}{oracle}&0.005&0.051&0.198&0.958&0.050&0.050\\
\cdashline{3-8}
\multirow{2}{*}{constant}&Lasso&0.077&0.090&0.143&0.438&0.047&0.036\\
&poly-Lasso&0.086&0.098&0.143&0.386&0.047&0.036\\
\cdashline{3-8}
\multirow{2}{*}{{\color{newblue} \bf log-Lasso}}&{\color{newblue} \bf Lasso}&\color{newblue}0.019&\color{newblue}0.053&\color{newblue}0.173&\color{newblue}0.894&\color{newblue}0.049&\color{newblue}0.044\\
&poly-Lasso&0.021&0.054&0.174&0.896&0.050&0.044\\
\bottomrule
\end{tabular}
}
\end{table}

\begin{table}[h!]
\centering
\caption{\underline{Simulatio\tcr{n} setting d} with $p=500$.
The rest of the caption details remain the same as those in Table \ref{table:a_p10}.}\label{table:d_p500}
\vspace{0.5em}
\resizebox{!}{1.7in}{%
\begin{tabular}{llccccccc}
\toprule
$\pihat_N(\cdot)$&$\mhat(\cdot)$&Bias&RMSE&Length&Coverage&ESD&ASD\\
\hline
\multicolumn{2}{c}{} & \multicolumn{6}{c}{  \cellcolor{gray!50}  $ N=10000,\bar\pi_N=0.01 ~(N\bar\pi_N=100)$}\\
\cline{3-8}
\multicolumn{2}{c}{$\bar Y_\mathrm{labeled}$}&1.875&1.917&1.620&0.000&0.401&0.413\\
\multicolumn{2}{c}{oracle}&-0.002&0.157&0.613&0.970&0.157&0.156\\
\cdashline{3-8}
\multirow{2}{*}{constant}&Lasso&-0.183&0.535&1.225&0.756&0.503&0.313\\
&poly-Lasso&0.273&0.360&0.625&0.566&0.235&0.160\\
\cdashline{3-8}
\multirow{2}{*}{{\color{newblue} \bf log-Lasso}}&Lasso&-0.229&0.621&1.511&0.736&0.578&0.386\\
&{\color{newblue}\bf poly-Lasso}&\color{newblue}0.090&\color{newblue}0.263&\color{newblue}0.708&\color{newblue}0.824&\color{newblue}0.247&\color{newblue}0.181\\
\hline
\multicolumn{2}{c}{} & \multicolumn{6}{c}{  \cellcolor{gray!50}  $ N=50000,\bar\pi_N=0.01 ~(N\bar\pi_N=500)$}\\
\cline{3-8}
\multicolumn{2}{c}{$\bar Y_\mathrm{labeled}$}&1.890&1.900&0.731&0.000&0.191&0.187\\
\multicolumn{2}{c}{oracle}&-0.006&0.074&0.286&0.954&0.074&0.073\\
\cdashline{3-8}
\multirow{2}{*}{constant}&Lasso&-0.657&0.686&0.481&0.020&0.200&0.123\\
&poly-Lasso&0.086&0.113&0.194&0.562&0.074&0.049\\
\cdashline{3-8}
\multirow{2}{*}{{\color{newblue} \bf log-Lasso}}&Lasso&-0.254&0.377&0.968&0.684&0.279&0.247\\
&{\color{newblue}\bf poly-Lasso}&\color{newblue}0.010&\color{newblue}0.076&\color{newblue}0.247&\color{newblue}0.888&\color{newblue}0.075&\color{newblue}0.063\\
\hline
\multicolumn{2}{c}{} & \multicolumn{6}{c}{  \cellcolor{gray!50}   $ N=10000,\bar\pi_N=0.1 ~(N\bar\pi_N=1000)$}\\
\cline{3-8}
\multicolumn{2}{c}{$\bar Y_\mathrm{labeled}$}&1.355&1.360&0.470&0.000&0.118&0.120\\
\multicolumn{2}{c}{oracle}&0.000&0.058&0.224&0.964&0.058&0.057\\
\cdashline{3-8}
\multirow{2}{*}{constant}&Lasso&-0.551&0.566&0.343&0.002&0.128&0.088\\
&poly-Lasso&0.055&0.080&0.175&0.706&0.057&0.045\\
\cdashline{3-8}
\multirow{2}{*}{{\color{newblue} \bf log-Lasso}}&Lasso&-0.179&0.250&0.649&0.706&0.175&0.165\\
&{\color{newblue}\bf poly-Lasso}&\color{newblue}0.008&\color{newblue}0.058&\color{newblue}0.198&\color{newblue}0.910&\color{newblue}0.058&\color{newblue}0.051\\
\bottomrule
\end{tabular}
}
\end{table}

We consider several different choices on the outcome and propensity \tco{estimators}. In low dimensions ($p=10$), we consider two
\tcr{parametric} outcome \tcr{model} estimators: least squares (LS) \tcr{linear regression} and a polynomial (poly) regression with degree 2 (without interaction terms), and two non-parametric outcome estimators: random forest (RF) and reproducing kernel Hilbert space (RKHS) regression \tcr{using a Gaussian kernel}.
 In high dimensions ($p=500$), we
\tcr{consider two \tco{$\ell_1$}-regularized parametric} outcome \tcr{model} estimators: Lasso and a degree-2 polynomial regression with a Lasso-type penalty (poly-Lasso). As for the  PS estimators, we consider a constant estimator that essentially corresponds to a MCAR estimator, and an offset based logistic estimator (or it's \tco{$\ell_1$}-regularized version, log-Lasso, when $p=500$).
{\crev
Note that the DRSS estimators based on a constant PS estimate are essentially the same as the {\crevmag SS mean} 
estimators proposed by \cite{zhang2022high}, except here, we use a cross-fitted version of $\pihat_N(\cdot)$. In addition, the
{\crevmag SS mean estimators} proposed by \cite{zhang2019semi} {\crevmag based on a linear outcome model estimated via least squares} can be further seen as a special case of our proposed DRSS estimator essentially, where a (non-cross-fitted) constant PS and a (non-cross-fitted) least squares {\crevmag estimate for a linear} outcome {\crevmag model}
are used; see further discussions in Appendix \ref{sec:MCAR} of the \hyperref[supp_mat]{Supplement}.} \tco{The \jelena{tuning parameters} in the regularized estimators are chosen 
\tcr{via} 5-fold cross-validation. The hyperparameters in the RF
are chosen by minimizing the out-of-bag (OOB) error.}
The bandwidth parameter \tcr{for the Gaussian kernel} in RKHS regression is set to be $p$.


The simulations are {\crev repeated} 500 times and the results are presented in Tables \ref{table:a_p10}-\ref{table:d_p500}.  We report the \tcr{b}ias, \tcr{the} root mean square error (RMSE), \tcr{the} average length  and average coverage of the $95\%$ confidence intervals,  \tcr{the} empirical standard error and \tcr{the} averaged estimated standard error \tcr{for all settings}.
The {\color{newblue}\bf blue} 
color in the tables denotes the ``smallest'' \tcr{(i.e.\tco{,} most parsimonious)} and correctly specified parametric model for each of the settings.

We \tcr{first} check the proposed estimator's double robustness \tcr{in terms} of inference. As per Theorem \ref{t4}, the asymptotic normality results hold when the product rate condition is satisfied and when both of \tco{$\pi_N(\cdot)$ and $m(\cdot)$} are correct,
in which case,
 the proposed estimator is $(N\bar\pi_N)^{1/2}$-consistent with the asymptotic efficiency  matching that of  the oracle estimator.
  In low dimensions ($p=10$), the product rate condition always holds since $\pihat_N(\cdot)$ has an estimation error of rate $(N\bar\pi_N)^{-1/2}$ and $\mhat(\cdot)$ (parametric or non-parametric) is consistent. As in Tables \ref{table:a_p10}-\ref{table:d_p10}, the coverage of $\thetahat_\mathrm{DRSS}$ based on correct \tco{$\pi_N(\cdot)$ and $m(\cdot)$} is close to $95\%$ even with a small $N\bar\pi_N=100$. In high dimensions ($p=500$), the product rate condition depends on the true PS  model. When the true PS  model is P1 (MCAR), the corresponding $\pihat_N(\cdot)$ still has an estimation error of rate $(N\bar\pi_N)^{-1/2}$ and hence the product rate condition holds; see Tables \ref{table:a_p500} and \ref{table:b_p500}. When the true PS model is P2 (\tco{offset} logistic),  the product rate condition  requires $s_ms_\pi=o(N\bar\pi_N\{\log(p)\}^{-2})$\tco{, where $s_m:=\|\bbeta_0\|_0$ and $s_\pi:=\|\bgamma_0\|_0$}. We can see the coverages are slowly growing towards $95\%$ as $N\bar\pi_N$ increases in Tables \ref{table:c_p500} and \ref{table:d_p500}. More results with different sparsity levels in the high dimensions can be found in Section \ref{sim:hd}.  Here, in Tables \ref{table:b_p10}, \ref{table:d_p10}, and \ref{table:b_p500}, we can see fairly good coverages \emph{even if} the outcome model is misspecified and the confidence interval is constructed without a modification. This coincides with the Remarks \ref{remark:RAL_MCAR} and \ref{remark:RAL_logsitic}; see more details in Appendix \ref{sim:modCI} of the \hyperref[supp_mat]{Supplement}. {\crev On the other hand, as shown in Tables \ref{table:c_p10}, \ref{table:d_p10}, \ref{table:c_p500}, and \ref{table:d_p500}, all the semi-supervised mean estimators based on a constant PS model provide coverages below the desired $95\%$ when the true PS model is logistic. Recall that estimators based on constant PS estimators are essentially the estimators proposed by \cite{zhang2022high} (the ones with least squares  {\crevmag linear} outcome {\crevmag model} estimators are further essentially the estimators proposed by \cite{zhang2019semi}), where an MCAR setting is considered -- this is why we can see poor coverages as selection bias is ignored.}

  Regarding  efficiency, \tco{as in Tables \ref{table:a_p10}-\ref{table:d_p500},}  \jelena{we observe that}  the  \jelena{proposed} estimators based on correct parametric models provide ``optimal'' RMSEs that are close to the oracle estimator. \tco{In Tables \ref{table:a_p10}-\ref{table:d_p10}, the} RMSEs based on non-parametric (RF and RKHS) $\mhat(\cdot)$ are worse than those  based on (correctly specified) parametric \tco{models} with the difference  arising from the product rate of the estimation errors of $\pihat_N(\cdot)$ and $\mhat(\cdot)$. For a (correctly specified) \tco{$\pi_N(\cdot)$} with an estimation error $O_p((N\bar\pi_N)^{-1/2})$, such a difference is not significant and the RMSE is first order insensitive to 
  estimation error of $\mhat(\cdot)$.



%

We also check the double robustness \tcr{in terms} of consistency of the proposed estimators, when only one of \tco{$\pi_N(\cdot)$ and $m(\cdot)$} is correct. \tco{As \tcr{seen} in Tables \ref{table:c_p10}-\ref{table:b_p500}, the naive mean estimator, $\bar Y_\mathrm{labeled}$, is \tcr{\emph{not}} consistent when the selection bias occurs, i.e., the PS is not a constant. Nevertheless, a}s suggested by Theorem \ref{t4}, the proposed \tco{DRSS} estimator is still consistent, and it's consistency rate depends on the estimation error \tcr{rate} of the correct one among $\pihat_N(\cdot)$ and $\mhat(\cdot)$. The proposed  $\thetahat_\mathrm{DRSS}$ can still be $(N\bar\pi_N)^{1/2}$-consistent when the correct estimator has an estimation error of rate $(N\bar\pi_N)^{-1/2}$, which is typically true when the correct model is a low dimensional parametric model: \tcr{s}ee  Tables \ref{table:b_p10}, \ref{table:d_p10}, and \ref{table:b_p500} for correct \tco{$\pi_N(\cdot)$}; see Tables \ref{table:c_p10} and \ref{table:d_p10} for correct \tco{$m(\cdot)$}. If the correct estimator is linear (\tco{offset} logistic) in high dimensions, $\thetahat_\mathrm{DRSS}-\theta_0$ is of the order $O_p(\{(N\bar\pi_N)^{-1}s\log(p)\}^{1/2})$, where $s$ is the sparsity of the correct model; see results in Tables \ref{table:c_p500} and \ref{table:d_p500}. If the correct estimator is non-parametric, we would expect a non-parametric rate on the proposed estimator, matching results in Tables \ref{table:c_p10} and \ref{table:d_p10}.

\subsubsection{Results \tcr{under the} stratified labeling \tcr{model}}\label{sim:stratified}

Now we work on a special \tco{PS} model, \tcr{that of stratified labeling,} as discussed in Section \ref{stratified}:
\begin{enumerate}
\item[P3.] (Stratified \tco{PS}) Suppose we further observe the stratum indicators $\delta_i\in\{0,1\}$, with the following model: $\delta_i|\bX_i\sim\mathrm{Bernoulli}(p_\delta(\bX_i))$ and $R_i|\delta_i\sim\mathrm{Bernoulli}(0.5\bar\pi_N\delta_i+1.5\bar\pi_N(1-\delta_i))$, where $p_\delta(\bx)=g(\bx(1))$ and $g(u)=\exp(u)/\{1+\exp(u)\}$.
\end{enumerate}
We consider the same choices of $p$, $N$, and $\bar\pi_N$ as in Section \ref{sim:main}, and we focus on the following DGP \jelena{of}
Setting e\jelena{, i.e.,} P3+O2. Furthermore, we consider an additional PS estimator based on the stratified labeling of Section \ref{stratified}, where $p_\delta(\cdot)$ is estimated by a logistic regression (with a Lasso-type regularization when $p=500$). The results are shown in Tables \ref{table:f_p10} and \ref{table:f_p500} \tco{for the case $p=10$ and $p=500$, re\jelena{s}pectively}.


By Theorem \ref{t4}, the  estimators based on \tcr{a} correctly specified \tco{$\pi_N(\cdot)$ (stratified) and $m(\cdot)$}
(poly/RF/RKHS) provide $(N\bar\pi_N)^{1/2}$-consistent estimations and valid asymptotic confidence intervals. \jelena{Additionally}, when \tco{$\pi_N(\cdot)$} is correctly specified (stratified) and \tco{$m(\cdot)$} is misspecified (linear), the proposed DR\tcr{SS} mean estimator has a consistency rate \jelena{of} $O_p((N\bar\pi_N)^{-1/2}+r_{\delta,N})$, with $r_{\delta,N}$, defined in \eqref{rdeltaN}, satisfying $r_{\delta,N}=O(N^{-1/2})$ in low dimensions and $r_{\delta,N}=O(\{\tcr{s_{\delta}}\log(p)\}^{1/2}N^{-1/2})$, 
in high dimensions, \tcr{as discussed in Example \ref{ex:2}}. The simulation results in Tables \ref{table:f_p10} and \ref{table:f_p500} support our theoretical arguments: we can see the stratified+poly/RF/RKHS estimators provide coverages close to $95\%$, and all the   estimators based on a stratified $\pihat_N(\cdot)$ provide RMSEs of a similar magnitude.

\begin{table}[h!]
\centering
\caption{\underline{Simulatio\tcr{n} setting e} with $p=10$.
The rest of the caption details remain the same as those in Table \ref{table:a_p10}.} \label{table:f_p10}
\vspace{0.5em}
\resizebox{!}{3.4in}{%
\begin{tabular}{llccccccc}
\toprule
$\pihat_N(\cdot)$&$\mhat(\cdot)$&Bias&RMSE&Length&Coverage&ESD&ASD\\
\hline
\multicolumn{2}{c}{} & \multicolumn{6}{c}{  \cellcolor{gray!50}  $ N=10000,\bar\pi_N=0.01 ~(N\bar\pi_N=100)$}\\
\cline{3-8}
\multicolumn{2}{c}{$\bar Y_\mathrm{labeled}$}&-0.205&0.360&1.203&0.876&0.297&0.307\\
\multicolumn{2}{c}{oracle}&-0.002&0.101&0.415&0.964&0.101&0.106\\
\cdashline{3-8}
\multirow{4}{*}{constant}&LS&-0.082&0.318&1.149&0.916&0.308&0.293\\
&poly&-0.009&0.125&0.493&0.954&0.125&0.126\\
&RF&-0.100&0.259&0.919&0.908&0.239&0.234\\
&RKHS&-0.087&0.248&0.909&0.918&0.232&0.232\\
\cdashline{3-8}
\multirow{4}{*}{logistic}&LS&0.129&0.419&1.639&0.960&0.399&0.418\\
&poly&-0.011&0.146&0.633&0.976&0.146&0.162\\
&RF&0.075&0.328&1.346&0.960&0.320&0.343\\
&RKHS&0.109&0.341&1.337&0.974&0.323&0.341\\
\cdashline{3-8}
\multirow{4}{*}{{\color{newblue}\bf stratified}}&LS&0.009&0.324&1.256&0.948&0.324&0.320\\
&{\color{newblue}\bf poly}&\color{newblue}-0.010&\color{newblue}0.126&\color{newblue}0.510&\color{newblue}0.958&\color{newblue}0.126&\color{newblue}0.130\\
&RF&0.021&0.268&1.062&0.956&0.268&0.271\\
&RKHS&0.016&0.262&1.042&0.960&0.262&0.266\\
\hline
\multicolumn{2}{c}{} & \multicolumn{6}{c}{  \cellcolor{gray!50}  $ N=50000,\bar\pi_N=0.01 ~(N\bar\pi_N=500)$}\\
\cline{3-8}
\multicolumn{2}{c}{$\bar Y_\mathrm{labeled}$}&-0.210&0.251&0.532&0.652&0.138&0.136\\
\multicolumn{2}{c}{oracle}&-0.001&0.048&0.187&0.970&0.048&0.048\\
\cdashline{3-8}
\multirow{4}{*}{constant}&LS&-0.078&0.150&0.469&0.882&0.128&0.120\\
&poly&-0.001&0.050&0.188&0.942&0.050&0.048\\
&RF&-0.053&0.101&0.296&0.832&0.086&0.075\\
&RKHS&-0.053&0.098&0.299&0.866&0.083&0.076\\
\cdashline{3-8}
\multirow{4}{*}{logistic}&LS&-0.014&0.135&0.531&0.952&0.135&0.135\\
&poly&-0.001&0.051&0.199&0.946&0.051&0.051\\
&RF&-0.011&0.091&0.349&0.944&0.090&0.089\\
&RKHS&-0.002&0.090&0.353&0.942&0.090&0.090\\
\cdashline{3-8}
\multirow{4}{*}{{\color{newblue}\bf stratified}}&LS&-0.001&0.132&0.508&0.950&0.132&0.130\\
&{\color{newblue}\bf poly}&\color{newblue}-0.001&\color{newblue}0.050&\color{newblue}0.193&\color{newblue}0.944&\color{newblue}0.050&\color{newblue}0.049\\
&RF&-0.006&0.089&0.333&0.956&0.089&0.085\\
&RKHS&-0.005&0.088&0.336&0.940&0.088&0.086\\
\hline
\multicolumn{2}{c}{} & \multicolumn{6}{c}{  \cellcolor{gray!50}   $ N=10000,\bar\pi_N=0.1 ~(N\bar\pi_N=1000)$}\\
\cline{3-8}
\multicolumn{2}{c}{$\bar Y_\mathrm{labeled}$}&-0.202&0.224&0.377&0.428&0.097&0.096\\
\multicolumn{2}{c}{oracle}&0.001&0.041&0.173&0.970&0.041&0.044\\
\cdashline{3-8}
\multirow{4}{*}{constant}&LS&-0.076&0.120&0.334&0.836&0.093&0.085\\
&poly&0.001&0.042&0.172&0.968&0.042&0.044\\
&RF&-0.033&0.066&0.215&0.868&0.057&0.055\\
&RKHS&-0.034&0.067&0.219&0.906&0.057&0.056\\
\cdashline{3-8}
\multirow{4}{*}{logistic}&LS&-0.018&0.098&0.370&0.928&0.096&0.094\\
&poly&0.001&0.042&0.176&0.968&0.042&0.045\\
&RF&-0.004&0.060&0.240&0.954&0.060&0.061\\
&RKHS&0.001&0.061&0.245&0.952&0.061&0.063\\
\cdashline{3-8}
\multirow{4}{*}{{\color{newblue}\bf stratified}}&LS&0.002&0.095&0.363&0.938&0.095&0.093\\
&{\color{newblue}\bf poly}&\color{newblue}0.001&\color{newblue}0.042&\color{newblue}0.175&\color{newblue}0.968&\color{newblue}0.042&\color{newblue}0.045\\
&RF&0.002&0.059&0.236&0.952&0.059&0.060\\
&RKHS&0.002&0.060&0.241&0.948&0.060&0.061\\
\bottomrule
\end{tabular}
}
\end{table}

\begin{table}[h!]
\centering
\caption{\underline{Simulatio\tcr{n} setting e} with $p=500$.
The rest of the caption details remain the same as those in Table \ref{table:a_p10}.} \label{table:f_p500}
\vspace{0.5em}
\resizebox{!}{2.1in}{%
\begin{tabular}{llccccccc}
\toprule
$\pihat_N(\cdot)$&$\mhat(\cdot)$&Bias&RMSE&Length&Coverage&ESD&ASD\\
\hline
\multicolumn{2}{c}{} & \multicolumn{6}{c}{  \cellcolor{gray!50}  $ N=10000,\bar\pi_N=0.01 ~(N\bar\pi_N=100)$}\\
\cline{3-8}
\multicolumn{2}{c}{$\bar Y_\mathrm{labeled}$}&-0.189&0.363&1.193&0.880&0.311&0.304\\
\multicolumn{2}{c}{oracle}&0.005&0.098&0.417&0.960&0.098&0.106\\
\cdashline{3-8}
\multirow{2}{*}{constant}&Lasso&-0.167&0.352&1.186&0.888&0.310&0.303\\
&poly-Lasso&-0.103&0.220&0.696&0.894&0.195&0.178\\
\multirow{2}{*}{log-Lasso}&Lasso&-0.160&0.351&1.215&0.904&0.312&0.310\\
&poly-Lasso&-0.096&0.218&0.717&0.894&0.196&0.183\\
\multirow{2}{*}{{\color{newblue}\bf stratified}}&Lasso&0.005&0.339&1.318&0.926&0.339&0.336\\
&{\color{newblue}\bf poly-Lasso}&\color{newblue}0.010&\color{newblue}0.200&\color{newblue}0.780&\color{newblue}0.948&\color{newblue}0.200&\color{newblue}0.199\\
\hline
\multicolumn{2}{c}{} & \multicolumn{6}{c}{  \cellcolor{gray!50} $ N=50000,\bar\pi_N=0.01 ~(N\bar\pi_N=500)$}\\
\cline{3-8}
\multicolumn{2}{c}{$\bar Y_\mathrm{labeled}$}&-0.196&0.239&0.535&0.702&0.136&0.136\\
\multicolumn{2}{c}{oracle}&0.000&0.048&0.187&0.950&0.048&0.048\\
\cdashline{3-8}
\multirow{2}{*}{constant}&Lasso&-0.134&0.187&0.478&0.776&0.130&0.122\\
&poly-Lasso&-0.032&0.061&0.195&0.896&0.052&0.050\\
\multirow{2}{*}{log-Lasso}&Lasso&-0.101&0.163&0.491&0.844&0.128&0.125\\
&poly-Lasso&-0.022&0.056&0.196&0.914&0.052&0.050\\
\multirow{2}{*}{{\color{newblue}\bf stratified}}&Lasso&0.001&0.137&0.530&0.946&0.137&0.135\\
&{\color{newblue}\bf poly-Lasso}&\color{newblue}0.000&\color{newblue}0.052&\color{newblue}0.202&\color{newblue}0.944&\color{newblue}0.052&\color{newblue}0.051\\
\hline
\multicolumn{2}{c}{} & \multicolumn{6}{c}{  \cellcolor{gray!50}$ N=10000,\bar\pi_N=0.1 ~(N\bar\pi_N=1000)$}\\
\cline{3-8}
\multicolumn{2}{c}{$\bar Y_\mathrm{labeled}$}&-0.205&0.226&0.377&0.428&0.095&0.096\\
\multicolumn{2}{c}{oracle}&-0.001&0.044&0.173&0.944&0.044&0.044\\
\cdashline{3-8}
\multirow{2}{*}{constant}&Lasso&-0.124&0.151&0.335&0.688&0.087&0.086\\
&poly-Lasso&-0.024&0.051&0.173&0.900&0.045&0.044\\
\multirow{2}{*}{log-Lasso}&Lasso&-0.082&0.119&0.347&0.854&0.087&0.088\\
&poly-Lasso&-0.013&0.047&0.174&0.928&0.045&0.044\\
\multirow{2}{*}{{\color{newblue}\bf stratified}}&Lasso&-0.013&0.090&0.367&0.956&0.089&0.094\\
&{\color{newblue}\bf poly-Lasso}&\color{newblue}-0.002&\color{newblue}0.045&\color{newblue}0.176&\color{newblue}0.948&\color{newblue}0.045&\color{newblue}0.045\\
\bottomrule
\end{tabular}
}
\end{table}

\subsubsection{Results \tcr{for} high dimensional sparse models\tcr{: Investigating performance under varying sparsity levels}}\label{sim:hd}

Here we focus on the high dimension\tcr{al case} ($p=500$) with different sparsity levels. We consider the following \tco{PS} and outcome models:
\begin{enumerate}
\item[P2'.] (\tco{Offset} logistic \tco{PS} with different sparsity levels) Let $\pi_N(\bx)=g(\bxv^T\bgamma_{s_\pi}+\log(\bar\pi_N))$ and $R_i|\bX_i\sim\mathrm{Bernoulli}(\pi_N(\bX_i))$, where $g(u)=\exp(u)/\{1+\exp(u)\}$.
\item[O1'.] (Linear outcome with different sparsity levels) Let $Y_i=\bXv_i^T\bbeta_{s_m}+\varepsilon_i$.
\end{enumerate}
The parameter values are:
$$\bbeta_{s_m}=(-0.5,\sqrt{3/s_m}\boldsymbol{1}_{1\times s_m},\boldsymbol{0}_{1\times(p-s_m)})^T\;\;\mbox{and}\;\;\bgamma_{s_\pi}=(\bgamma_0(1),\sqrt{1/s_\pi}\boldsymbol{1}_{1\times s_\pi},\boldsymbol{0}_{1\times(p-s_\pi)})^T.$$
We consider \tcr{a DGP,} 
{Setting c': P2'+O1'},
with the following choices of $p$, $N$, $\bar\pi_N$, $s_m$ and $s_\pi$:
$$p~=~500,\;\;N~\in~\{50000,\;200000\},\;\;~\bar\pi_N~=0.01,\;\;(s_m,s_\pi)~\in~\{(3,15),\;(15,3)\}.$$
\tco{We illustrate the performance of the same estimators that we considered in Section \ref{sim:main} (for $p=500$); \tcr{the results are presented in} 
Table \ref{table:c'_p500}.}

\tco{In Table \ref{table:c'_p500}, we observe that the RMSEs of $\thetahat_\mathrm{DRSS}$ based on log-Lasso PS estimators are smaller than those based on constant PS estimators. This coincides with our Remark \ref{remark:conds}, as well as Theorems \ref{t4} and \ref{thm:high-dim} - if both of the nuisance functions are correctly specified, we have $\thetahat_\mathrm{DRSS}-\theta_0=O_p((N\bar\pi_N)^{-1/2}+\sqrt{s_ms_\pi}\log(p)/(N\bar\pi_N))$; if only the outcome model is correctly specified, we have a slower upper bound $\thetahat_\mathrm{DRSS}-\theta_0=O_p(\sqrt{s_m\log(p)/(N\bar\pi_N)})$.
\jelena{F}or the DRSS estimators based on log-Lasso PS estimators, we can see that the biases of the   estimators, originat\tcr{ing} from the product rate $\sqrt{s_ms_\pi}\log(p)/(N\bar\pi_N)$, are non-ignorable \tcr{compared to} the RMSEs, especially for smaller $N$. As $N$ grows, \tcr{however,} the coverages of the confidence intervals \tcr{start getting} closer to the desired $95\%$ \tcr{level}. This also coincides with our Remark \ref{remark:conds} that we expect a valid inference result when $s_ms_\pi\log(p)=o(N\bar\pi_N)$ for correctly specified models.}

\begin{table}[h!]
\centering
\caption{\underline{Simulatio\tcr{n} setting c'} with $p=500$.
The rest of the caption details remain the same as those in Table \ref{table:a_p10}.}\label{table:c'_p500}
\vspace{0.5em}
\resizebox{!}{2.2in}{%
\begin{tabular}{llccccccc}
\toprule
$\pihat_N(\cdot)$&$\mhat(\cdot)$&Bias&RMSE&Length&Coverage&ESD&ASD\\
\hline
\multicolumn{2}{c}{} & \multicolumn{6}{c}{  \cellcolor{gray!50}   $s_m=3, s_\pi=15, N=50000, \bar\pi_N=0.01 ~(N\bar\pi_N=500)$}\\
\cline{3-8}
\multicolumn{2}{c}{$\bar Y_\mathrm{labeled}$}&0.755&0.760&0.351&0.000&0.090&0.090\\
\multicolumn{2}{c}{oracle}&0.000&0.074&0.284&0.944&0.074&0.072\\
\cdashline{3-8}
\multirow{2}{*}{constant}&Lasso&0.089&0.103&0.184&0.522&0.052&0.047\\
&poly-Lasso&0.098&0.111&0.185&0.458&0.052&0.047\\
\multirow{2}{*}{{\color{newblue} \bf log-Lasso}}&{\color{newblue} \bf Lasso}&\color{newblue}0.038&\color{newblue}0.071&\color{newblue}0.203&\color{newblue}0.860&\color{newblue}0.060&\color{newblue}0.052\\
&poly-Lasso&0.042&0.073&0.204&0.848&0.060&0.052\\
\hline
\multicolumn{2}{c}{} & \multicolumn{6}{c}{  \cellcolor{gray!50}   $s_m=3, s_\pi=15,  N=200000, \bar\pi_N=0.01 ~(N\bar\pi_N=2000)$}\\
\cline{3-8}
\multicolumn{2}{c}{$\bar Y_\mathrm{labeled}$}&0.754&0.756&0.175&0.000&0.045&0.045\\
\multicolumn{2}{c}{oracle}&0.000&0.035&0.143&0.960&0.035&0.037\\
\cdashline{3-8}
\multirow{2}{*}{constant}&Lasso&0.044&0.051&0.090&0.506&0.025&0.023\\
&poly-Lasso&0.049&0.055&0.090&0.430&0.025&0.023\\
\multirow{2}{*}{{\color{newblue} \bf log-Lasso}}&{\color{newblue} \bf Lasso}&\color{newblue}0.012&\color{newblue}0.032&\color{newblue}0.111&\color{newblue}0.914&\color{newblue}0.030&\color{newblue}0.028\\
&poly-Lasso&0.013&0.033&0.111&0.914&0.030&0.028\\
\hline
\multicolumn{2}{c}{} & \multicolumn{6}{c}{  \cellcolor{gray!50}  $s_m=15, s_\pi=3, N=50000, \bar\pi_N=0.01 ~(N\bar\pi_N=500)$}\\
\cline{3-8}
\multicolumn{2}{c}{$\bar Y_\mathrm{labeled}$}&0.752&0.757&0.349&0.000&0.085&0.089\\
\multicolumn{2}{c}{oracle}&0.001&0.068&0.283&0.966&0.068&0.072\\
\cdashline{3-8}
\multirow{2}{*}{constant}&Lasso&0.155&0.169&0.196&0.200&0.068&0.050\\
&poly-Lasso&0.184&0.197&0.201&0.118&0.070&0.051\\
\multirow{2}{*}{{\color{newblue} \bf log-Lasso}}&{\color{newblue} \bf Lasso}&\color{newblue}0.049&\color{newblue}0.086&\color{newblue}0.237&\color{newblue}0.824&\color{newblue}0.071&\color{newblue}0.060\\
&poly-Lasso&0.058&0.094&0.245&0.794&0.074&0.063\\
\hline
\multicolumn{2}{c}{} & \multicolumn{6}{c}{  \cellcolor{gray!50}   $s_m=15, s_\pi=3,   N=200000, \bar\pi_N=0.01 ~(N\bar\pi_N=2000)$}\\
\cline{3-8}
\multicolumn{2}{c}{$\bar Y_\mathrm{labeled}$}&0.753&0.754&0.175&0.000&0.044&0.045\\
\multicolumn{2}{c}{oracle}&0.000&0.036&0.144&0.964&0.036&0.037\\
\cdashline{3-8}
\multirow{2}{*}{constant}&Lasso&0.076&0.082&0.091&0.172&0.032&0.023\\
&poly-Lasso&0.089&0.094&0.091&0.094&0.032&0.023\\
\multirow{2}{*}{{\color{newblue} \bf log-Lasso}}&{\color{newblue} \bf Lasso}&\color{newblue}0.013&\color{newblue}0.037&\color{newblue}0.124&\color{newblue}0.912&\color{newblue}0.034&\color{newblue}0.032\\
&poly-Lasso&0.015&0.038&0.125&0.904&0.035&0.032\\
\bottomrule
\end{tabular}
}
\end{table}

{\crev
\subsubsection{Results when both {\crevmag nuisance} models are misspecified}\label{sec:both-mis}

In {\crevmag this section, w}e further investigate the case that both nuisance models {\crevmag (i.e., the outcome and PS models)} are misspecified {\crevmag and compare the performance of various SS estimators in such cases}. We consider the following nuisance models:
\begin{enumerate}
\item[P3.] (Offset logistic PS with quadratic signals) $\pi_N(\bx)=g(\bxv^T\bgamma_0+\sum_{j=1}^{p+1}\bdelta_0(j)\bXv_i(j)^2+\log(\bar\pi_N))$, where $g(\cdot)$ is defined in \eqref{divmod-def}.
\item[O2.] (Quadratic outcome) $Y_i=\bXv_i^T\bbeta_0+\sum_{j=1}^{p+1}\balpha_0(j)\bXv_i(j)^2+\varepsilon_i$.
\end{enumerate}
Under P3+O2, three new DGPs, g1, g2, and g3, are considered. The DGPs g1-g3 only varies from the parameter values, and we choose the parameters as follows:
\begin{enumerate}
\item[g1.] $\bgamma_0=(\gamma_1,0.3,0.3,0.3,\boldsymbol0_{1\times(p-3)})^T$, $\bdelta_0=(0,0.3,0.3,0.3,\boldsymbol0_{1\times(p-3)})^T$, $\bbeta_0=(2,1,1,1,\boldsymbol0_{1\times(p-3)})^T+r_1\bgamma_0$, $\balpha_0=r_1\bdelta_0$.
\item[g2.] $\bgamma_0=(\gamma_2,0.5,0.5,0.5,\boldsymbol0_{1\times(p-3)})^T$, $\bdelta_0=r_2(0,1,1,1,\boldsymbol0_{1\times(p-3)})^T$, $\bbeta_0=(2,0.3,0.3,0.3,\boldsymbol0_{1\times(p-3)})^T$, $\balpha_0=(0,0.3,0.3,0.3,\boldsymbol0_{1\times(p-3)})^T$.
\item[g3.] $\bgamma_0=(\gamma_1,1,1,1,\boldsymbol0_{1\times(p-3)})^T$, $\bdelta_0=\balpha_0=r_3(0,1,1,1,\boldsymbol0_{1\times(p-3)})^T$, $\bbeta_0=(2,1,1,1,\boldsymbol0_{1\times(p-3)})^T$.
\end{enumerate}
In the above, $\gamma_1$, $\gamma_2$, and $\gamma_3$ are chosen so that $\E(R)=\bar\pi_N=0.05$ for each DGP; $r_1$, $r_2$, and $r_3$ denote the misspecification levels. For Setting g1, we choose $r_1\in\{1,0.5,0.3,0.1\}$; the PS model has a non-ignorable misspecification error, and the outcome model's misspecification error decays as $r_1$ decreases. For Setting g2, we choose $r_2\in\{0.3,0.2,0.1,0.05\}$; the outcome model has a non-ignorable misspecification error, and the PS model's misspecification error decays as $r_2$ decreases. For Setting g3, we choose $r_3\in\{0.3,0.2,0.1,0.05\}$; both model's misspecification error decays as $r_3$ decreases.

As in Section \ref{sim:main}, we also consider the naive labeled mean estimator $\bar Y_\mathrm{labeled}$, the proposed estimator $\thetahat_\mathrm{DRSS}$
(based on {\crevmag a} Lasso {\crevmag estimate for the} outcome {\crevmag model} and {\crevmag a} log-Lasso/ constant {\crevmag estimate for the} PS {\crevmag model}) and its oracle version. Here we use ``DRSS'' to refer in particular to the estimator $\thetahat_\mathrm{DRSS}$ based on Lasso and log-Lasso estimates {\crevmag of the respective nuisance models}; ``SS'' denotes the one based on {\crevmag a} Lasso outcome and constant PS estimates, as it is the same as the semi-supervised (SS) estimator {\crevmag of} \citet{zhang2022high}, except here we use cross-fitted PS estimates. In addition, we further implement the following estimators that have been considered in \cite{kang2007demystifying}: 1) two variants of the IPW estimator, IPW-POP (re-weighting the respondents to resemble the full population) and IPW-NR (re-weighting the respondents to resemble the nonrespondents' population), where the PS models are estimated by log-Lasso, {\crevmag and} 2) the regression-based (Reg) estimator $\thetahat_\mathrm{Reg}$ using a Lasso {\crevmag based} outcome {\crevmag model} estimate, which can be seen as a regularized version of the OLS regression estimate {\crevmag used} therein. We choose $N=2000$, $p=100$, and repeat the simulations 500 times. In Table \ref{table:g1-g3}, we report the bias and the root mean square error (RMSE) based on the considered estimators.

Similarly to Section \ref{sim}, the naive estimator $\bar Y_\mathrm{labeled}$ has a poor performance since it converges to $\E(Y\mid R=1)\neq\E(Y)$. When both models have non-ignorable misspecification errors, none of the estimators has a good performance, and as also noticed by \cite{kang2007demystifying}, a doubly-robust-type estimator may perform even worse than both the IPW and Reg estimators under such a case. As in Table \ref{table:g1-g3}, DRSS has a larger RMSE than the IPW estimators and the Reg estimator when $r_1=1$ under Setting g1, $r_2\in\{0.3,0.2\}$ under Setting g2, and $r_3=0.3$ under Setting g3; it is also worse than the Reg estimator when $r_3=0.2$ under Setting g3. However, when at least one misspecification level is relatively small, the DRSS estimator outperforms all the others in the sense of the RMSEs (except the oracle one); see Setting g1 with $r_1\in\{0.5,0.3,0.1\}$, Setting g2 with $r_2\in\{0.1,0.05\}$, and Setting g3 with $r_3\in\{0.1,0.05\}$. In addition, similarly to the DRSS estimator, the SS estimator can also be seen as a de-biased version of the Reg estimator. However, since the bias correction is constructed based on a constant PS model, which is far away from the truth, such a correction seems to move the estimate in the wrong direction -- SS always performs worse than Reg in Table \ref{table:g1-g3}.}

\definecolor{ao(english)}{rgb}{0.0, 0.5, 0.0}
\begin{table}[h!]
\centering
{\crev
\caption{Simulations under Settings g1, g2, and g3, with $p=100, N=2000$ and $\bar\pi_N=0.05 ~(N\bar\pi_N=100)$. Bias: empirical bias; RMSE: root mean square error. The words ``bad'' and ``good'' indicate that the models have large and small misspecification errors, respectively. The symbol ``$\to$'' indicates a change in the model misspecification levels, which is characterized by $r_1$, $r_2$, and $r_3$ -- the smaller the values are, the smaller the misspecification error is. The {\color{ao(english)}green} color denotes the smallest RMSE among the estimators (except the oracle one) for each setting.} \label{table:g1-g3}
\vspace{0.5em}
\resizebox{!}{2.35in}{
\begin{tabular}{cccccccccccc}
\toprule
$\thetahat$&Bias&RMSE&&Bias&RMSE&&Bias&RMSE&&Bias&RMSE\\
\hline
&\multicolumn{11}{c}{\cellcolor{gray!10} Setting g1 (``bad'' outcome + ``bad'' PS $\to$ ``good'' outcome + ``bad'' PS)}\\
&\multicolumn{2}{c}{\cellcolor{gray!60} $r_1=1$}&&\multicolumn{2}{c}{\cellcolor{gray!60} $r_1=0.5$}&&\multicolumn{2}{c}{\cellcolor{gray!60} $r_1=0.3$}&&\multicolumn{2}{c}{\cellcolor{gray!60} $r_1=0.1$}\\
\cline{2-3}\cline{5-6}\cline{8-9}\cline{11-12}
$\bar Y_\mathrm{labeled}$&2.404&2.430&&1.812&1.835&&1.575&1.597&&1.337&1.361\\
oracle&0.005&0.152&&0.005&0.149&&0.005&0.148&&0.005&0.147\\
IPW-POP&0.767&0.861&&0.371&0.526&&0.212&0.426&&0.054&0.373\\
IPW-NR&0.758&{\color{ao(english)}0.854}&&0.363&0.520&&0.205&0.422&&0.047&0.371\\
Reg&0.914&0.937&&0.523&0.543&&0.377&0.401&&0.240&0.274\\
SS&0.965&0.988&&0.561&0.581&&0.408&0.432&&0.263&0.297\\
DRSS&0.891&0.938&&0.429&{\color{ao(english)}0.472}&&0.242&{\color{ao(english)}0.303}&&0.054&{\color{ao(english)}0.198}\\
\hline
&\multicolumn{11}{c}{\cellcolor{gray!10} Setting g2 (``bad'' outcome + ``bad'' PS $\to$ ``bad'' outcome + ``good'' PS)}\\
&\multicolumn{2}{c}{\cellcolor{gray!60} $r_2=0.3$}&&\multicolumn{2}{c}{\cellcolor{gray!60} $r_2=0.2$}&&\multicolumn{2}{c}{\cellcolor{gray!60} $r_2=0.1$}&&\multicolumn{2}{c}{\cellcolor{gray!60} $r_2=0.05$}\\
\cline{2-3}\cline{5-6}\cline{8-9}\cline{11-12}
$\bar Y_\mathrm{labeled}$&1.405&1.415&&1.147&1.160&&0.853&0.867&&0.709&0.727\\
oracle&-0.003&0.173&&0.002&0.153&&-0.005&0.143&&-0.002&0.146\\
IPW-POP&0.735&0.797&&0.477&0.531&&0.295&0.337&&0.228&0.278\\
IPW-NR&0.730&{\color{ao(english)}0.793}&&0.469&{\color{ao(english)}0.525}&&0.287&0.331&&0.222&0.274\\
Reg&0.943&0.984&&0.585&0.636&&0.295&0.367&&0.178&0.273\\
SS&1.027&1.061&&0.664&0.709&&0.362&0.421&&0.246&0.317\\
DRSS&0.921&1.099&&0.537&0.658&&0.228&{\color{ao(english)}0.328}&&0.133&{\color{ao(english)}0.254}\\
\hline
&\multicolumn{11}{c}{\cellcolor{gray!10} Setting g3 (``bad'' outcome + ``bad'' PS $\to$ ``good'' outcome + ``good'' PS)}\\
&\multicolumn{2}{c}{\cellcolor{gray!60} $r_3=0.3$}&&\multicolumn{2}{c}{\cellcolor{gray!60} $r_3=0.2$}&&\multicolumn{2}{c}{\cellcolor{gray!60} $r_3=0.1$}&&\multicolumn{2}{c}{\cellcolor{gray!60} $r_3=0.05$}\\
\cline{2-3}\cline{5-6}\cline{8-9}\cline{11-12}
$\bar Y_\mathrm{labeled}$&3.635&3.641&&3.147&3.154&&2.721&2.728&&2.524&2.530\\
oracle&-0.007&0.349&&0.008&0.362&&-0.010&0.348&&0.028&0.317\\
IPW-POP&0.959&1.313&&0.779&1.048&&0.661&0.896&&0.768&0.920\\
IPW-NR&0.896&1.237&&0.708&0.977&&0.597&0.838&&0.709&0.865\\
Reg&0.957&{\color{ao(english)}1.042}&&0.569&{\color{ao(english)}0.650}&&0.444&0.519&&0.480&0.535\\
SS&1.102&1.176&&0.693&0.758&&0.527&0.589&&0.557&0.603\\
DRSS&0.780&1.638&&0.396&0.773&&0.191&{\color{ao(english)}0.399}&&0.228&{\color{ao(english)}0.426}\\
\bottomrule
\end{tabular}
}
}
\end{table}

\subsection{Application to the NHEFS data}\label{sec:NHEFS}

We now apply our imbalanced ATE estimator proposed in Section \ref{ATE} to assess the effect of smoking and alcohol drinking on weight gain, using a subset of data from the National Health and Nutrition Examination Survey Data I Epidemiologic Follow-up Study (NHEFS).  As per \cite{hernan2023causal}, the NHEFS was jointly initiated by the National Center for Health Statistics and the National Institute on Aging in collaboration with other agencies of the United States Public Health Service. The NHEFS dataset has been studied by \cite{hernan2023causal} and \cite{ertefaie2020nonparametric}. The subset of the NHEFS data we consider consists of $N=1561$ cigar\tcr{e}tte smokers aged $25-74$ years, who had a baseline visit and a follow-up visit approximately 10 years later. We consider two types of product (joint) treatment indicators $R_1^{(1)},R_1^{(2)}\in\{0,1\}$: $R^{(1)}=1$ denotes that the individual has not \tcr{quit} 
smoking before the follow-up visit and has not drunk alcohol before the baseline visit, and $R^{(1)}=0$ otherwise; $R^{(2)}=1$ denotes that the individual has \tcr{quit} 
smoking before the follow-up visit and has not drunk alcohol before the baseline visit, and $R^{(2)}=0$ otherwise. We omit 5 individuals whose alcohol drinking information was missing. The weight gain (in kg), $Y\in\R$, was measured as the body weight at the follow-up visit minus the body weight at the baseline visit. Same as in \cite{hernan2023causal} and \cite{ertefaie2020nonparametric}, the following 9 confounding variables, $\bX$ are considered: sex (0: male, 1: female), age (in years), race (0: white, 1: other), education (5 categories), intensity and duration of smoking (number of cigarettes per day and years of smoking), physical activity in daily life (3 categories), recreational exercise (3 categories), and weight (in kg).

We estimate the ATE of the product (joint) treatments $R^{(1)}$ and $R^{(2)}$ on weight gain. The ATE estimators $\thetahat_\mathrm{ATE}^{(1)}$ and $\thetahat_\mathrm{ATE}^{(2)}$ are constructed using \eqref{ATEhat}, based on samples $\S^{(1)}:=(Y_i,R_i^{(1)},\bX_i)_{i=1}^N$ and $\S^{(2)}:=(Y_i,R_i^{(2)},\bX_i)_{i=1}^N$, respectiv\tcr{e}ly. Recall that the sample size is $N=1561$, and the \tcr{dimension} 
(after converting categorical variables) is $p=12$. The estimated proportions of the treated groups are $\pihat_N^{(1)}:=N^{-1}\sum_{i=1}^NR_i^{(1)}=0.088$ and $\pihat_N^{(2)}:=N^{-1}\sum_{i=1}^NR_i^{(2)}=0.037${\crevyz, which clearly indicates that the treatment groups are very unbalanced}. 
We consider \jelena{three} PS estimators: a constant estimator, an offset based logistic estimator with a Lasso-type penalty (log-Lasso), and a random forest (RF). We consider \jelena{four} outcome models: a Lasso estimator, a degree-2 polynomial estimator \jelena{without interactions} and with a Lasso-type penalty (poly-Lasso), a random forest (RF), and a reproducing kernel Hilbert space (RKHS) estimator.  We co\tcr{m}pare the proposed estimators with naive emp\tcr{i}rical difference (empdiff) estimators, $(\sum_{i=1}^NR_i^{(j)})^{-1}\sum_{i=1}^NR_i^{(j)}Y_i-\{\sum_{i=1}^N(1-R_i^{(j)})\}^{-1}\sum_{i=1}^N(1-R_i^{(j)})Y_i$, for $j=1,2$. To reduce the randomness coming from the sample splitting, we repeat the sample splitting for $B=10$ times and report the median of the ATE estimators based on each split. The asymptotic variance is then estimated by \tco{a plugged-in}
version using the mean estimators as well as the asymptotic variance estimators based on each split\tcr{; see} more details \tcr{of this technique} in Definition 3.3 of \cite{chernozhukov2018double}.

We report the ATE estimators, the corresponding $95\%$ confidence intervals, and the length of the confidence intervals in the Table  \ref{NHEFS00}. We can see negative estimated ATEs for $\theta_\mathrm{ATE}^{(1)}$ and positive estimated ATEs for $\theta_\mathrm{ATE}^{(2)}$. Moreover,  our proposed ATE estimators are close to each other and fairly different from the empirical difference estimator, especially  for $\theta_\mathrm{ATE}^{(2)}$. Therein,  all  our confidence intervals  do not include $0$ while the \tcr{one} 
based on the empirical difference does. The difference between our proposed ATE estimators and the empirical difference estimator
\tcr{seems to suggest presence of substantial confounding via $\bX$, and a significant causal effect of the treatment on the response after adjusting for the confounding.}

\begin{table}[h!]
\setlength{\tabcolsep}{3.5pt}
\def\arraystretch{1.5}
\setlength{\belowrulesep}{0.5pt}
\centering
\caption{\tco{\underline{Real data analysis}: estimation and inference of $\theta_\mathrm{ATE}^{(1)}$ and $\theta_\mathrm{ATE}^{(2)}$. We compare a naive empirical difference (empdiff) estimator with our proposed estimators based on various choices of nuisance estimators. ATE: the estimated average treatment effect; CI: a $95\%$ confidence interval; Leng\tcr{t}h: length of the $95\%$ confidence interval.}} \label{NHEFS00}
\vspace{0.6em}
\resizebox{!}{1.8in}{%
\begin{tabular}{llcccllcccc}
\toprule
  \multicolumn{3}{c}{ } &\multicolumn{3}{c}{  \cellcolor{gray!50}  $\theta_\mathrm{ATE}^{(1)}$} & & \multicolumn{3}{c}{  \cellcolor{gray!50}  $\theta_\mathrm{ATE}^{(2)}$} \\
  \cline{4-6}  \cline{8-10}
$\pihat_N(\cdot)$&$\mhat(\cdot)$&&ATE&CI&Length & &ATE&CI&Length \\
\hline
\multicolumn{2}{c}{empdiff}&&-2.003&(-3.282,-0.725)&2.558&&1.867&(-0.642,4.377)&5.019\\
\cdashline{1-2} \cdashline{4-6}\cdashline{8-10}
\multirow{4}{*}{constant}&Lasso&&-1.935&(-3.219,-0.651)&2.568&&4.209&(1.743,6.676)&4.933\\
&poly-Lasso&&-1.865&(-3.152,-0.578)&2.574&&3.291&(0.719,5.864)&5.145\\
&RF&&-1.729&(-2.992,-0.466)&2.526&&3.095&(0.607,5.584)&4.977\\
&RHKS&&-1.941&(-3.227,-0.655)&2.573&&3.183&(0.642,5.723)&5.081\\
\cdashline{2-2}  \cdashline{4-6}\cdashline{8-10}
\multirow{4}{*}{log-Lasso}&Lasso&&-1.967&(-3.518,-0.416)&3.102&&4.780&(1.954,7.605)&5.650\\
&poly-Lasso&&-1.717&(-3.321,-0.113)&3.207&&3.890&(0.804,6.976)&6.172\\
&RF&&-1.873&(-3.424,-0.322)&3.102&&3.890&(0.955,6.825)&5.870\\
&RHKS&&-2.051&(-3.663,-0.440)&3.223&&4.221&(1.068,7.375)&6.307\\
\cdashline{2-2}  \cdashline{4-6}\cdashline{8-10}
\multirow{4}{*}{RF}&Lasso&&-1.727&(-2.970,-0.484)&2.486&&4.932&(1.518,8.345)&6.827\\
&poly-Lasso&&-1.612&(-2.878,-0.346)&2.532&&4.693&(0.763,8.622)&6.827\\
&RF&&-1.608&(-2.845,-0.371)&2.474&&4.456&(0.942,7.970)&7.027\\
&RHKS&&-1.772&(-3.016,-0.528)&2.487&&4.411&(0.621,8.200)&7.579\\
\bottomrule
\end{tabular}
}
\end{table}


\section{Discussion}\label{discussion}

In this paper, we study the mean estimation problem in the semi-supervised setting with a decaying \tco{PS} \jelena{while} \tcr{allowing for selection bias in the labeling mechanism. To our knowledge this is one of the first full-hearted attempts in extending the SS inference literature to the case of selection bias, and that too in a very general way, as well as the MAR literature to the case of a (uniformly) decaying PS.} The \tcr{proposed DRSS} mean estimator is based on estimators of the outcome and the decaying \tcr{PS} 
models. We establish estimation and inference results  under different cases of the correctness of the models, \jelena{while allowing}  flexible \jelena{model} choices, including high-dimensional and non-parametric methods. \tcr{The subtleties of the problem setting and the non-standard asymptotics, among others, make the method and its analyses challenging and our results reveal several novel insights in the process. In particular, we} find that the consistency rate of the proposed estimator depends on the (expected) size of the labeled sample and the tail of the \tco{PS} distribution. Throughout the paper, $Na_N$ (recall that $a_N=[\E\{\pi_N^{-1}(\bX)\}]^{-1}$) is a crucial value, \tcr{in} that it serves as the ``effective sample size'' in our \tcr{MAR-SS setting with a decaying PS.} 
\jelena{This paper provides details as to why this happens.}

\tcr{As a necessary component of analyzing the MAR-SS setting, we} further propose 
estimators \tcr{of the decaying PS under three different} 
models: MCAR, stratified labeling, and a \tcr{novel} offset logistic model, \tcr{under both high and low dimensional settings.} The consistency rates of the \tcr{PS} 
models are established\tco{,} \jelena{which are }\tcr{of independent interest}.
We also extend our methods to an ATE estimation problem where the treatment groups can be extremely imbalanced. We provide extensive numerical studies to illustrate our results in finite-sample simulations, \tcr{as well as a real data analysis using the NHEFS data}.

The semi-supervised decaying \tco{PS} setting is an interesting scenario that occurs in numerous applications \tcr{in the modern era, and yet has been largely under-studied so far}. We provide a detailed analysis of the mean estimation problem under such  setting. \tcr{We hope it serves as a start towards understanding this practically very relevant, and yet technically challenging, scenario in all its subtleties}, \jelena{therefore opening doors to many new questions where inferential results need to be adjusted for the ``effective sample size''.}

\appendix

\phantomsection
\addcontentsline{toc}{section}{Supplementary Material}

\section*{Supplementary Material}\label{supp_mat}

\textbf{Supplement to ``Double Robust Semi-supervised Inference for the Mean: Selection Bias under MAR Labeling with Decaying Overlap''}.
In the supplement (Appendices \ref{sec:offset}-\ref{Ssec:proof}), we present additional results and discussions, as well as all proofs of the main theoretical results. \tco{We provide some fu\tcr{r}ther discussions on the offset based \tco{PS} model in Appendix \ref{sec:offset}. In Appendix \ref{sec:MCAR}, we provide additional results on the MCAR \tco{PS} model. We illustrate additional numerical simulation results for the \tco{adjusted} confidence intervals in Appendix \ref{sim:modCI}. The proofs of the theoretical results are included in Appendix \ref{Ssec:proof}.}

\phantomsection
\addcontentsline{toc}{section}{Data Availability Statement}
\section*{Data Availability Statement}



 
The NHEFS data are available at \href{https://www.hsph.harvard.edu/miguel-hernan/causal-inference-book/}{https://www.hsph.harvard.edu/miguel-hernan/causal-inference-book/}. This data set, along with the R codes for all the simulation studies as well as real data analysis presented in this article, are available separately at \href{https://github.com/yuqianruc/DRSS-code}{https://github.com/yuqianruc/DRSS-code}.

\phantomsection
\addcontentsline{toc}{section}{Funding Acknowledgement}
{\crevmag
\section*{Funding Acknowledgement}\label{funding}

AC's research was supported in part by the National Science Foundation grant NSF DMS-2113768.
JB's research was supported in part by the National Science Foundation grant NSF DMS-1712481.

\par\medskip
\noindent
The authors would also like to thank the Editor, the anonymous Associate Editor and the two Reviewers for their constructive
comments and useful suggestions that helped significantly improve the article.
}

\phantomsection
\addcontentsline{toc}{section}{References}



\appendix
\setcounter{section}{0}

\renewcommand\theequation{\thesection.\arabic{equation}}  

\section*{Supplementary Material to ``Double Robust Semi-Supervised Inference for the Mean: Selection Bias under MAR Labeling with Decaying Overlap''}

\tco{This \tcr{supplementary} document (Appendices~\ref{sec:offset}--\ref{Ssec:proof}) 
 contains additional results and \tcr{discussions} \tcr{that could not be accommodated in the main manuscript}, as well as the proofs of the main theoretical results.}  All results and notation are numbered and used as in the main text unless stated otherwise.


\paragraph*{\jelena{Organization}}
\phantomsection
\tco{The rest of the document is organized as follows. 
\tcr{First, we} introduce the (additional) notations we will use in the document.
In Appendix \ref{sec:offset}, we provide some fu\tcr{r}ther discussions on the offset based propensity score \tco{(PS)} model. In Appendix \ref{sec:MCAR}, we provide additional results on the Missing completely at random (MCAR) \tco{PS} model. In Appendix \ref{sim:modCI}, we illustrate additional numerical simulation results for the \tco{adjusted} confidence intervals. The proofs of the theoretical results are included in Appendix \ref{Ssec:proof}.}

\paragraph*{Notation}\label{Ssec:notation}

\tco{Constants $c,C>0$, independent of $N$ and $p$, may change values from one line to the other.
For any $\tilde{\S}\subseteq\S=(\bZ_i)_{i=1}^N$, define $\P_{\tilde{\S}}$ as the joint distribution of $\tilde{\S}$ and $\E_{\tilde{\S}}(f)=\int fd\P_{\tilde{\S}}$. For any $r > 0$, let $ \| f(\cdot) \|_{r,\P} := \{\E | f(\bZ)|^r\}^{1/r}$. We abbreviate ``with probability approaching \jelena{one}'' and ``almost surely'' by ``w.p.a. 1'' and ``a.s.'', respectively.}

\section{Further discussions on the offset based \tco{PS} model} \label{sec:offset}

\tco{We provide further discussions \tcr{here} on the offset based \tcr{PS} 
model proposed in Section \ref{logistic} of the main file}.
\begin{remark}[\tcr{Rationale behind the offset model \eqref{offset_model} and its connections with a diverging intercept model}] \label{remark:offset}
Instead of an offset based model \tcr{as in} \eqref{offset_model}, let us now directly consider a standard logistic model for $\P(R=1|\bX) = \pi_N(\bX)$ but allowing (necessarily) for a diverging intercept, given by:
\begin{align}
\pi_N(\bX)~=~g(\bXv^T\bbeta)~=~ \frac{\exp(\bXv^T\bbeta)}{1+\exp(\bX^T\bbeta)}, \;\; \mbox{where} \label{divmod-def-supp} 
\end{align}
$\bbeta=(\bbeta(1),\bbeta(-1)^T)^T\in\R^{p+1}$ is a vector allowed to depend on $N$; \tcr{e.g.,} see \cite{owen2007infinitely} and \cite{wang2020logistic}. For further simplification, let us 
\tcr{assume} that the slope $\bbeta(-1)$, while allowed to depend on $N$, is  finite, i.e.\tco{,} $\| \bbeta(-1)\|_2 < C < \infty$ for some $C$ independent of $N$.

Under the model \eqref{divmod-def-supp}, the following holds.
Let $\mbox{MGF}_{\bX}(\bv) := \E\{ \exp(\bv^{T} \bX)\}$ denote the moment generating function (MGF) of $\bX$ at $\bv \in \R^p$ and assume $\mbox{MGF}_{\bX}(\bv)$ exists (i.e.\tco{,} finite) at $\bv = \bbeta(-1)$ and $\bv = -\bbeta(-1)$. Then, the following holds for the intercept $\bbeta(1)$:
\begin{align}
& \frac{1}{\bar\pi_N} \frac{1-\bar\pi_N}{\mbox{MGF}_{\bX}(-\bbeta(-1))} \; \leq \; \exp(-\bbeta(1)) \; \leq \;   \frac{1}{\bar\pi_N} \mbox{MGF}_{\bX}(\bbeta(-1)), \;\; \mbox{and consequently}, \label{divmod-int-bound1} \\
& \frac{1}{\bar\pi_N} \frac{1-\bar\pi_N}{\E\{ \exp (\|\bbeta(-1)\|_2 \|\bX\|_2) \}} \; \leq \; \exp(-\bbeta(1)) \; \leq \;   \frac{1}{\bar\pi_N} \E\{ \exp (\|\bbeta(-1)\|_2 \|\bX\|_2) \}. \label{divmod-int-bound2}
\end{align}
\tcr{For the special case of a Gaussian $\bX$, i.e.}\tco{,} $\bX \sim \Nsc_p(\bzero,\bSigma)$,
$$\mbox{MGF}_{\bX}(-\bbeta(-1)) = \mbox{MGF}_{\bX}(\bbeta(-1)) \leq \exp\{\|\bbeta(-1) \|_2^2 \lambda_{\max}(\bSigma)\}.$$ Hence, as long as $\|\bbeta(-1) \|_2^2 < C < \infty$ and $\lambda_{\max}(\bSigma) < \infty$, then using \eqref{divmod-int-bound1}, $\exp(-\bbeta(1)) \asymp \bar\pi_N^{-1} \rightarrow \infty$.
More generally, if $\|\bbeta(-1) \|_2^2 < C < \infty$ and $\E\{ \exp(C\| \bX \|_2)\} < \infty$ (\tcr{e.g.}\tco{,} if $\bX$ is sub-Gaussian), then using \eqref{divmod-int-bound2}, we will have  $\exp(-\bbeta(1)) \asymp \bar\pi_N^{-1} \rightarrow \infty$.

\emph{Rationale \tcr{for} the offset model \eqref{offset_model}.} The result clearly shows that the intercept $\bbeta(1)$ diverges to $-\infty$ and does so \emph{precisely} at a rate of $\log (\bar\pi_N)$, i.e.\tco{,} $c_1 + \log\tcr{(\bar\pi_N)} \leq \bbeta(1) \leq C_1 + \log\tcr{(\bar\pi_N)}$.
This provides a clear justification for our offset based model \eqref{offset_model} where we precisely extract out this $\log\tcr{(\bar\pi_N)}$ as an offset (to be estimated separately and plugged in apriori to the sample likelihood equation), and then treat the  intercept $\alpha_0$ and the slope parameter $\bbeta_0$ to be well-behaved, i.e.\tco{,} finite and independent of $N$ (or at least bounded in $N$).

This makes the parameter space 
\tcr{more} amenable to theoretical analysis where it is common practice to assume that the truths (the true \emph{unconstrained} minimizers) lie as interior points of some compact set. Such assumptions are commonplace in most of empirical process and \tcr{$M$-}estimation theory, and these results won't be applicable without this assumption, something that has clear justification under the offset model but not under the diverging intercept model.
\end{remark}

\begin{remark}[Connections with density ratio estimation]\label{rem:densityratio-supp}
\tcr{It is interesting (though elementary) to note that t}he \tco{PS} is also related to the density ratio \tcr{of $\bX$ (given $R = 0$ or $1$), in} that
$$\Lambda_N(\bX)~:=~\frac{f(\bX|R=0)}{f(\bX|R=1)}~=~\frac{\P(R=0|\bX)\P(R=1)}{\P(R=1|\bX)\P(R=0)}~=~\frac{\{1-\pi_N(\bX)\}\bar\pi_N}{\pi_N(\bX)(1-\bar\pi_N)},$$
where $f(\cdot|R=\cdot)$ is the conditional density function 
of $\bX$ \tcr{given} 
$R$. The density ratio is usually used in the \tcr{so-called} ``covariate shift'' setting in \tco{semi-supervised learning (SS\tcr{L) and missing data}}, where $R_i$'s are treated as fixed (or condition\tcr{ed} 
 on) and $\P_{\bX}\neq\P_{\bX|R=1}$ is allowed; see for example \cite{kawakita2013semi}, \cite{liu2020doubly}, and Section 4 of \cite{kallus2020role}.

Here, we discuss a simple and fairly obvious connection of the offset model \eqref{offset_model} to a corresponding model for density ratio estimation. The analysis here can actually be seen to be model-free and non-parametric. Observe that
$$\mbox{logit}\{\pi_N(\bX)\}~=~\log(\bar\pi_N) -\log(1-\bar\pi_N)-\log\{\Lambda_N(\bX)\}.$$
The standard approach to modeling the density ratio is to model $\log \{\Lambda_{\tcr{N}}(\bX) \}$ through 
basis function expansion based on \tcr{some} basis functions $\{\phi_j(\bX)\}_{j=1}^d$ (e.g.\tco{,} the linear bases will lead to standard parametric forms). But this in general can be difficult to implement in practice. However, the above representation suggests that the \emph{same} model can be fitted by simply using a logistic regression model for $R|\bX$ with covariates as the \emph{same} basis functions, and further using $\log\{\bar\pi_N/(1-\bar\pi_N)\}$ as an \emph{offset} (which can be estimated separately and plugged in apriori into the likelihood equation). This provides a simple and flexible regression modelling approach to estimate the density ratio. Our offset based model \eqref{offset_model} precisely implements such a model (albeit we had different motivations to consider it), and therefore provides a way to estimate the density ratio as well. This is a key quantity involved (as a nuisance function) in the semi-parametric efficiency bound for our parameter; see Theorem 4.1 of \cite{kallus2020role}. Our approach provides an automated and agnostic way of bypassing its estimation through a theoretically equivalent but practically more flexible regression modeling approach. 

A discussion similar to above can be found in Section 1 of \cite{Qin_1998} who further proves that the estimation approach as above corresponds to \tcr{an} optimal choice of the estimating equation for estimating the density ratio among the class of all such equations. It is also interesting to note that the \tco{semi-supervised (SS)} setting actually bears a very close relation to so-called \emph{case-control study designs} (which are \emph{retrospective} designs, as opposed to the \emph{prospective} cohort studies that we usually consider), since here the labeling indicator $R$ is typically treated as \emph{non-random (or conditioned)}, which is similar in spirit to case-control designs (with $R$ being replaced by case/control status). For statistical analyses of these kind of studies, density ratio estimation models are often required and an estimation strategy via a logistic regression model of the \tco{PS}, similar as above, is often employed; see  Section 1 of \cite{Qin_1998} for more discussions. 

\end{remark}

\begin{remark}[Connection with the maximum likelihood estimate (MLE) of the model \eqref{divmod-def-supp}]
In fact, there is an one-one correspond\tcr{en}ce between $\bgammahat$ (we suppress the dependence on $k$ for a moment) and the MLE of the model \eqref{divmod-def-supp}: if $(\bbetahat(1), \bbetahat(-1))$ denotes a sample MLE, i.e.\tco{,} a solution (assuming it exists) to the (sample) likelihood equation for the model  \eqref{divmod-def-supp}, then $\bgammahat = (\bbetahat(1) - \log \pihat_N, \bbetahat(-1))$ is a sample MLE for the model \eqref{offset_model}. Conversely, if $\bgammahat$ is a sample MLE for the model \eqref{offset_model}, then $(\bbetahat(1),\bbetahat(-1)) = (\bgammahat(1) + \log(\pihat_N), \bgammahat(-1))$ is a sample MLE for the model \eqref{divmod-def-supp}. All these claims are straightforward to show by means of direct verification.
\end{remark}

\begin{remark}[Existence and uniqueness of $\bgammahat$]\label{remark:existence}
The uniqueness of $\bgammahat$ is a direct consequence of the convexity of the sample log-lik\tcr{eli}hood. As for the existence, we appeal to the one-one correspondence between $\bgammahat$ and the sample MLE of the model \eqref{divmod-def-supp}. We further use the results of \cite{owen2007infinitely} who demonstrated the existence of the sample MLE for the model \eqref{divmod-def-supp} under a fairly mild (sample) overlap condition; see Lemma 5 therein. Note that \cite{owen2007infinitely} shows this result for a slightly modified version of the log-likelihood wherein the empirical average over unlabeled data is replaced by an expectation (assuming $N$ is very large). But the same proof technique could be applied to the actual log-likelihood along with a corresponding appropriate modification of the (sample) overlap condition to conclude the existence \tcr{of} 
the sample MLE for model \eqref{divmod-def-supp}. Consequently, this also establishes the existence of the sample MLE for the offset model \eqref{offset_model}.
\end{remark}

\def\tcr{} 
\def\tco{} 

\section{Missing completely at random \jelena{(MCAR)} {\crevmag labeling: Theory and comparisons}}\label{MCAR} \label{sec:MCAR}

\tco{Apart from the offset based model and the stratified labeling model discussed in Sections \ref{logistic} and \ref{stratified}, a simple but commonly used \tco{PS} model would be a \tco{MCAR} mechanism. In this section, we consider \tcr{this} special \tcr{MCAR} mechanism \tcr{with}  
 $\pi_N(\cdot)\equiv\bar\pi_N$, \tcr{and derive the properties of $\thetahat_\mathrm{DRSS}$ including an adjusted \tco{regular and asymptotically linear (RAL)} expansion allowing for misspecification of $\mhat(\cdot)$.}}
In this case, a cross-fitted estimator of the \tco{PS} is proposed as \tco{$\pihat_N(\bX_i)=N_{-k}^{-1}\sum_{i\in\mathcal I_{-k}}R_i$ for any $i\in\mathcal I_k$, where $N_{-k}:=|\mathcal I_{-k}|$ and $\mathcal I_{-k}:=\mathcal I\setminus\mathcal I_k$}. Based on such a MCAR \tco{PS} estimator, we have the following result on the conditions and conclusions in Theorem \ref{t4}.
\begin{theorem}\label{thm:ex1}
Assume $\pi_N(\bX)\equiv\bar\pi_N$, $N\bar\pi_N\to\infty$ as $N\to\infty$, $\|m(\cdot)-\mu(\cdot)\|_{2,\P}<\infty$ and $\|\mhat(\cdot)-\mu(\cdot)\|_{2,\P}=o_p(1)$. Then, $a_N=\bar\pi_N$ and
$$
\E\left[ \frac{a_N}{\pi_N(\bX)} \left\{1 - \frac{\pi_N(\bX)}{\pihat_N(\bX)} \right\}^2\right]~=~\E\left\{1 - \frac{\pi_N(\bX)}{\pihat_N(\bX)} \right\}^2 ~=~ O_p\left((N\bar\pi_N)^{-1}\right).
$$
Furthermore,
\begin{align}
\thetahat_\mathrm{DRSS}-\theta_0~=~&N^{-1}\sum_{i=1}^N \Psi(\bZ_i) + o_p\left( (Na_N)^{-1/2}\right), \quad \mbox{where} \;\; \Psi(\bZ) ~:=~ \psi_{\mu}(\bZ) + \mathrm{IF}_{\pi}(\bZ),\label{def:Psi_Z}\\
\psi_{\mu}(\bZ) ~=~& \frac{R}{\bar\pi_N}\{Y - \mu(\bX)\} + \mu(\bX) - \theta_0,\nonumber\\
\mathrm{IF}_{\pi}(\bZ) ~:=~& \left(\frac{R - \bar\pi_N}{\bar\pi_N}\right)\Delta_{\mu},\quad\Delta_{\mu} ~:=~ \E \{ \mu(\bX) - m(\bX) \}.\nonumber
\end{align}
\end{theorem}
Note that Theorem \ref{thm:ex1} still holds if the \tco{PS} is estimated without cross-fitting that $\pihat_N(\bX)\equiv\pihat_N=n/N$, where $n=\sum_{i=1}^NR_i$.
\begin{remark}\label{remark:RAL_MCAR}
The modification on the RAL expansion of the mean estimator is needed only when $\Delta_{\mu}= \E \{ \mu(\bX) - m(\bX) \}\neq0$. Recall Remark \ref{remark:RAL_logsitic}; if the outcome model is fitted by a linear model that $\mu(\bX)=\bXv^T\bbeta^*$, where $\bXv=(1,\bX^T)^T$, $\bbeta^*=\arg\min_{\bbeta\in\R^{p+1}}\E\{(Y-\bXv^T\bbeta)^2\}=\{\E(\bXv\bXv^T)\}^{-1}\E(\bXv Y)$ is the optimal population slope. Then, we have $\Delta_{\mu}= \E \{ \mu(\bX) - m(\bX) \}=\E(\bXv^T)\{\E(\bXv\bXv^T)\}^{-1}\E(\bXv Y)-\E(Y)=0$. This suggests that, the RAL modification is unnecessary when $\mhat(\cdot)$ converges to the linear projection. In other words, for such cases, the original asymptotic normality \eqref{norm:main} still holds even if $\mu(\cdot)\neq m(\cdot)$ and it coincides with the results in \cite{zhang2019semi,zhang2022high}. Classical examples for such $\mhat(\cdot)$ include least squ\tcr{a}res (LS) estimator and regularized least squ\tcr{a}res such as Lasso and ridge under appropriate conditions.
\end{remark}
\paragraph*{\tcr{Reconciliation with ``traditional'' SS inference literature under MCAR}} Now, we consider the ``traditional'' \tco{SS} setting \tcr{where} 
all the $R_i$'s are considered \emph{deterministic} (or condition\tcr{ed}) apart from an underlying MCAR assumption. \tcr{Under this} 
\tco{SS} setting, we consider the \tco{SS} mean estimator proposed in \cite{zhang2022high}. In fact, their estimator is a special case of our \tco{double robust SS (\tcr{DRSS})} mean estimator \tcr{$\thetahat_\mathrm{DRSS}$} except that the \tco{PS} is estimated without cross-fitting, \tcr{i.e.}\tco{,} $\pihat_N(\bX)\equiv\pihat_N=n/N$. {\crev Moreover, the {\crevmag SS mean estimator}
proposed by \cite{zhang2019semi} {\crevmag based on a linear outcome model estimated via least squares} can be further seen as a special case of the SS estimator of \cite{zhang2022high} -- the latter estimator allows the usage of flexible outcome models (including high-dimensional or non-parametric models), whereas the former one is restricted to 
{\crevmag a linear} outcome nuisance model in low dimensions {\crevmag estimated via least squares}. In addition, since \cite{zhang2019semi} only focus on low-dimensional parametric estimators, their outcome function is also estimated without cross-fitting, whereas a cross-fitting technique is required in \cite{zhang2022high} to estimate the outcome function. In the following, we compare our proposed mean estimator{\crevyz s} with the latest SS mean estimator proposed by \cite{zhang2022high}.}

{\crevyz Under this SS setting, the semi-supervised mean estimator proposed by \cite{zhang2022high}, denoted as $\thetahat_\mathrm{SS}$, has the following RAL expansion: 
\begin{align}
\thetahat_\mathrm{SS}-\theta_0~=~&N^{-1}\sum_{i=1}^N\psi_{\mu,\mathrm{SS}}(\bZ_i)+o_p(n^{-1/2}),\;\;\text{where}\label{RAL_muhat}\\
\psi_{\mu,\mathrm{SS}}(\bZ)~=~&\frac{NR}{n}\{Y-\mu(\bX)\}+\mu(\bX)- \theta_0.\label{def:psi_mu_SS}
\end{align}
Here, conditional on $R_i$'s, $\{\psi_{\mu,\mathrm{SS}}(\bZ_i)\}_{i=1}^N$ are independent and identically distributed\tcr{, with mean zero}.

\paragraph{Variance comparison} In the following, we present an interesting comparison of the mean estimators' asymptotic variances under the following three cases:
\begin{enumerate}
\item[(a)] $R_i$'s are considered as random (MCAR), $\bar\pi_N$ is known, and the mean estimator $\tilde\theta$, defined as \eqref{theta:knownpi}, is based on the true \tco{PS} $\bar\pi_N$; 
\item[(b)] $R_i$'s are considered as random (MCAR), the mean estimator $\thetahat_\mathrm{DRSS}$, defined as \eqref{theta:unknownpi} and studied in Theorem \ref{thm:ex1}, is based on the cross-fitted constant estimate that $\pihat_N(\bX_i)=|\S_{-k}|^{-1}\sum_{i\not\in\mathcal I_k}R_i$ for $i\in\mathcal I_k$; and
\item[(c)] $R_i$'s are considered as fixed (SS) \tcr{and} the semi-supervised mean estimator $\thetahat_\mathrm{SS}$ is as defined in \cite{zhang2022high}.
\end{enumerate}
Recall that $\psi_{\mu,\pi}(\bZ)$, $\Psi(\bZ)$, and $\psi_{\mu,\mathrm{SS}}(\bZ)$, defined in \eqref{psi1}, \eqref{def:Psi_Z}, and \eqref{def:psi_mu_SS}, are the IFs for the estimators considered in the above cases (a), (b), and (c), respectively; see the RAL expansions provided in Theorems \ref{t3}, \ref{thm:ex1}, and \eqref{RAL_muhat}. Hence, the considered estimators have the following asymptotic variances:
\begin{align*}
\text{(a)}&\;\;\Var\{\psi_{\mu,\pi}(\bZ)\}~=~\frac{\E\{Y-\mu(\bX)\}^2}{\bar\pi_N}-\Delta_\mu^2+\Var\{\mu(\bX)\}+2\mathrm{Cov}\{Y-\mu(\bX),\mu(\bX)\},\\
\text{(b)}&\;\;\Var\{\Psi(\bZ)\}~=~\frac{\E\{Y-\mu(\bX)\}^2}{\bar\pi_N}-\frac{\Delta_\mu^2}{\bar\pi_N}+\Var\{\mu(\bX)\}+2\mathrm{Cov}\{Y-\mu(\bX),\mu(\bX)\},\;\;\text{and}\\
\text{(c)}&\;\;\Var\{\psi_{\mu,\mathrm{SS}}(\bZ)|(R_i)_{i\in\mathcal I}\}~=~\frac{\E\{Y-\mu(\bX)\}^2}{\pihat_N}-\frac{\Delta_\mu^2}{\pihat_N}+\Var\{\mu(\bX)\}+2\mathrm{Cov}\{Y-\mu(\bX),\mu(\bX)\}.
\end{align*}
We can see that, $\Var\{\psi_{\mu,\pi}(\bZ)\}=\Var\{\Psi(\bZ)\}+(\bar\pi_N^{-1}-1)\Delta_\mu^2\geq\Var\{\Psi(\bZ)\}$, i.e., the asymptotic variance of $\tilde\theta$ is larger than (or equal to) the asymptotic variance of $\thetahat_\mathrm{DRSS}$. It suggests that, under the MCAR setting, even if $\bar\pi_N$ is known, it is s\tcr{ti}ll worth \tcr{estimating} 
$\bar\pi_N$ instead of directly plug\tcr{ging} in the true value $\bar\pi_N$ as long as $\Delta_\mu\neq0$. As for the asymptotic variance of $\thetahat_\mathrm{SS}$ under the \tco{SS} setting, notice the fact that
$$\frac{\bar\pi_N}{\pihat_N}-1~=~O_p\left((N\bar\pi_N)^{-1/2}\right).$$
Hence, $\Var\{\psi_{\mu,\mathrm{SS}}(\bZ)|(R_i)_{i\in\mathcal I}\}=\Var\{\Psi(\bZ)\}\{1+O_p((N\bar\pi_N)^{-1/2})\}=\Var\{\Psi(\bZ)\}\{1+o_p(1)\}$.}


\section{Inference results based on \tco{adjusted} confidence intervals}\label{sim:modCI}

In Section \ref{numerical}, we illustrate\tcr{d} the simulation performance of the proposed confidence interval \eqref{CI}, which requires both the outcome and \tco{PS} models \tcr{to be} 
correctly specified, as in part (a) of Theorem \ref{t4}. In this section, we compare \eqref{CI} with \tco{an adjusted} version based on \tco{the asymptotic 
expansion} in part (b) of Theorem \ref{t4} \tco{and the RAL expansion in Remark \ref{remark:RAL_mis}.}
The \tco{adjusted} confidence interval allows \tcr{inference via $\thetahat_\mathrm{DRSS}$ even under} misspecified outcome models, and the \tco{adjusted} RAL expansions based on different \tco{PS} models are provided in Theorems \ref{thm:ex3}, \ref{thm:ex2}, and \ref{thm:ex1}. Here, we only focus on the \tcr{results for the} sk\tcr{e}wed \tcr{offset} logistic \tcr{PS} model as discussed in Theorem \ref{thm:ex3}, \tcr{and we present numerical results to validate the inference provided by the \tco{adjusted} RAL expansion \eqref{modRAL} of $\thetahat_\mathrm{DRSS}$ given therein}.

Apart from the settings c and d in Section \ref{sim:main}, an additional DGP, Setting f: P2+O3, is considered. Here, P2 is the \tco{offset} logistic \tco{PS} model as in Section \ref{sim:main}, and O3 is a cubic outcome model defined as follows:
\begin{enumerate}
\item[] O3. (Cubic outcome) $Y_i=\bXv_i^T\bbeta_0+\sum_{j=1}^{p+1}\balpha_0(j)\bXv_i(j)^2+\sum_{j=1}^{p+1}\bzeta_0(j)\bXv_i(j)^3+\varepsilon_i$.
\end{enumerate}
The parameter value is defined as:
$$\bzeta_0~=~(0,0.2,0.2,0.2,\boldsymbol{0}_{1\times(p-3)})^T.$$

We illustrate the behavior of the original confidence interval \eqref{CI} and the \tco{adjusted} confidence interval based \tcr{on} the RAL expansion \eqref{modRAL}. We consider the Settings c, d, and f, \tcr{where} 
the outcome models are polynomial (without interaction) with degrees 1, 2, and 3, \tcr{respectively}. Apart from the emp\tcr{i}rical estimator $\bar Y_\mathrm{labeled}$ and the oracle estimator as in Section \ref{sim:main}, we also consider the proposed mean estimators $\thetahat_\mathrm{DRSS}$ based on an offset \tcr{logistic model based PS} 
estimator, and polyn\tcr{o}m\tcr{i}al \tcr{model based} outcome \tcr{regression} estimators with degrees 1, 2, and 3. The simulation results are \tcr{presented} in Table \ref{table:cde_p10}.

We can clearly see the improvement of the coverage based on the \tco{adjusted} confidence intervals, especially for polynomial estimators $\mhat(\cdot)$ with degrees 2 and 3. As mentioned in Remark \ref{remark:RAL_logsitic}, 
a latent misspecification arises \tcr{here} since the effective sample size $N\bar\pi_N=100$ is comparable with the dimension of the working model: for polynomial regression with degrees 2 and 3, the dimensions of the design matrix are 21 and 31, respectively. Under such a circumstance, $\mhat(\cdot)$ \tcr{tends to be} 
\tcr{a} 
biased estimate and a (latent) misspecification arises, \tcr{in}
that \tcr{its (effective) target (or limit) becomes some} $\mu(\cdot)\neq m(\cdot)$.

Such an example suggests that,  the \tco{adjusted} confidence intervals, when \tco{$\pi_N(\cdot)$} is correctly specified,   \tcr{allow us to better capture the model complexity} of $\mhat(\cdot)$ \jelena{and} improve the efficiency of the \tco{DRSS} estimator. \jelena{\tcr{The m}odified confidence intervals \tco{can \emph{still}}}  provide   valid inference  even when a degree  of freedom \tcr{of the model} becomes comparable with the effective sample size.

In Table \ref{table:cde_p10}, one can see that neither of the  \tco{averages of \tcr{the} estimated standard deviations (ASDs)} or the \tco{adjusted} ASDs are close to the \tco{emp\tcr{i}rical standard deviations (ESDs)} for \tcr{the DRSS} mean estimators based on polynomial regressions with degrees 2 and 3, while we can still \tcr{achieve} fairly acceptable coverages for the confidence intervals. This is not contradicted with our theory: we only obtain asymptotic results in terms of convergence in distribution \tcr{or} 
 probability, whereas $\mbox{ASD}=\mbox{ESD}+o(1)$ requires a convergence in mean \tcr{(i.e.\tco{,} $L_1$ convergence)}. Such a difference is \tcr{possibly related to} 
the instability of the LS-type outcome estimator, when the dimension of the working model is comparable with the sample size.

\begin{table}[h!]
\centering
\caption{Simulations \tcr{under Settings c, d and f}, with $p=10, N=10000$ and $\bar\pi_N=0.01 ~(N\bar\pi_N=100)$. Bias: emp\tcr{i}rical bias; RMSE: root mean square error; Length: average length of the $95\%$ confidence intervals; Coverage: average coverage of the $95\%$ confidence intervals; ESD: emp\tcr{i}rical standard deviation; ASD: average of estimated standard deviations. \tco{The results for \tco{adjusted (adj)} confidence intervals based on  \tcr{the} %
RAL expansion \tcr{\eqref{modRAL}} in Theorem \ref{thm:ex3} are provided in parentheses.} } \label{table:cde_p10}
\vspace{0.5em}
\resizebox{!}{1.5in}{%
\begin{tabular}{llccccccc}
\toprule
$\pihat_N(\cdot)$&$\mhat(\cdot)$&Bias&RMSE&Length(\tco{adj})&Coverage(\tco{adj})&ESD&ASD(\tco{adj})\\
\hline
\multicolumn{2}{c}{} & \multicolumn{6}{c}{  \cellcolor{gray!50} Setting c }\\
\cline{3-8}
\multicolumn{2}{c}{$\bar Y_\mathrm{labeled}$}&0.979&0.998&0.784&0.002&0.197&0.200\\
\multicolumn{2}{c}{oracle}&-0.007&0.164&0.607&0.972&0.164&0.155\\
\cdashline{3-8}
\multirow{3}{*}{{\color{blue}\bf logistic}}&{\color{blue}\bf poly1(LS)}&-0.009&0.227&0.865(0.881)&0.952(0.964)&0.227&0.221(0.225)\\
&poly2&0.002&0.277&1.017(1.039)&0.940(0.964)&0.277&0.260(0.265)\\
&poly3&-0.013&0.491&1.436(1.465)&0.922(0.956)&0.492&0.366(0.374)\\
\hline
\multicolumn{2}{c}{} & \multicolumn{6}{c}{  \cellcolor{gray!50} Setting d }\\
\cline{3-8}
\multicolumn{2}{c}{$\bar Y_\mathrm{labeled}$}&1.923&1.968&1.628&0.002&0.418&0.415\\
\multicolumn{2}{c}{oracle}&0.011&0.158&0.615&0.960&0.158&0.157\\
\cdashline{3-8}
\multirow{3}{*}{{\color{blue}\bf logistic}}&poly1(LS)&0.493&1.638&4.352(4.202)&0.908(0.936)&1.564&1.110(1.072)\\
&{\color{blue}\bf poly2}&-0.006&0.401&1.058(1.080)&0.916(0.954)&0.401&0.270(0.275)\\
&poly3&-0.013&0.562&1.457(1.483)&0.918(0.942)&0.563&0.372(0.378)\\
\hline
\multicolumn{2}{c}{} & \multicolumn{6}{c}{  \cellcolor{gray!50} Setting f }\\
\cline{3-8}
\multicolumn{2}{c}{$\bar Y_\mathrm{labeled}$}&2.613&2.670&2.182&0.000&0.549&0.557\\
\multicolumn{2}{c}{oracle}&-0.003&0.164&0.623&0.966&0.164&0.159\\
\cdashline{3-8}
\multirow{3}{*}{{\color{blue}\bf logistic}}&poly1(LS)&0.302&1.267&4.406(4.163)&0.914(0.918)&1.232&1.124(1.062)\\
&poly2&-0.018&0.584&1.752(1.800)&0.862(0.900)&0.584&0.447(0.459)\\
&{\color{blue}\bf poly3}&-0.005&0.410&1.279(1.316)&0.894(0.926)&0.410&0.326(0.336)\\
\bottomrule
\end{tabular}
}
\end{table}

\section{Proofs of main results}\label{Ssec:proof}
\subsection{Auxiliary lemmas}

The following Lemmas \jelena{will} be useful in the proofs.
\begin{lemma}\label{l1}
Let $(X_N)_{N\geq1}$ and $(Y_N)_{N\geq1}$ be sequences of random variables in $\mathbb R$. If $\E(|X_N|^r|Y_N)=O_p(1)$ for any $r\geq1$, then $X_N=O_p(1)$.
\end{lemma}

\begin{proof}[Proof of Lemma \ref{l1}]
For any $c>0$, there exists $C>0$ such that, for large enough $N$,
$$\mathbb P\left\{\E\left(X_N^r\bigr|Y_N\right)>C\right\}~<~c/2.$$
Let $\delta=(2C/c)^{1/r}$, then
\begin{align*}
&\mathbb P(|X_N|>\delta)=\E\left(\E\left[\mathbbm{1}{\{|X_N|>\delta\}}\bigr|Y_N\right]\right)\\
&\qquad=\E\left[\mathbbm{1}{\{\E(|X_N|^r\bigr|Y_N)\leq C\}}\E\left(\mathbbm{1}{\{|X_N|>\delta\}}\bigr|Y_N\right)\right]\\
&\qquad\qquad+\E\left[\mathbbm{1}{\{\E(|X_N|^r\bigr|Y_N)> C\}}\E\left(\mathbbm{1}{\{|X_N|>\delta\}}\bigr|Y_N\right)\right]\\
&\qquad\leq\E\left[\mathbbm{1}{\{\E(|X_N|^r\bigr|Y_N)\leq C\}}\E\left(\delta^{-r}|X_N|^r\bigr|Y_N\right)\right]+\E\left(\mathbbm{1}{\{\E(|X_N|^r\bigr|Y_N)> C\}}\right)\\
&\qquad=\delta^{-r}\E\left[\mathbbm{1}{\{\E(|X_N|^r\bigr|Y_N)\leq C\}}\E\left(|X_N|^r\bigr|Y_N\right)\right]+\mathbb P\left[\E(|X_N|^r\bigr|Y_N)> C\right]\\
&\qquad\leq c/2+c/2=c.
\end{align*}
That is, $X_N=O_p(1)$.
\end{proof}

\begin{lemma}[Lemma 6.1 of \cite{chernozhukov2018double}]\label{l2}
Let $(X_N)_{N\geq1}$ and $(Y_N)_{N\geq1}$ be sequences of random variables in $\R$. If for any $c>0$, $\mathbb P(|X_N|>c|Y_N)=o_p(1)$, then $X_N=o_p(1)$.
\end{lemma}

In particular, Lemma \ref{l2} occurs if $\E(|X_N|^q|Y_N)=o_p(1)$ for some $q\geq1$. A typical example we used in our proofs is $X_N=\sum_{i=1}^NZ_{N,i}/N$, where $(Z_{N,i})_{N\geq1,i\leq N}$ is a row-wise independent and identically distributed triangular array with $\E(|Z_{N,i}||Y_N)=o_p(1)$.

\begin{lemma}\label{l3}
Let $(Z_{N,i})_{N\geq1,i\leq N}$ be a row-wise independent and identical distributed triangular array, suppose there exists a sequence $b_N$ such that $N^{-r}b_N^{-1-r}\E\left(|Z_{N,1}|^{1+r}\right)=o(1)$ with $0<r<1$ and $b_N>0$. Then,
$$
N^{-1}\sum_{i=1}^NZ_{N,i}-\E(Z_{N,1})~=~o_p(b_N).
$$
\end{lemma}

\begin{proof}[Proof of Lemma \ref{l3}]
Let $Y_{N,i}=Z_{N,i}\mathbbm{1}\{|Z_{N,i}|\leq Nb_N)\}$. For any $c>0$,
\begin{align*}
&\P\left(\left|N^{-1}\sum_{i=1}^NZ_{N,i}-\E(Y_{N,1})\right|\geq c b_N \right)\\
&\qquad\leq\P\left(\cup_{i=1}^N[Z_{N,i}\neq Y_{N,i}]\cup\left[\left|\sum_{i=1}^NY_{N,i}-\E(Y_{N,1})\right|\geq Nc b_N \right]\right)\\
&\qquad\leq\P\left(\cup_{i=1}^N[Z_{N,i}\neq Y_{N,i}]\right)+\P\left(\left|\sum_{i=1}^NY_{N,i}-\E(Y_{N,1})\right|\geq Nc b_N \right),
\end{align*}
where with a slight abuse of notation, here $\P$ denotes the joint distribution of $(Z_{N,i})_{N\geq1,i\leq N}$. By Markov's inequality,
\begin{align*}
&\P\left(\cup_{i=1}^N[Z_{N,i}\neq Y_{N,i}]\right)\leq N\P(Z_{N,1}\neq Y_{N,1})=N\P(|Z_{N,1}|> N b_N )\\
&\qquad\leq N(Nb_N)^{-1-r}\E\left(|Z_{N,1}|^{1+r}\right)=N^{-r}b_N^{-1-r}\E\left(|Z_{N,1}|^{1+r}\right)=o(1),
\end{align*}
\jelena{where the last equality follows from the assumptions. Moreover, by Chebyshev's inequality}
\begin{align*}
&\P\left(\left|\sum_{i=1}^NY_{N,i}-\E(Y_{N,1})\right|\geq Nc b_N \right)\leq(Nc b_N )^{-2}\E\left\{\left|\sum_{i=1}^NY_{N,i}-\E(Y_{N,1})\right|^2\right\}\\
&\qquad=c^{-2}N^{-1}b_N^{-2}\E\{Y_{N,i}-\E(Y_{N,i})\}^2\leq c^{-2}N^{-1}b_N^{-2}\E(Y_{N,i}^2)\\
&\qquad=c^{-2}N^{-1}b_N^{-2}\E[Z_{N,1}^2\mathbbm{1}\{|Z_{N,1}|\leq Nb_N\}]\leq c^{-2}N^{-1}b_N^{-2}(Nb_N)^{1-r}\E\left(|Z_{N,1}|^{1+r}\right)\\
&\qquad=c^{-2}N^{-r}b_N^{-1-r}\E\left(|Z_{N,1}|^{1+r}\right)=o(1),
\end{align*}
\jelena{where in the second to last inequality we used Markov's inequality on $Z_{N,1}$s.}
Hence,
$$
N^{-1}\sum_{i=1}^NZ_{N,i}-\E(Y_{N,1})=o_p(b_N).
$$
\jelena{In addition, by similar arguments}
\begin{align*}
&\E(Z_{N,1})-\E(Y_{N,1})=\E\left[Z_{N,1}\mathbbm{1}\{|Z_{N,1}|>Nb_N\}\right]=\E\left[|Z_{N,1}|^{1+r}|Z_{N,1}|^{-r}\mathbbm{1}\{|Z_{N,1}|>Nb_N\}\right]\\
&\qquad\leq(Nb_N)^{-r}\E\left(|Z_{N,1}|^{1+r}\right)=b_NN^{-r}b_N^{-1-r}\E\left(|Z_{N,1}|^{1+r}\right)=o(b_N).
\end{align*}
Therefore,
$$N^{-1}\sum_{i=1}^NZ_{N,i}-\E(Z_{N,1})=N^{-1}\sum_{i=1}^NZ_{N,i}-\E(Z_{N,1})+\E(Z_{N,1})-\E(Y_{N,1})=o_p(b_N).$$
\end{proof}

\begin{lemma}\label{l4}
\tco{For any function $g(\cdot)$ and $\theta\in\R$, define
$$\psi(\bZ,\theta)~:=~g(\bZ)-\theta.$$
Let $\theta^0:=\E\{g(\bZ)\}$. Assume
\begin{equation}\label{cond:l4}
\E\{\psi^2(\bZ,\theta^0)\}~\asymp~ b_N^{-1},\;\;N^{-r}b_N^{r+1}\E\{|\psi(\bZ,\theta^0)|^{2+2r}|\}~=~o(1),
\end{equation}
\jelena{for} some sequence $b_N$ and $0<r<1$. \jelena{Moreover, let} $\thetahat\in\R$ \jelena{be such that} $\thetahat-\theta^0=o_p(b_N^{-1/2})$. \jelena{Additionally,} for $k\leq\K$, \jelena{and} some (possibly random) function $g_{-k}(\cdot)\independent\mathbb S_k$, define $\psi_{-k}(\bZ,\theta):=g_{-k}(\bZ)-\theta$ and suppose that
$$\E\{\psi_{-k}(\bZ,\theta^0)-\psi(\bZ,\theta^0)\}^2~=~o_p(b_N^{-1}).$$
Then, as $N\to\infty$,} \jelena{we have}
$$N^{-1}\sum_{k=1}^{\mathbb K}\sum_{i\in\mathcal I_k}\psi_{-k}^2(\bZ_i,\thetahat)~=~\E\{\psi^2(\bZ,\theta^0)\}\{1+o_p(1)\}.$$
\end{lemma}

\begin{proof}[Proof of Lemma \ref{l4}]

\tco{\jelena{By Young's inequality} with $(a+b)^2\leq2a^2+2b^2$,
\begin{align}
&|\mathcal I_k|^{-1}\sum_{i\in\mathcal I_k}\{\psi_{-k}(\bZ_i,\thetahat)-\psi(\bZ_i,\theta^0)\}^2\nonumber\\
&\qquad\leq2(\thetahat-\theta^0)^2+2|\mathcal I_k|^{-1}\sum_{i\in\mathcal I_k}\{\psi_{-k}(\bZ_i,\theta^0)-\psi(\bZ_i,\theta^0)\}^2=o_p(b_N^{-1})\jelena{.}\label{eq:l4}
\end{align}
In what follows we will use the following equality which is a consequence of Lemma \ref{l3} and the condition in \eqref{cond:l4}:
\begin{equation}\label{eq:psi2}
|\mathcal I_k|^{-1}\sum_{i\in\mathcal I_k}\{\psi_{-k}(\bZ_i,\theta^0)-\psi(\bZ_i,\theta^0)\}^2=o_p(b_N^{-1}).
\end{equation}
Using the fact that $a^2-b^2=(a+b)(a-b)=(a-b)^2+2b(a-b)$, \jelena{and using the triangle and then Cauchy-Schwarz inequality}
\begin{align*}
&\left||\mathcal I_k|^{-1}\sum_{i\in\mathcal I_k}\psi_{-k}^2(\bZ_i,\thetahat)-|\mathcal I_k|^{-1}\sum_{i\in\mathcal I_k}\psi^2(\bZ_i,\theta^0)\right|\\
&\qquad=\bigr||\mathcal I_k|^{-1}\sum_{i\in\mathcal I_k}\{\psi_{-k}(\bZ_i,\thetahat)-\psi(\bZ_i,\theta^0)\}^2+2|\mathcal I_k|^{-1}\sum_{i\in\mathcal I_k}\{\psi_{-k}(\bZ_i,\thetahat)-\psi(\bZ_i,\theta^0)\}\psi(\bZ_i,\theta^0)\bigr|\\
&\qquad\leq|\mathcal I_k|^{-1}\sum_{i\in\mathcal I_k}\{\psi_{-k}(\bZ_i,\thetahat)-\psi(\bZ_i,\theta^0)\}^2\\
&\qquad\qquad+2|\mathcal I_k|^{-1}\left[\sum_{i\in\mathcal I_k}\{\psi_{-k}(\bZ_i,\thetahat)-\psi(\bZ_i,\theta^0)\}^2\sum_{i\in\mathcal I_k}\psi^2(\bZ_i,\theta^0)\right]^{1/2}\\
&\qquad\overset{(i)}{=}o_p(a_N^{-1})+o_p(a_N^{-1/2})[\E\{\psi^2(\bZ,\theta^0)\}\{1+o_p(1)\}]^{1/2}\overset{(ii)}{=}o_p(a_N^{-1}),
\end{align*}
where $(i)$ follows by \eqref{eq:l4} and \eqref{eq:psi2}, and in $(ii)$, \jelena{we utilized the assumption  $\E\{\psi^2(\bZ,\theta^0)\}\asymp a_N^{-1}$ to conclude  the asymptotic order of the quantities of interest. Then,} \jelena{by utilizing the result of Lemma \ref{l3}, i.e., \eqref{eq:psi2}},
we have
\begin{align*}
&N^{-1}\sum_{k=1}^{\mathbb K}\sum_{i\in\mathcal I_k}\psi_{-k}^2(\bZ_i,\thetahat)=|\mathcal I_k|^{-1}\sum_{i\in\mathcal I_k}\psi^2(\bZ_i,\theta^0)+o_p(a_N^{-1})\\
&\qquad=\E\{\psi^2(\bZ,\theta^0)\}\{1+o_p(1)\}+o_p(a_N^{-1})=\E\{\psi^2(\bZ,\theta^0)\}\{1+o_p(1)\},
\end{align*}
since $\E\{\psi^2(\bZ,\theta^0)\}\asymp a_N^{-1}$ by assumption.}
\end{proof}

\begin{lemma}\label{subG}
The following are some useful properties regarding sub-Gaussian variables.
\begin{itemize}
\item[(a)] If $|X|\leq|Y|$ a.s., then $\|X\|_{\psi_2}\leq\|Y\|_{\psi_2}$. If $|X|\leq M$ a.s. for some constant $M$, then $\|X\|_{\psi_2}\leq\{\log(2)\}^{-1/2}M$.
\item[(b)] If $\|X\|_{\psi_2}\leq\sigma$, then $\E(|X|^m)\leq2\sigma^m\Gamma(m/2+1)$, for all $m\geq1$, where $\Gamma(a):=\int_0^\infty x^{a-1}\exp(-x)dx$ denotes the Gamma function. Hence, $\E(|X|)\leq\sigma\sqrt\pi$ and $\E(|X|^m)\leq2\sigma^m(m/2)^{m/2}$ for $m\geq2$.
\item[(c)] If $\|X-\E(X)\|_{\psi_2}\leq\sigma$, then $\E(\exp[t\{X-\E(X)\}])\leq\exp(2\sigma^2t^2)$, for all $t\in\R$.
\item[(d)] Let $\bX\in\R^p$ be a random vector with $\sup_{1\leq j\leq p}\|\bX(j)\|_{\psi_2}\leq\sigma$. Then, $\|\|\bX\|_\infty\|_{\psi_2}\leq\sigma\{\log(p)+2\}^{1/2}$.
\item[(e)] Let $(X_i)_{i=1}^N$ be independent random variables with means $(\mu_i)_{i=1}^N$ such that $\|X_i-\mu_i\|_{\psi_2}\leq\sigma$. Then, $\|N^{-1}\sum_{i=1}^N(X_i-\mu_i)\|_{\psi_2}\leq4\sigma N^{-1/2}$.
\end{itemize}
\end{lemma}
Lemma \ref{subG} \jelena{is a simple consequence of}  Lemma\jelena{s} D.1 and D.2 of \cite{chakrabortty2019high}.

\begin{lemma}\label{RSC-supp} 
Assume $(\bX_i)_{i=1}^N$ are independent and identically distributed, $\lambda_{\min}\{\E(\bXv_i\bXv_i^T)\geq c>0$ and $\sup_{\|\bv\|_2=1}\E\{(\bXv_i^T\bv)^4\}<C<\infty$, with constants $c$ and $C$. Assume $\bgamma_0\in\R^{p+1}$ satisfy\jelena{es}$\|\bgamma_0\|_2<C<\infty$, $\bXv_i^T\bgamma_0$ is a sub-Gaussian random variable, and $\bX_i$ is a marginal sub-Gaussian random vector \jelena{with}
\begin{align}
&\|\bXv_i^T\bgamma_0\|_{\psi_2}~=~\inf\left\{t>0:\E[\exp\{t^{-2}(\bXv_i^T\bgamma_0)^2\}]\leq2\right\}~<~\infty,\\
&\sup_{1\leq j\leq p}\|\bX_i(j)\|_{\psi_2}~=~\inf\left\{t>0:\E[\exp\{t^{-2}\bX_i^2(j)\}]\leq2\right\}~<~\infty.
\end{align}
Recall that
\begin{align*}
\ell_N^{\mbox{\small bal}}(\bgamma)&~=~-N^{-1}\sum_{i=1}^N [R_i^*\bXv_i^T\bgamma-\log\{1+\exp(\bXv_i^T\bgamma)\}]\quad\forall\;\bgamma\in\R^{p+1},\\
\delta\ell(\bDelta;1;\bgamma)&~=~\ell_N^{\mbox{\small bal}}(\bgamma+\bDelta)-\ell_N^{\mbox{\small bal}}(\bgamma)-\bDelta^T \bnabla_{\bgamma} \ell_N^{\mbox{\small bal}}(\bgamma)\quad\forall\;\bgamma,\bDelta\in\R^{p+1}.
\end{align*}
where $(R_i^*)_{i=1}^N$ are i.i.d. \jelena{pseudo} binary random variables satisfying $\P(R_i^*=1|\bX)=g(\bXv^T\bgamma_0)$. Then, for some constants $c_1,c_2,c_3,c_4 > 0$,
$$\delta \ell(\bDelta;1;\bgamma_0)~\geq~ c_1\|\bDelta\|_2\left\{\|\bDelta\|_2-c_2\sqrt\frac{\log(p+1)}{N}\|\bDelta\|_1\right\}\;\; \forall \; \bDelta \in \R^{p+1},\;\; \| \bDelta \|_2 \leq 1,$$
with probability at least $1-c_3\exp(-c_4 N)$.
\end{lemma}
Lemma \ref{RSC-supp} is a slightly more general version of Proposition 2 of \cite{negahban2010unified}: instead of assuming $\bX$ to be joint sub-Gaussian with mean zero, one can repeat their proof by only requiring $\bX$ to be a marginal sub-Gaussian vector and $\bXv^T\bgamma_0$ be a sub-Gaussian variable, as well as an additional 4-th moment condition that $\sup_{\|\bv\|_2=1}\E\{(\bXv^T\bv)^4\}<C<\infty$. Unlike \cite{negahban2010unified}, the intercept term is also considered here: since we do not require zero-mean covariates, the intercept term $\bXv(1)=1$ can be seen as a sub-gaussian variable.

\begin{lemma}\label{Hoeffding}
(Theorem 3.26 of \cite{wainwright2019high}) Let $\mathcal F$ be a class of functions of the form $f:\mathcal X\to\R$, and let $(\bX_1,\dots,\bX_N)$ be drawn from a product distribution $\P=\bigotimes_{i=1}^N\P_i$, where each $\P_i$ is supported on some set $\mathcal X_i\subseteq\mathcal X$. For each $f\in\mathcal F$ and $i=1,\dots,N$, assume that there are real numbers $a_{i,f}\leq b_{i,f}$ such that $f(\bx)\in[a_{i,f},b_{i,f}]$ for all $\bx\in\mathcal X_i$. Let
$Z~=~\sup_{f\in\mathcal F}\left\{N^{-1}\sum_{i=1}^Nf(\bX_i)\right\}.$
Then for all $t\geq0$, we have
$\P\{Z\geq\E(Z)+t\}~\leq~\exp\left(- {Nt^2}/{4L^2}\right),$
where $L^2:=\sup_{f\in\mathcal F}\{N^{-1}\sum_{i=1}^N(b_{i,f}-a_{i,f})^2\}$.
\end{lemma}

\subsection{Proofs of the Main Statements}

\begin{proof}[Proof of Theorem \ref{t3}]

\tco{We prove Theorem \ref{t3} by  \jelena{decomposing the estimation error into two terms:}  $N^{-1}\sum_{i=1}^N\psi_{\mu,\pi}(\bZ_i)$ and $\Deltahat_{N,1,k}$ \jelena{defined below in} \eqref{eq:thetatil}. \jelena{We use}  Lemma \ref{l1} and the Lindeberg-Feller theorem} for \jelena{self-normalized partial sums}.
Observe that
\begin{align}
\tilde\theta-\theta_0=& N^{-1}\sum_{i=1}^{ N}\left\{\frac{R_i-\pi_N(\bX_i)}{\pi_N(\bX_i)}[Y_i-\mhat(\bX_i)]+Y_i-\theta_0\right\}=N^{-1}\sum_{i=1}^N\psi_{\mu,\pi}(\bZ_i)+\sum_{k=1}^{\mathbb K}\Deltahat_{N,1,k},\label{eq:thetatil}
\end{align}
where
\begin{equation}\label{deltahatN1k}
\Deltahat_{N,1,k}=-N^{-1}\sum_{i\in\mathcal I_k}\left\{\frac{R}{\pi_N(\bX_i)}-1\right\}\{\mhat(\bX_i;\mathbb S_{-k})-\mu(\bX_i)\}.
\end{equation}
Consider the remainder term $\Deltahat_{N,1,k}$. For each $k\leq\mathbb K$, notice that $\Deltahat_{N,1,k}$ is a summation of independent random variables conditional on the training sample $\mathbb S_{-k}$:
$$\Deltahat_{N,1,k} = - N^{-1} \sum_{i \in \mathcal{I}_k} \xi_i, \qquad \xi_i =\left\{\frac{R_i}{\pi_N(\bX_i)}-1\right\}\{\mhat(\bX_i;\mathbb S_{-k})-\mu(\bX_i)\},$$
with $\xi _i \perp \xi_j |  \mathbb S_{-k}  $ for $i,j \in \mathcal{I}_k$. Hence,
with
$$\xi =\left\{\frac{R}{\pi_N(\bX)}-1\right\}\{\mhat(\bX;\mathbb S_{-k})-\mu(\bX)\},$$
and recall that $\E_{\mathbb S_k}$ denotes the expectation \tco{with respect to (w.r.t.)} the samples in the $k$-th fold,
\begin{align}
\E_{\mathbb S_k}(\Deltahat_{N,1,k})=&-N^{-1}|\mathcal I_k|\E\left\{\E\left( \xi \bigr|\bX\right)\right\}=0,\label{delta1-1}\\
\E_{\mathbb S_k }(\Deltahat_{N,1,k}^2)=&N^{-2}|\mathcal I_k|\E\left(\E\left[\left\{\frac{R}{\pi_N(\bX)}-1\right\}^2\{\mhat(\bX;\mathbb S_{-k})-\mu(\bX)\}^2\bigr|\bX\right]\right)\label{delta1-2}\\
=&N^{-2}|\mathcal I_k|\E\left\{\left[\frac{1}{\pi_N(\bX)}-1\right][\mhat(\bX;\mathbb S_{-k})-\mu(\bX)]^2\right\}\label{delta1-3}\\
=&O_p((Na_N)^{-1}c_{\mu,N}^2),\label{delta1-4}
\end{align}
In the above equations, \eqref{delta1-1} and \eqref{delta1-2} used the fact that $\xi _i \perp \xi_j |  \mathbb S_{-k}  $ for $i,j \in \mathcal{I}_k$; \eqref{delta1-3} used the fact that $R^2=R$; \eqref{delta1-4} used the fact that $|\mathcal I_k|<N$ and the definition of $c_{\mu,N}$; the definition $\E(R|\bX)=\pi_N(\bX)$ is also used in \eqref{delta1-1} and \eqref{delta1-3}. These techniques will be used for multiple times throughout the proof, and we will not emphasis them in \jelena{again} in the following proofs.

By Lemma \ref{l1},
\begin{equation}\label{rate:deltahatN1k}
\Deltahat_{N,1,k}=O_p((Na_N)^{-1/2}c_{\mu,N}).
\end{equation}
As for the influence function $N^{-1}\sum_{i=1}^N\psi_{\mu,\pi}(\bZ_i)$,
\begin{align*}
\E_{\mathbb S}\left[N^{-1}\sum_{i=1}^N\psi_{\mu,\pi}(\bZ_i)\right]&=\E\left(\E\left\{\mu(\bX)-\theta_0+\frac{R[Y-\mu(\bX)]}{\pi_N(\bX)}\bigr|\bX\right\}\right)=0,\\
\E_{\mathbb S}\left[N^{-1}\sum_{i=1}^N\psi_{\mu,\pi}(\bZ_i)\right]^2&=N^{-1}\E\left[\E\left(\left\{\mu(\bX)-\theta_0+\frac{R[Y-\mu(\bX)]}{\pi_N(\bX)}\right\}^2\bigr|\bX\right)\right]=N^{-1}V_N(\mu),
\end{align*}
where
\begin{align*}
&V_N(\mu)=\E\left[\mu(\bX)-\theta_0+\frac{R\{Y-\mu(\bX)\}}{\pi_N(\bX)}\right]^2=\E\left[\frac{\{R-\pi_N(\bX)\}\{Y-\mu(\bX)\}}{\pi_N(\bX)}+Y-\theta_0\right]^2\\
&\qquad=\E\left[\left\{\frac{1-\pi_N(\bX)}{\pi_N(\bX)}\right\}^2\{Y-\mu(\bX)\}^2\right]+\Var(Y).
\end{align*}
To control the order of $V_N(\mu)$, we enforce uniform lower and upper bounds for $\E[\{Y-\mu(\bX)\}^2|\bX]$ and $\Var(Y)$. Under \tcr{A}ssumption \ref{a1},
$$\E[\{Y-m(\bX)\}^2|\bX]\geq\sigma_{\zeta,1}^2,\qquad\E[\{Y-\mu(\bX)\}^2|\bX]\leq\sigma_{\zeta,2}^2,\qquad\Var(Y)\leq\sigma_{\zeta,2}^2.$$
\tco{Additionally}, we have the following lower bounds \jelena{as} $m(\bX)=\E(Y|\bX)$,
\begin{align*}
\E[\{Y-\mu(\bX)\}^2|\bX]&=\E[\{Y-m(\bX)\}^2|\bX]+\E[\{m-\mu(\bX)\}^2|\bX]\geq\sigma_{\zeta,1}^2,\\
\Var(Y)&=\E\left(\E[\{Y-m(\bX)\}^2|\bX]+\E[\{m(\bX)-\theta_0\}^2|\bX]\right)\geq\sigma_{\zeta,1}^2.
\end{align*}
Recall that by definition, $a_N=\E\{\pi_N^{-1}(\bX)\}$. Therefore,
\begin{align}
&a_NV_N(\mu)\geq a_N\left\{\sigma_{\xi,1}^2\E\left[\frac{1-\pi_N(\bX)}{\pi_N(\bX)}\right]+\sigma_{\xi,1}^2\right\}= \sigma_{\xi,1}^2>0,\label{Vmu1}\\
&a_NV_N(\mu)\leq a_N\left\{\sigma_{\xi,2}^2\E\left[\frac{1-\pi_N(\bX)}{\pi_N(\bX)}\right]+\sigma_{\xi,2}^2\right\}=\sigma_{\xi,2}^2<\infty,\label{Vmu2}
\end{align}
and $V_N(\mu)\asymp a_N^{-1}$. Since
\begin{equation}\label{EIF2}
\E\left\{(Na_N)^{1/2}N^{-1}\sum_{i=1}^N\psi_{\mu,\pi}(\bZ_i)\right\}^2=a_NV_N(\mu)=O(1),
\end{equation}
by Lemma \ref{l1}, $N^{-1}\sum_{i=1}^N\psi_{\mu,\pi}(\bZ_i)=O_p((Na_N)^{-1/2})$. Therefore,
$$\thetahat_\mathrm{DRSS}-\theta_0=O_p((Na_N)^{-1/2}).$$
\tco{In addition}, under \tcr{A}ssumption \ref{a4}, for any $c>0$,
$$N^{-1}\sum_{i=1}^N\E[V_N^{-1}(\mu)\psi_{\mu,\pi}^2(\bZ_i)\mathbbm{1}\{V_N^{-1/2}(\mu)|\psi_{\mu,\pi}(\bZ_i)|>cN^{1/2}\}]=o(1).$$
By Proposition 2.27 (Lindeberg-Feller theorem) of \cite{van2000asymptotic},
\begin{equation}\label{normal:psi}
 V_N^{-1/2}(\mu)N^{-1\jelena{/2}}\sum_{i=1}^N\psi_{\mu,\pi}(\bZ_i)\xrightarrow{d}\mathcal N(0,1).
\end{equation}
Recall that
\begin{align*}
&N^{1/2}V_N^{-1/2}(\mu)(\tilde\theta-\theta_0)=V_N^{-1/2}(\mu)N^{-1/2}\sum_{i=1}^N\psi_{\mu,\pi}(\bZ_i)+N^{1/2}V_N^{-1/2}(\mu)\sum_{k=1}^{\mathbb K}\Deltahat_{N,1,k}\\
&\qquad=V_N^{-1/2}(\mu)N^{-1/2}\sum_{i=1}^N\psi_{\mu,\pi}(\bZ_i)+O_p(N^{1/2}a_N^{1/2}(Na_N)^{-1/2}c_{\mu,N})\\
&\qquad=V_N^{-1/2}(\mu)N^{-1/2}\sum_{i=1}^N\psi_{\mu,\pi}(\bZ_i)+O_p(c_{\mu,N})=V_N^{-1/2}(\mu)N^{-1/2}\sum_{i=1}^N\psi_{\mu,\pi}(\bZ_i)+o_p(1).
\end{align*}
By Lemma 2.8 (Slutsky) of \cite{van2000asymptotic},
$$N^{1/2}V_N^{-1/2}(\mu)(\tilde\theta-\theta_0)\xrightarrow{d}\mathcal N(0,1).$$
\end{proof}

\begin{proof}[Proof of Theorem \ref{t4}]
\tco{We prove Theorem \ref{t4} \jelena{by considering} two cases: (a) the nuisance models are both correctly specified, and (b) only one of the nuisance models is correctly specified. For case (a), we \jelena{design a suitable decomposition, \eqref{eq:DRSSa}, and apply Lemma \ref{l1} and} the Lindeberg-Feller theorem \jelena{for self-normalized sums} to obtain asymptotic normality. For case (b), \jelena{we design two different decompositions of the estimation error: one suitable for the case when PS model is correct \eqref{eq:DRSSbi}  and the other suitable for the case when the outcome model is correct  \eqref{eq:DRSSbii}}}. 

\jelena{\bf Case (a): $\mu(\cdot)=m(\cdot)$ and $e_N(\cdot)=\pi_N(\cdot)$.} Observe that
\begin{align}
&\thetahat_\mathrm{DRSS}-\theta_0= N^{-1}\sum_{i=1}^{ N}\left[\frac{R_i-\pihat_N(\bX_i;\S_{-k})}{\pihat_N(\bX_i;\S_{-k})}\{Y_i-\mhat(\bX_i;\S_{-k})\}+Y_i-\theta_0\right]\nonumber\\
&\qquad=N^{-1}\sum_{i=1}^N\psi_{\mu,e}(\bZ_i)+\sum_{k=1}^{\mathbb K}(\Deltahat_{N,1,k}+\Deltahat_{N,2,k}+\Deltahat_{N,3,k}),\label{eq:DRSSa}
\end{align}
where $\Deltahat_{N,1,k}$ is defined as \eqref{deltahatN1k} and we further define
\begin{align}
&\Deltahat_{N,2,k}=N^{-1}\sum_{i\in\mathcal I_k}\left\{\frac{R_i}{\pihat_N(\bX_i;\mathbb S_{-k})}-\frac{R_i}{e_N(\bX_i)}\right\}\{Y_i-m(\bX_i)\},\label{deltahatN2k}\\
&\Deltahat_{N,3,k}=-N^{-1}\sum_{i\in\mathcal I_k}\left\{\frac{R_i}{\pihat_N(\bX_i;\mathbb S_{-k})}-\frac{R_i}{e_N(\bX_i)}\right\}\{\mhat(\bX_i;\mathbb S_{-k})-\mu(\bX_i)\}.\label{deltahatN3k}
\end{align}
Recall from \eqref{rate:deltahatN1k}, $\Deltahat_{N,1,k}=O_p((Na_N)^{-1/2}c_{\mu,N})$. 
As for the remainder term $\Deltahat_{N,2,k}$,
\begin{align}
\E_{\mathbb S_k}(\Deltahat_{N,2,k})=&N^{-1}|\mathcal I_k|\E\left(\E\left[\left\{\frac{R}{\pihat_N(\bX;\mathbb S_{-k})}-\frac{R}{e_N(\bX)}\right\}\{Y-m(\bX)\}\bigr|\bX\right]\right)=0,\nonumber\\
\E_{\mathbb S_k }(\Deltahat_{N,2,k}^2)=&N^{-2}|\mathcal I_k|\E\left(\E\left[\left\{\frac{R}{\pihat_N(\bX;\mathbb S_{-k})}-\frac{R}{e_N(\bX)}\right\}^2\{Y-m(\bX)\}^2\bigr|\bX\right]\right)\nonumber\\
=&N^{-2}|\mathcal I_k|\E\left[\frac{\pi_N(\bX)}{e_N^2(\bX)}\left\{1-\frac{e_N(\bX)}{\pihat_N(\bX;\mathbb S_{-k})}\right\}^2\{Y-m(\bX)\}^2\right]\label{piNeN}\\
\overset{(i)}{\leq}&N^{-1}\sigma_{\xi,2}^2c_{e,N}^2a_N^{-1}=O_p((Na_N)^{-1}c_{e,N}^2),\label{EDeltahatN2k2}
\end{align}
\tco{where $(i)$ holds under the Assumption \ref{a1} and the condition in \eqref{drthm:ratecond4} with $e_N(\cdot)=\pi_N(\cdot)$ and also noting the fact that $|\mathcal I_k|\leq N$.}
By Lemma \ref{l1},
\begin{equation}\label{rate:deltahatN2k}
\Deltahat_{N,2,k}=O_p((Na_N)^{-1/2}c_{e,N}).
\end{equation}
Now, consider the last remainder term $\Deltahat_{N,3,k}$, \tco{by the triangular inequality and the tower rule,}
\begin{align*}
\E_{\mathbb S_k}(|\Deltahat_{N,3,k}|)\leq&N^{-1}|\mathcal I_k|\E\left[\E\left\{\left|\frac{R}{\pihat_N(\bX;\mathbb S_{-k})}-\frac{R}{e_N(\bX)}\right||\mhat(\bX_i;\mathbb S_{-k})-\mu(\bX)|\bigr|\bX\right\}\right]\\
=&N^{-1}|\mathcal I_k|\E\left\{\left|1-\frac{e_N(\bX)}{\pihat_N(\bX;\mathbb S_{-k})}\right||\mhat(\bX_i;\mathbb S_{-k})-\mu(\bX)|\right\}=O_p(r_{\mu,N}r_{e,N}).
\end{align*}
By Lemma \ref{l1},
\begin{equation}\label{rate:deltahatN3k}
\Deltahat_{N,3,k}=O_p(r_{\mu,N}r_{e,N}).
\end{equation}
Lastly, for the influence function $N^{-1}\sum_{i=1}^N\psi_{\mu,e}(\bZ_i)$,
\begin{align*}
\E_{\mathbb S}\left\{N^{-1}\sum_{i=1}^N\psi_{\mu,e}(\bZ_i)\right\}&=\E\left(\E\left[\mu(\bX)-\theta_0+\frac{R\{Y-m(\bX)\}}{\pi_N(\bX)}\bigr|\bX\right]\right)=0,\\
\E_{\mathbb S}\left\{N^{-1}\sum_{i=1}^N\psi_{\mu,e}(\bZ_i)\right\}^2&=N^{-1}\E\left\{\E\left(\left[\mu(\bX)-\theta_0+\frac{R\{Y-m(\bX)\}}{\pi_N(\bX)}\right]^2\bigr|\bX\right)\right\}=N^{-1}V_N(\mu,e).
\end{align*}
Now we control the rate of the variance, $V_N(\mu,e)$. Under \tcr{A}ssumption \ref{a1}, $\Var(Y)\geq\E[\{Y-m(\bX)\}^2|\bX]\geq\sigma_{\zeta,1}^2$, $\E[\{Y-m(\bX)\}^2|\bX]\leq\sigma_{\zeta,2}^2$ and $\Var(Y)\leq\sigma_{\zeta,2}^2$. Hence,
\begin{align*}
&a_NV_N(\mu,e)\geq a_N\left[\sigma_{\xi,1}^2\E\left\{\frac{1-\pi_N(\bX)}{\pi_N(\bX)}\right\}+\sigma_{\xi,1}^2\right]\geq\sigma_{\zeta,1}^2>0,\\
&a_NV_N(\mu,e)\leq a_N\left[\sigma_{\xi,2}^2\E\left\{\frac{1-\pi_N(\bX)}{\pi_N(\bX)}\right\}+\sigma_{\xi,2}^2\right]\leq\sigma_{\zeta,2}^2<\infty.
\end{align*}
It follows that $V_N(\mu,e)\asymp a_N^{-1}$.

\tco{\jelena{Recall} the definition of $\psi_{\mu,e}$ in \eqref{def:psimue}. }
By Lemma \ref{l1}, $N^{-1}\sum_{i=1}^N\psi_{\mu,e}(\bZ_i)=O_p((Na_N)^{-1/2})$. Therefore, $\thetahat_\mathrm{DRSS}-\theta_0=O_p((Na_N)^{-1/2})$. \tco{Moreover}, by Proposition 2.27 (Lindeberg-Feller theorem) of \cite{van2000asymptotic},
\begin{equation}
N^{1/2}V_N^{1/2}(\mu,e)N^{-1}\sum_{i=1}^N\psi_{\mu,e}(\bZ_i)\xrightarrow{d}\mathcal N(0,1).
\end{equation}
By  Lemma 2.8 (Slutsky) of \cite{van2000asymptotic},
$$
N^{1/2}V_N^{-1/2}(\mu,e)(\thetahat_\mathrm{DRSS}-\theta_0)\xrightarrow{d}\mathcal N(0,1).
$$

\jelena{\bf Case (b.i): $e_N(\cdot)=\pi_N(\cdot)$.} Observe that
\begin{equation}\label{eq:DRSSbi}
\thetahat_\mathrm{DRSS}-\theta_0=N^{-1}\sum_{i=1}^N\psi_{\mu,e}(\bZ_i)+\sum_{k=1}^{\mathbb K}(\Deltahat_{N,1,k}+\Deltahat_{N,2,k}+\Deltahat_{N,3,k}+\Deltahat_{N,4,k}),
\end{equation}
where $\Deltahat_{N,1,k}$, $\Deltahat_{N,2,k}$, and $\Deltahat_{N,3,k}$ are defined as \eqref{deltahatN1k}, \eqref{deltahatN2k}, and \eqref{deltahatN3k}, respectively, and we further define
$$\Deltahat_{N,4,k}=N^{-1}\sum_{i\in\mathcal I_k}\left\{\frac{R_i}{\pihat_N(\bX_i;\mathbb S_{-k})}-\frac{R_i}{e_N(\bX_i)}\right\}\{m(\bX_i)-\mu(\bX_i)\}.$$
As shown in \eqref{rate:deltahatN1k}, \eqref{rate:deltahatN2k}, and \eqref{rate:deltahatN3k}, we have $\Deltahat_{N,1,k}=O_p((Na_N)^{-1/2}c_{\mu,N})$, $\Deltahat_{N,2,k}=O_p((Na_N)^{-1/2}c_{e,N})$ and $\Deltahat_{N,2,k}=O_p(r_{\mu,N}r_{e,N})$. \jelena{In addition}, for the remainder term $\Deltahat_{N,4,k}$,
\begin{align*}
&\E_{\mathbb S_k}(|\Deltahat_{N,4,k}|)\leq N^{-1}|\mathcal I_k|\E\left[\E\left\{\left|\frac{R}{\pihat_N(\bX;\mathbb S_{-k})}-\frac{R}{e_N(\bX)}\right||m(\bX_i)-\mu(\bX)|\bigr|\bX\right\}\right]\\
&\qquad=N^{-1}|\mathcal I_k|\E\left\{\left|1-\frac{e_N(\bX)}{\pihat_N(\bX;\mathbb S_{-k})}\right||m(\bX_i)-\mu(\bX)|\right\}\\
&\qquad\leq\left\|1-\frac{e_N(\cdot)}{\pihat_N(\cdot)}\right\|_{2,\mathbb P_{\bX}}\|m(\cdot)-\mu(\cdot)\|_{2,\mathbb P_{\bX}}=O_p(r_{e,N}).
\end{align*}
By Lemma \ref{l1},
$$\Deltahat_{N,4,k}=O_p(r_{e,N}).$$
Lastly, for the \jelena{influence function} $N^{-1}\sum_{i=1}^N\psi_{\mu,e}(\bZ_i)$, similarly as in \eqref{EIF2} and by Lemma \ref{l1},
$$N^{-1}\sum_{i=1}^N\psi_{\mu,e}(\bZ_i)=O_p((Na_N)^{-1/2}).$$

\jelena{\bf Case (b.ii)\tco{: $\mu(\cdot)=m(\cdot)$.} }Observe that
\begin{equation}\label{eq:DRSSbii}
\thetahat_\mathrm{DRSS}-\theta_0=N^{-1}\sum_{i=1}^N\psi_{\mu,e}(\bZ_i)+\sum_{k=1}^{\mathbb K}(\Deltahat_{N,1,k}+\Deltahat_{N,2,k}+\Deltahat_{N,3,k}+\Deltahat_{N,4,k}),
\end{equation}
where $\Deltahat_{N,1,k}$, $\Deltahat_{N,2,k}$, and $\Deltahat_{N,3,k}$ are defined as \eqref{deltahatN1k}, \eqref{deltahatN2k}, and \eqref{deltahatN3k}, respectively, and we further define
$$\Deltahat_{N,5,k}=N^{-1}\sum_{i\in\mathcal I_k}\left\{\frac{R_i}{\pi_N(\bX_i)}-\frac{R_i}{ e_N(\bX_i)}\right\}\{\mhat(\bX_i;\mathbb S_{-k})-\mu(\bX_i)\}.$$
Similarly as shown in \eqref{rate:deltahatN1k}, \eqref{rate:deltahatN2k}, and \eqref{rate:deltahatN3k}, we have $\Deltahat_{N,1,k}=O_p((Na_N)^{-1/2}c_{\mu,N})$, $\Deltahat_{N,2,k}=O_p((Na_N)^{-1/2}c_{e,N})$ and $\Deltahat_{N,2,k}=O_p(r_{\mu,N}r_{e,N})$. Here, the only difference from the previous proofs is that, in \eqref{piNeN}, instead of obtaining $\pi_N(\bX)/e_N^2(\bX)=\pi_N^{-1}(\bX)$ using $e_N(\cdot)=\pi_N(\cdot)$, here we bound $\pi_N(\bX)/e_N^2(\bX)\leq c^{-2}\pi_N^{-1}(\bX)$ by assuming that, a.s., $\pi_N(\bX)/e_N(\bX)\geq c$ . For the remainder term $\Deltahat_{N,5,k}$,
\begin{align*}
&\E_{\mathbb S_k}(|\Deltahat_{N,5,k}|)\leq N^{-1}|\mathcal I_k|\E\left[\E\left\{\left|\frac{R}{\pi_N(\bX)}-\frac{R}{ e_N(\bX)}\right||\mhat(\bX_i;\mathbb S_{-k})-\mu(\bX)|\bigr|\bX\right\}\right]\\
&\qquad=N^{-1}|\mathcal I_k|\E\left\{\left|1-\frac{\pi_N(\bX)}{ e_N(\bX)}\right||\mhat(\bX_i;\mathbb S_{-k})-\mu(\bX)|\right\}\\
&\qquad\leq\|1-\pi_N(\cdot)/ e_N(\cdot)\|_{2,\mathbb P_\bX}\|\mhat(\cdot)-\mu(\cdot)\|_{2,\mathbb P_\bX}=O_p(r_{\mu,N}).
\end{align*}
By Lemma \ref{l1},
$$\Deltahat_{N,5,k}=O_p(r_{\mu,N}).$$
Lastly, for the influence function $N^{-1}\sum_{i=1}^N\psi_{\mu,e}(\bZ_i)$, we have
\begin{align*}
\E_{\mathbb S}\left\{N^{-1}\sum_{i=1}^N\psi_{\mu,e}(\bZ_i)\right\}&=\E\left(\E\left[m(\bX)-\theta_0+\frac{R\{Y-m(\bX)\}}{ e_N(\bX)}\bigr|\bX\right]\right)=0,\\
\E_{\mathbb S}\left\{N^{-1}\sum_{i=1}^N\psi_{\mu,e}(\bZ_i)\right\}^2&=N^{-1}\E\left\{\E\left(\left[m(\bX)-\theta_0+\frac{R\{Y-m(\bX)\}}{ e_N(\bX)}\right]^2\bigr|\bX\right)\right\}=N^{-1}V_N(\mu,e).
\end{align*}
Here,
\begin{align*}
&V_N(\mu,e)=\Var\{m(\bX)\}+\E\{\pi_N(\bX)\{Y-m(\bX)\}^2/\{\pi_N(\bX)\}^2\}\\
&\qquad\leq\sigma_{\zeta,2}^2(1+\E[\pi_N(\bX)/\{\pi_N(\bX)\}^2])\leq\sigma_{\zeta,2}^2(1+C^2a_N^{-1})=O(a_N^{-1}).
\end{align*}
By Lemma \ref{l1},
$$N^{-1}\sum_{i=1}^N\psi_{\mu,e}(\bZ_i)=O_p((Na_N)^{-1/2}).$$
\end{proof}

\begin{proof}[Proof of Theorem \ref{thm:var}]
\tco{We prove the consistency results of the asymptotic variance estimators for the two cases (known PS and unknown PS). The results follows from Lemma \ref{l4} after we validate the conditions therein.}

\jelena{\bf Case (a).} By Lemma \ref{l4}, it is sufficient to show $a_N\E(\delta_{N,1,k}^2)=o_p(1)$, where
$$\delta_{N,1,k}=-\left\{\frac{R}{\pi_N(\bX)}-1\right\}\{\mhat(\bX;\mathbb S_{-k})-\mu(\bX)\}.$$
Recall from \eqref{delta1-4}, we have
$$a_N\E(\deltahat_{N,1,k}^2)=O_p(c_{\mu,N}^2)=o_p(1).$$

\jelena{\bf Case (b).} By Lemma \ref{l4}, it is sufficient to show $a_N\E(\delta_{N,1,k}+\delta_{N,2,k}+\delta_{N,3,k})^2=o_p(1)$, where
\begin{align*}
\delta_{N,1,k}&=-\left\{\frac{R}{\pi_N(\bX)}-1\right\}\{\mhat(\bX;\mathbb S_{-k})-\mu(\bX)\},\\
\delta_{N,2,k}&=\left\{\frac{R}{\pihat_N(\bX;\mathbb S_{-k})}-\frac{R}{e_N(\bX)}\right\}\{Y-m(\bX)\},\\
\delta_{N,3,k}&=-\left\{\frac{R}{\pihat_N(\bX;\mathbb S_{-k})}-\frac{R}{e_N(\bX)}\right\}\{\mhat(\bX;\mathbb S_{-k})-\mu(\bX)\}.
\end{align*}
Recall \eqref{delta1-4} and \eqref{EDeltahatN2k2}, we have
$$a_N\E(\deltahat_{N,1,k}^2)=O_p(c_{\mu,N}^2)=o_p(1),\qquad a_N\E(\deltahat_{N,2,k}^2)=O_p(c_{e,N}^2)=o_p(1).$$
Besides, by the condition \eqref{cond:drift2},
$$a_N\E(\deltahat_{N,3,k}^2)=\E\left[\frac{a_N}{\pi_N(\bX)}\left\{1-\frac{\pi_N(\bX)}{\pihat(\bX;\mathbb S_{-k})}\right\}^2\{\mhat(\bX;\mathbb S_{-k})-m(\bX)\}^2\right]=o_p(1).$$
Therefore,
$$a_N\E(\Deltahat_{N,1,k}+\Deltahat_{N,2,k}+\Deltahat_{N,3,k})^2\leq 3a_N\E(\Deltahat_{N,1,k}^2+\Deltahat_{N,2,k}^2+\Deltahat_{N,3,k}^2)=o_p(1).$$
\end{proof}

\begin{proof}[Proof of Theorem \ref{thm:ex3}]
\tco{In the proof of Theorem \ref{thm:ex3}, we work directly on the cross-fitted version of $\bgammahat$. The results for a non cross-fitted $\bgammahat$ can be obtained analogously by repeating the procedure using the full sample $\S$. Here, we first obtain an RAL expansion of the offset logistic regression estimator. Then,   we establish the RAL expansion of the DRSS estimator.}

For any $k\leq\K$, $a\in(0,1]$, and $\bgamma\in\R^{p+1}$, let
$$\ell_N(\bgamma;a)=-N_{-k}^{-1}\sum_{i\in\mathcal I_{-k}}\left[R_i\bXv_i^T\bgamma-\log\{1+a\exp(\bXv_i^T\bgamma)\}\right].$$
Define $g(u)=\exp(u)/\{1+\exp(u)\}$, then $\dot{g}(u)=g(u)\{1-g(u)\}$ and $\ddot{g}(u)=g(u)\{1-g(u)\}\{1-2g(u)\}$. We have
\begin{align}
&g(u+\log(a))=\frac{a\exp(u)}{1+a\exp(u)}\geq\frac{a\exp(u)}{1+\exp(u)}=ag(u),\quad\forall u\in\R,\;a\in(0,1],\label{g:lower}\\
&g(u)\leq\exp(u),\quad\dot{g}(u)\leq g(u)\leq\exp(u),\quad|\ddot{g}(u)|\leq g(u)\leq\exp(u),\quad\forall u\in\R.\label{g:upper}
\end{align}
For any $\bu\in\R^{p+1}$, define
$$\ell_N(\bu)=N_{-k}\left\{\ell_N(\bgamma_0+(N_{-k}\bar\pi_N)^{-1/2}\bu,\pihat_N)-\ell_N(\bgamma_0,\bar\pi_N)\right\}-N_{-k}^{-1}\sum_{i\in\mathcal I_{-k}}R_i\log(\pihat_N/\bar\pi_N).$$
Since $\bgamma=\bgammahat$ minimizes $\ell_N(\bgamma;\pihat_N)$, the terms $\ell_N(\bgamma_0,\bar\pi_N)$ and $N_{-k}^{-1}\sum_{i\in\mathcal I_{-k}}R_i\log(\pihat_N/\bar\pi_N)$ are both independent of $\bgamma$, we know that $\bu_N=(N_{-k}\bar\pi_N)^{1/2}(\bgammahat-\bgamma_0)$ minimizes $\ell_N(\bu)$. Here, $\pihat_N=N_{-k}^{-1}\sum_{i\in\mathcal I_{-k}}R_i$ is the cross-fitted estimate of $\bar\pi_N$. By Taylor's Theorem,\jelena{where} some $(\tilde{\bgamma}_1,\log(\pitil_{N,1}))$ lies between $(\bgamma_0,\log(\bar\pi_N))$ and $(\bgamma_0+(N_{-k}\bar\pi_N)^{-1/2}\bu,\log(\pihat_N))$,
$$\ell_N(\bu)=\frac{1}{2}\bu^T\mathbf{A}_N(\tilde{\bgamma}_1,\pitil_{N,1})\bu+\mathbf{B}_{N,1}^T(\tilde{\bgamma}_1,\pitil_{N,1})\bu+C_N(\tilde{\bgamma}_1,\pitil_{N,1}),$$
where
\begin{align*}
\mathbf{A}_N(\tilde{\bgamma}_1,\pitil_{N,1})&=(N_{-k}\bar\pi_N)^{-1}\sum_{i\in\mathcal I_{-k}}\dot{g}(\bXv_i^T\tilde{\bgamma}_1+\log(\pitil_{N,1}))\bXv_i\bXv_i^T,\\
\mathbf{B}_{N,1}(\tilde{\bgamma}_1,\pitil_{N,1})&=-(N_{-k}\bar\pi_N)^{-1/2}\sum_{i\in\mathcal I_{-k}}\biggr\{R_i-g(\bXv_i^T\bgamma_0+\log(\bar\pi_N))\\
&\qquad-\dot{g}(\bXv_i^T\tilde{\bgamma}_1+\log(\pitil_{N,1}))\log(\pihat_N/\bar\pi_N)\biggr\}\bXv_i,\\
C_N(\tilde{\bgamma}_1,\pitil_{N,1})&=\frac{1}{2}\sum_{i\in\mathcal I_{-k}}\left\{\dot{g}(\bXv_i^T\tilde{\bgamma}_1+\log(\pitil_{N,1}))-\dot{g}(\bXv_i^T\bgamma_0+\log(\bar\pi_N))\right\}\left\{\log(\pihat_N/\bar\pi_N)\right\}^2.
\end{align*}
Define
\begin{align}
\mathcal J(\bgamma_0,\bar\pi_N)&=\E\left\{\bXv\bXv^T\dot{g}(\bXv^T\bgamma_0+\log(\bar\pi_N))\right\},\nonumber\\
\mathbf{B}_{N,2}&=-(N_{-k}\bar\pi_N)^{-1/2}\sum_{i\in\mathcal I_{-k}}\biggr\{R_i-g(\bXv_i^T\bgamma_0+\log(\bar\pi_N))\nonumber\\
&\qquad-\dot{g}(\bXv_i^T\bgamma_0+\log(\bar\pi_N))\log(\pihat_N/\bar\pi_N)\biggr\}\bXv_i,\nonumber\\
\boldsymbol{\zeta}_N&=(N_{-k}\bar\pi_N)^{1/2}\mathcal J^{-1}(\bgamma_0,\bar\pi_N)N_{-k}^{-1}\sum_{i\in\mathcal I_{-k}}\biggr\{R_i-g(\bXv_i^T\bgamma_0+\log(\bar\pi_N))\nonumber\\
&\qquad-\dot{g}(\bXv_i^T\bgamma_0+\log(\bar\pi_N))\log(\pihat_N/\bar\pi_N)\biggr\}\bXv_i.\label{def:zetaN}
\end{align}
Then, $\boldsymbol{\zeta}_N$ is the unique minimizer of
$$Z_N(\bu)=\bu^T\bar\pi_N^{-1}\mathcal J(\bgamma_0,\bar\pi_N)\bu/2+\mathbf{B}_{N,2}^T\bu.$$
By Lemma 2 of \cite{hjort2011asymptotics}, for each $\delta>0$,
$$\P_{\S_{-k}}(\|\bu_N-\boldsymbol{\zeta}_N\|_2\geq\delta)\leq\P\left\{\Delta_N(\delta)\geq\frac{1}{2}h_N(\delta)\right\},$$
where $\S_{-k}=\S\setminus\S_k$ and
$$\Delta_N(\delta)=\sup_{\|\bu-\boldsymbol{\zeta}_N\|_2\leq\delta}|\ell_N(\bu)-Z_N(\bu)|,\quad h_N(\delta)=\inf_{\|\bu-\boldsymbol{\zeta}_N\|_2=\delta}Z_N(\bu)-Z_N(\boldsymbol{\zeta}_N).$$
Hence, to prove
\begin{equation}\label{IF_gamma}
\left\|\bgammahat-\bgamma_0-(N_{-k}\bar\pi_N)^{-1/2}\boldsymbol{\zeta}_N\right\|_2=(N_{-k}\bar\pi_N)^{-1/2}\|\bu_N-\boldsymbol{\zeta}_N\|_2=o_p\left((N\bar\pi_N)^{-1/2}\right),
\end{equation}
it suffices to show that, for each $\delta>0$, $\Delta_N(\delta)=o_p(1)$ and $h_N(\delta)>c(\delta)$ with some constant $c(\delta)>0$ independent of $N$. First notice that
\begin{align*}
&h_N(\delta)=\inf_{\|\bu-\boldsymbol{\zeta}_N\|_2=\delta}\frac{1}{2}(\bu-\boldsymbol{\zeta}_N)^T\bar\pi_N^{-1}\mathcal J(\bgamma_0,\bar\pi_N)(\bu-\boldsymbol{\zeta}_N)\\
&\qquad\geq\frac{1}{2}\delta^2\bar\pi_N^{-1}\lambda_{\min}\{\mathcal J(\bgamma_0,\bar\pi_N)\}\geq\frac{1}{2}\delta^2\lambda_{\min}\left[\E\{\bXv\bXv^T\dot{g}(\bXv^T\bgamma_0)\}\right].
\end{align*}
Now, it remains to show $\Delta_N(\delta)=o_p(1)$. A sufficient condition we would like to show is the following:
\begin{align}
\sup_{\|\bu-\boldsymbol{\zeta}_N\|_2\leq\delta}\left|\bu^T\{\mathbf{A}_N(\tilde{\bgamma}_1,\pitil_{N,1})-\bar\pi_N^{-1}\mathcal J(\bgamma_0,\bar\pi_N)\}\bu\right|&=o_p(1),\label{AN}\\
\sup_{\|\bu-\boldsymbol{\zeta}_N\|_2\leq\delta}\left|(\mathbf{B}_{N,1}(\tilde{\bgamma}_1,\pitil_{N,1})-\mathbf{B}_{N,2})^T\bu\right|&=o_p(1),\label{BN}\\
\tco{\left|C_N(\tilde{\bgamma}_1,\pitil_{N,1})\right|}&=o_p(1).\label{CN}
\end{align}
To prove \eqref{AN}-\eqref{CN}, we first analyze some basic properties of $\pitil_{N,1}$ and $\boldsymbol{\zeta}_N$. \jelena{With} \eqref{piest}, we \jelena{have}
\begin{equation}\label{pitilde}
\pitil_N=\bar\pi_N\left\{1+O_p\left((N\bar\pi_N)^{-1/2}\right)\right\},\;\;\mbox{for any $\pitil_N$ lies between $\bar\pi_N$ and $\pihat_N$}.
\end{equation}
\tco{In addition}, by the fact that $\log(u)\leq1-u$ for all $u>0$ and \eqref{pitilde} we have
\begin{equation}\label{logpitilde}
|\log(\pitil_N/\bar\pi_N)|\leq|1-\pitil_N/\bar\pi_N|=O_p\left((N\bar\pi_N)^{-1/2}\right),\;\;\mbox{for any $\pitil_N$ lies between $\bar\pi_N$ and $\pihat_N$}.
\end{equation}
For any $t<\infty$ and $r<\infty$,
\begin{align}
&\E\left\{\exp(t\|\bXv\|_2)\|\bXv\|_2^r\right\}\leq r!\E\left[\exp\{(t+1)\|\bXv\|_2\}\right]\nonumber\\
&\qquad\leq r!\exp(t+1)\E\left[\exp\{(t+1)\|\bX\|_2\}\right]<\infty.\label{ineq:Eexp}
\end{align}
\tco{Now, to control the supremum over $\|\bu-\boldsymbol{\zeta}_N\|_2\leq\delta$ in \eqref{AN} and \eqref{BN}, we analyse asymptotic properties for $\boldsymbol{\zeta}_N$ defined in \eqref{def:zetaN}. We consider the following representation:}
\begin{align}
\boldsymbol{\zeta}_N&=\boldsymbol{\zeta}_{N,1}-\boldsymbol{\zeta}_{N,2},\quad\mbox{where}\label{def:zeta}\\
\boldsymbol{\zeta}_{N,1}&=(N_{-k}\bar\pi_N)^{1/2}\mathcal J^{-1}(\bgamma_0,\bar\pi_N)N_{-k}^{-1}\sum_{i\in\mathcal I_{-k}}\left\{R_i-g(\bXv_i^T\bgamma_0+\log(\bar\pi_N))\right\}\bXv_i,\label{def:zeta1}\\
\boldsymbol{\zeta}_{N,2}&=\log(\pihat_N/\bar\pi_N)(N_{-k}\bar\pi_N)^{1/2}\mathcal J^{-1}(\bgamma_0,\bar\pi_N)N_{-k}^{-1}\sum_{i\in\mathcal I_{-k}}\dot{g}(\bXv_i^T\bgamma_0+\log(\bar\pi_N))\bXv_i.\label{def:zeta2}
\end{align}
\jelena{Moreover}, define
\begin{align}
\boldsymbol{\zeta}_{N,3}&=\log(\pihat_N/\bar\pi_N)(N_{-k}\bar\pi_N)^{1/2}\mathcal J^{-1}(\bgamma_0,\bar\pi_N)\E\left\{\dot{g}(\bXv^T\bgamma_0+\log(\bar\pi_N))\bXv\right\}\nonumber\\
&=(N_{-k}\bar\pi_N)^{1/2}\log(\pihat_N/\bar\pi_N)\mathbf{e}_1.\label{def:zeta3}
\end{align}
\jelena{In the above we used}  $\mathcal J^{-1}(\bgamma_0,\bar\pi_N)\E\{\dot{g}(\bXv^T\bgamma_0+\log(\bar\pi_N))\bXv\}=\mathbf{e}_1$. \tco{Note that,
$$\dot{g}(\bXv^T\bgamma_0+\log(\bar\pi_N))=\frac{\bar\pi_N\exp(\bXv^T\bgamma_0)}{\{1+\bar\pi_N\exp(\bXv^T\bgamma_0)\}^2}\geq\frac{\bar\pi_N\exp(\bXv^T\bgamma_0)}{\{1+\exp(\bXv^T\bgamma_0)\}^2}=\bar\pi_N\dot{g}(\bXv^T\bgamma_0).$$
Hence,
\begin{equation}\label{rate:J_inv}
\|\mathcal J^{-1}(\bgamma_0,\bar\pi_N)\|_2\leq\bar\pi_N^{-1}\left\|\left[\E\{\dot{g}(\bXv^T\bgamma_0)\bXv\bXv^T\}\right]^{-1}\right\|_2=O(\bar\pi_N^{-1}).
\end{equation}
Then,}
\begin{align*}
\E_{\S_{-k}}\|\boldsymbol{\zeta}_{N,1}\|_2^2&\overset{(i)}{=}\bar\pi_N\E\left\{\dot{g}(\bXv^T\bgamma_0+\log(\bar\pi_N))\|\mathcal J^{-1}(\bgamma_0,\bar\pi_N)\bXv\|_2^2\right\}\\
&\overset{(ii)}{\leq}\bar\pi_N\|\mathcal J^{-1}(\bgamma_0,\bar\pi_N)\|_2^2\E\left\{\dot{g}(\bXv^T\bgamma_0+\log(\bar\pi_N))\|\bXv\|_2^2\right\}\\
&\overset{(iii)}{\leq}\bar\pi_N^2\|\mathcal J^{-1}(\bgamma_0,\bar\pi_N)\|_2^2\E\left\{\exp(\|\bXv\|_2\|\bgamma_0\|_2)\|\bXv\|_2^2\right\}\overset{(iv)}{=}O(1),
\end{align*}
\tco{where $(i)$ holds by the tower rule with the fact $\E[\{R-g(\bXv^T\bgamma_0+\log(\bar\pi_N))\}^2|\bX]=\dot{g}(\bXv^T\bgamma_0+\log(\bar\pi_N))$, $(ii)$ holds by the fact that $|\boldsymbol{A}\boldsymbol{a}|\leq\|\boldsymbol{A}\|_2\|\boldsymbol{a}\|_2$ for any $\boldsymbol{a}\in\R^{p+1}$ and $\boldsymbol{A}\in\R^{(p+1)\times(p+1)}$, $(iii)$ follows by the fact that $\dot{g}(\bXv^T\bgamma_0+\log(\bar\pi_N))\leq g(\bXv^T\bgamma_0+\log(\bar\pi_N))\leq\bar\pi_N\exp(\|\bXv\|_2\|\bgamma_0\|_2)$, and $(iv)$ holds holds by \eqref{ineq:Eexp} and \eqref{rate:J_inv}. Besides,}
\begin{align*}
&\E_{\S_{-k}}\left\|\{\log(\pihat_N/\bar\pi_N)\}^{-1}(\boldsymbol{\zeta}_{N,2}-\boldsymbol{\zeta}_{N,3})\right\|_2^2=\bar\pi_N\Var\left\{\dot{g}(\bXv^T\bgamma_0+\log(\bar\pi_N))\|\mathcal J^{-1}(\bgamma_0,\bar\pi_N)\bXv\|_2\right\}\\
&\qquad\leq\bar\pi_N\E\left\{\dot{g}^2(\bXv^T\bgamma_0+\log(\bar\pi_N))\|\mathcal J^{-1}(\bgamma_0,\bar\pi_N)\bXv\|_2^2\right\}\\
&\qquad\overset{(i)}{\leq}\bar\pi_N\|\mathcal J^{-1}(\bgamma_0,\bar\pi_N)\|_2^2\E\left\{\dot{g}^2(\bXv^T\bgamma_0+\log(\bar\pi_N))\|\bXv\|_2^2\right\}\\
&\qquad\overset{(ii)}{\leq}\bar\pi_N^3\|\mathcal J^{-1}(\bgamma_0,\bar\pi_N)\|_2^2\E\left\{\exp(2\|\bXv\|_2\|\bgamma_0\|_2)\|\bXv\|_2^2\right\}\overset{(iii)}{=}O(\bar\pi_N),
\end{align*}
\tco{where $(i)$ holds by the fact that $|\boldsymbol{A}\boldsymbol{a}|\leq\|\boldsymbol{A}\|_2\|\boldsymbol{a}\|_2$ for any $\boldsymbol{a}\in\R^{p+1}$ and $\boldsymbol{A}\in\R^{(p+1)\times(p+1)}$, $(ii)$ follows by the fact that $\dot{g}(\bXv^T\bgamma_0+\log(\bar\pi_N))\leq\bar\pi_N\exp(\|\bXv\|_2\|\bgamma_0\|_2)$, and $(iii)$ holds holds by \eqref{ineq:Eexp} and \eqref{rate:J_inv}.} By Chebyshev's Inequality,
$$\|\boldsymbol{\zeta}_{N,1}\|=O_p(1),\qquad\left\|\{\log(\pihat_N/\bar\pi_N)\}^{-1}(\boldsymbol{\zeta}_{N,2}-\boldsymbol{\zeta}_{N,3})\right\|_2=O_p(\bar\pi_N^{1/2}).$$
Hence, by \eqref{logpitilde},
\begin{equation}\label{diff:zeta2-3}
\left\|(\boldsymbol{\zeta}_{N,2}-\boldsymbol{\zeta}_{N,3})\right\|_2=\{\log(\pihat_N/\bar\pi_N)\}O_p(\bar\pi_N^{1/2})=O_p(N^{-1/2}),
\end{equation}
\tco{with}
\begin{align*}
\left\|\boldsymbol{\zeta}_{N,3}\right\|_2&\leq|\log(\pihat_N/\bar\pi_N)|(N_{-k}\bar\pi_N)^{1/2}\|\mathcal J^{-1}(\bgamma_0,\bar\pi_N)\|_2\bar\pi_N\E\left\{\exp(\|\bXv\|_2\|\bgamma_0\|_2)\|\bXv\|_2\right\}=O_p(1).
\end{align*}
Therefore,
$$\|\boldsymbol{\zeta}_N\|_2\leq\|\boldsymbol{\zeta}_{N,1}\|_2+\|\boldsymbol{\zeta}_{N,2}-\boldsymbol{\zeta}_{N,3}\|_2+\|\boldsymbol{\zeta}_{N,3}\|_2=O_p(1).$$
It follows that,
\begin{align}
&\sup_{\|\bu-\boldsymbol{\zeta}_N\|_2\leq\delta}\|\bu\|_2\leq\sup_{\|\bu-\boldsymbol{\zeta}_N\|_2\leq\delta}\|\bu-\boldsymbol{\zeta}_N\|_2+\|\boldsymbol{\zeta}_N\|_2\leq\delta+\|\boldsymbol{\zeta}_N\|_2=O_p(1),\label{2norm:u}\\
&\sup_{\|\bu-\boldsymbol{\zeta}_N\|_2\leq\delta}\|\tilde{\bgamma}_1-\bgamma_0\|_2\leq\sup_{\|\bu-\boldsymbol{\zeta}_N\|_2\leq\delta}(N_{-k}\bar\pi_N)^{-1/2}\|\bu\|_2=O_p\left((N_{-k}\bar\pi_N)^{-1/2}\right),\label{2norm:gammadiff}\\
&\sup_{\|\bu-\boldsymbol{\zeta}_N\|_2\leq\delta}\|\tilde{\bgamma}_1\|_2\leq\sup_{\|\bu-\boldsymbol{\zeta}_N\|_2\leq\delta}\|\tilde{\bgamma}_1-\bgamma_0\|_2+\|\bgamma_0\|_2<M,\;\;\text{w.p.a. 1},\label{2norm:gamma}
\end{align}
where $M>0$ is a constant independent of $N$.

Now, we prove \eqref{AN}. For any $\bu$ satisfying $\|\bu-\boldsymbol{\zeta}_N\|_2\leq\delta$,
\begin{align*}
&\left|\bu^T\{\mathbf{A}_N(\tilde{\bgamma}_1,\pitil_{N,1})-\bar\pi_N^{-1}\mathcal J(\bgamma_0,\bar\pi_N)\}\bu\right|\\
&\qquad\leq\left|(N_{-k}\bar\pi_N)^{-1}\|\bu\|_2^2\sum_{i\in\mathcal I_{-k}}\dot{g}(\bXv_i^T\tilde{\bgamma}_1+\log(\pitil_{N,1}))-\dot{g}(\bXv_i^T\bgamma_0+\log(\bar\pi_N))\|\bXv_i\|_2^2\right|\\
&\qquad\qquad+\|\bu\|_2^2\left\|(N_{-k}\bar\pi_N)^{-1}\sum_{i\in\mathcal I_{-k}}\dot{g}(\bXv_i^T\bgamma_0+\log(\bar\pi_N))\bXv_i\bXv_i^T-\bar\pi_N^{-1}\mathcal J(\bgamma_0,\bar\pi_N)\right\|_2.
\end{align*}
By Taylor's Theorem, with some $(\tilde{\bgamma}_2,\pitil_{N,2})$ lies between $(\bgamma_0,\bar\pi_N)$ and $(\tilde{\bgamma}_1,\pitil_{N,1})$, uniformly on $\|\bu-\boldsymbol{\zeta}_N\|_2\leq\delta$,
\begin{align}
&\left|N_{-k}^{-1}\sum_{i\in\mathcal I_{-k}}\dot{g}(\bXv_i^T\tilde{\bgamma}_1+\log(\pitil_{N,1}))\|\bXv_i\|_2^2-N_{-k}^{-1}\sum_{i\in\mathcal I_{-k}}\dot{g}(\bXv_i^T\bgamma_0+\log(\bar\pi_N))\|\bXv_i\|_2^2\right|\nonumber\\
&\qquad\overset{(i)}{=}\left|N_{-k}^{-1}\sum_{i\in\mathcal I_{-k}}\ddot{g}(\bXv_i^T\tilde{\bgamma}_2+\log(\pitil_{N,2}))\left\{\bXv_i^T(\tilde{\bgamma}_1-\bgamma_0)+\log(\pitil_{N,1}/\bar\pi_N)\right\}\|\bXv_i\|_2^2\right|\nonumber\\
&\qquad\overset{(ii)}{\leq}\pitil_{N,2}N_{-k}^{-1}\sum_{i\in\mathcal I_{-k}}\exp(\bXv_i^T\tilde{\bgamma}_2)\left|\bXv_i^T(\tilde{\bgamma}_1-\bgamma_0)+\log(\pitil_{N,1}/\bar\pi_N)\right|\|\bXv_i\|_2^2\nonumber\\
&\qquad\overset{(iii)}{\leq}\pitil_{N,2}N_{-k}^{-1}\sum_{i\in\mathcal I_{-k}}\exp(\|\bXv_i\|_2M)\left\{\|\bXv_i\|_2\|\tilde{\gamma}_1-\bgamma_0\|_2+|\log(\pitil_{N,1}/\bar\pi_N)|\right\}\|\bXv_i\|_2^2,\label{ANp1}
\end{align}
with probability approaching 1. Here, $(i)$ holds by Taylor's Theorem, $(ii)$ holds by \eqref{g:upper}, $(iii)$ holds by \eqref{2norm:gamma}. \tco{Recall \eqref{ineq:Eexp}, b}y Markov's Inequality,
\begin{equation}\label{markov}
N_{-k}^{-1}\sum_{i\in\mathcal I_{-k}}\exp(\|\bXv_i\|_2M)\|\bXv_i\|_2^r=O_p(1).
\end{equation}
Hence,
\begin{align}
&\sup_{\|\bu-\boldsymbol{\zeta}_N\|_2\leq\delta}(N_{-k}\bar\pi_N)^{-1}\|\bu\|_2^2\sum_{i\in\mathcal I_{-k}}\left|\dot{g}(\bXv_i^T\tilde{\bgamma}_1+\log(\pitil_{N,1}))-\dot{g}(\bXv_i^T\bgamma_0+\log(\bar\pi_N))\right|\|\bXv_i\|_2^2\nonumber\\
&\qquad\overset{(i)}{\leq}\sup_{\|\bu-\boldsymbol{\zeta}_N\|_2\leq\delta}\bar\pi_N^{-1}\|\bu\|_2^2\pitil_{N,2}\left\{\|\tilde{\bgamma}_1-\bgamma_0\|_2O_p(1)+|\log(\pitil_{N,1}/\bar\pi_N)|O_p(1)\right\}\nonumber\\
&\qquad\overset{(ii)}{=}O_p\left((N\bar\pi_N)^{-1/2}\right)=o_p(1).\label{ANp1op1}
\end{align}
where $(i)$ holds by \eqref{ANp1} and \eqref{markov}, $(ii)$ holds by \eqref{pitilde}, \eqref{logpitilde}, \eqref{2norm:u} and \eqref{2norm:gammadiff}. Notice that
\begin{align*}
\E\{\dot{g}(\bXv^T\bgamma_0+\log(\bar\pi_N))\|\bXv\|_2^2\}&\leq\bar\pi_N\E\{\exp(\bXv^T\bgamma_0)\|\bXv\|_2^2\}.\\
\|\mathcal J(\bgamma_0,\bar\pi_N)\|_2&\leq\bar\pi_N\|\E\{\exp(\bXv^T\bgamma_0)\bXv\bXv^T\}\|_2\leq\bar\pi_N\E\left\{\exp(\bXv^T\bgamma_0)\|\bXv\|_2^2\right\}.
\end{align*}
Recall that $p$ is fixed, by Theorem 5.48 of \cite{vershynin2010introduction}, with some constant $C>0$,
\begin{align*}
&\E_{\S_{-k}}\left\|N_{-k}^{-1}\sum_{i\in\mathcal I_{-k}}\dot{g}(\bXv_i^T\bgamma_0+\log(\bar\pi_N))\bXv_i\bXv_i^T-\mathcal J(\bgamma_0,\bar\pi_N)\right\|_2\\
&\qquad\leq\max\left[\|\mathcal J(\bgamma_0,\bar\pi_N)\|_2^{1/2}C\sqrt\frac{\bar\pi_N\log\{\min(N,p+1)\}}{N},\frac{C^2\bar\pi_N\log\{\min(N,p+1)\}}{N}\right]\\
&\qquad=O\left(\max\left(N^{-1/2}\bar\pi_N,N^{-1}\bar\pi_N\right)\right)=O\left(N^{-1/2}\bar\pi_N\right).
\end{align*}
By Markov's Inequality,
$$\left\|N_{-k}^{-1}\sum_{i\in\mathcal I_{-k}}\dot{g}(\bXv_i^T\bgamma_0+\log(\bar\pi_N))\bXv_i\bXv_i^T-\mathcal J(\bgamma_0,\bar\pi_N)\right\|_2=O_p\left(N^{-1/2}\bar\pi_N\right).$$
It follows that
\begin{align}
&\sup_{\|\bu-\boldsymbol{\zeta}_N\|_2\leq\delta}\|\bu\|_2^2\left\|(N_{-k}\bar\pi_N)^{-1}\sum_{i\in\mathcal I_{-k}}\dot{g}(\bXv_i^T\bgamma_0+\log(\bar\pi_N))\bXv_i\bXv_i^T-\bar\pi_N^{-1}\mathcal J(\bgamma_0,\bar\pi_N)\right\|_2\nonumber\\
&\qquad=O_p\left((N\bar\pi_N)^{-1/2}\right)=o_p(1).\label{ANp2op1}
\end{align}
Hence, by \eqref{ANp1op1} and \eqref{ANp2op1},
\begin{equation}\label{ANop1}
\sup_{\|\bu-\boldsymbol{\zeta}_N\|_2\leq\delta}\left|\bu^T\{\mathbf{A}_N-\bar\pi_N^{-1}\mathcal J(\bgamma_0,\bar\pi_N)\}\bu\right|=O_p\left((N\bar\pi_N)^{-1/2}\right)=o_p(1).
\end{equation}

Now, we show \eqref{BN}. By Taylor's Theorem, \jelena{where} some $(\tilde{\bgamma}_3,\pitil_{N,3})$ lies between $(\bgamma_0,\bar\pi_N)$ and $(\tilde{\bgamma}_1,\pitil_{N,1})$,
\begin{align}
&|(\mathbf{B}_{N,1}-\mathbf{B}_{N,2})^T\bu|\nonumber\\
&\qquad\overset{(i)}{=}\biggr|\log\left(\frac{\pihat_N}{\bar\pi_N}\right)(N_{-k}\bar\pi_N)^{-1/2}\sum_{i\in\mathcal I_{-k}}\ddot{g}(\bXv_i^T\tilde{\bgamma}_3+\log(\pitil_{N,3}))\left\{\bXv_i^T(\tilde{\bgamma}_1-\bgamma_0)+\log\left(\frac{\pitil_{N,1}}{\bar\pi_N}\right)\right\}\bXv_i^T\bu\biggr|\nonumber\\
&\qquad\overset{(ii)}{\leq}\left|\log\left(\frac{\pitil_{N,1}}{\bar\pi_N}\right)\right|(N_{-k}\bar\pi_N)^{-1/2}\pitil_{N,3}\sum_{i\in\mathcal I_{-k}}\exp(\|\bXv_i\|_2M)\|\bXv_i\|_2^2\|\tilde{\bgamma}_1-\bgamma_0\|_2\|\bu\|_2\nonumber\\
&\qquad\qquad+\left|\log\left(\frac{\pitil_{N,1}}{\bar\pi_N}\right)\right|(N_{-k}\bar\pi_N)^{-1/2}\pitil_{N,3}\sum_{i\in\mathcal I_{-k}}\exp(\|\bXv_i\|_2M)\left|\log\left(\frac{\pitil_{N,1}}{\bar\pi_N}\right)\right|\|\bXv_i\|_2\|\bu\|_2\quad\mbox{w.p.a. 1}\nonumber\\
&\qquad\overset{(iii)}{\leq}\left|\log\left(\frac{\pihat_N}{\bar\pi_N}\right)\right|N_{-k}^{1/2}\bar\pi_N^{-1/2}\|\bu\|_2\pitil_{N,3}\left\{\|\tilde{\bgamma}_1-\bgamma_0\|_2O_p(1)+\left|\log\left(\frac{\pitil_{N,1}}{\bar\pi_N}\right)\right|O_p(1)\right\}\quad\mbox{w.p.a. 1}\nonumber\\
&\qquad\overset{(iv)}{=}O_p\left((N\bar\pi_N)^{-1/2}\right)=o_p(1),\qquad\mbox{uniformly on $\|\bu-\boldsymbol{\zeta}_N\|_2\leq\delta$,}\label{BNop1}
\end{align}
where $(i)$ holds by Taylor's Theorem, $(ii)$ holds by \eqref{g:upper} and \eqref{2norm:gamma}, $(iii)$ holds by \eqref{markov}, $(iv)$ holds by \eqref{pitilde}, \eqref{logpitilde} \eqref{2norm:u} and \eqref{2norm:gammadiff}.

As for \eqref{CN}, by Taylor's Theorem, with some $(\tilde{\bgamma}_4,\pitil_{N,4})$ lies between $(\bgamma_0,\bar\pi_N)$ and $(\tilde{\bgamma}_1,\pitil_{N,1})$,
\begin{align}
&|C_N(\tilde{\bgamma}_1,\pitil_{N,1})|\overset{(i)}{=}\left|\frac{1}{2}\sum_{i\in\mathcal I_{-k}}\ddot{g}(\bXv_i^T\tilde{\bgamma}_4+\log(\pitil_{N,4}))\left\{\bXv^T(\tilde{\bgamma}_1-\bgamma_0)+\log\left(\frac{\pitil_{N,1}}{\bar\pi_N}\right)\right\}\left\{\log\left(\frac{\pihat_N}{\bar\pi_N}\right)\right\}^2\right|\nonumber\\
&\qquad\overset{(ii)}{\leq}\frac{1}{2}\pitil_{N,4}\sum_{i\in\mathcal I_{-k}}\exp(\|\bXv_i\|M)\left\{\|\bXv\|_2\|\tilde{\bgamma}_1-\bgamma_0\|_2+\left|\log\left(\frac{\pitil_{N,1}}{\bar\pi_N}\right)\right|\right\}\left\{\log\left(\frac{\pihat_N}{\bar\pi_N}\right)\right\}^2\;\;\mbox{w.p.a. 1}\nonumber\\
&\qquad\overset{(iii)}{\leq}\frac{1}{2}\left|\log\left(\frac{\pihat_N}{\bar\pi_N}\right)\right|^2\pitil_{N,4}N_{-k}\left\{\|\tilde{\bgamma}_1-\bgamma_0\|_2O_p(1)+\left|\log\left(\frac{\pitil_{N,1}}{\bar\pi_N}\right)\right|O_p(1)\right\}\;\;\mbox{w.p.a. 1}\nonumber\\
&\qquad\overset{(iv)}{=}O_p\left((N\bar\pi_N)^{-1/2}\right)=o_p(1).\label{CNop1}
\end{align}
where $(i)$ holds by Taylor's Theorem, $(ii)$ holds by \eqref{g:upper} and \eqref{2norm:gamma}, $(iii)$ holds by \eqref{markov}, $(iv)$ holds by \eqref{pitilde}, \eqref{logpitilde} \eqref{2norm:u} and \eqref{2norm:gammadiff}.

Combining \eqref{ANop1}, \eqref{BNop1} and \eqref{CNop1}, we have
$$\Delta_N(\delta)=o_p(1),\qquad\mbox{for any $\delta>0$},$$
and hence \eqref{IF_gamma} holds. \tco{Recall the definition of $\boldsymbol{\zeta}_{N,3}$ in \eqref{def:zeta3}, we have}
\begin{align*}
&\left\|\boldsymbol{\zeta}_{N,3}-N_{-k}^{-1/2}\bar\pi_N^{1/2}\sum_{i\in\mathcal I_{-k}}(\bar\pi_N^{-1}R_i-1)\mathbf{e}_1\right\|_2=\left\|\boldsymbol{\zeta}_{N,3}-(N_{-k}\bar\pi_N)^{1/2}\frac{\pihat_N-\bar\pi_N}{\bar\pi_N}\mathbf{e}_1\right\|_2\\
&\qquad\overset{(i)}{=}\left\|(N_{-k}\bar\pi_N)^{1/2}\frac{(\pihat_N-\bar\pi_N)^2}{\pitil_{N,5}^2}\mathbf{e}_1\right\|_2=(N_{-k}\bar\pi_N)^{1/2}\frac{(\pihat_N-\bar\pi_N)^2}{\pitil_{N,5}^2}=O_p\left((N\bar\pi_N)^{-1/2}\right)=o_p(1).
\end{align*}
\tco{where $(i)$ follows from the Taylor's Theorem with some $\pitil_{N,5}$ \jelena{lying} between $\bar\pi_N$ and $\pihat_N$.}
Hence, with $\mathrm{IF}_{\bgamma}(\bZ)=\mathcal J^{-1}(\bgamma_0,\bar\pi_N)\{R-g(\bXv^T\bgamma_0+\log(\bar\pi_N))\}\bXv-(\bar\pi_N^{-1}R-1)\mathbf{e}_1$,
\begin{align*}
&\left\|\bgammahat-\bgamma_0-N_{-k}^{-1}\sum_{i\in\mathcal I_{-k}}\mathrm{IF}_{\bgamma}(\bZ_i)\right\|_2=(N_{-k}\bar\pi_N)^{-1/2}\left\|\boldsymbol{\zeta_N}-N_{-k}^{-1/2}\bar\pi_N^{1/2}\sum_{i\in\mathcal I_{-k}}\mathrm{IF}_{\bgamma}(\bZ_i)\right\|_2\\
&\qquad=(N_{-k}\bar\pi_N)^{-1/2}\left\|N_{-k}^{-1/2}\bar\pi_N^{1/2}\sum_{i\in\mathcal I_{-k}}(\bar\pi_N^{-1}R_i-1)\mathbf{e}_1-\boldsymbol{\zeta}_{N,3}+(\boldsymbol{\zeta}_{N,3}-\boldsymbol{\zeta}_{N,2})\right\|_2\\
&\qquad\leq(N_{-k}\bar\pi_N)^{-1/2}\left\|N_{-k}^{-1/2}\bar\pi_N^{1/2}\sum_{i\in\mathcal I_{-k}}(\bar\pi_N^{-1}R_i-1)\mathbf{e}_1-\boldsymbol{\zeta}_{N,3}\right\|_2+(N_{-k}\bar\pi_N)^{-1/2}\left\|\boldsymbol{\zeta}_{N,3}-\boldsymbol{\zeta}_{N,2}\right\|_2\\
&\qquad=(N_{-k}\bar\pi_N)^{-1/2}O_p\left((N\bar\pi_N)^{-1/2}+N^{-1/2}\right)=o_p\left((N_{-k}\bar\pi_N)^{-1/2}\right).
\end{align*}
Now, it remains to analy{\jelena{ze}} the IF of the \tco{PS} $\pihat_N(\bX)=g(\bXv^T\bgammahat+\log(\pihat_N))$. \tcr{For this,} 
define
\begin{align}
&\widehat{\bbeta}=\bgammahat+\log(\pihat_N)\mathbf{e}_1,\quad\bbeta_0=\bgamma_0+\log(\bar\pi_N)\mathbf{e}_1,\label{def:beta}\\
&\mathrm{IF}_{\bbeta}(\bZ)=\mathcal J^{-1}(\bgamma_0,\bar\pi_N)\{R-g(\bXv^T\bgamma_0+\log(\bar\pi_N))\}\bXv.\label{def:IF_beta}
\end{align}
Then,
\begin{align}
&\left\|\widehat{\bbeta}-\bbeta_0-N_{-k}^{-1}\sum_{i\in\mathcal I_{-k}}\mathrm{IF}_{\bbeta}(\bZ_i)\right\|_2\overset{(i)}{=}\left\|\bgammahat-\bgamma_0+\log(\pihat_N/\bar\pi_N)\mathbf{e}_1-(N_{-k}\bar\pi_N)^{-1/2}\boldsymbol{\zeta}_{N,1}\right\|_2\nonumber\\
&\qquad\overset{(ii)}{=}\left\|\bgammahat-\bgamma_0-(N_{-k}\bar\pi_N)^{-1/2}\boldsymbol{\zeta}_N+(N_{-k}\bar\pi_N)^{-1/2}(\boldsymbol{\zeta}_{N,3}-\boldsymbol{\zeta}_{N,1}-\boldsymbol{\zeta}_N)\right\|_2\nonumber\\
&\qquad\overset{(iii)}{\leq}\left\|\bgammahat-\bgamma_0-(N_{-k}\bar\pi_N)^{-1/2}\boldsymbol{\zeta}_N\right\|_2+(N_{-k}\bar\pi_N)^{-1/2}\left\|\boldsymbol{\zeta}_{N,3}-\boldsymbol{\zeta}_{N,2}\right\|_2\nonumber\\
&\qquad\overset{(iv)}{=}o_p\left((N\bar\pi_N)^{-1/2}\right)+(N_{-k}\bar\pi_N)^{-1/2}O_p(N^{-1/2})=o_p\left((N\bar\pi_N)^{-1/2}\right),\label{IF_beta}
\end{align}
where $(i)$ holds by \eqref{def:zeta1}, \eqref{def:beta} and \eqref{def:IF_beta}, $(ii)$ holds by \eqref{def:zeta3}, $(iii)$ holds by \eqref{def:zeta} and the triangular inequality, $(iv)$ holds by \eqref{IF_gamma} and \eqref{diff:zeta2-3}. It follows that
\begin{align*}
\left\|\widehat{\bbeta}-\bbeta_0\right\|_2&\overset{(i)}{\leq}\left\|\widehat{\bbeta}-\bbeta_0-N_{-k}^{-1}\sum_{i\in\mathcal I_{-k}}\mathrm{IF}_{\bbeta}(\bZ_i)\right\|_2+(N_{-k}\bar\pi_N)^{-1/2}\left\|\boldsymbol{\zeta}_{N,1}\right\|_2\\
&\overset{(ii)}{=}o_p\left((N\bar\pi_N)^{-1/2}\right)+O_p\left((N\bar\pi_N)^{-1/2}\right)=O_p\left((N\bar\pi_N)^{-1/2}\right),
\end{align*}
where $(i)$ holds by \eqref{def:zeta1} and the triangular inequality, $(ii)$ holds by \eqref{IF_beta} and \eqref{diff:zeta2-3}. \tco{Furthermore},
\begin{align*}
\left\|\bgammahat-\bgamma_0\right\|_2&\leq\left\|\bgammahat-\bgamma_0-(N_{-k}\bar\pi_N)^{-1/2}\boldsymbol{\zeta}_N\right\|_2+(N_{-k}\bar\pi_N)^{-1/2}\left\|\boldsymbol{\zeta}_N\right\|_2\\
&=o_p\left((N\bar\pi_N)^{-1/2}\right)+O_p\left((N\bar\pi_N)^{-1/2}\right)=O_p\left((N\bar\pi_N)^{-1/2}\right)=o_p(1),
\end{align*}
and hence $\left\|\bgammahat-\bgamma_0\right\|_2<1$ w.p.a. 1. By Taylor's Theorem, for any $\bx\in\mathcal X$, \jelena{where} some $(\tilde{\bgamma}_6,\pitil_{N,6})$ (depending on $\bx$) lies between $(\bgamma_0,\bar\pi_N)$ and $(\bgammahat,\pihat_N)$ and $\tilde{\bbeta}=\tilde{\bgamma}_6+\log(\pitil_{N,6})\mathbf{e}_1$,
\begin{align*}
&1-\frac{g(\bxv^T\bgamma_0+\log(\bar\pi_N))}{g(\bxv^T\bgammahat+\log(\pihat_N))}-\left\{1-g(\bxv^T\bbeta_0)\right\}(\widehat{\bbeta}-\bbeta_0)^T\bxv\\
&\qquad=g(\bxv^T\bbeta_0)\left\{g^{-1}(\bxv^T\tilde{\bbeta})-1\right\}\left\{(\widehat{\bbeta}-\bbeta_0)^T\bxv\right\}^2\\
&\qquad=g(\bxv^T\bbeta_0)\exp(-\bxv^T\tilde{\bbeta})\left\{(\widehat{\bbeta}-\bbeta_0)^T\bxv\right\}^2\\
&\qquad=\pitil_{N,6}^{-1}g(\bxv^T\bbeta_0)\exp(-\bxv^T\tilde{\bgamma})\left\{(\widehat{\bbeta}-\bbeta_0)^T\bxv\right\}^2\\
&\qquad\leq\max(\bar\pi_N^{-1},\pihat_N^{-1})g(\bxv^T\bbeta_0)\exp\left\{\|\bxv\|_2(\|\bgamma_0\|_2+\|\tilde{\bgamma}-\bgamma_0\|_2)\right\}\left\{(\widehat{\bbeta}-\bbeta_0)^T\bxv\right\}^2\\
&\qquad\leq\max(\bar\pi_N^{-1},\pihat_N^{-1})g(\bxv^T\bbeta_0)\exp\left\{\|\bxv\|_2(\|\bgamma_0\|_2+\|\bgammahat-\bgamma_0\|_2)\right\}\left\{(\widehat{\bbeta}-\bbeta_0)^T\bxv\right\}^2.
\end{align*}
Therefore, for any fixed $r>0$, with $\bX$ independent of $\bgammahat$ and $\pihat_N$,
\begin{align}
&\left\|1-\frac{g(\bXv^T\bbeta_0)}{g(\bXv^T\widehat{\bbeta})}-\left\{1-g(\bXv^T\bbeta_0)\right\}(\widehat{\bbeta}-\bbeta_0)^T\bXv\right\|_{r,\P}\nonumber\\
&\qquad\leq\max(\bar\pi_N^{-1},\pihat_N^{-1})\left\|g(\bXv^T\bbeta_0)\exp\left\{\|\bXv\|_2(\|\bgamma_0\|_2+\|\bgammahat-\bgamma_0\|_2)\right\}\left\{(\widehat{\bbeta}-\bbeta_0)^T\bXv\right\}^2\right\|_{r,\P}\nonumber\\
&\qquad\leq\max(\bar\pi_N^{-1},\pihat_N^{-1})\left\|g(\bXv^T\bbeta_0)\exp\left\{\|\bXv\|_2(\|\bgamma_0\|_2+1)\right\}\left\{(\widehat{\bbeta}-\bbeta_0)^T\bXv\right\}^2\right\|_{r,\P}\;\;\mbox{w.p.a. 1}\nonumber\\
&\qquad\leq\max(1,\bar\pi_N\pihat_N^{-1})\|\widehat{\bbeta}-\bbeta_0\|_2^2\left\|\exp\left\{\|\bXv\|_2(2\|\bgamma_0\|_2+1)\right\}\|\bXv\|_2^2\right\|_{r,\P}\;\;\mbox{w.p.a. 1}\nonumber\\
&\qquad=\{1+o_p(1)\}O_p\left((N\bar\pi_N)^{-1}\right)O(1)=O_p\left((N\bar\pi_N)^{-1}\right)=o_p\left((N\bar\pi_N)^{-1/2}\right).\label{IF_pip1}
\end{align}
Define
\begin{equation}\label{def:IF_pi-k}
\mathrm{IF}_\pi(\bZ;S_{-k})=\left\{1-g(\bXv^T\bbeta_0)\right\}\bXv^TN_{-k}^{-1}\sum_{i\in\mathcal I_k}\mathrm{IF}_{\bbeta}(\bZ_i),
\end{equation}
where $\bZ$ is independent of $(\bZ_i)_{i\in\mathcal I_k}$. Then,
\begin{align}
&\left\|\left\{1-g(\bXv^T\bbeta_0)\right\}(\widehat{\bbeta}-\bbeta_0)^T\bXv-\mathrm{IF}_\pi(\bZ;S_{-k})\right\|_{r,\P}\nonumber\\
&\qquad\leq\left\|\|\bXv\|_2\right\|_{r,\P}\left\|\widehat{\bbeta}-\bbeta_0-N_{-k}^{-1}\sum_{i\in\mathcal I_{-k}}\mathrm{IF}_{\bbeta}(\bZ_i)\right\|_2=o_p\left((N\bar\pi_N)^{-1/2}\right),\label{IF_pip2}
\end{align}
and
$$\left\|\left\{1-g(\bXv^T\bbeta_0)\right\}(\widehat{\bbeta}-\bbeta_0)^T\bXv\right\|_{r,\P}\leq\|\widehat{\bbeta}-\bbeta_0\|_2\left\|\|\bXv\|_2\right\|_{r,\P}=O_p\left((N\bar\pi_N)^{-1/2}\right).$$
Hence,
$$\left\|1-\frac{g(\bXv^T\bbeta_0)}{g(\bXv^T\widehat{\bbeta})}\right\|_{r,\P}=O_p\left((N\bar\pi_N)^{-1/2}\right).$$
For any fixed $r>0$,
$$\left\|\pi_N^{-1}(\bX)\right\|_{r,\P}=\left\|1+\bar\pi_N^{-1}\exp(-\bXv^T\bgamma_0)\right\|_{r,\P}\leq1+\bar\pi_N^{-1}\left\|\exp(\|\bXv\|_2\|\bgamma_0\|_2)\right\|_{r,\P}=O(\bar\pi_N^{-1}).$$
\tco{Additionally}, by Jensen's Inequality,
\begin{equation}\label{a_N}
\left\|\pi_N^{-1}(\bX)\right\|_{r,\P}=\left[\E\{\bar\pi_N^{-r}(\bX)\}\right]^{1/r}\geq\E\{\pi_N^{-1}(\bX)\}=\bar\pi_N^{-1},
\end{equation}
and hence $\left\|\pi_N^{-1}(\bX)\right\|_{r,\P}\asymp\bar\pi_N^{-1}$, which implies that $a_N\asymp\bar\pi_N$. It follows that, with $r,s>0$ satisfying $1/r+1/s=1$ and $2s=2+c$,
\begin{align}
\E\left[ \frac{a_N}{\pi_N(\bX)} \{\mhat(\bX) - \mu(\bX) \}^2 \right]&\leq a_N\left\|\pi_N^{-1}(\bX)\right\|_{r,\P}\|\mhat(\cdot)-\mu(\cdot)\|_{2s,\P}^2=o_p(1),\label{mhataN}\\
\E\left[ \frac{a_N}{\pi_N(\bX)} \left\{1 - \frac{\pi_N(\bX)}{\pihat_N(\bX)} \right\}^2\right]&\leq a_N\left\|\pi_N^{-1}(\bX)\right\|_{r,\P}\left\|1 - \frac{g(\bXv^T\bbeta_0)}{g(\bXv^T\widehat{\bbeta})}\right\|_{2s,\P}^2=O_p\left((N\bar\pi_N)^{-1}\right)=o_p(1),\nonumber
\end{align}
where \eqref{mhataN} requires an additional assumption that $\|\hat m(\cdot)-\mu(\cdot)\|_{2+c,\P}=o_p(1)$.

Now, \tcr{we} \jelena{analyze $\thetahat_\mathrm{DRSS}- \theta_0$, where we } further assume that $\|m(\cdot)-\mu(\cdot)\|_{2+c,\P}<\infty$. Apply\tcr{ing} part (b) of Theorem \ref{t4}, we have
$$(\thetahat_\mathrm{DRSS}- \theta_0) = \frac{1}{N} \sum_{i=1}^N \psi_{\mu}(\bZ_i) + o_p\left((N\bar\pi_N)^{-1/2}\right) + \Deltahat_N,$$
where $\psi_{\mu,e}(\bZ)=\mu(\bX) - \theta_0 + R/\pi_N(\bX) \{ Y - \mu(\bX)\}$ and
\begin{align*}
\Deltahat_N&=N^{-1}\sum_{k=1}^{\K}\sum_{i\in\mathcal I_k}\frac{R_i}{\pi_N(\bX_i)}\left\{1-\frac{\pi_N(\bX_i)}{\pihat_N(\bX_i)}\right\}\{\mu(\bX_i)-m(\bX_i)\}\\
&=N^{-1}\sum_{k=1}^{\K}\sum_{i\in\mathcal I_k}\frac{R_i}{\pi_N(\bX_i)}\{\mu(\bX_i)-m(\bX_i)\}\mathrm{IF}_\pi(\bZ;S_{-k})\\
&\qquad+N^{-1}\sum_{k=1}^{\K}\sum_{i\in\mathcal I_k}\frac{R_i}{\pi_N(\bX_i)}\{\mu(\bX_i)-m(\bX_i)\}\left\{1-\frac{g(\bXv_i^T\bbeta_0)}{g(\bXv_i^T\widehat{\bbeta})}-\mathrm{IF}_\pi(\bZ;S_{-k})\right\}.
\end{align*}
For each $k\leq\K$,
\begin{align*}
&\E_{\S_k}\left||\mathcal I_k|^{-1}\sum_{i\in\mathcal I_k}\frac{R_i}{\pi_N(\bX_i)}\{\mu(\bX_i)-m(\bX_i)\}\left\{1-\frac{g(\bXv_i^T\bbeta_0)}{g(\bXv_i^T\widehat{\bbeta})}-\mathrm{IF}_\pi(\bZ;S_{-k})\right\}\right|\\
&\qquad\leq\E\left|\{\mu(\bX)-m(\bX)\}\left\{1-\frac{g(\bXv^T\bbeta_0)}{g(\bXv^T\widehat{\bbeta})}-\mathrm{IF}_\pi(\bZ;S_{-k})\right\}\right|\\
&\qquad\leq\|\mu(\cdot)-m(\cdot)\|_{2,\P}\left\|1-\frac{g(\bXv^T\bbeta_0)}{g(\bXv^T\widehat{\bbeta})}-\mathrm{IF}_\pi(\bZ;S_{-k})\right\|_{2,\P}\overset{(i)}{=}o_p\left((N\bar\pi_N)^{-1/2}\right),
\end{align*}
where $(i)$ holds by \eqref{IF_pip1} and \eqref{IF_pip2}. By Lemma \ref{l2},
\begin{align*}
&|\mathcal I_k|^{-1}\sum_{i\in\mathcal I_k}\frac{R_i}{\pi_N(\bX_i)}\{\mu(\bX_i)-m(\bX_i)\}\left\{1-\frac{g(\bXv_i^T\bbeta_0)}{g(\bXv_i^T\widehat{\bbeta})}-\mathrm{IF}_\pi(\bZ_i;S_{-k})\right\}=o_p\left((N\bar\pi_N)^{-1/2}\right),
\end{align*}
and hence
\begin{align*}
&N^{-1}\sum_{k=1}^{\K}\sum_{i\in\mathcal I_k}\frac{R_i}{\pi_N(\bX_i)}\{\mu(\bX_i)-m(\bX_i)\}\left\{1-\frac{g(\bXv_i^T\bbeta_0)}{g(\bXv_i^T\widehat{\bbeta})}-\mathrm{IF}_\pi(\bZ_i;S_{-k})\right\}=o_p\left((N\bar\pi_N)^{-1/2}\right).
\end{align*}
Besides, for each $k\leq\K$, with $r,s>0$ satisfying $1/r+1/s=1$ and $2s=2+c$, \tco{and recall\tcr{ing} the definition of $\mathrm{IF}_\pi(\bZ;S_{-k})$ in \eqref{def:IF_pi-k}, we have}
\begin{align*}
&\Var_{\S_k}\left[|\mathcal I_k|^{-1}\sum_{i\in\mathcal I_k}\frac{R_i}{\pi_N(\bX_i)}\{\mu(\bX_i)-m(\bX_i)\}\mathrm{IF}_\pi(\bZ_i;S_{-k})\right]\\
&\qquad\overset{(i)}{\leq} |\mathcal I_k|^{-1}\left\|\E\left[\pi_N^{-1}(\bX)\{\mu(\bX)-m(\bX)\}^2\|\bXv\|_2\right]\right\|_2\left\|N_{-k}^{-1}\sum_{j\in\mathcal I_{-k}}\mathrm{IF}_{\bbeta}(\bZ_j)\right\|_2^2\\
&\qquad\leq |\mathcal I_k|^{-1}\|\pi_N^{-1}(\bX)\|_{2r,\P}\|\mu(\bX)-m(\bX)\|_{2s,\P}^2\left\|\|\bXv\|_2\right\|_{2r,\P}(N_{-k}\bar\pi_N)^{-1}\left\|\boldsymbol{\zeta}_{N,1}\right\|_2^2=O\left((N\bar\pi_N)^{-2}\right).
\end{align*}
By Lemma \ref{l1} and recall\tcr{ing} the definition \eqref{def:IF_pi-k},
\begin{align*}
&|\mathcal I_k|^{-1}\sum_{i\in\mathcal I_k}\frac{R_i}{\pi_N(\bX_i)}\{\mu(\bX_i)-m(\bX_i)\}\mathrm{IF}_\pi(\bZ_i;S_{-k})=\E_\bX\left[\{\mu(\bX)-m(\bX)\}\mathrm{IF}_\pi(\bZ;S_{-k})\right]+O_p\left((N\bar\pi_N)^{-1}\right)\\
&\qquad=N_{-k}^{-1}\sum_{j\in\mathcal I_{-k}}\mathrm{IF}_\pi(\bZ_j)+O_p\left((N\bar\pi_N)^{-1}\right)=N_{-k}^{-1}\sum_{j\in\mathcal I_{-k}}\mathrm{IF}_\pi(\bZ_j)+o_p\left((N\bar\pi_N)^{-1/2}\right),
\end{align*}
where $\mathrm{IF}_\pi(\bZ)=\E\left[\{1-\pi_N(\bX)\}\{\mu(\bX)-m(\bX)\}\bXv^T\right]J^{-1}(\bar\pi_N,\bgamma_0)\bXv\{R-\pi_N(\bX)\}$. Therefore,
\begin{align*}
\Deltahat_N&=\K^{-1}\sum_{k=1}^{\K}N_{-k}^{-1}\sum_{j\in\mathcal I_{-k}}\mathrm{IF}_\pi(\bZ_j)+o_p\left((N\bar\pi_N)^{-1/2}\right)=N^{-1}\sum_{i=1}^N\mathrm{IF}_\pi(\bZ_i)+o_p\left((N\bar\pi_N)^{-1/2}\right).
\end{align*}
\end{proof}

\begin{proof}[Proof of Lemma \ref{lemma:RSC}]
For any $a\in(0,1]$, we have the corresponding Jacobian (or Hessian) matrices of $\ell_N(\bgamma;a)$ and $\ell_N(\bgamma,1)$ \tco{w.r.t.} $\bgamma \in \R^{p+1}$ satisfy the following (analytical) inequality:
\begin{align*}
&\frac{\partial^2}{\partial \bgamma \partial \bgamma^T} \{\ell_N(\bgamma;a)\}=N^{-1}\sum_{i=1}^N\dot{g}(\bXv_i^T\bgamma+\log(a))\bXv_i\bXv_i^T\\
&\qquad\succeq aN^{-1}\sum_{i=1}^N\dot{g}(\bXv_i^T\bgamma)\bXv_i\bXv_i^T=a\frac{\partial^2}{\partial \bgamma \partial \bgamma^T} \{\ell_N(\bgamma;1)\},
\end{align*}
since for any $a\in(0,1]$ and $u\in\R$,
$$\dot{g}(u+\log(a))=\frac{a\exp(u)}{\{1+a\exp(u)\}^2}\geq\frac{a\exp(u)}{\{1+\exp(u)\}^2}=a\dot{g}(u).$$
Let $\mathcal G(\bgamma;a):=\ell_N(\bgamma;a)-a\ell(\bgamma;1)$. Then,
$$\frac{\partial^2}{\partial \bgamma \partial \bgamma^T}\{\mathcal G(\bgamma;a)\}\succeq\bzero,\qquad\forall\;\bgamma\in\R^{p+1}.$$
That is, the function $\mathcal{G}(\bgamma,a)$ is convex in $\bgamma \in \R^{p+1}$, and hence by the basic properties of convex functions, we have: for any $\bgamma, \bDelta \in \R^{p+1}$,
$$\mathcal G(\bgamma+\bDelta;a)-\mathcal G(\bgamma;a)-\bDelta^T\{\nabla_{\bgamma}\mathcal G(\bgamma;a)\}\geq0$$
and hence
\begin{align*}
&\delta\ell(\bDelta;a;\bgamma)=\ell_N(\bgamma+\bDelta;a)-\ell_N(\bgamma;a)-\bDelta^T\{\nabla_{\bgamma}\ell_N(\bgamma;a)\}\\
&\qquad\geq a\left[\ell_N(\bgamma+\bDelta;1)-\ell_N(\bgamma;1)-\bDelta^T\{\nabla_{\bgamma}\ell_N(\bgamma;1)\}\right]=a\left\{\delta\ell(\bDelta;1;\bgamma)\right\}.
\end{align*}
Therefore,
$$\delta\ell(\bDelta;a;\bgamma)\geq a\kappa\|\bDelta\|_2^2,\qquad\forall\;\bDelta\in A,$$
if $\delta\ell(\bDelta;1;\bgamma)\geq \kappa\|\bDelta\|_2^2$ for all $\bDelta\in A$.
\end{proof}

\begin{proof}[Proof of Lemma \ref{lemma:gradient}]
We first \jelena{derive} the following useful properties: define $\mu_{\bgamma_0}=\E(\bXv^T\bgamma_0)$, then by Lemma \ref{subG} and some calculation, for all $t\in\R$, $j\leq p+1$, and $r\geq1$,
\begin{align}
|\mu_{\bgamma_0}|&\leq\E(|\bXv^T\bgamma_0|)\leq\sigma_{\bgamma_0}\sqrt\pi<2\sigma_{\bgamma_0},\nonumber\\
\|\bXv^T\bgamma_0-\mu_{\bgamma_0}\|_{\psi_2}&\leq\|\bXv^T\bgamma_0\|_{\psi_2}+\|\mu_{\bgamma_0}\|_{\psi_2}\leq\sigma_{\bgamma_0}+\{\log(2)\}^{-1/2}|\mu_{\bgamma_0}|<4\sigma_{\bgamma_0},\nonumber\\
\E\{\exp(t\bXv^T\bgamma_0)\}&=\exp(t\mu_{\bgamma_0})\E[\exp\{t(\bXv^T\bgamma_0-\mu_{\bgamma_0})\}]\leq\exp\{2\sigma_{\bgamma_0}|t|+20\sigma_{\bgamma_0}^2t^2\},\label{upper:mgf}\\
\|\bXv(j)-\E\{\bXv(j)\}\|_{\psi_2}&\leq\|\bXv(j)\|_{\psi,2}+\|\E\{\bXv(j)\}\|_{\psi_2}\leq\sigma+\{\log(2)\}^{-1/2}\sqrt\pi\sigma,\nonumber\\
\max_{1\leq j\leq p+1}\E\{|\bXv(j)|^r\}&\leq r!\max_{1\leq j\leq p+1}\E[\exp\{|\bXv(j)|\}]\leq r!\exp(20\sigma^2).\nonumber
\end{align}
Notice that $\||\cdot|\|_{\psi_2}$ is a monotone increasing function \jelena{leading to} $\|X_2\|_{\psi_2}\geq\|X_1\|_{\psi_2}$ if $|X_2|\geq|X_1|$. Hence,
$$\max_{1\leq j\leq p+1}\|\{R_i-\pi_N(\bX_i)\}\bXv_{ij}\|_{\psi_2}\leq\max_{1\leq j\leq p+1}\|\bXv_{ij}\|_{\psi_2}\leq\sigma.$$
\tco{In addition},
\begin{align*}
&\max_{1\leq j\leq p+1}\E\left[\{R-\pi_N(\bX)\}^2\bXv^2(j)\right]\\
&\qquad=\max_{1\leq j\leq p+1}\E\left[\pi_N(\bX)\{1-\pi_N(\bX)\}\bXv^2(j)\right]\leq\max_{1\leq j\leq p+1}\bar\pi_N\E\left\{\exp(\bXv^T\bgamma_0)\bXv^2(j)\right\}\\
&\qquad\leq\bar\pi_N\left[\E\{\exp(2\bXv^T\bgamma_0)\}\max_{1\leq j\leq p+1}\E\{\bXv^4(j)\}\right]^{1/2}\leq2\exp(2\sigma_{\bgamma_0}+40\sigma_{\bgamma_0}^2+10\sigma^2)\bar\pi_N.
\end{align*}
Now, apply Theorem 3.4 of \cite{kuchibhotla2018moving}, for any $t_1\geq0$, with probability at least $1-3\exp(-t_1)$,
\begin{align}
&\left\|\nabla_{\bgamma}\ell_N(\bgamma_0;\bar\pi_N)\right\|_\infty=\left\|N_{-k}\sum_{i\in\mathcal I_{-k}}\{R_i-\pi_N(\bX_i)\}\bXv_i\right\|_\infty\nonumber\\
&\qquad\leq7\sqrt{\frac{2\exp(2\sigma_{\bgamma_0}+40\sigma_{\bgamma_0}^2+10\sigma^2)\bar\pi_N\{t_1+\log(p+1)\}}{N}}\\
&\qquad\qquad+\frac{c_6\sigma\sqrt{\log(2N)}\{t_1+\log(p+1)\}}{N},\label{event:B}
\end{align}
with some constant $c_6$ independent of $N$. Define $\mathcal B=\mathcal B(t_1)$ \jelena{to} be an event that \eqref{event:B} holds, then $\P(\mathcal B)\geq1-3\exp(-t_1)$.

Now, we consider the error \jelena{that originated} from the first step estimation $\pihat_N$. By Taylor's Theorem, for each $i\leq N$, there exists $\bar\pi_N'$ (depends on $i$) lies between $\bar\pi_N$ and $\pihat_N$, such that
\begin{align}
&\left|g(\bXv_i^T\bgamma_0+\log(\pihat_N))-g(\bXv_i^T\bgamma_0+\log(\bar\pi_N))\right|=\frac{|\pihat_N-\bar\pi_N|}{\bar\pi_N'}|\phi(\bXv_i^T\bgamma+\log(\bar\pi_N'))|\\
&\qquad\leq\frac{|\pihat_N-\bar\pi_N|}{\bar\pi_N'}g(\bXv_i^T\bgamma+\log(\bar\pi_N'))\leq\frac{|\pihat_N-\bar\pi_N|}{\min(\bar\pi_N,\pihat_N)}g(\bXv_i^T\bgamma+\log(\max\{\bar\pi_N,\pihat_N\})),\label{unifromg}
\end{align}
since function $g(\cdot)$ is monotone increasing. Observe that, for each $r\geq2$,
\begin{align*}
\E|R-\bar\pi_N|^r&=\E\left[|1-\bar\pi_N|^r\pi_N(\bX)+|-\bar\pi_N|^r\{1-\pi_N(\bX)\}\right]\\
&=(1-\bar\pi_N)^r\bar\pi_N+\bar\pi_N^r(1-\bar\pi_N)\leq2\bar\pi_N\leq \frac{r!}{2}1^{r-2}\cdot2\bar\pi_N.
\end{align*}
By Theorem 1 of \cite{Van_de_Geer_2013}, for any $t_2>0$,
$$\P_{\S}\left(|\pihat_N-\bar\pi_N|\geq2\sqrt\frac{t_2\bar\pi_N}{N}+\frac{t_2}{N}\right)\leq2\exp(-t_2).$$
Define event
\begin{equation}\label{prob:pihat}
\mathcal A=\mathcal A(t_2):=\{|\pihat_N-\bar\pi_N|<2\sqrt{t_2\bar\pi_N/N}+t_2/N\}.
\end{equation}
Then, $\P_{\S}(\mathcal A)\geq 1-2\exp(-t_2)$. Define
$$\pi_{N,\mathrm{min}}=\bar\pi_N-2\sqrt{t_2\bar\pi_N/N}-t_2/N,\quad\pi_{N,\mathrm{max}}=\bar\pi_N+2\sqrt{t_2\bar\pi_N/N}+t_2/N.$$
Suppose $t_2<N\bar\pi_N/9$, then $2\sqrt{t_2\bar\pi_N/N}+t_2/N<7\bar\pi_N/9$, $\pi_{N,\mathrm{min}}>2\bar\pi_N/9>0$ and $\pi_{N,\mathrm{max}}<16\bar\pi_N/9<16/9$. Recall \eqref{unifromg}, on event $\mathcal A$, we have for each $i\leq N$,
$$
\left|g(\bXv_i^T\bgamma_0+\log(\pihat_N))-g(\bXv_i^T\bgamma_0+\log(\bar\pi_N))\right|\leq\frac{|\pihat_N-\bar\pi_N|}{\pi_{N,\mathrm{min}}}g(\bXv_i^T\bgamma_0+\log(\pi_{N,\mathrm{max}})),
$$
and
\begin{align*}
&\left\|\nabla_{\bgamma}\ell_N(\bgamma_0;\pihat_N)-\nabla_{\bgamma}\ell_N(\bgamma_0;\bar\pi_N)\right\|_\infty=\left\|N^{-1}\sum_{i=1}^N\{g(\bXv_i^T\bgamma_0+\log(\pihat_N))-g(\bXv_i^T\bgamma_0+\log(\bar\pi_N))\}\bXv_i\right\|_\infty\\
&\qquad\leq\frac{\pihat_N-\bar\pi_N}{\pi_{N,\mathrm{min}}}\left\|N^{-1}\sum_{i=1}^Ng(\bXv_i^T\bgamma_0+\log(\pi_{N,\mathrm{max}}))\bXv_i\right\|_\infty\\
&\qquad\leq\frac{\pihat_N-\bar\pi_N}{\pi_{N,\mathrm{min}}}\left\|N^{-1}\sum_{i=1}^N\bV_i\right\|_\infty+\frac{\pihat_N-\bar\pi_N}{\pi_{N,\mathrm{min}}}\left\|\E\left\{g(\bXv^T\bgamma_0+\log(\pi_{N,\mathrm{max}}))\bXv\right\}\right\|_\infty,
\end{align*}
where
$$\bV_i=g(\bXv_i^T\bgamma_0+\log(\pi_{N,\mathrm{max}}))\bXv_i-\E\left\{g(\bXv^T\bgamma_0+\log(\pi_{N,\mathrm{max}}))\bXv\right\}.$$
For any vector $\bv$, let $\bv(j)$ denotes the j-th element of the vector $\bv$. Notice that, on event $\mathcal A$,
\begin{align*}
&\max_{1\leq j\leq p+1}\left|\E\left\{g(\bXv^T\bgamma_0+\log(\pi_{N,\mathrm{max}}))\bXv(j)\right\}\right|\leq\max_{1\leq j\leq p+1}\pi_{N,\mathrm{max}}\E\{\exp(\bXv^T\bgamma_0)|\bXv(j)|\}\\
&\qquad\leq\pi_{N,\mathrm{max}}\left[\E\{\exp(2\bXv^T\bgamma_0)\}\max_{1\leq j\leq p+1}\E\{\bXv^2(j)\}\right]^{1/2}\leq\pi_{N,\mathrm{max}}\sqrt2\exp(2\sigma_{\bgamma_0}+40\sigma_{\bgamma_0}^2+10\sigma^2),
\end{align*}
and hence
\begin{align*}
\max_{1\leq j\leq p+1}\|\bV(j)\|_{\psi_2}&\leq\max_{1\leq j\leq p+1}\left\|g(\bXv_i^T\bgamma_0+\log(\pi_{N,\mathrm{max}}))\bXv_i(j)\right\|_{\psi_2}\\
&\qquad+\max_{1\leq j\leq p+1}\left\|\E\left\{g(\bXv^T\bgamma_0+\log(\pi_{N,\mathrm{max}}))\bXv(j)\right\}\right\|_{\psi_2}\\
&\leq\max_{1\leq j\leq p+1}\left\|\bXv_i(j)\right\|_{\psi_2}+\pi_{N,\mathrm{max}}\sqrt2\exp(2\sigma_{\bgamma_0}+40\sigma_{\bgamma_0}^2+10\sigma^2)\\
&\leq\sigma+\frac{16\bar\pi_N}{9}\sqrt2\exp(2\sigma_{\bgamma_0}+40\sigma_{\bgamma_0}^2+10\sigma^2).
\end{align*}
\tco{Additionally},
\begin{align*}
&\max_{1\leq j\leq p+1}\E(\bV_i^2)\leq\max_{1\leq j\leq p+1}\E\left\{g^2(\bXv^T\bgamma_0+\log(\pi_{N,\mathrm{max}}))\bXv^2(j)\right\}\\
&\qquad\leq\pi_{N,\mathrm{max}}^2\max_{1\leq j\leq p+1}\E\left\{\exp(2\bXv^T\bgamma_0)\bXv^2(j)\right\}\\
&\qquad\leq\pi_{N,\mathrm{max}}^2\left[\E\{\exp(4\bXv^T\bgamma_0)\}\max_{1\leq j\leq p+1}\E\{\bXv^4(j)\}\right]^{1/2}\\
&\qquad\leq\pi_{N,\mathrm{max}}^22\exp(4\sigma_{\bgamma_0}+160\sigma_{\bgamma_0}^2+10\sigma^2).
\end{align*}
Define
\begin{equation}\label{event:C}
\mathcal C=\left\{\left\|N^{-1}\sum_{i=1}^N\bV_i\right\|_\infty\leq7c_8\pi_{N,\mathrm{max}}\sqrt{\frac{t_1+\log(p+1)}{N}}+\frac{c_9\sqrt{\log(2N)}\{t_1+\log(p+1)\}}{N}\right\},
\end{equation}
where $c_8=\sqrt2\exp(2\sigma_{\bgamma_0}+80\sigma_{\bgamma_0}^2+5\sigma^2)$, $c_9=c_6\{\sigma+16\bar\pi_N\exp(2\sigma_{\bgamma_0}+40\sigma_{\bgamma_0}^2+10\sigma^2)/9\}$. By Theorem 3.4 of \cite{kuchibhotla2018moving}, $\P(\mathcal C)\geq1-3\exp(-t_1)$. It follows that, on events $\mathcal A$ and $\mathcal C$,
\begin{align*}
&\left\|\nabla_{\bgamma}\ell_N(\bgamma_0;\pihat_N)-\nabla_{\bgamma}\ell_N(\bgamma_0;\bar\pi_N)\right\|_\infty\\
&\qquad\leq\frac{|\pihat_N-\bar\pi_N|}{\pi_{N,\mathrm{min}}}\left\|N\sum_{i=1}^N\bV_i\right\|_\infty+\frac{|\pihat_N-\bar\pi_N|}{\pi_{N,\mathrm{min}}}\left\|\E\left\{g(\bXv^T\bgamma_0+\log(\pi_{N,\mathrm{max}}))\bXv\right\}\right\|_\infty\\
&\qquad\leq\frac{2\sqrt{t_2\bar\pi_N/N}+t_2/N}{\pi_{N,\mathrm{min}}}\left\{7c_8\pi_{N,\mathrm{max}}\sqrt{\frac{t_1+\log(p+1)}{N}}+\frac{c_9\sqrt{\log(2N)}\{t_1+\log(p+1)\}}{N}\right\}\\
&\qquad\qquad+\frac{2\sqrt{t_2\bar\pi_N/N}+t_2/N}{\pi_{N,\mathrm{min}}}\pi_{N,\mathrm{max}}\sqrt2\exp(2\sigma_{\bgamma_0}+40\sigma_{\bgamma_0}^2+10\sigma^2).
\end{align*}
Recall that, when $t_2<N\bar\pi_N/9$,
$$\frac{2\sqrt{t_2\bar\pi_N/N}+t_2/N}{\pi_{N,\mathrm{min}}}<\frac{7}{2},\qquad\pi_{N,\min}>\frac{2}{9}\bar\pi_N,\qquad\pi_{N,\max}<\frac{16}{9}\bar\pi_N.$$
Hence, when $t_2<N\bar\pi_N/9$, on events $\mathcal A$, $\mathcal B$ and $\mathcal C$,
\begin{align*}
&\left\|\nabla_{\bgamma}\ell_N(\bgamma_0;\pihat_N)\right\|_\infty\leq\left\|\nabla_{\bgamma}\ell_N(\bgamma_0;\bar\pi_N)\right\|_\infty+\left\|\nabla_{\bgamma}\ell_N(\bgamma_0;\pihat_N)-\nabla_{\bgamma}\ell_N(\bgamma_0;\bar\pi_N)\right\|_\infty\\
&\qquad\leq7\sqrt{\frac{2\exp(2\sigma_{\bgamma_0}+40\sigma_{\bgamma_0}^2+10\sigma^2)\bar\pi_N\{t_1+\log(p+1)\}}{N}}+\frac{c_6\sigma\sqrt{\log(2N)}\{t_1+\log(p+1)\}}{N}\\
&\qquad\qquad+\frac{2\sqrt{t_2\bar\pi_N/N}+t_2/N}{\pi_{N,\mathrm{min}}}\left\{7c_8\pi_{N,\mathrm{max}}\sqrt{\frac{t_1+\log(p+1)}{N}}+\frac{c_9\sqrt{\log(2N)}\{t_1+\log(p+1)\}}{N}\right\}\\
&\qquad\qquad+\frac{2\sqrt{t_2\bar\pi_N/N}+t_2/N}{\pi_{N,\mathrm{min}}}\pi_{N,\mathrm{max}}\sqrt2\exp(2\sigma_{\bgamma_0}+40\sigma_{\bgamma_0}^2+10\sigma^2)\\
&\qquad\leq C_1(\bar\pi_N+\bar\pi_N^{1/2})\sqrt\frac{\{t_1+\log(p+1)\}}{N}+(C_2+C_3\bar\pi_N)\frac{\sqrt{\log(2N)}\{t_1+\log(p+1)\}}{N}\\
&\qquad\qquad+C_4\left\{\sqrt\frac{t_2\bar\pi_N}{N}+\frac{t_2}{N}\right\}.
\end{align*}
where $\P_{\S}(\mathcal A\cap\mathcal B\cap\mathcal C)\geq1-6\exp(-t_1)-2\exp(-t_2)$,
\begin{align}
& C_1=62\exp(2\sigma_{\gamma_0}+80\sigma_{\gamma_0}^2+5\sigma^2),\qquad C_2=\frac{9}{2}c_6\sigma,\label{constants1}\\
&C_3=\frac{56}{9}\exp(2\sigma_{\bgamma_0}+40\sigma_{\bgamma_0}^2+10\sigma^2),\qquad C_4=16\sqrt2\exp(2\sigma_{\bgamma_0}+40\sigma_{\bgamma_0}^2+10\sigma^2).\label{constants2}
\end{align}
\end{proof}

\begin{proof}[Proof of Theorem \ref{thm:high-dim}]
\tco{Here, we establish a non-asymptotic property of the offset logistic regression estimator based on the full sample $\bS$. The result follows from the Lemmas \ref{lemma:RSC} and \ref{lemma:gradient}, where we obtained the RSC property and controlled the gradient $\|\nabla_{\bgamma}\ell_N(\bgamma_0;\pihat_N)\|_\infty$, respectively. After that, we validate the conditions required in Theorem \ref{t4} for the proposed offset logistic PS estimator.}

For any $t\in\R$, set $t_1=t_2=t\log(p+1)$. By Lemma \ref{lemma:gradient}, on events $\mathcal A$, $\mathcal B$ and $\mathcal C$, with $\P_{\S}(\mathcal A\cap\mathcal B\cap\mathcal C)\geq1-8(p+1)^{-t}$,
\begin{align*}
&\left\|\nabla_{\bgamma}\ell_N(\bgamma_0;\pihat_N)\right\|_\infty\\
&\qquad\leq(t+1)\left\{C_1(\bar\pi_N+\bar\pi_N^{1/2})\sqrt\frac{\log(p+1)}{N}+(C_2+C_3\bar\pi_N)\frac{\sqrt{\log(2N)}\log(p+1)}{N}\right\}\\
&\qquad\qquad+C_4t^{1/2}\bar\pi_N^{1/2}\sqrt\frac{\log(p+1)}{N}+C_4t\frac{\log(p+1)}{N}\\
&\qquad\leq(t+1)\left\{(C_1+C_4)(\bar\pi_N+\bar\pi_N^{1/2})\sqrt\frac{\log(p+1)}{N}+(C_2+C_4+C_3\bar\pi_N)\frac{\sqrt{\log(2N)}\log(p+1)}{N}\right\}\\
&\qquad\leq(t+1)M_N,
\end{align*}
where
$$M_N=2(C_1+C_4)\bar\pi_N^{1/2}\sqrt\frac{\log(p+1)}{N}+(C_2+C_3+C_4)\frac{\sqrt{\log(2N)}\log(p+1)}{N}.$$
Hence, for any $\lambda_N\geq2(1+c)M_N$ with constant $c>0$,
$$2\left\|\nabla_{\bgamma}\ell_N(\bgamma_0;\pihat_N)\right\|_\infty\leq\lambda_N,\qquad\mbox{on events}\;\;\mathcal A,\mathcal B\;\mbox{and}\;\mathcal C.$$
Define event
\begin{equation}\label{RSC:leq1}
\mathcal D:=\left\{\delta\ell(\bDelta;\pihat_N;\bgamma)\geq\pihat_N\kappa\|\bDelta\|_2^2,\qquad\forall\;\bDelta\in\C_\delta(S;3)\;\;\mbox{and}\;\;\delta\leq1\right\}.
\end{equation}
By Lemma \ref{lemma:RSC}, $\P(\mathcal D)\geq1-\alpha_N$. Let $\delta_N^*=2\lambda_Ns^{1/2}(\pihat_N\kappa)^{-1}$. Then, the RSC condition holds for $\ell_N(\cdot;\pihat_N)$ with parameter $\pihat_N\kappa$ over $\C_{\delta_N^*}(S;3)$. By Theorem 1 of \cite{negahban2010unified}, when $\lambda_N\geq2\left\|\nabla_{\bgamma}\ell_N(\bgamma_0;\pihat_N)\right\|_\infty$, $2\lambda_Ns^{1/2}(\pihat_N\kappa)^{-1}\leq1$ and on event $\mathcal D$,
$$\|\bgammahat-\bgamma_0\|_2\leq\delta_N^*\leq2\lambda_Ns^{1/2}(\pihat_N\kappa)^{-1}.$$
Recall \eqref{prob:pihat}, for any $t>0$, on event $\mathcal A=\mathcal A(t)$,
\begin{align*}
\pihat_N&\geq\bar\pi_N-2\sqrt\frac{t\log(p+1)\bar\pi_N}{N}-\frac{t\log(p+1)}{N}\geq\frac{2}{9}\bar\pi_N,\\
2\lambda_Ns^{1/2}(\pihat_N\kappa)^{-1}&\leq\frac{1}{9}\lambda_Ns^{1/2}\bar\pi_N^{-1}\kappa^{-1}\leq1,
\end{align*}
when $t<N\bar\pi_N\{\log(p+1)\}^{-1}/9$ and $\lambda_N\leq9\kappa\bar\pi_Ns^{-1/2}$. Hence, if $N\bar\pi_N>9c\log(p+1)$,
$$\|\bgammahat-\bgamma_0\|_2\leq\frac{1}{9}\lambda_Ns^{1/2}\bar\pi_N^{-1}\kappa^{-1},\quad\mbox{on events $\mathcal A$, $\mathcal B$, $\mathcal C$ and $\mathcal D$},$$
where $\P_{\S}(\mathcal A\cap\mathcal B\cap\mathcal C\cap\mathcal D)\geq1-8(p+1)^{-c}-\alpha_N$.

Now, consider the asymptotic performance that as $N\to\infty$, $\log(p)\log(N)=O(N\bar\pi_N)$ and $s\log(p)=o(N\bar\pi_N)$. Then,
$$M_N\asymp\bar\pi_N^{1/2}\sqrt\frac{\log(p)}{N}.$$
Hence, with some $\lambda_N\asymp\{N^{-1}\bar\pi_N\log(p)\}^{1/2}$,
\begin{equation}\label{rate:gammahat}
\|\bgammahat-\bgamma_0\|_2=O_p\left(\sqrt\frac{s\log(p)}{N\bar\pi_N}\right)=o_p(1).
\end{equation}
Now we analyze the consistency rate of the \tco{PS} estimator $\pihat_N(\cdot)$. For any $r>0$,
$$\left\|1-\frac{\pi_N(\cdot)}{\pihat_N(\cdot)}\right\|_{r,\P}\leq\left\|\pihat_N(\cdot)-\pi_N(\cdot)\right\|_{2r,\P}\|\pihat_N^{-1}(\cdot)\|_{2r,\P}.$$
Let $u_0=\bxv^T\bgamma_0+\log(\bar\pi_N)$ and $\Delta_u=\bxv^T\bgammahat+\log(\pihat_N)-\{\bxv^T\bgamma_0+\log(\bar\pi_N)\}$. By mean value theorem, and notice that $g'(u)=g(u)\{1-g(u)\}$, for some $v'\in(0,1)$,
\begin{align*}
&|g(u_0+\Delta_u)-g(u_0)|=g'(u_0+v'\Delta_u)|\Delta_u|\leq g(u_0+v'\Delta_u)|\Delta_u|\\
&\qquad\leq\max\{g(u_0),g(u_0+\Delta_u)\}|\Delta_u|\leq\{g(u_0)+g(u_0+\Delta_u)\}|\Delta_u|,
\end{align*}
since the function $g(\cdot)>0$ is monotone increasing. Besides, notice that, on $\mathcal A$ and $\mathcal E:=\{\|\bgammahat-\bgamma_0\|_2\leq1\}$,
\begin{align*}
|\log(\pihat_N)-\log(\bar\pi_N)|&\leq\frac{|\pihat_N-\bar\pi_N|}{\min(\pihat_N,\bar\pi_N)}\leq\frac{2\sqrt{t_2\bar\pi_N/N}+t_2/N}{2\bar\pi_N/9}\leq\frac{21}{2}\sqrt\frac{t_2}{N\bar\pi_N},\\
g(u_0)&=\frac{\bar\pi_N\exp(-\bxv^T\bgamma_0)}{1+\bar\pi_N\exp(-\bxv^T\bgamma_0)}\leq\bar\pi_N\exp(-\bxv^T\bgamma_0),\\
g(u_0+\Delta_u)&=\frac{\pihat_N\exp(-\bxv^T\bgammahat)}{1+\pihat_N\exp(-\bxv^T\bgammahat)}\leq\pihat_N\exp(-\bxv^T\bgammahat)\leq\frac{16}{9}\bar\pi_N\exp(-\bxv^T\bgammahat),
\end{align*}
when $t_2<N\bar\pi_N/9$. Hence, on $\mathcal A\cap\mathcal E$,
\begin{align*}
&\left\|\pihat_N(\cdot)-\pi_N(\cdot)\right\|_{2r,\P}\leq\bar\pi_N\left\|\left\{\exp(-\bXv^T\bgamma_0)+\frac{16}{9}\exp(-\bXv^T\bgammahat)\right\}\left\{|\bXv^T(\bgammahat-\bgamma_0)|+\frac{21}{2}\sqrt\frac{t_2}{N\bar\pi_N}\right\}\right\|_{2r,\P}\\
&\qquad\leq\left[\left\|\exp(-\bXv^T\bgamma_0)\right\|_{4r,\P}+\frac{16}{9}\left\|\exp(-\bXv^T\bgamma_0)\right\|_{8r,\P}\left\|\exp\{-\bXv^T(\bgammahat-\bgamma_0)\}\right\|_{8r,\P}\right]\\
&\qquad\qquad\cdot\left\{\left\|\bXv^T(\bgammahat-\bgamma_0)\right\|_{4r,\P}+\frac{21}{2}\sqrt\frac{t_2}{N\bar\pi_N}\right\}\\
&\qquad\leq C\left\{\|\bgammahat-\gamma_0\|_2+\sqrt\frac{t_2}{N\bar\pi_N}\right\},
\end{align*}
with some constant $C>0$. Here, $\P(\mathcal A)\geq 1-\exp(-t_2)$ and recall \eqref{rate:gammahat}. Hence,
\begin{equation}\label{pihatdiff}
\left\|\pihat_N(\cdot)-\pi_N(\cdot)\right\|_{2r,\P}=O_p\left(\sqrt\frac{s\log(p)}{N\bar\pi_N}\right).
\end{equation}
\tco{Additionally}, observe that
\begin{align*}
\|\pihat_N^{-1}(\cdot)\|_{2r,\P}&=\|1+\pihat_N^{-1}\exp(-\bXv^T\bgammahat)\|_{2r,\P}\leq1+\pihat_N^{-1}\|\exp(-\bXv^T\bgammahat)\|_{2r,\P}\\
&\leq1+\pihat_N^{-1}\|\exp(-\bXv^T\bgamma_0)\|_{4r,\P}\|\exp(-U)\|_{4r,\P}.
\end{align*}
By \eqref{upper:mgf}, $\|\exp(-\bXv^T\bgamma_0)\|_{4r,\P}=O(1)$. By \eqref{piest}, $\pihat_N^{-1}=\bar\pi_N^{-1}\{1+o_p(1)\}.$ By Lemma part (b) of \ref{subG},
$$|\E(U)|\leq\E(|U|)\leq\sigma\sqrt{\pi}\|\bgammahat-\bgamma_0\|_2.$$
Hence, by triangular inequality and part (a) of \ref{subG},
$$\|U-\E(U)\|_{\psi_2}\leq\|U\|_{\psi_2}+\|\E(U)\|_{\psi_2}\leq\{1+\sqrt{\pi/\log(2)}\}\sigma\|\tilde{\bgamma}-\bgamma\|_2\leq4\sigma\|\bgammahat-\bgamma_0\|_2.$$
By part (c) of Lemma \ref{subG},
$$\|\exp\{-U+\E(U)\}\|_{4r,\P}=\left(\E\exp[-4r\{U-\E(U)\}]\right)^{1/(4r)}\leq\exp\left(128r\sigma^2\|\bgammahat-\bgamma_0\|_2^2\right).$$
Hence,
\begin{align*}
\|\exp(-U)\|_{4r,\P}&=\|\exp\{-U+\E(U)\}\|_{4r,\P}\exp\{-\E(U)\}\\
&\leq\exp\left(128r\sigma^2\|\bgammahat-\bgamma_0\|_2^2\right)\exp(\sigma\sqrt{\pi}\|\bgammahat-\bgamma_0\|_2)=1+o_p(1).
\end{align*}
It follows that
\begin{equation}\label{pihatr}
\|\pihat_N^{-1}(\cdot)\|_{2r,\P}\leq1+\bar\pi_N^{-1}\{1+o_p(1)\}\cdot O(1)\cdot\{1+o_p(1)\}=O_p(\bar\pi_N^{-1}).
\end{equation}
Therefore, by \eqref{pihatdiff} and \eqref{pihatr},
$$\left\|1-\frac{\pi_N(\cdot)}{\pihat_N(\cdot)}\right\|_{r,\P}\leq\left\|\pihat_N(\cdot)-\pi_N(\cdot)\right\|_{2r,\P}\|\pihat_N^{-1}(\cdot)\|_{2r,\P}=O_p\left(\sqrt\frac{s\log(p)}{N\bar\pi_N}\right).$$
Besides, recall \eqref{a_N} and notice that
\begin{align*}
\|\pi_N^{-1}(\bX)\|_{r,\P}&\leq1+\bar\pi_N^{-1}\|\exp(-\bXv^T\bgamma_0)\|_{r,\P}\leq1+\bar\pi_N^{-1}\exp(2\sigma_{\bgamma_0}+20\sigma_{\bgamma_0}r)=O(\bar\pi_N^{-1}).
\end{align*}
Therefore,
\begin{align*}
\E\left[ \frac{a_N}{\pi_N(\bX)} \left\{1 - \frac{\pi_N(\bX)}{\pihat_N(\bX)} \right\}^2\right]&\leq a_N\left\|\pi_N^{-1}(\bX)\right\|_{2,\P}\left\|1 - \frac{\pi_N(\cdot)}{\pihat_N(\cdot)}\right\|_{4,\P}^2=O_p\left(\frac{s\log(p)}{N\bar\pi_N}\right).
\end{align*}
If further assume $\|\mhat(\cdot)-m(\cdot)\|_{2+c,\P}=o_p(1)$, then
\begin{align*}
\E\left[ \frac{a_N}{\pi_N(\bX)} \{\mhat(\bX) - m(\bX) \}^2 \right]&\leq a_N\left\|\pi_N^{-1}(\bX)\right\|_{1+c/2,\P}\|\mhat(\cdot)-m(\cdot)\|_{2+c,\P}^2=o_p(1).
\end{align*}
\end{proof}

\begin{proof}[Proof of Lemma \ref{RSC:general}]
The proof of Lemma \ref{RSC:general} is based on the proof of Proposition 2 in \cite{negahban2010unified}. Here, we only provide the details that are different from their proof and we will use our notations in the folloing proof. As a reminder, $N$ denotes the number of samples, $\bXv\in\R^{p+1}$ is the covariate containing the intercept term and $\bgamma_0\in\R^{p+1}$ is the coefficient of the balanced logistic model (the dimension $p$ in their proof will be replaced by $p+1$ everywhere because of the usage of the intercept term).

The proof consists of 3 main steps: 1) show that (71) of \cite{negahban2010unified} holds under our assumptions and the parameter $K_3$ we choose, 2) prove a slightly different version of (72) in \cite{negahban2010unified}, 3) conclude the RSC property result.

{\bf Step 1.} For the inequality (71), similarly as in their proof, we notice that $\E\{(\bXv^T\bDelta)^2\}\geq\kappa_l\|\bDelta\|_2^2=\kappa_l$ for any $\|\bDelta\|_2=1$. Hence, it suffices to show their inequality (73). Ins\jelena{t}ead of assuming $\bXv$ to be a zero-mean jointly sub-Gaussian random vector (which is not ture since we have the intercept term here), we only assume a $(2+c)$-th moment condition that $\sup_{\|\bv\|_2\leq1}\|\bXv^T\bv\|_{2+c,\P}\leq M<\infty$ and a $c$-th moment condition that $\|\bXv^T\bgamma_0\|_{c,\P}\leq\mu_c<\infty$, with our choice on the constant $K_3$. We have
\begin{align*}
\sup_{\|\bDelta\|_2\leq1}\P(|\bXv^T\bDelta|\geq\tau/2)&\leq(\tau/2)^{-2-c}\sup_{\|\bv\|_2\leq1}\E|\bXv^T\bDelta|^{2+c}\leq M^{2+c}(\tau/2)^{-2-c},\\
\P(|\bXv^T\bgamma_0|\geq T)&\leq\E|\bXv^T\bgamma_0|^cT^{-c}=\mu_c^cT^{-c}.
\end{align*}
Hence, by H\"older's Inequality, for any $\|\bDelta\|_2\leq1$,
\begin{align*}
\E\left\{(\bXv^T\bDelta)^21_{|\bXv^T\bgamma_0|\geq T}\right\}&\leq\|\bXv^T\bDelta\|_{2+c,\P}^2\{\P(|\bXv^T\bgamma_0|\geq T)\}^{\frac{c}{2+c}}\leq M^2\mu_c^{\frac{c^2}{2+c}}T^{-\frac{c^2}{2+c}},\\
\E\left\{(\bXv^T\bDelta)^21_{|\bXv^T\bDelta|\geq\tau/2}\right\}&\leq\|\bXv^T\bDelta\|_{2+c,\P}^2\{\P(|\bXv^T\bDelta|\geq\tau/2)\}^{\frac{c}{2+c}}\leq M^{2+c}(\tau/2)^{-c}.
\end{align*}
It follows that, for $\tau^2=T^2=K_3\geq1$,
\begin{align*}
&\E\left\{(\bXv^T\bDelta)^2-g_{\bDelta}(\bX)\right\}\leq\E\left\{(\bXv^T\bDelta)^21_{|\bXv^T\bgamma_0|\geq T}\right\}+\E\left\{(\bXv^T\bDelta)^21_{|\bXv^T\bDelta|\geq\tau/2}\right\}\\
&\qquad\leq M^2\mu_c^{\frac{c^2}{2+c}}T^{-\frac{c^2}{2+c}}+M^{2+c}(\tau/2)^{-c}\leq(M^2\mu_c^{\frac{c^2}{2+c}}+M^{2+c}2^c)K_3^{-\frac{c^2}{4+2c}}.
\end{align*}
Hence, (73) of \cite{negahban2010unified} holds when we set
$$K_3=\max\left[1,\left\{2\kappa_l^{-1}(M^2\mu_c^{\frac{c^2}{2+c}}+M^{2+c}2^c)\right\}^{\frac{4+2c}{c}}\right].$$

{\bf Step 2.} We will demonstrate a slightly different version of (72) in \cite{negahban2010unified} that
\begin{equation}\label{eq72}
\P_{\S}\left\{Z(t)\geq\frac{\kappa_l}{4}+66K_3\sigma\sqrt\frac{\log(p+1)}{N}t\right\}\leq\exp\left\{-\frac{N\kappa_l^2}{64K_3^2}-\sigma^2t^2\log(p+1)\right\}.
\end{equation}
Set $z^*(t)=\kappa_l/4+2K_3\sigma\sqrt{\log(p+1)/N}t$ and let $\mathcal F:=\{\pm f(\cdot):f(\bu)=g_{\bDelta}(\bu)-\E\{g_{\bDelta}(\bX)\},\|\bDelta\|_2=1,\|\bDelta\|_1=t\}$. Since $0\leq g_{\bDelta}(\bu)\leq K_3$ for all $\bu$, we have $|f(\bX_i)|\leq K_3$ for all $f\in\mathcal F$.
By Lemma \ref{Hoeffding}, we have a slightly different version of their (76):
\begin{align}
\P_{\S}[Z(t)\geq\E\{Z(t)\}+z^*(t)]&\leq\exp\left[-\frac{N\{z^*(t)\}^2}{4K_3^2}\right]\leq\exp\left\{-\frac{N\kappa_l^2}{64K_3^2}-\sigma^2t^2\log(p+1)\right\}.\label{eq76}
\end{align}
Now, we need to obtain an upper bound for $\E_{\S}\|N^{-1}\sum_{i=1}^N\varepsilon_i\bu_i\|_\infty$ only using the marginal sub-Gaussianity of $\bXv$. Firstly, since $|\varepsilon_i\bu_i(j)|\leq |X_i(j)|$, by part (a) of Lemma \ref{subG},
$$\sup_{1\leq j\leq p+1}\|\varepsilon_i\bu_i(j)\|_{\psi_2}\leq\sup_{1\leq j\leq p+1}\|\bXv_i(j)\|_{\psi_2}\leq\sigma.$$
Notice that $\E(\varepsilon\bu)=0$ since $\varepsilon$ is independent with $\bu$ and $\E(\varepsilon)=0$, by part (e) of Lemma \ref{subG}, for any $1\leq j\leq p+1$,
$$\left\|N^{-1}\sum_{i=1}^N\varepsilon_i\bu_i(j)\right\|_{\psi_2}\leq4\sigma/\sqrt N.$$
By part (d) of Lemma \ref{subG},
$$\left\|\left\|N^{-1}\sum_{i=1}^N\varepsilon_i\bu_i\right\|_\infty\right\|_{\psi_2}\leq4\{\log(p+1)+2\}^{1/2}\sigma/\sqrt N\leq8\{\log(p+1)\}^{1/2}\sigma/\sqrt N,$$
for any $p\geq1$. Hence, by part (b) of Lemma \ref{subG},
$$\E_{\S}\left\|N^{-1}\sum_{i=1}^N\varepsilon_i\bu_i\right\|_\infty\leq8\sqrt\pi\sigma\sqrt\frac{\log(p+1)}{N}.$$
Combining the upper bound with (78) of \cite{negahban2010unified}, we have a slightly different version of their inequality (77):
$$\E_{\S}\{Z(t)\}\leq64K_3t\sqrt\pi\sigma\sqrt\frac{\log(p+1)}{N}.$$
and recall \eqref{eq76}, hence \eqref{eq72} follows. Notice that the statements in Step 2 are all independent of the choice of $K_3$, so our choice on the constant $K_3$ does not affect the validity of the resutlts.

{\bf Step 3.} By inequality (71) of \cite{negahban2010unified} and our \eqref{eq72}, we conclude that: for any $t>0$,
\begin{align*}
&\P_{\S}\left[\widehat{\E}_N\{g_{\bDelta}(\bX)\}<\frac{\kappa_l}{4}-66K_3\sigma\sqrt\frac{\log(p+1)}{N}t,\;\;\exists\bDelta\in\R^{p+1},\;\mbox{with}\;\|\bDelta\|_1=t,\|\bDelta\|_2=1\right]\\
&\qquad\leq\exp\left\{-\frac{N\kappa_l^2}{64K_3^2}-\sigma^2t^2\log(p+1)\right\}.
\end{align*}
Let $\S(1,t)=\{\bDelta\in\R^{p+1}:\|\bDelta\|_2\leq1,\|\bDelta\|_1/\|\bDelta\|_2=t\}$. By their inequality (66) and the technique in (69),
\begin{align*}
&\P_{\S}\left[\delta\ell(\bDelta;1;\bgamma_0)<L_{\psi}(K_3^{1/2})\left\{\frac{\kappa_l}{4}\|\bDelta\|_2^2-66K_3\sigma\sqrt\frac{\log(p+1)}{N}\|\bDelta\|_2t\right\},\;\;\exists\bDelta\in\S(1,t)\right]\\
&\qquad\leq\exp\left\{-\frac{N\kappa_l^2}{64K_3^2}-\sigma^2t^2\log(p+1)\right\},
\end{align*}
where for a logistic model, $L_\psi(K_3^{1/2})=\dot{g}(2K_3^{1/2})>0$.

By a peeling argument as in \cite{raskutti2010restricted}, \eqref{RSC:finite} holds. If further assume that $s\log(p)=o(N)$. For any $\bDelta\in\mathcal C(S,3)$,
$$\|\bDelta\|_1=\|\bDelta_S\|_1+\|\bDelta_{S^c}\|_1\leq4\|\bDelta_S\|_1\leq4\sqrt s\|\bDelta_S\|_2\leq4\sqrt s\|\bDelta\|_2.$$
and hence \eqref{RSC:asymp} holds.
\end{proof}

\begin{proof}[Proof of Theorem \ref{thm:ex2}]
\tco{We establish the asymptotic properties of the stratified PS estimator and the DRSS estimator based on the stratified PS estimator.}

Let $\bar\pi_N=\E(R)$, then $\pi_{1,N}+\pi_{0,N}\in(\bar\pi_N/(1-C),\bar\pi_N/C)$ and $\pi(\bX)\in(C\bar\pi_N/(1-C),(1-C)\bar\pi_N/C)$ for all $\bX\in\mathcal X$. Let $N_1=\sum_{i\in\mathcal I_{-k}}\delta_i$ and $N_0=\sum_{i\in\mathcal I_{-k}}(1-\delta_i)$. Similarly as in \eqref{piest}, for $j\in\{0,1\}$, $N_j^{-1}=O_p(N^{-1})$, $\pihat_j(\S_{-k})-\pi_{j,N}=O_p(\sqrt{\bar\pi_N/N})$ and $1-\pi_{j,N}/\hat\pi_j(\S_{-k})=O_p(1/\sqrt{N\bar\pi_N})$. Hence, $\pihat_j^{-1}(\S_{-k})=\pi_{j,N}^{-1}\{1+O_p(1/\sqrt{N\bar\pi_N})\}$. It follows that there exists $c>0$ such that
$$\P_{\S_{-k}}(\pihat_j(\S_{-k})>c\bar\pi_N)\to1,\quad\mbox{for}\;\;j\in\{0,1\}.$$
Hence, with probability approaching 1,
\begin{align*}
\pihat_N(\bX;\S_{-k})&=\pihat_1(\S_{-k})\widehat p_\delta(\bX;\S_{-k})+\pihat_0(\S_{-k})\{1-\widehat p_\delta(\bX;\S_{-k})\}\\
&\geq c\bar\pi_N\widehat p_\delta(\bX;\S_{-k})+c\bar\pi_N\{1-\widehat p_\delta(\bX;\S_{-k})\}=c\bar\pi_N.
\end{align*}
Observe that
\begin{align*}
\pihat_N(\bX;\S_{-k})-\pi_N(\bX)&=\{\pihat_1(\S_{-k})-\pihat_0(\S_{-k})\}\{\widehat p_\delta(\bX;\S_{-k})-p_\delta(\bX)\}\\
&\qquad+\{\pihat_1(\S_{-k})-\pi_{1,N}\}p_\delta(\bX)+\{\pihat_0(\S_{-k})-\pi_{0,N}\}\{1-p_\delta(\bX)\}.
\end{align*}
Hence,
\begin{align*}
\left\|\frac{\pihat_N(\cdot;\S_{-k})-\pi_N(\cdot)}{\pi_N(\cdot)}\right\|_{2,\mathbb P_\bX}&=O_p\left(r_{p_\delta,N}+(N\bar\pi_N)^{-1/2}\right),\\
\left\|\frac{\pihat_N(\cdot;\S_{-k})-\pi_N(\cdot)}{\pihat_N(\cdot;\S_{-k})}\right\|_{2,\mathbb P_\bX}&=O_p\left(r_{p_\delta,N}+(N\bar\pi_N)^{-1/2}\right).
\end{align*}
Following the case (b) in Theorem \ref{t4} that $\pi_N(\cdot)= e_N(\cdot)$ being correctly specified,
$$\thetahat_\mathrm{DRSS}-\theta_0=\frac{1}{N}\sum_{i=1}^N\psi_{\mu,e}(\bZ_i)+\Deltahat_N+O_p\left(\frac{c_{\mu,N}}{\sqrt{Na_N}}+\frac{c_{e,N}}{\sqrt{Na_N}}+r_{\pi,m.N}\right),$$
where $\psi_{\mu,e}(\bZ)=\mu(X)-\theta_0+R/\bar\pi_N(X)[Y-\mu(X)]$, $\Deltahat_N=\sum_{k=1}^{\mathbb K}\Deltahat_{N,k}$ and
$$\Deltahat_{N,k}=N^{-1}\sum_{i\in\mathcal I_k} \frac{R_i}{\pi_N(\bX_i)}\left\{1 - \frac{\pi_N(\bX_i)}{\pihat_N(\bX_i;\S_{-k})}\right\}\{\mu(\bX_i) - m(\bX_i)\}.$$
With a slight abuse of notation, let $\bZ=(\delta,R,\bX)$, $\bZ_i=(\delta_i,R_i,\bX_i)$ and $\mathbb S_k=\{\bZ_i:i\in\mathcal I_k\}$. Then,
\begin{align*}
\Var_{\mathbb S_k}(\Deltahat_{N,k})&=N^{-2}|\mathcal I_k|\Var\left[\frac{R}{\pi_N(\bX)}\left\{1 - \frac{\pi_N(\bX)}{\pihat_N(\bX;\S_{-k})}\right\}\{\mu(\bX) - m(\bX)\}\right]\\
&\leq N^{-1}\E\left[\frac{1}{\pi_N(\bX)}\left\{1 - \frac{\pi_N(\bX)}{\pihat_N(\bX;\S_{-k})}\right\}^2\{\mu(\bX) - m(\bX)\}^2\right]\\
&\leq \frac{1-C}{C}(N\bar\pi_N)^{-1}\left\|\frac{\pihat_N(\cdot;\S_{-k})-\pi_N(\cdot)}{\pihat_N(\cdot;\S_{-k})}\right\|_{2,\mathbb P_\bX}^2\|\mu(\cdot)-m(\cdot)\|_{\infty,\mathbb P_\bX}^2\\
&=O_p\left((N\bar\pi_N)^{-1}r_{p_\delta,N}^2+(N\bar\pi_N)^{-2}\right),
\end{align*}
and by Lemma \ref{l1},
$$
\Deltahat_{N,k}=\E_{\S_k}(\Deltahat_{N,k})+O_p\left((N\bar\pi_N)^{-1/2}r_{p_\delta,N}+(N\bar\pi_N)^{-1}\right).
$$
\tco{In addition},
\begin{align*}
&N|\mathcal I_k|^{-1}\E_{\mathbb S_k}(\Deltahat_{N,k})=\E\left[\frac{R}{\pi_N(\bX)}\left\{1 - \frac{\pi_N(\bX)}{\pihat_N(\bX;\S_{-k})}\right\}\{\mu(\bX) - m(\bX)\}\right]\\
&\qquad=\E\left[\left\{1 - \frac{\pi_N(\bX)}{\pihat_N(\bX;\S_{-k})}\right\}\{\mu(\bX) - m(\bX)\}\right]\\
&\qquad=\E\left[\frac{\pihat_N(\bX;\S_{-k})-\pi_N(\bX)}{\pi_N(\bX)}\{\mu(\bX) - m(\bX)\}\right]\\
&\qquad\qquad-\E\left[\frac{\pihat_N(\bX;\S_{-k})-\pi_N(\bX)}{\pi_N(\bX)}\cdot\frac{\pihat_N(\bX;\S_{-k})-\pi_N(\bX)}{\pihat_N(\bX;\S_{-k})}\{\mu(\bX) - m(\bX)\}\right]\\
&\qquad=\E\left[\frac{\pihat_N(\bX;\S_{-k})-\pi_N(\bX)}{\pi_N(\bX)}\{\mu(\bX) - m(\bX)\}\right]\\
&\qquad\qquad+O_p\left(\left\|\frac{\pihat_N(\cdot;\S_{-k})-\pi_N(\cdot)}{\pi_N(\cdot)}\right\|_{2,\mathbb P_\bX}\left\|\frac{\pihat_N(\cdot;\S_{-k})-\pi_N(\cdot)}{\pihat_N(\cdot;\S_{-k})}\right\|_{2,\mathbb P_\bX}\|\mu(\cdot)-m(\cdot)\|_\infty\right)\\
&\qquad=\E\left[\frac{\pihat_N(\bX;\S_{-k})-\pi_N(\bX)}{\pi_N(\bX)}\{\mu(\bX) - m(\bX)\}\right]+O_p\left(\|\widehat p_\delta(\cdot)-p_\delta(\cdot)\|_{2,\mathbb P_\bX}^2+(N\bar\pi_N)^{-1}\right).
\end{align*}
Let $\pitil_N(\cdot;\S_{-k})=\pihat_1(\S_{-k})p_\delta(\cdot)+\pihat_0(\S_{-k})\{1-p_\delta(\cdot)\}$. Then,
\begin{align*}
&\E\left[\frac{\pihat_N(\bX;\S_{-k})-\pi_N(\bX)}{\pi_N(\bX)}\{\mu(\bX) - m(\bX)\}\right]\\
&\qquad=\E\left[\frac{\pitil_N(\bX;\S_{-k})-\pi_N(\bX)}{\pi_N(\bX)}\{\mu(\bX) - m(\bX)\}\right]+\E\left[\frac{\pihat_N(\bX;\S_{-k})-\pitil_N(\bX;\S_{-k})}{\pi_N(\bX)}\{\mu(\bX) - m(\bX)\}\right]\\
&\qquad=\E\left[\frac{\pitil_N(\bX;\S_{-k})-\pi_N(\bX)}{\pi_N(\bX)}\{\mu(\bX) - m(\bX)\}\right]+O_p\left(r_{p_\delta,N}\|\mu(\cdot)-m(\cdot)\|_{2,\mathbb P_\bX}\right)\\
&\qquad=\{\pihat_1(\S_{-k})-\pi_{1,N}\}\E\left[\frac{p_\delta(\bX)}{\pi_N(\bX)}\{\mu(\bX)-m(\bX)\}\right]\\
&\qquad\qquad+\{\pihat_0(\S_{-k})-\pi_{0,N}\}\E\left[\frac{1-p_\delta(\bX)}{\pi_N(\bX)}\{\mu(\bX)-m(\bX)\}\right]+O_p\left(r_{p_\delta,N}\right).
\end{align*}
Let $\widehat p_\delta(\S_{-k})=N_{-k}^{-1}\sum_{i\in\mathcal I_{-k}}\delta_i$ and $p_\delta=\E[p_\delta(\bX)]$. Then, similarly as \eqref{piest}, $\widehat p_\delta^{-1}(\S_{-k})=p_\delta^{-1}\{1+O_p(1/\sqrt N)\}$. Hence,
\begin{align*}
&\pihat_1(\S_{-k})-\pi_{1,N}=N_{-k}^{-1}\sum_{i\in\mathcal I_{-k}}\frac{\delta_iR_i}{\widehat p_\delta(\S_{-k})}-\pi_{1,N}\\
&\qquad=N_{-k}^{-1}\sum_{i\in\mathcal I_{-k}}\frac{\delta_iR_i}{p_\delta}-\pi_{1,N}+O_p(N^{-1/2})N_{-k}^{-1}\sum_{i\in\mathcal I_{-k}}\frac{\delta_iR_i}{p_\delta}=N_{-k}^{-1}\sum_{i\in\mathcal I_{-k}}\frac{\delta_iR_i}{p_\delta}-\pi_{1,N}+O_p(N^{-1/2}\bar\pi_N).
\end{align*}
Similarly,
$$\pihat_0(\S_{-k})-\pi_{0,N}=N_{-k}^{-1}\sum_{i\in\mathcal I_{-k}}\frac{(1-\delta_i)R_i}{1-p_\delta}-\pi_{0,N}+O_p(N^{-1/2}\bar\pi_N).$$
Let
\begin{align*}
\mathrm{IF}_\pi(\bZ)&=\left\{\frac{\delta R}{p_\delta}-\pi_{1,N}\right\}\E\left[\frac{p_\delta(\bX)}{\pi_N(\bX)}\{\mu(\bX)-m(\bX)\}\right]\\
&\qquad+\left\{\frac{(1-\delta)R}{1-p_\delta}-\pi_{0,N}\right\}\E\left[\frac{1-p_\delta(\bX)}{\pi_N(\bX)}\{\mu(\bX)-m(\bX)\}\right].
\end{align*}
Then,
$$
\E\left[\frac{\pihat_N(\bX;\S_{-k})-\pi_N(\bX)}{\pi_N(\bX)}\{\mu(\bX) - m(\bX)\}\right]=N_{-k}^{-1}\sum_{i\in\mathcal I_{-k}}\mathrm{IF}_\pi(\bZ_i)+O_p\left(N^{-1/2}\right).
$$
Hence,
\begin{align*}
\Deltahat_N&=\sum_{k=1}^{\K}\Deltahat_{N,k}=(\K N_{-k})^{-1}\sum_{k=1}^{\K}\sum_{i\in\mathcal I_{-k}}\mathrm{IF}_\pi(\bZ_i)+O_p\left(\|\widehat p_\delta(\cdot)-p_\delta(\cdot)\|_{2,\mathbb P_\bX}^2+(N\bar\pi_N)^{-1}\right)\\
&\qquad+O_p\left(r_{p_\delta,N}\right)+O_p\left(N^{-1/2}\right)+O_p\left((N\bar\pi_N)^{-1/2}r_{p_\delta,N}+(N\bar\pi_N)^{-1}\right)\\
&=N^{-1}\sum_{i=1}^N\mathrm{IF}_\pi(\bZ_i)+O_p\left(r_{p_\delta,N}+(N\bar\pi_N)^{-1}+N^{-1/2}\right).
\end{align*}
By part (b) of Theorem \ref{t4},
\begin{align*}
\thetahat_\mathrm{DRSS}- \theta_0&= \frac{1}{N} \sum_{i=1}^N \Psi(\bZ_i) + O_p\left(r_{p_\delta,N}+(N\bar\pi_N)^{-1}+N^{-1/2}\right)\\
&\qquad+ O_p\left(r_{\mu,N}(N\bar\pi_N)^{-1/2}+\left\{r_{p_\delta,N}+(N\bar\pi_N)^{-1/2}\right\}\left\{(N\bar\pi_N)^{-1/2}+r_{\mu,N}\right\}\right),\\
&= \frac{1}{N} \sum_{i=1}^N \Psi(\bZ_i) + O_p\left((N\bar\pi_N)^{-1}+N^{-1/2}+r_{\mu,N}(N\bar\pi_N)^{-1/2}+r_{p_\delta,N}\right),
\end{align*}
where $\Psi(\bZ) := \psi_{\mu}(\bZ) + \mathrm{IF}_{\pi}(\bZ)$ and $\E\{\Psi(\bZ)\} = 0$ with
\begin{align*}
\psi_{\mu}(\bZ) &= \frac{R}{\pi_N(\bX)}\{Y - \mu(\bX)\} + \mu(\bX) - \theta_0,\\
\mathrm{IF}_\pi(\bZ)&=\left\{\frac{\delta R}{p_\delta}-\pi_{1,N}\right\}\E\left[\frac{p_\delta(\bX)}{\pi_N(\bX)}\{\mu(\bX)-m(\bX)\}\right]\\
&\qquad+\left\{\frac{(1-\delta)R}{1-p_\delta}-\pi_{0,N}\right\}\E\left[\frac{1-p_\delta(\bX)}{\pi_N(\bX)}\{\mu(\bX)-m(\bX)\}\right].
\end{align*}
If further  $r_{p_\delta,N}=o_p((N\bar\pi_N)^{-1/2})$,
$$
\thetahat_\mathrm{DRSS}- \theta_0=\frac{1}{N} \sum_{i=1}^N \Psi(\bZ_i) + o_p\left((N\bar\pi_N)^{-1/2}\right).
$$
\end{proof}

\begin{proof}[Proof of Theorem \ref{t6}]
Since $\pi_N(\bX)>c\bar\pi_N$, we have
$$a_N=1/\E\{\pi_N^{-1}(\bX)\}\geq c^{-1}\bar\pi_N.$$
\tco{Additionally}, by Jensen's inequality, $a_N\leq\bar\pi_N$. Hence, $a_N\asymp\bar\pi_N$. \tco{For each $k\leq\K$, define the following event
$$\mathcal E_{-k}:=\{\pihat_N(\bx;\S_{-k})<2C\bar\pi_N,\;\;\forall\bx\in\mathcal X\}.$$
Then, under conditions $e_N(\bX)<C\bar\pi_N$ for all $\bx\in\mathcal X$, \eqref{4.4}, and $r_{e,N}=o(1)$, we have
\begin{equation}\label{E-k}
\P_{\S_{-k}}(\mathcal E_{-k})\geq\P\left(\sup_{x\in\mathcal X}\left|\frac{\pihat_N(\bx;\S_{-k})-e_N(\bX)}{\bar\pi_N}\right|>C\right)=1-o(1).
\end{equation}
Recall that $\varepsilon=Y-Rm_1(\bX)-(1-R)m_0(\bX)$. W}e have $\E(\varepsilon|R,\bX)=0$. Observe that
$$
\hat\theta^0-\theta^0=N^{-1}\sum_{i=1}^N\psi_0(\bZ_i)+\sum_{k=1}^{\mathbb K}(\Deltahat_{N,1,k}'+\Deltahat_{N,2,k}'+\Deltahat_{N,3,k}'+\Deltahat_{N,4,k}'+\Deltahat_{N,5,k}'),
$$
where
\begin{align*}
\psi_0(\bZ)&= \mu_0(\bX) - \theta^0 + \frac{1-R}{1- e_N(\bX)} \{ Y - \mu_0(\bX)\}\\
&=\frac{ e_N(\bX)-R}{1- e_N(\bX)}\{m_0(\bX)-\mu_0(\bX)\}+m_0(\bX)-\theta^0+\frac{\varepsilon(1-R)}{1- e_N(\bX)},\\
\Deltahat_{N,1,k}'&=-N^{-1}\sum_{i\in\mathcal I_k}\left\{\frac{1-R_i}{1-\pi_N(\bX_i)}-1\right\}\{\mhat_0(\bX_i;\mathbb S_{-k})-\mu_0(\bX_i)\},\\
\Deltahat_{N,2,k}'&=N^{-1}\sum_{i\in\mathcal I_k}\left\{\frac{1-R_i}{1-\pihat_N(\bX_i;\mathbb S_{-k})}-\frac{1-R_i}{1- e_N(\bX_i)}\right\}\{Y_i-m_0(\bX_i)\},\\
\Deltahat_{N,3,k}'&=-N^{-1}\sum_{i\in\mathcal I_k}\left\{\frac{1-R_i}{1-\pihat_N(\bX_i;\mathbb S_{-k})}-\frac{1-R_i}{1- e_N(\bX_i)}\right\}\{\mhat_0(\bX_i;\mathbb S_{-k})-\mu_0(\bX_i)\},\\
\Deltahat_{N,4,k}'&=N^{-1}\sum_{i\in\mathcal I_k}\left\{\frac{1-R_i}{1-\pihat_N(\bX_i;\mathbb S_{-k})}-\frac{1-R_i}{1- e_N(\bX_i)}\right\}\{m_0(\bX_i)-\mu_0(\bX_i)\},\\
\Deltahat_{N,5,k}'&=N^{-1}\sum_{i\in\mathcal I_k}\left\{\frac{1-R_i}{1-\pi_N(\bX_i)}-\frac{1-R_i}{1- e_N(\bX_i)}\right\}\{\mhat_0(\bX_i;\mathbb S_{-k})-\mu_0(\bX_i)\}.
\end{align*}
We first obtain the rates for the terms $N^{-1}\sum_{i=1}^N\psi_0(\bZ_i)$, $\Deltahat_{N,1,k}'$, and $\Deltahat_{N,2,k}'$ for each $k\leq\K$. Observe the following properties for the first moments:
\begin{align*}
\E\{\psi_0(\bZ)\}&=\E\left[\frac{ e_N(\bX)-\pi_N(\bX)}{1- e_N(\bX)}\{m_0(\bX)-\mu_0(\bX)\}\right]=\mathbbm{1}\{ e_N(\cdot)\neq\pi_N(\cdot),\;\mu(\cdot)\neq m(\cdot)\}O_p(\bar\pi_N),\\
\E_{\S_k}(\Deltahat_{N,1,k}')&=\E_{\S_k}(\Deltahat_{N,2,k}')=0.
\end{align*}
\tco{For the second moments, we have
\begin{align*}
&\Var\{\psi_0(\bZ)\}=\Var\left[\frac{\{ e_N(\bX)-R\}\{m_0(\bX)-\mu_0(\bX)\}}{1- e_N(\bX)}+m_0(\bX)-\theta^0+\frac{\varepsilon(1-R)}{1- e_N(\bX)}\right]\\
&\qquad\overset{(i)}{=}\Var\left[\frac{\{ e_N(\bX)-R\}\{m_0(\bX)-\mu_0(\bX)\}}{1- e_N(\bX)}+m_0(\bX)-\theta^0\right]+\Var\left\{\frac{\varepsilon(1-R)}{1- e_N(\bX)}\right\}\\
&\qquad\leq\left\|\frac{\{ e_N(\bX)-R\}\{m_0(\bX)-\mu_0(\bX)\}}{1- e_N(\bX)}+m_0(\bX)-\theta^0\right\|_{2,\P}^2+\left\|\frac{\varepsilon(1-R)}{1- e_N(\bX)}\right\|_{2,\P}^2\\
&\qquad\leq2\left\|\frac{\{ e_N(\bX)-R\}\{m_0(\bX)-\mu_0(\bX)\}}{1- e_N(\bX)}\right\|_{2,\P}+2\left\|m_0(\bX)-\theta^0\right\|_{2,\P}^2+\left\|\frac{\varepsilon(1-R)}{1- e_N(\bX)}\right\|_{2,\P}^2\\
&\qquad\overset{(ii)}{=}2\E\left(\frac{[\{ e_N(\bX)-\pi_N(\bX)\}^2+\pi_N(\bX)\{1-\pi_N(\bX)\}]\{m_0(\bX)-\mu_0(\bX)\}^2}{\{1- e_N(\bX)\}^2}\right)+2\left\|m_0(\bX)-\theta^0\right\|_{2,\P}^2\\
&\qquad\qquad+\left\|\frac{\varepsilon(1-R)}{1- e_N(\bX)}\right\|_{2,\P}^2\\
&\qquad\overset{(iii)}{\leq}2\left[(1-C\bar\pi_N)^{-2}\{(2C\bar\pi_N)^2+C\bar\pi_N\}+1\right]\left\|m_0(\bX)-\theta^0\right\|_{2,\P}^2+(1-C\bar\pi_N)^{-2}\|\varepsilon\|_{2,\P}^2=O(1).
\end{align*}
where $(i)$ holds by the fact that $\E(\varepsilon|R,\bX)=0$, $(ii)$ holds by the tower rule with $\E(R|\bX)=\pi_N(\bX)$, and $(iii)$ follows by the assumption that $\bar\pi_N(\bx),e_N(\bx)<C\bar\pi_N$ for all $\bx\in\mathcal X$. Besides,
\begin{align*}
&\E_{\S_k}(\Deltahat_{N,1,k}'^2)=N^{-2}|\mathcal I_k|\E\left[\left\{\frac{\pi_N(\bX)-R}{1-\pi_N(\bX)}\right\}^2\{\mhat_0(\bX;\S_{-k})-\mu_0(\bX)\}^2\right]\\
&\qquad=N^{-2}|\mathcal I_k|\E\left[\frac{\pi_N(\bX)}{1-\pi_N(\bX)}\{\mhat_0(\bX;\S_{-k})-\mu_0(\bX)\}^2\right]\\
&\qquad\overset{(i)}{\leq} N^{-1}(1-C\bar\pi_N)^{-1}C\bar\pi_N\|\mhat_0(\cdot;\S_{-k})-\mu_0(\cdot)\|_{2,\P_X}^2=O_p(N^{-1}\bar\pi_Nr_{\mu,0,N}^2),
\end{align*}
\tco{where $(i)$ holds by the fact that $\bar\pi_N(\bx)<C\bar\pi_N$ for all $\bx\in\mathcal X$.} On the event $\mathcal E_{-k}$, \jelena{with} \eqref{E-k}, we have
\begin{align*}
&\E_{\S_k}(\Deltahat_{N,2,k}'^2)=N^{-2}|\mathcal I_k|\E\left[\left\{\frac{1-R}{1-\pihat_N(\bX;\mathbb S_{-k})}-\frac{1-R}{1- e_N(\bX)}\right\}^2\{Y-m_0(\bX)\}^2\right]\\
&\qquad=N^{-2}|\mathcal I_k|\E\left[\frac{\{1-\pi_N(\bX)\}\{\pihat_N(\bX;\S_{-k})- e_N(\bX)\}^2\varepsilon^2}{\{1- e_N(\bX)\}^2\{1-\pihat_N(\bX;\S_{-k})\}^2}\right]\\
&\qquad\overset{(i)}{\leq} N^{-1}(1-2C\bar\pi_N)^{-4}(1-c\bar\pi_N)\bar\pi_N^2\sup_{\bx\in\mathcal X}\left|\frac{\pihat_N(\bx;\S_{-k})- e_N(\bx)}{\bar\pi_N}\right|^2\|\varepsilon\|_{2,\P}^2=O_p(N^{-1}\bar\pi_N^2r_{e,N}^2),
\end{align*}
\tco{where (i) holds by the fact that $c\bar\pi_N<\bar\pi_N(\bx),e_N(\bx)<C\bar\pi_N$ and $\pihat_N(\bx;\S_{-k})<2C\bar\pi_N$ for all $\bx\in\mathcal X$ on $\mathcal E_{-k}$. Here, if we fix (or conditional on) $\S_{-k}$, on the event $\mathcal E_{-k}$, the inequality $\pihat_N(\bx;\S_{-k})<2C\bar\pi_N$ holds \jelena{almost surely,} w.r.t. the probability measure $\P$; if $\S_{-k}$ is treated as random, recall \eqref{E-k}, the inequality holds w.p.a. 1 \jelena{,} w.r.t. the joint probability measre of $\P$ and $\P_{\S_{-k}}$. As a result, we have $\E_{\S_k}(\Deltahat_{N,2,k}'^2)=O_p(N^{-1}\bar\pi_N^2r_{e,N}^2)$ w.r.t. the joint probability meas\jelena{u}re of $\P$ and $\P_{\S_{-k}}$.}
By Lemma \ref{l1},}
\begin{align*}
N^{-1}\sum_{i=1}^N\psi_0(\bZ_i)&=\mathbbm{1}\{ e_N(\cdot)\neq\pi_N(\cdot),\;\mu(\cdot)\neq m(\cdot)\}O_p(\bar\pi_N)+O_p(N^{-1/2}),\\
\Deltahat_{N,1,k}'&=O_p(N^{-1/2}\bar\pi_N^{1/2}r_{\mu,0,N}),\\
\Deltahat_{N,2,k}'&=O_p(N^{-1/2}\bar\pi_Nr_{e,N}).
\end{align*}
Now we consider the terms $\Deltahat_{N,3,k}$, $\Deltahat_{N,4,k}$, and $\Deltahat_{N,5,k}$. \tco{On the event $\mathcal E_{-k}$, we have}
\begin{align*}
&\E_{\S_k}|\Deltahat_{N,3,k}|\overset{(i)}{\leq} N^{-1}|\mathcal I_k|\E\left\{\left|\frac{1-R}{1-\pihat_N(\bX;\mathbb S_{-k})}-\frac{1-R}{1- e_N(\bX)}\right||\mhat_0(\bX;\mathbb S_{-k})-\mu_0(\bX)|\right\}\\
&\qquad\overset{(ii)}{=}N^{-1}|\mathcal I_k|\E\left\{\frac{\{1-\pi_N(\bX)\}|\pihat_N(\bX;\S_{-k})- e_N(\bX)|}{\left\{1-\pihat_N(\bX;\mathbb S_{-k})\right\}\left\{1- e_N(\bX)\right\}}|\mhat_0(\bX;\mathbb S_{-k})-\mu_0(\bX)|\right\}\\
&\qquad\overset{(iii)}{\leq}\frac{1-c\bar\pi_N}{(1-2C\bar\pi_N)^{2}}\bar\pi_N\sup_{\bx\in\mathcal X}\left|\frac{\pihat_N(\bx;\S_{-k})- e_N(\bx)}{\bar\pi_N}\right|\|\mhat_0(\cdot;\S_{-k})-\mu_0(\cdot)\|_{2,\P_X}=O_p(\bar\pi_Nr_{e,N}r_{\mu,0,N}),
\end{align*}
\tco{where $(i)$ holds by the triangular inequality, $(ii)$ follows by the tower rule with the fact that $E(R|\bX)=\pi_N(\bX)$, and $(iii)$ holds by the fact that $c\bar\pi_N<\bar\pi_N(\bx),e_N(\bx)<C\bar\pi_N$ and $\pihat_N(\bx;\S_{-k})<2C\bar\pi_N$ for all $\bx\in\mathcal X$ on $\mathcal E_{-k}$. Similarly, on the event $\mathcal E_{-k}$,}
\begin{align*}
&\E_{\S_k}|\Deltahat_{N,4,k}|\leq N^{-1}|\mathcal I_k|\E\left\{\left|\frac{1-R}{1-\pihat_N(\bX;\mathbb S_{-k})}-\frac{1-R}{1-\pi_N(\bX)}\right||m_0(\bX)-\mu_0(\bX)|\right\}\\
&\qquad=N^{-1}|\mathcal I_k|\E\left\{\frac{\{1-\pi_N(\bX)\}|\pihat_N(\bX;\S_{-k})-\pi_N(\bX)|}{\{1-\pihat_N(\bX;\mathbb S_{-k})\}\{1- e_N(\bX)\}}|m_0(\bX)-\mu_0(\bX)|\right\}\\
&\qquad\overset{(i)}{\leq}\frac{1-c\bar\pi_N}{(1-2C\bar\pi_N)^{2}}\bar\pi_N\sup_{\bx\in\mathcal X}\left|\frac{\pihat_N(\bx;\S_{-k})- e_N(\bx)}{\bar\pi_N}\right|\|m_0(\cdot)-\mu_0(\cdot)\|_{2,\P_X}\\
&\qquad=\mathbbm{1}\{m_0(\cdot)\neq\mu_0(\cdot)\}O_p(\bar\pi_Nr_{e,N}),
\end{align*}
\tco{where $(i)$ holds by the assumption that $c\bar\pi_N<\bar\pi_N(\bx),e_N(\bx)<C\bar\pi_N$ and $\pihat_N(\bx;\S_{-k})<2C\bar\pi_N$ for all $\bx\in\mathcal X$ on $\mathcal E_{-k}$.} \tco{Additionally}, we also have
\begin{align*}
&\E_{\S_k}|\Deltahat_{N,5,k}|\leq N^{-1}|\mathcal I_k|\E\left\{\left|\frac{1-R}{1-\bar\pi_N(\bX;\mathbb S_{-k})}-\frac{1-R}{1-\pi_N(\bX)}\right||\mhat_0(\bX;\mathbb S_{-k})-\mu_0(\bX)|\right\}\\
&\qquad=N^{-1}|\mathcal I_k|\E\left\{\frac{|\pi_N(\bX)- e_N(\bX)|}{1- e_N(\bX)}|\mhat_0(\bX;\mathbb S_{-k})-\mu_0(\bX)|\right\}\\
&\qquad\leq(1-C\bar\pi_N)^{-1}\sup_{\bx\in\mathcal X}\left|\bar\pi_N(\bx)- e_N(\bx)\right|\|\mhat_0(\cdot;\S_{-k})-\mu_0(\cdot)\|_{2,\P_X}\\
&\qquad=\mathbbm{1}\{ e_N(\cdot)\neq\pi_N(\cdot)\}O_p(\bar\pi_Nr_{\mu,0,N}),
\end{align*}
\tco{since $\bar\pi_N(\bx),e_N(\bx)<C\bar\pi_N$ for all $\bx\in\mathcal X$ by assumption.}
By Lemma \ref{l1},
\begin{align*}
\Deltahat_{N,3,k}'&=O_p(\bar\pi_Nr_{e,N}r_{\mu,N}),\\
\Deltahat_{N,4,k}'&=\mathbbm1\{m_0(\cdot)\neq\mu_0(\cdot)\}O_p(\bar\pi_Nr_{e,N}),\\
\Deltahat_{N,5,k}'&=\mathbbm{1}\{e_N(\cdot)\neq\pi_N(\cdot)\}O_p(\bar\pi_Nr_{\mu,0,N}).
\end{align*}
Therefore,
\begin{align*}
\thetahat_\mathrm{DRSS}^0-\theta^0&=N^{-1}\sum_{i=1}^N\psi_0(\bZ_i)+O_p(N^{-1/2}\bar\pi_N^{1/2}r_{\mu,0,N}+N^{-1/2}\bar\pi_Nr_{e,N}+\bar\pi_Nr_{e,N}r_{\mu,0,N})\\
&\qquad+\mathbbm1\{m_0(\cdot)\neq\mu_0(\cdot)\}O_p(\bar\pi_Nr_{e,N})+\mathbbm{1}\{ e_N(\cdot)\neq\pi_N(\cdot)\}O_p(\bar\pi_Nr_{\mu,0,N}).
\end{align*}
\end{proof}

Corollary \ref{c4.2} is a direct consequence of Theorems \ref{t4} and \ref{t6}.

\subsection{Proof of Theorem \ref{thm:ex1}}
\begin{proof}[Proof of Theorem \ref{thm:ex1}]
\tco{Here we provide asymptotic results of the DRSS estimator based on a MCAR PS.}

Under MCAR, neither $\pi_N(\bX) \equiv \bar\pi_N$ nor $\pihat_N(\bX;\S_{-k})\equiv\pihat_N(\S_{-k})$ depend on $\bX$, and we recall that $\bar\pi_N = \P(R=1)$. For each $k\leq\K$, $\pihat_N(\S_{-k})=N_{-k}^{-1}\sum_{i\in\mathcal I_{-k}}R_i$, where $N_{-k}=N-N/\K$. Notice that
\begin{align*}
\E_{\S_{-k}}\left[\left\{\frac{\pihat_N(\S_{-k})-\bar\pi_N}{\bar\pi_N}\right\}^2\right]&=\bar\pi_N^{-2}N_{-k}^{-1}\E(R-\bar\pi_N)^2=N_{-k}^{-1}\bar\pi_N^{-1}(1-\bar\pi_N)=O\left((N\bar\pi_N)^{-1}\right).
\end{align*}
By Lemma \ref{l1},
\begin{equation}
\label{piest1}\frac{\pihat_N(\S_{-k})-\bar\pi_N}{\bar\pi_N}=O_p\left((N\bar\pi_N)^{-1/2}\right).
\end{equation}
By the fact that
$$\frac{\pihat_N(\S_{-k}) - \bar\pi_N}{\pihat_N(\S_{-k})} \left\{1 + \frac{\pihat_N(\S_{-k}) - \bar\pi_N}{\bar\pi_N}\right\} = \frac{\pihat_N(\S_{-k}) - \bar\pi_N}{\bar\pi_N},$$
we have
\begin{equation}\label{piest}
\frac{\pihat_N(\S_{-k})-\bar\pi_N}{\pihat_N(\S_{-k})}=\left\{1 + \frac{\pihat_N(\S_{-k}) - \bar\pi_N}{\bar\pi_N}\right\}^{-1}\frac{\pihat_N(\S_{-k}) - \bar\pi_N}{\bar\pi_N}=O_p\left((N\bar\pi_N)^{-1/2}\right).
\end{equation}
Hence,
$$\frac{\pihat_N(\S_{-k})-\bar\pi_N}{\pihat_N(\S_{-k})}-\frac{\pihat_N(\S_{-k})-\bar\pi_N}{\bar\pi_N}=\frac{\pihat_N(\S_{-k})-\bar\pi_N}{\bar\pi_N}\cdot\frac{\pihat_N(\S_{-k})-\bar\pi_N}{\pihat_N(\S_{-k})}=O_p\left((N\bar\pi_N)^{-1}\right).$$
\tco{Additionally}, notice that
\begin{align*}
\E_{\S_k}\left[|\mathcal I_k|^{-1}\sum_{i\in\mathcal I_k}\frac{R_i}{\bar\pi_N}\{ \mu(\bX_i) - m(\bX_i)\}\right]&=\E\{\mu(\bX)-m(\bX)\},\\
\E_{\S_k}\left[|\mathcal I_k|^{-1}\sum_{i\in\mathcal I_k}\frac{R_i}{\bar\pi_N}\{ \mu(\bX_i) - m(\bX_i)\}\right]^2&=|\mathcal I_k|^{-1}\bar\pi_N^{-1}\E\left[\{\mu(\bX)-m(\bX)\}^2\right]=O\left((N\bar\pi_N)^{-1}\right).
\end{align*}
\tco{Let $\Delta_{\mu}= \E \{ \mu(\bX) - m(\bX) \}$. By Lemma \ref{l1},
\begin{equation}\label{eq:Deltamuest}
|\mathcal I_k|^{-1}\sum_{i\in\mathcal I_k}\frac{R_i}{\bar\pi_N}\{ \mu(\bX_i) - m(\bX_i)\}-\Delta_{\mu}=O_p\left((N\bar\pi_N)^{-1/2}\right).
\end{equation}
Using the definition of $\Deltahat_N$ from Theorem \ref{t4} and adapting it to the MCAR setting,
\begin{align*}
{\Deltahat_N} &= N^{-1}\sum_{i=1}^N\frac{R_i}{\pi_N(\bX_i)} \left\{ \frac{\pihat_N(\S_{-k})  - \bar\pi_N}{ \pihat_N(\S_{-k}) } \right\} \{ \mu(\bX_i) - m(\bX_i)\} \\
& = \K^{-1}\sum_{k=1}^{\K}\frac{\pihat_N(\S_{-k}) - \bar\pi_N}{\pihat_N(\S_{-k})} \left[|\mathcal I_k|^{-1}\sum_{i\in\mathcal I_k}\frac{R_i}{\bar\pi_N}\{ \mu(\bX_i) - m(\bX_i)\} \right].
\end{align*}
Recall \eqref{piest} and \eqref{eq:Deltamuest}, since $\K<\infty$ is a fixed number, we have
\begin{align*}
\sup_{k\leq\K}\left|\frac{\pihat_N(\S_{-k})-\bar\pi_N}{\pihat_N(\S_{-k})}\right|&=O_p\left((N\bar\pi_N)^{-1/2}\right),\\
\sup_{k\leq\K}\left||\mathcal I_k|^{-1}\sum_{i\in\mathcal I_k}\frac{R_i}{\bar\pi_N}\{ \mu(\bX_i) - m(\bX_i)\}-\Delta_{\mu}\right|&=O_p\left((N\bar\pi_N)^{-1/2}\right).
\end{align*}
Hence,
\begin{align*}
&\left|\Deltahat_N-\K^{-1}\sum_{k=1}^{\K}\frac{\pihat_N(\S_{-k}) - \bar\pi_N}{\pihat_N(\S_{-k})}\Delta_\mu\right|\leq\sup_{k\leq\K}\left|\frac{\pihat_N(\S_{-k})-\bar\pi_N}{\pihat_N(\S_{-k})}\right|\sup_{k\leq\K}\left||\mathcal I_k|^{-1}\sum_{i\in\mathcal I_k}\frac{R_i}{\bar\pi_N}\{ \mu(\bX_i) - m(\bX_i)\}-\Delta_{\mu}\right|\\
&\qquad=O_p\left((N\bar\pi_N)^{-1}\right).
\end{align*}
It follows that,
\begin{align*}
\Deltahat_N& = \K^{-1}\sum_{k=1}^{\K}\left(\frac{\pihat_N(S_{-k}) - \bar\pi_N}{\bar\pi_N} \right)\Delta_\mu + O_p\left( (N\bar\pi_N)^{-1}\right)\\
& \overset{(i)}{=} \K^{-1}\sum_{k=1}^{\K}N_{-k}^{-1}\sum_{i\in\mathcal I_{-k}}\frac{R_i - \bar\pi_N}{\bar\pi_N} \Delta_{\mu} + O_p\left( (N\bar\pi_N)^{-1}\right) \\
& = \K^{-1}\sum_{k=1}^{\K}N_{-k}^{-1}\left(\sum_{i=1}^N\frac{R_i - \bar\pi_N}{\bar\pi_N}-\sum_{i\in\mathcal I_k}\frac{R_i - \bar\pi_N}{\bar\pi_N}\right) \Delta_{\mu} + O_p\left( (N\bar\pi_N)^{-1}\right) \\
& = \left\{N_{-k}^{-1}-(\K N_{-k})^{-1}\right\}\sum_{i=1}^N\frac{R_i - \bar\pi_N}{\bar\pi_N} \Delta_{\mu} + O_p\left( (N\bar\pi_N)^{-1}\right) \\
& \overset{(ii)}{=} N^{-1}\sum_{i=1}^N \mathrm{IF}_{\pi}(\bZ_i) + O_p\left( (N\bar\pi_N)^{-1}\right),
\end{align*}
where $\mathrm{IF}_{\pi}(\bZ)= (\bar\pi_N^{-1}R - 1)\Delta_{\mu}$. Here, $(i)$ holds by the definition that $\pihat_N(S_{-k})=N_{-k}^{-1}\sum_{i\in\mathcal I_{-k}}R_i$ and $(ii)$ follows by the fact that $N_{-k}^{-1}-(\K N_{-k})^{-1}=\K/\{(\K-1)N\}-1/\{(\K-1)N\}=N^{-1}$.}
\end{proof}

\phantomsection
\addcontentsline{toc}{section}{References}



\end{document}